# In-situ Growth of Ultrathin Magnetic Films and Tuning the Magnetic Properties by Ion-sculpting

*A Thesis Submitted to*

Devi Ahilya Vishwavidyalaya, Indore

for the Degree of

**Doctor of Philosophy in Physics**

**Faculty of Science**

*by*

## Anup Kumar Bera

Supervised by

## Dr. Dileep Kumar

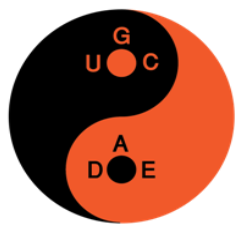

UGC-DAE Consortium for Scientific Research,

University Campus, Khandwa Road,

Indore-452001, M.P., India.

August 2022

# CERTIFICATE OF THE SUPERVISOR

It is certified that the work contained in the thesis titled "*In-situ Growth of Ultrathin Magnetic Films and Tuning the Magnetic Properties by Ion-sculpting*" by **Mr. Anup Kumar Bera** is worthy of consideration for the award of the degree of Doctor of Philosophy of Devi Ahilya Vishwavidyalaya, Indore (M.P.) and is a record of the original research work carried out under my guidance and supervision. The results contained in it have not been submitted in part or full to any other university or institute for the award of any degree/diploma. The thesis fulfils the requirement of the ordinance relating to the Ph.D. degree of the university. It is up to the standard both in respect of contents and language for being referred to the examiner. I also certify that the candidate has put an attendance of more than 200 days with me.

**Dr. Dileep Kumar**
Signature of the supervisor

Forwarded

Signature of Head U.T.D./Principle
Dr. Vasant Sathe

# DECLARATION BY THE CANDIDATE

I, Anup Kumar Bera, hereby declare that the thesis entitled "**In-situ Growth of Ultrathin Magnetic Films and Tuning the Magnetic Properties by Ion-sculpting**" is my own work conducted under the supervision of **Dr. Dileep Kumar**, Scientist -F, **UGC DAE CSR, Indore** and approved by the Research Degree Committee. I have put more than 200 days of attendance with the supervisor at center.

I further declare that to the best of my knowledge, the thesis does not contain any part of the work submitted for the award of any degree either in this University or in any other University/Deemed University without proper citation.

Besides this-

- ❖ I have successfully completed the coursework as per UGC regulation 2009 norms.
- ❖ I have also given a pre-Ph.D. presentation and successfully incorporated the changes suggested on the basis of feedback and comments received.
- ❖ I have also published more than one research paper in an ISSN/referred journal from the research work of the thesis and have produced evidence of the same in the form of reprints.

   Dr. Dileep Kumar                                                             Anup Kumar Bera  
Signature of the supervisor                                   Signature of the Candidate

Forwarded

Signature of Head U.T.D./Principal

Dr. Vasant Sathe

# *Contents*

Contents

Contents

*Dedicated to My Parents*

*for their endless love, support and encouragement*

# *Acknowledgements*

This thesis would not have been possible for me to complete without the support, close association and guidance of many people. I take this opportunity to extend my sincere gratitude to all of them who made this thesis possible.

First and foremost, I offer my profound gratitude to my thesis supervisor, Dr. Dileep Kumar, for his constant support, unreserved help and efficient guidance throughout the past few years. From finding an appropriate research topic at the beginning to the finishing of this thesis, his expert and steady hands steered me through the unavoidable difficulties and hurdles. His motivational words and novel ideas always inspired me. I am grateful to him for providing excellent experimental facilities, insightful discussions, valuable suggestions and critical review of my academic papers and this thesis. Without his kind and patient instructions, it would not have been possible for me to finish this thesis.

I extend my sincere thanks to Prof. Ajay Gupta for his advice and suggestions. The insightful discussions during his visit to our lab are gratefully acknowledged. I am grateful to Dr. V. R. Reddy for his instantaneous support in Kerr Microscopy and XRR measurement, as well as for fruitful scientific discussions. I also sincerely thank Dr. Mukul Gupta for thin film deposition and Dr. R. Venkatesh for FESEM measurements. My special thanks go to Dr. Pooja Gupta and her colleagues, Dr. P. Nageswararao, Dr. Mahesh K. Swami at RRCAT for MOKE, AFM and beamline measurement. Their prompt response and friendly behavior are gratefully acknowledged. The experimental support from Dr. Sujay Chakravarty in the CSR Kalpakkam node is gratefully acknowledged. I sincerely acknowledge Prof. Stephan V. Roth and Dr. Matthias Schwartzkopf (P03, Desy, Germany) for SAXS measurement, valuable discussions and insightful comments. I am thankful to Dr. Mukesh Ranjan (IPR, Gujrat) for providing rippled substrate. I would also like to thank Mr. Nitin Patil in CSR Workshop, Er. Layanta Bahera (XRD LAB), Er. Satish Potdar (FESEM LAB), Er. Mohan Gangrade (LT Lab) and Er. Gome (GIXRD LAB) for their help in various measurements.

I express my whole heartedly thanks to my senior Ph.D. fellow Dr. Sadhana Singh for her guidance, moral support and for making me familiar with the lab equipment. I am grateful to another senior Ph.D. fellow Mr. Arun Singh Dev, for his kind assistance during any experiment-related works, care, motivation and encouragement. I  sincerely acknowledge the help, support, pleasant company and scientific discussions with Md. Shahid Jamal and Mr. Avinash G. Khanderao. I really enjoyed the nice company at tea breaks of other Ph.D. fellows Mr. Sharanjeet Singh and Ms. Manisha Priyadarshini during this journey. My heartful thanks to all of them for patiently dealing with my childish acts, listening to the serious talks and always giving me good suggestions. I am very much thankful to all of them for maintaining a friendly, cheerful environment in the lab and making this long journey enjoyable and memorable.

I am indebted to the Department of Science and Technology (DST), India for providing financial assistantship to perform experiments at DESY, Germany and KEK, Japan. In addition, I want to thank UGC DAE CSR, Indore, for providing financial support for attending national conferences.

I am fortunate to have lovely friends at UGC DAE CSR, Indore, with whom I built sweet memories to carry forward throughout my entire life. I would like to thank Mr. Koushik, Mr. Manik, Mr. Sudip, Mr. Krishnendu, Mr. Sourav, Mr. Akash, Mr. Arup, Mr. Supriyo, Mr. Binoy, Mr. Vinoy, Mr. Satish, Mr. Shubham, Mr. Sumit, Ms. Priyanka, Mr. Vivek, Anjali mam, Gyanendra sir, Zainab mam and Biswajit da, in no particular order for their love, spontaneous support and care whenever needed. I warmly thank the company of Mr. Manik and Mr. Koushik with whom I have spent countless hours in and outside of lab. They have enriched me by their scientific aptitude. With all of them, the 6 years spent at CSR, Indore did not feel like a very long time.

Finally, I would like to acknowledge my family, who means the world to me. My parents - Baba and Maa, whose endless love, sacrifice, valuable prayers, encouragement and continuous support throughout my life bring me to this level. Without their blessings, it was impossible for me to complete this thesis. I owe my deepest gratitude to my better half Monika, for her patience and understanding of my goals and aspirations. Her eternal support has always been my strength. I acknowledge the support and comfort from my younger brother Arup. I would like to extend my gratitude to my late grandfather and grandmother. I would also like to thank other members of my extended family.

Anup

# Preface

Magnetism in low dimensional systems such as magnetic nanostructures, ultrathin film and multilayers is of great interest from both fundamental and applied points of view and have become an active field of research. In particular, magnetic anisotropy (MA) and magnetization reversal are key properties that find their application in high-density memories [1], magnetic sensors [2], spintronic devices [3], and wireless communication devices [4]. In epitaxial thin-film, magnetic anisotropy arises due to the periodic arrangement of lattices that give rise to magnetocrystalline anisotropy (MCA). The origin of MCA is related to spin-orbit coupling [5]. However, due to the absence of long-range structural ordering in the polycrystalline and amorphous thin film, MCA is absent [6]. Therefore, it is always challenging to induce MA in polycrystalline and amorphous thin films. In this kind of system, there are two possible methods of inducing MA. One is by controlling morphology, which gives rise to shape anisotropy. Another method is the generation of artificial stress that gives magnetoelastic anisotropy. Therefore, researchers have utilized several approaches, for example, oblique angle deposition [7][8], magnetic field annealing [9], stress anneal [10][11], insertion of antiferromagnetic underlayer [12] etc. for flexible and fine tailoring of magnetic anisotropy, although the strength of MA is limited. As magnetism is a collective phenomenon, it is strongly sensitive to interfacial and morphological symmetry breaking effects. Therefore, the artificial engineering of MA by manipulating surface and interface morphology is a fascinating field to research. In past decades, mainly multistep lithography-based techniques have been utilized for patterning the surface. However, due to high cost, low yield, limitations to fabricating the large-scale surface areas or a restriction to isotropic patterns restricts its application [13]. Therefore, cheaper, high throughput and time-saving approaches for synthesizing nano-dimensional morphological features with the well-controlled distribution of size and shape are thus extensively searched. In this respect, ion beam erosion (IBE) has been demonstrated as an easy and handy tool in particular way to induce MA by engineering surface and interface morphology through the self-assembled formation of nanometric patterns on the surface. Bradley and Harper first proposed the commonly accepted mechanism for pattern formation in 1988 [14]. It is based on two competitive counter acting processes like curvature-dependent sputtering, which tries to roughen the surface and surface self-diffusivity, which smoothen the surface. IBE is a versatile technique as several parameters like ion energy, ion flux, angle of incidence, ion species, and the target material can be varied. Ferromagnetic films deposited on the pre-patterned nanostructured substrate [15–17] (bottom-up) or post-growth IBE of the film surface [18–20] (top-down) deposited on planner substrate are found to imprint an uniaxial magnetic anisotropy (UMA). A theoretical study has also been done by Schlömann [21], who proposed a model on the demagnetization field associated with the morphological parameters of film responsible for this induced UMA. Several experiments have recently been performed on epitaxial, polycrystalline and amorphous systems establishing the relationship between UMA with film morphology and thickness.

Despite theoretical and experimental reports in the last ten years and technological applications, a clear understanding of the phenomenon and its related properties has not been developed to date. Apart from this, the methodology for inducing UMA suffers from its limitations. For example, in the case of direct ion beam erosion of thin film, ion beam-generated modifications happen only in the top surface and subsurface regions of the film due to the limited penetration depth of ions [20]. Also, because of sputter removal of material, the starting film thickness must be high for homogeneous pattern formation and continuity of the film keeping intact [22]. On the other hand, in the bottom-up approach, the strength of UMA falls off after a critical film thickness due to the merging of ripples crests with their nearest neighbours [23],[24],[25], which hinders its practical applications. Therefore, the new and novel route with the in-depth realization of the process involved in nanoscale control of magnetic properties is highly desired. Finally, yet importantly, despite being aware of the fact that the interaction between the ion beam and thin-film can alter the structure or crystallinity of material [26], which in turn can induce magneto crystalline anisotropy as demonstrated in our recent study [19], no other studies in literature have not taken this contribution into account in the induced magnetic anisotropy. This is an important factor and cannot be neglected in practice. All the studies in this field have explained the induced magnetic anisotropy in connection with morphological anisotropy or shape anisotropy. This may be owing to the use of ex-situ characterization techniques where surface oxidation and contamination or use of capping layer destroys film surface structure. Therefore, in order to extract genuine and unambiguous information about surface structure, film morphology and its correlation with magnetic anisotropy use of surface sensitive techniques and more importantly in-situ characterization is highly required.

Preface

In the present thesis, a detailed investigation of IBE-induced morphology, surface structure and related induced magnetic anisotropy property in polycrystalline and epitaxial thin films has been performed in-situ. Thin films were grown in an ultra-high vacuum (UHV) chamber and characterized simultaneously for their magnetic, transport, structural and morphological properties during and after the different stages of growth and IBE. The techniques such as magneto-optical Kerr effect (MOKE), reflection high energy electron diffraction (RHEED, x-ray reflectivity (XRR) and x-ray diffuse scattering (XDS) were used for in-situ characterization. In addition, various ex-situ structural, morphological and magnetic characterization techniques such as Small and Wide-angle x-ray scattering, Atomic Force Microscopy, Field emission scanning electron microscope, KERR microscopy, and nuclear forward scattering were used to complement in-situ experiments for thorough study.

We have divided our study into different sections to address and study the following points in the field of ion beam erosion-induced magnetic anisotropy in thin-film.

- ❖ Understanding the origin of UMA in polycrystalline and epitaxial thin film induced by IBE.
- ❖ The role of IBE on texture and structure of the polycrystalline and epitaxial thin film.
- ❖ Role of film structure and texture on the induced UMA.
- ❖ Possibilities of combining film morphology and texture for enhancement of the strength of UMA.
- ❖ Stability of UMA in higher film thickness.
- ❖ Use of shadowing effect on the pre-patterned substrate for controlling morphology induced UMA.
- ❖ Role of concurrent material deposition and IBE on morphology, structure and magnetic anisotropy.

Co and Fe polycrystalline and epitaxial thin films are selected as model systems for the present study to fulfil the objectives, as mentioned earlier. The work in the present thesis is presented by dividing it into seven chapters, including one appendix. A summary of the individual chapter and an appendix are given in the following sections.

**Chapter 1:** The first chapter of the thesis presents a brief introduction to thin-film, magnetic energetics of thin-film systems, the basics of ion beam erosion related to morphology, texture modification, and its importance in the field of surface science, physics and engineering.

A detailed review of IBE-induced UMA in crystalline, polycrystalline and amorphous thin films has also been provided. It also briefly describes the importance of an in-situ investigation to study ion beam erosion-induced phenomena. Other deposition methods like OAD to get magnetic anisotropy have also been mentioned briefly. At the end of the chapter motivation and objective of the present thesis have been presented.

**Chapter 2:** This chapter describes the various experimental techniques used for thin film deposition and its characterization. A brief description of the versatile UHV chamber used for in-situ synthesis and characterization of thin films is included in this chapter. Thin films have been grown inside the chamber by electron beam evaporation. Its magnetic, structural, morphological and transport characterizations are done by MOKE, XRR, XDS, RHEED and two probe resistivity measurements. The whole set-up is connected to a Cu-k$_\alpha$ x-ray source, which delivers a beam with adequate collimation and intensity for in-situ x-ray scattering investigations. All the measurements can be done simultaneously during the deposition of the film, thus making it possible to study the evolution of magnetic, transport and structural properties with film thickness under UHV conditions. The sample is attached to a 5-axis manipulator to perform an angle-dependent study with an option to cool and heat the sample from 50 K to 1200 K. This chapter also includes a brief description of some of the ex-situ techniques which have been used as complementary techniques to the in-situ study such as ion beam sputtering (IBS) for deposition, atomic force microscopy and field emission scanning electron microscope for imaging surface morphology, KERR microscopy to study magnetic domain dynamics. Furthermore, synchrotron-based small and wide-angle x-ray scattering, nuclear resonance scattering (NRS) techniques have also been used for detailed morphology, structure and magnetic characterization.

**Chapter 3:** This chapter studies the growth, structure and morphology of magnetic thin films deposited obliquely on the nanopatterned substrate. This work has been performed with the aim of achieving strong magnetic anisotropy in polycrystalline thin films. Co films are selected as a model system for the present study. Thin films were deposited obliquely on ripple patterned substrate along and across the ripple direction. The idea behind this unconventional approach was to combine shape anisotropy contribution originating from both columnar and rippled structures so that a stronger in-plane magnetic anisotropy can be obtained. A clear anisotropy in their growth behaviour has been observed. Detailed characterization reveals the formation of tilted columnar structure, and the films are in the polycrystalline state. All the films exhibit UMA with an easy axis of magnetization parallel to the substrate ripple direction with a striking difference in their strength and domain configuration. The increased shadowing effect in the films deposited obliquely normal to the direction of the ripples causing preferential coalescence of the columns along the substrate ripples, resulting in strong in-plane magnetic anisotropy in the film. The induced UMA is also found one order magnitude higher compared to the reference samples deposited obliquely on a flat substrate and deposited normally on ripple substrate. This peculiarity in magnetic behaviour is addressed by considering the morphological anisotropy governed by the enhanced shadowing effect, the shape anisotropy, and the dipolar interactions among the magneto statically coupled nanocolumnar structure. The present observation explores the crucial role of initial surface topography together with the shadowing effect on the formation of self-organized nanostructured thin film and hence, UMA. In our view, these results provide a new, cost-effective and straightforward avenue of fine tailoring the strength of morphology-induced UMA in the polycrystalline thin film.

**Chapter 4:** A detailed in-situ investigation to understand the origin of IBE-induced magnetic anisotropy in polycrystalline thin films is presented in this chapter. IBE was directly applied on polycrystalline Fe thin film to study its evolution of magnetic anisotropy. This is an unique study as film deposition, IBE and all characterization are done in-situ. Compared to the earlier studies, where IBE was applied to modify surface morphology, the present study demonstrated the use of low-energy ions to modify surface structure. Therefore, a combined role of structure and morphology of polycrystalline Fe thin film on the evolution of UMA subject to several cycles of the IBE process is investigated through a set of self-consistent in-situ experiments. In the initial stages of IBE, surface texturing of Fe film imprints a recognizable magneto-crystalline anisotropy (MCA) with the easy axis of magnetization along the IBE direction. Further erosion results in the development of well-defined correlated morphology that incorporates reinforcement of shape anisotropy and switches the easy axis of magnetization by 90°. The observed unusual crossover is ascribed to be originated from an interplay between the relative weight of MCA and shape anisotropy, which is successfully explained by considering modified surface structure, morphology and the relative strength of the dipolar stray fields. The present phenomenological understanding provides a promising option for the fine tailoring of UMA strength by controlling surface crystallinity and fine tailoring UMA strength by controlling surface crystallinity and anisotropic morphology.

**Chapter 5:** This chapter describes a superior approach for inducing and enhancing the strength of UMA by combining crystalline and shape anisotropy. We have seen in the previous chapter that IBE on the film surface can lead to textured surface topography, which can induce magnetic anisotropy. However, limited penetration depth restricts ion beam-induced modification to the surface and sub-surface region. Therefore, a scientific strategy would be to induce crystallographic texture throughout the whole layer of film to enhance the strength of UMA. Thus, the film was deposited by sequential deposition and subsequent IBE method. For the present study, a cobalt thin-film system is selected as a model system due to its advantage of easy growth and magnetocrystalline easy anisotropy axis along its "c" axis. This chapter deals with a detailed in-situ investigation of the structure, morphology, and magnetic anisotropy of Co thin film deposited by the sequential deposition and subsequent IBE method. A thorough in-situ investigation indicates that the film grows in the biaxially textured polycrystalline state with unidirectional correlated surface morphology. The film also exhibits pronounced UMA of one order magnitude higher than the earlier work on post-growth IBE of thin or film deposited over patterned ripple substrate. The huge difference between the UMA constant estimated from the MOKE hysteresis loop and Schlömann's formula suggests that the induced UMA not only originates from the shape anisotropy and an in-plane component of MCA to crystallographic texture is also simultaneously contributing to enhancing the strength of UMA. The formation of correlated surface morphology on the film surface was confirmed by AFM measurement from an independent film prepared in identical conditions. The work has been further extended to study the evolution of UMA and both the crystallographic texture as well as surface morphology, the sample was heated from room temperature up to 350ºC. It is observed that at 300ºC, film texture and UMA starts disappearing. Finally, at 350ºC, both RHEED and XDS measurements indicate isotropic film structure and morphology. At this point, UMA completely disappears, which confirms that both film texture and morphology were contributing to enhancing the strength of UMA. The present chapter demonstrates the novelty of the current approach, which can be utilized to tune in-plane UMA precisely by modifying film surface morphology and texture throughout the whole film layer.

**Chapter 6:** Compared to the previous chapters, this chapter deals with an epitaxial thin film, bringing some exciting and significant results. This chapter aims to study the different kinds of magnetic switching behavior arising from the interplay of intrinsic crystalline anisotropy and induced uniaxial magnetic anisotropy. This chapter is divided into two parts. The first part describes the growth and magnetic anisotropy of 20nm thick Fe film grown on ion eroded Ag(001) substrate that has been studied in-situ. The real-time growth process was monitored by RHEED. The film grows epitaxially on Ag surface with crystallographic orientations (001)[110]$_{Fe}$ || (001)[100]$_{Ag}$ via Volmer–Weber mechanism, where islands grow larger to form a continuous film at around 1 nm thickness. The induced UMA arises due to morphological symmetry breaking of Ag surface, couples with the intrinsic cubic four-fold anisotropy of bcc-Fe, and leads to the multiple-step jump in magnetic hysteresis loops. Magnetic switching and its correlation with the domain dynamics of the film, as studied by the Kerr microscopy technique, revealed a striking difference in domain structure depending upon the direction of the applied magnetic field with respect to crystallographic direction. It is found that the magnetization reversal takes place via a "single-step jump" mechanism through sweeping of 180° domain walls. In contrast, the coherent rotation and the sweeping of 90° domain walls are responsible for a "double step jump" magnetization reversal.

On the other hand, in the second part, IBE was directly applied on epitaxial Fe thin film is grown on MgO(001) substrate. The film grows epitaxially in 3D island mode and exhibits intrinsic bi-axial magnetic anisotropy. After that, to study the evolution of its surface structure and magnetic anisotropy, its surface was eroded in steps with Ar+ ion of energy one keV at an angle of incidence 50º from the surface normal. Simultaneously, the evolution of its surface structure and magnetic anisotropy is recorded after each IBE step. It is observed that after a certain amount of IBE cycle, the film surface is converted into (2×2) reconstructed surface. An UMA is also induced in the film and has converted the film from purely biaxial to almost uniaxial in nature. Furthermore, thermal annealing has also been observed to remove surface reconstruction and induce UMA. The experiment has been repeated with higher film thickness. In this case, only the film surface gets reconstructed without affecting magnetic anisotropy as IBE generated structure as well as morphology modification limited to the only surface and subsurface region. Therefore, we have demonstrated that apart from traditional tailoring of film morphology and related engineering of UMA, IBE can be further extended for modification of surface structure by proper selection of ion flux and energy. Furthermore, this type of system can be used as a potential candidate where competition or interplay of magnetocrystalline anisotropy (MCA) and shape anisotropy can be studied. The understanding acquired from the present study, responsible for different kinds of magnetic switching behaviour, can be further extended for multilayer systems where a separation of interlayer exchange coupling and anisotropy is often needed.

**Chapter 7:** The overall conclusion of the research work carried out in this thesis has been given in this chapter. Further, the future scope of the studies presented in this thesis is also discussed in this chapter.

**Appendix**

A unique strategy to control the growth of self-organized nanostructures and modulate nanopatterns along with concurrent material deposition for inducing UMA is presented. The experiment was performed by varying several ion beam parameters such as angle of incidence, irradiation time, and ion energy to study its effect on morphology, texture, and magnetic anisotropy. Furthermore, the Fe$^{57}$ isotope was used to perform nuclear forward Scattering (NFS) measurement to extract information about interfacial magnetism. Detailed study insights that the films grow in the polycrystalline state with a certain crystallographic texture directly related to ion incidence angle. Small-angle x-ray scattering (SAXS) pattern taken along and across the IBE direction reveals the formation of unidirectional correlated nanometric morphology on the film surface. The surface morphology of the film as imaged by atomic force microscope indicates the formation of unidirectional elongated grains/islands on the film surface. The films exhibit an UMA, however, the strength and orientation of the anisotropy axis are dependent on the angle of incidence of the ion beam on the sample surface. The origin of UMA is explained by considering the shape anisotropy of self-assembled nano-structures and magneto-crystalline anisotropy originating from crystallographic texture.

# Chapter 1:

# Introduction

The field of nano-magnetism explores the magnetic characteristics of objects with submicron-sized dimensions. Magnetism, despite being one of the oldest scientific fields, the physics of surfaces, interfaces, and nanostructures has emerged over the past few decades as one of the primary research areas due to the technological and scientific trend toward the miniaturization of physical systems. This chapter will briefly discuss fundamental concepts of magnetic interactions, the origin of magnetic anisotropy, and the magnetization dynamics of the nanostructured thin film. It also provides an overview of the low-energy ion-solid interaction and its application in tailoring morphology, structure, and magnetic anisotropy. At the end of this introduction, the motivations and objectives of this thesis are outlined.

# Chapter 1: Introduction

## 1.1 Magnetic thin film nanostructures

## 1.2 Magnetic interaction

    1.2.1 Exchange Interactions

    1.2.2 Dipole-Dipole interactions

    1.2.3 Spin-orbit interactions

    1.2.4 Zeeman energy

## 1.3 Magnetic Anisotropy

    1.3.1 Magneto-crystalline Anisotropy

    1.3.2 Shape Anisotropy

    1.3.3 Magnetoelastic Anisotropy

    1.3.4 Perpendicular magnetic anisotropy and In-plane magnetic anisotropy.

## 1.4 Origin of magnetic anisotropy in thin films

## 1.5 Ion sculpting and tuning of magnetic anisotropy/ surface modification

    1.5.1 Ion solid interaction.

    1.5.2 Pattern formation by IBE.

    1.5.3 Ion Beam Erosion: A tool for tuning magnetic anisotropy

## 1.6 Aim and scope of the present thesis

## 1.1 Magnetic Thin Film Nanostructures

Magnetic thin films are tremendously important technologically since most electronic devices that exploit magnetic behaviours use thin-film architectures. In addition to their technological importance, they are of interest because they show novel physics due to their reduced size and dimensionality. Thus, thin film magnetism is an important field to study in magnetism because of the distinctive magnetic characteristics provided by the two-dimensional structure. In-plane and perpendicular magnetic anisotropies [27], Interlayer coupling [28], Exchange bias [29], Exchange spring [30], Spin valve [31], Enhanced magnetic moment, Current induced domain wall motion [32], Domain wall resistance [33], Giant magnetoresistance [34], Tunneling magneto resistance[35] which usually does not have a bulk equivalent. Surface/interface roughness, stresses, strain, and morphology of ultra-thin magnetic films control many important physical and chemical properties of such structures. Therefore, these two-dimensional structures have immense possibilities to tailor their properties in order to achieve the desired functionality.

## 1.2 Magnetic interactions

### 1.2.1 Exchange interaction

The exchange interaction causes the collective ordering in magnetic materials. In its most basic version, it considers the direct overlap of the wave functions of localized electrons with spins $s_i$. The Pauli exclusion principle and the short-range Coulomb interaction are responsible for this contribution. Its most fundamental implication is that neighboring magnetic moments tend to align collinearly. Any departures from the ideal scenario result in a cost in energy. The exchange Hamiltonian of two atoms bearing spin $s_i$ and $s_j$ reads:

$$H_{exch} = -2 \sum_{i<j} J_{ij} \boldsymbol{s}_i \cdot \boldsymbol{s}_j \qquad (1.1)$$

Here, $J_{ij}$ is the exchange integral describing the overlap of the $i^{th}$ the wave functions of spin $s_i$ with the $j^{th}$ of spin $s_j$. An antiparallel or antiferromagnetic alignment is obtained for $J_{ij} < 0$, whereas, $J_{ij} > 0$ aligns neighbouring spins parallel, i.e., ferromagnetic ordering.

However, the scalar product of **s_i.s_j** is independent of the choice of coordinate system. Therefore, the energy involved in the spin-spin exchange is isotropic in space, and it does not contribute in any way to magnetic anisotropy.

### 1.2.2 Dipole-dipole interaction

A dipolar magnetic field $H_{dipole}$ is linked with each magnetic moment. The magnetic field associated with a dipole of magnetic moment $m_i$ at a distance $r_{ij}$ is given by:

$$H_{dipole} = \frac{\mu_0}{4\pi} \sum_i \frac{3(\vec{m_i}.\hat{r_i})\hat{r_i} - \vec{m_i}}{r_i^3} \qquad 1.2$$

In an ensemble of magnetic moments, there are dipolar interactions between the moments and the field that is made by the other moments. Adding up all of the interactions between the moments and the fields around them, one can find the dipolar energy as given by:

$$E_{dipole} = \sum_{i>j} \frac{\vec{m_i}.\vec{m_j} - 3(\vec{m_i}.\hat{r_{ij}})(\vec{m_j}.\hat{r_{ij}})}{r_{ij}^3} \qquad 1.3$$

The dipole energy is clearly dependent on the orientation of the magnetic moments $m_i$, $m_j$ with respect to $r_{ij}$, as shown by the second term in the dipolar interaction: $E_{dip}$ is lowest when the magnetization M points parallel to $r_{ij}$, and it costs energy to rotate the two dipole moments perpendicular to the $r_{ij}$-axis. The dipolar energy decreases as the cube of the distance and depends on the magnetization of the material. At an atomic distance, the dipolar energy is negligible compared to the exchange coupling, but it is dominating at a relatively larger distance. Therefore, it is a long-range interaction.

### 1.2.3 Spin-orbit interaction

The spin-orbit interaction illustrates how the orbital motion of an electron affects its spin. In a basic classical model, this is regarded as a coupling between an electron's spin momentum and the magnetic field created by the electron's motion around the atom. This coupling results in an energetically advantageous alignment of magnetization M with respect to the sample's crystallographic axes. The energy contribution of spin-orbit interaction is expressed as:

$$E_{LS} = \xi \mathbf{L}.\mathbf{S} \qquad 1.4$$

Here, L and S are orbital and spin angular momentum, $\xi$ is the spin-orbit coupling constant that depends on the potential energy seen by the electrons and generated by surrounding charges of electrons and atoms. The equation shows the consequences of spin-orbit interaction. The angular momentum of the electron tends to align its spin along favoured crystal axes, consequently, the crystal structure and symmetry affect magnetic anisotropy, adding a magneto-crystalline anisotropy component. The spin-orbit interaction determines the magnitude of the anisotropy.

### 1.2.4 Zeeman energy

When an external magnetic field $H_{ext}$ is applied to a specimen, the magnetization M experience a torque. This $H_{ext}$ tries to align M parallel to the applied field direction. It is basically the interaction/potential energy of a system in the presence of a magnetic field. The corresponding energy is termed as Zeeman energy and is expressed as

$$E_z = - H_{ext}.M \qquad 1.5$$

When a strong magnetic field is applied in a direction other than the easy axes, the local magnetization vector rotates or switches via domain wall motion to align with the applied field, this lowers the Zeeman energy but increases the magnetic anisotropy energy.

**Therefore, the macroscopic behavior of magnetization is a compromise between four interactions:**

- ❖ **Exchange interaction:** favors uniform magnetization. It is very strong but short-ranged.

- ❖ **Dipolar interaction:** tends to avoid the formation of the magnetic poles. It is weak but long-ranged.

- ❖ **Magnetocrystalline anisotropy:** orients magnetic moments along privileged directions.

- ❖ **Zeeman energy:** interaction with an external magnetic field and alignment of magnetic moments along the field.

Thus, the state of equilibrium will be established by a compromise between four competing energies. The total energy will be given by:

$$E_T = E_{ex} + E_H + E_{dip} + E_{an}$$

$$E_T = -2\sum_{i\neq j} J_{ij} S_i.Sj - \vec{m_i}.\vec{H}_{ex} + \sum_{i>j} \frac{\vec{m_i}.\vec{m_j} - 3(\vec{m_i}.\hat{r}_{ij})(\vec{m_j}.\hat{r}_{ij})}{r_{ij}^3} - k_i \left(\frac{\vec{m_i}}{m_i}.\hat{r}_i\right)^2 \qquad 1.6$$

## 1.3 Magnetic Anisotropy

In any application involving a magnetic substance, we are not only interested in the size or magnitude of the magnetization but also its direction. We must know the orientation of the magnetization vector in the absence of any external effects (such as applied fields), how simple it is to modify that magnetization vector's orientation with an applied field and the orientation of the magnetization vector after the applied field is removed. All of these concerns are intrinsically linked to the magnetic anisotropy of the material. Anisotropy refers to the direction dependence of the physical properties of a material. When it comes to magnetic materials, magnetic anisotropy refers to the phenomenon in which the magnetization prefers to align itself along a certain axis or plane. The existence of magnetic anisotropy results in a preferential magnetization direction, known as the easy direction, since magnetizing the material along this direction is "easier." The magnetic characteristics of a material are highly influenced by the energy associated with magnetization and its relationship to the easy direction.

A uniaxial anisotropy can usually be represented by a unidirectional anisotropy term in the total energy calculation of the system:

$$E_u = K_u Sin^2 \phi \qquad 1.7$$

where $\phi$ is the angle between the magnetization and the easy axis. $Ku$ is the anisotropy constant whose value accounts for the strength of the uniaxial anisotropy. Magnetic anisotropy can be divided into three categories: Magneto-crystalline anisotropy, Shape anisotropy and Magneto-elastic anisotropy.

## 1.3.1 Magneto-crystalline anisotropy

Magneto-crystalline anisotropy is the most common kind of anisotropy that is found in single crystal materials and epitaxial thin films. Due to magneto-crystalline anisotropy, the free energy of a single crystal varies in the direction of M with respect to the crystallographic axes. This anisotropy is inherent to the material and reflects the crystalline symmetry of the material. Spin-orbit coupling is the physical mechanism responsible for crystallographic anisotropy. The orientation of orbit is fixed with respect to the lattice. Even when huge fields are applied, the orientations of the orbits cannot be altered because they are so firmly fixed to the lattice. On the other hand, the spin and orbital motion of each electron is also interlinked. When an external field attempts to realign an electron's spin, the orbit of that electron tends to be reoriented as well. However, since the orbit is firmly linked to the lattice, attempts to move the spin axis are resisted. We call the energy needed to rotate the spin axis away from the easy direction as anisotropy energy. Therefore, the anisotropy energy is simply the energy required to overcome the spin-orbit interaction.

For example, in the case of Fe, <100> direction is the easy axis direction and the <111> direction are hard axis. For nickel, it is the other way around. Cobalt has the hexagonal "c" axis [001] as the easy direction.

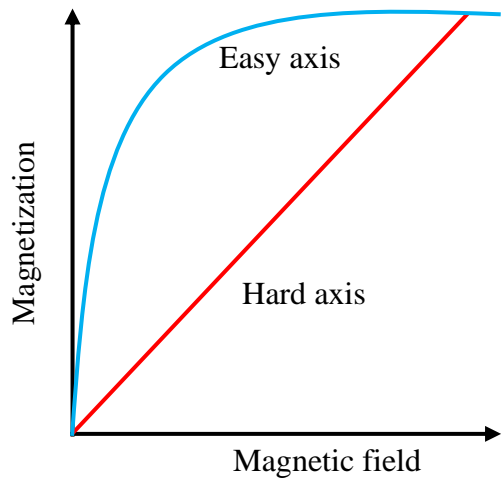

**Figure 1.1:** Representative magnetization curve along the easy and hard directions.

For a cubic symmetry, the energy density can be expressed in terms of the direction cosines and the anisotropy constants as,

$$E^C_{crys} = K_0 + K_1(\alpha_1^2\alpha_2^2 + \alpha_2^2\alpha_3^2 + \alpha_3^2\alpha_1^2) + K_2(\alpha_1^2\alpha_2^2\alpha_3^2) + \cdots \qquad 1.8$$

Where $K_i$ are empirically determined magneto-crystalline anisotropy constants, $\alpha_i$ are the directions cosines with respect to the x, y and z-axis, respectively

For a hexagonal system, the energy density is given by

$$E^C_{crys} = K_0 + K_1 Sin^2\theta + K_2 Sin^4\theta + K_3 Sin^6\theta + \ldots \qquad 1.9$$

Magneto crystalline anisotropy is not exhibited by polycrystalline samples, which have no preferred orientation of the grains. It has an overall isotropic behaviour concerning the energy being needed to magnetize it along an arbitrary direction.

### 1.3.2 Shape anisotropy

Shape anisotropy arises due to the asymmetric shape of the material. It is significant for thin films and nanostructures, where one dimension is much smaller than the other two. If the shape of a sample deviates from spherical symmetry, one or more directions of the specimen exhibit an easy magnetization axis due to shape anisotropy. The long-range nature of the dipolar interaction results in a contribution that varies with the geometry of the specimen, resulting in shape anisotropy in thin films. When a sample of finite size is magnetized, free charges appear at its surfaces or edges. These poles themselves act as a source of the additional magnetic field. The magnetic fields outside and inside the sample are known as stray and demagnetizing fields, respectively.

Assume that an external magnetic field ($H_{ext}$) applied from bottom to top of the page has magnetized our prolate spheroid from Fig. 1.2. Consequently, the prolate spheroid has a north pole on the top end and a south pole on the bottom edge. By definition, the lines of force emanate from the north pole and terminate at the south pole, resulting in the arrangement of field lines seen in Fig. 1.2. The field within the sample, as shown in the diagram, points from top to bottom, i.e., in the opposite direction of the applied external field. Because this internal field tends to demagnetize the magnet, it is referred to as the demagnetizing field, $H_d$.

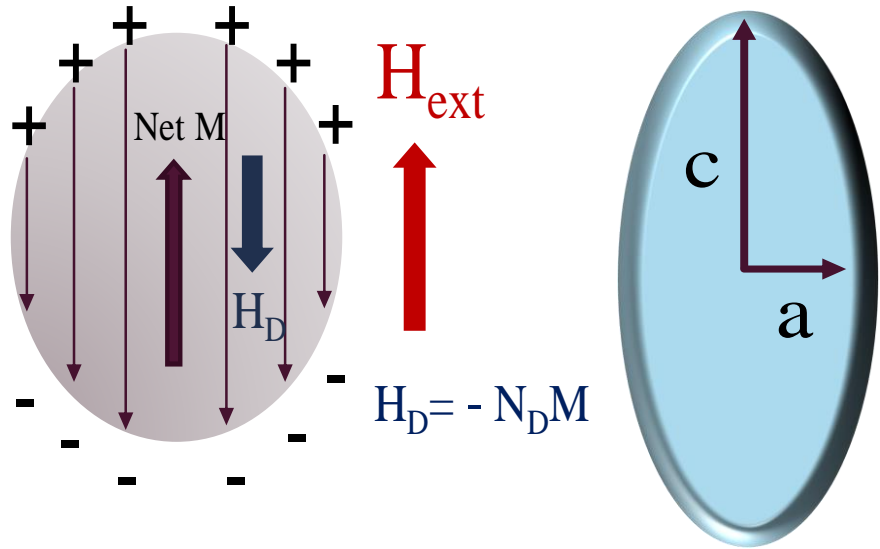

**Figure 1.2:** Demagnetization field produced by surface pole distribution in the presence of an external magnetic field of a prolate spheroid.

The demagnetizing field $H_d$ produced by a body is proportional to the magnetization M of the body itself. Therefore, it can be expressed as:

$$H_d = -N_d M. \qquad 1.10$$

Here, $N_d$ is the demagnetization factor and depends on the shape of the sample. The demagnetizing factor $N_d$ is, in general, a tensor element. The calculations show that for elongated samples, $N_d$ is least along the long axis and most along the short axis. Therefore, the effective magnetic field inside a material in the presence of an external magnetic field is

$$H_{eff} = H_{ext} - N_d M \qquad 1.11$$

The majority of the applied field is used to overcome the demagnetizing field along the short axis. Because of this, magnetizing the sample along the long axis is easier than the short axis.

The energy of a sample in its own stray field is given by the stray field energy, which is also termed as magnetostatic energy: $E_{ms} = \frac{1}{2} N_D M_s^2$. Therefore, the corresponding shape anisotropy constant can be expressed as:

$$K_{shape} = \frac{1}{2}(Na - Nc)M_s^2 \qquad 1.12$$

$N_D$ = Na-Nc; As Na>Nc; therefore, it pushes the magnetization M along the longer side of the nanostructure. For this reason, a cylindrical shape has an easy magnetization axis parallel to the axis of the cylinder and a disk-shaped structure has easy magnetization axis parallel to the disk surface.

Polycrystalline samples without a preferred orientation of the grains do not show any magneto crystalline anisotropy. If the sample or grains is not spherical, the preferred direction of magnetic moments determines the magnetostatic energy of the system. The phenomenon is completely magnetostatic in nature and is linked to demagnetizing fields.

### 1.3.3 Magneto elastic anisotropy

Stress or strain can be another source of magnetic anisotropy, which gives rise to magneto-elastic anisotropy. Magnetism relies on the configurations of electrons inside each atom and the structural arrangement of atoms. Therefore, putting stress on a crystal will cause changes in its magnetism. Stress or strain can be induced by external mechanical force or can be intrinsic due to substrate lattice mismatch etc. Similarly, an externally applied field can change the shape of a crystal by distorting its structure. This phenomenon is known as magnetostriction. The anisotropy energy associated with the stress or strain is expressed as:

$$E_{me} = \frac{3}{2} \lambda \sigma Sin^2\theta \qquad 1.13$$

Here, $\lambda$ is the magnetostriction constant, $\sigma$ is mechanical stress, and $\theta$ is the angle between magnetization and the strain axis.

Mechanical stress may change the domain structure and provide a new source of magnetic anisotropy.

### 1.3.4 Perpendicular magnetic anisotropy and In-plane magnetic anisotropy.

Magnetic anisotropy, in which the preferred orientation of the spins or magnetic moments is perpendicular to the film surface, is known as perpendicular magnetic anisotropy (PMA). PMA is generally observed in ultra-thin magnetic films (≈1nm). The film quality is a critical feature of the occurrence of perpendicular magnetic anisotropy. Many variables, including roughness, interfacial spin-orbit coupling, and the formation of interface alloying, may induce perpendicular magnetic anisotropy. Materials having perpendicular anisotropy have recently received a lot of interest because of the potential uses in spin-polarized current devices and high-density data storage devices. The other kind of magnetic anisotropy is In-plane magnetic anisotropy. Here, spins or magnetic moments are preferentially oriented parallel to the film plane. Because of the demagnetizing effects in the thin film, magnetization in thin films tends to lay in the plane of the film. The demagnetizing effect in thin films with perpendicular anisotropy is countered by uniaxial anisotropy, which maintains the moments aligned perpendicular to the film plane. A short summary of different types of anisotropy and their origin is described below in the table.

| Types of anisotropy | Origins (energy responsible) |
|---|---|
| Crystal or magneto crystalline anisotropy | Crystal symmetry |
| Shape anisotropy | Sample shape (magnetostatic energy) |
| Stress anisotropy | Stress (magneto elastic energy) |
| Induced anisotropy | Magnetic annealing (directional order in SS) |
| | Stress annealing (anisotropic distribution of solute) |
| | Plastic deformation(residual stress) |
| | Magnetic irradiation (related to defects generated) |
| | Exchange coupling at the interface through fine particle or thin layer |
| | Oblique angle deposition |

## 1.4 Anisotropy in thin film

The characteristics of thin films are often determined by surface or interface effects, which is a major difference between bulk and thin-film materials. This distinction is particularly apparent in magnetic anisotropy. When comparing thin films to bulk materials, additional factors to magnetic anisotropy must be taken into account [36]. Néel [37] suggested that a surface might generate substantial anisotropies owing to its lower symmetry compared to that of the bulk and that this anisotropy dominates the behaviour of the film as its thickness decreases. These are known as magnetic surface anisotropies (MSA), and they occur when a large portion of a thin film is at the surface (film/vacuum interface) or the film/substrate interface. Additional anisotropies arise because atoms at any of these surfaces are in substantially different local environments than atoms in bulk. Due to the short dimensions of thin films and the interactions between the film and substrate, other phenomena become significant. In certain situations, the MSA is sufficient to overcome the magnetostatic energy of the film and generate an easy axis normal to the film surface.

The overall anisotropy of thin films may be represented by two effective bulk and interface anisotropy terms [38]:

$$K_{eff} = K_v + 2K_s/t \qquad 1.14$$

Where, Kv is the volume anisotropy term, Ks is the surface anisotropy term, factor 2 counts for two interfaces, and *t* is the film thickness. Kv tends to align magnetization parallel to the film plane and depends on demagnetization energy, magnetoelastic energy and magneto crystalline anisotropy [39].

Magneto crystalline anisotropy is inherently present in epitaxial thin films. The MCA averages out for thin films made from polycrystalline and amorphous materials. However, because of the preferred orientation of grains, a textured polycrystalline film may show MCA in principle. A few research groups have found that the magnetic characteristics of an epitaxial layer are influenced by steps on the substrate's surface. Furthermore, as shown by Bruno [40], magnetostatic anisotropy for an ultrathin film may prefer magnetization perpendicular to the film, depending on the degree of surface and interface roughness.

Strain is expected to exist in thin films for a number of causes, and the accompanying magnetoelastic anisotropies must be considered. A major source of strain in epitaxial films is a mismatch in the lattice constants of the film and substrate. There are other types of strains that are induced during deposition, such as the strain that develops around defects [41].

The shape of material gives rise to magnetostatic anisotropy or shape anisotropy. For a thin film or nano-meter sized object, the shape, size, distribution, and mutual dipolar interaction between them decides the overall shape anisotropy. It is observed that nanowire ripples exhibit magnetic anisotropy with an easy axis aligned parallel to the long axis.

## 1.5. Ion sculpting and tuning of magnetic anisotropy/ surface modification

Research, development, and manufacturing of modern materials have come to rely on ion beam technologies, which are used to characterise and modify solid surfaces and thin films. These applications depend on the fundamental interactions between an energetic ion and the atoms' electrons and nuclei. Interactions between ion beam and materials result in nanometer-scale modifications, such as localized atomic rearrangement. The energy delivered from the ion to the substance may vary from meV volts to hundreds of keV. Therefore, ion beam has become an is a multifunctional tool for thin film deposition, depth profiling, sputter-etching, morphology modification, ion beam assisted growth, control of crystallographic texture, grain growth, sputter cleaning, micromachining and sputter deposition, recrystallization, formation of intermetallic compounds, chemically modified surface layers [42]. It is also beneficial to the performance of thin films and coatings for improving adhesion, densification of films, modification of residual stresses, grain size and morphology [43]. XPS, AES, and SIMS are the most widely used techniques in surface analysis and depth profiling, where the ion beam is used as a primary tool.

### 1.5.1 Ion-Solid interaction

Sputtering is the removal of material from the surface of a solid by the impact of energetic particles. This mechanism has already formulated by Sigmund in the 1960s [44]. The ions penetrating the target material slows down by losing its kinetic energy and momentum in elastic and inelastic collisions with target nuclei and electrons, respectively. However, at kinetic energies of a few keV and lower, nuclear collisions are the primary means of transferring the ions' momentum and kinetic energy to the target atoms, with inelastic collisions playing only a small role [45][46]. In one of these collisions, a target atom gets some of the ion's kinetic energy and momentum and may therefore be started in motion. If an atom's kinetic energy is strong enough to overcome the surface binding energy, it will sputter away from the surface. So, sputter erosion can have different effects on the surface, depending on many things like the energy of the ions that hit it, the angle at which they hit it, the ratio of masses, the temperature of the sputtered substrate, the materials it is made of, etc

Low-energy ion beam erosion (IBE) is a powerful technology for creating diverse self-organized nanostructures, such as ripples, dots, pits and triangular-shaped prisms on different materials, including amorphous, single-crystalline as well as compound semiconductors and polycrystalline materials. The IBE technology offers a single-shot, cost-effective, scalable method for the fabrication of large areas beyond the limits of conventional optical lithography. Furthermore, it is a versatile technique as ion species, ion energy, angle of incidence, ion flux, surface temperature can vary widely and applicable to all kinds of crystalline, polycrystalline, amorphous, metal, semiconductor, insulators systems etc.

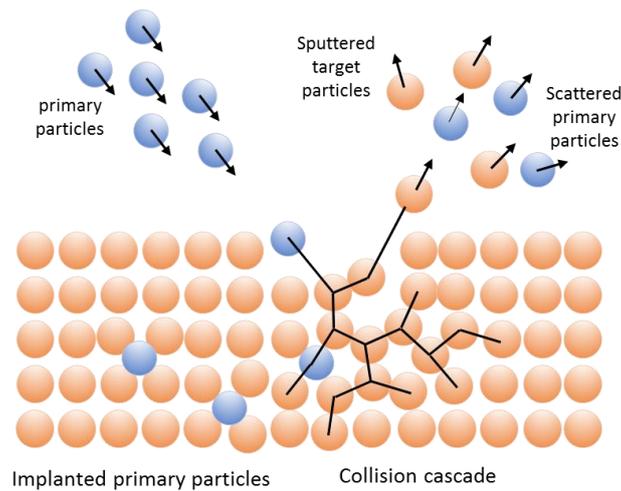

**Figure 1.3:** Sketch of various fundamental processes that occur due to ion-solid interaction.

The interaction of energetic ions with solids decides ion rages, ion stopping process and ion-induced damages via channelling and collision cascades. These fundamental interactions consist of binary elastic particle collisions, ion stopping and recoil processes, channelling and sputtering processes, ion implantation, and specimen modifications such as radiation damage, preferential sputtering, and ion mixing. These interactions are the basis for ion damage in materials caused by implantation, sputtering, and mixing operations.

Ion implantation is a process for implanting controlled amounts of chosen dopants into targets, particularly in semiconductors, in the form of an accelerated beam of ions to modify chemical, electronic, mechanical and electrical properties. Ion implantation involves the acceleration, volatilization, ionization, and separation of dopant atoms by their mass-to-charge ratios before they are directed towards a target. The ion beam also causes damage to solid objects. Typical ion energy varies from 10 to 500KeV, even up to a few MeV. In this energy range, ion penetration depth ranges from ≈ 10nm to 10μm depending upon target materials, ion energy, and dopant [45][46].

Ion irradiation involves the use of particle accelerators to fire energetic ions at a substance. Ion implantation is a type of ion irradiation. The primary distinction between ion implantation and swift heavy ion irradiation methods is that the impinging ions get stuck in the material during ion implantation. In contrast, in swift heavy ion irradiation, the ions that hit the material don't get stuck because they have a wider range. In addition, in ion irradiation, the melting point of the materials may be surpassed in an area surrounding the ion track, depending on the energy density. As a result, crystalline material may be converted to amorphous material and vice versa. Furthermore, it can create damaged tracks along the trajectory of the ion beam and interface mixing [45][46].

**1.5.2 Pattern formation by IBE**

Back in 1962, Navez et al.[47] conducted research on the morphology of glass surfaces by bombarding them with a 4 keV ionized air. They discovered that the surface developed periodic patterns while the sputtering was taking place, and the lateral dimensions of these structures varied from 30 to 120 nm, depending on the angle of incidence. However, the establishment of surface nanopatterning by IBS in the form we are familiar with today stems mostly from the late 1980s. Ion beam erosion-induced pattern formation was first described theoretically by the linear continuum Bradley and Harper (BH) model in 1988 [14], which combined surface diffusion with Sigmund's sputtering theory [44]. According to this theory, periodic ripple-like patterns are produced by the interplay between two competitive counteracting roughening and smoothening processes. Sigmund had already shown that the local erosion rate of a surface under an ion bombardment is higher in valleys than on hills. This curvature dependency of the sputter yield creates a surface instability, which amplifies all initial modulations. However, in the presence of a competing smoothing process such as surface self-diffusion activated thermally or by the ion itself, a mass current also exists simultaneously stimulated by the gradient of surface energy assumed to be proportional to the surface local curvature. During the interplay between these two competitive processes, a wavelength selection is seen, with the most unstable mode growing at the quickest rate. The linear continuum equation describes this surface height evolution known as the Bradley-Harper (BH) equation. The surface height evolution $h$(r, t) as described by BH theory stands as:

$$\partial_t h(x, y, t) = v_x(\theta)\partial_x^2 h + v_y(\theta)\partial_y^2 h - D\nabla^4 h$$

where h is a function describing the surface topography, D is a coefficient describing the surface diffusion and $v_{x,y}$ are coefficients describing the dependence of the erosion speed on the surface curvature—these coefficients depend on the angle of incidence θ. Since the coefficients $v_{x,y}$ depends on the angle of ion incidence θ, the pattern periodicity and orientation are also expected to depend on the angle. Ion bombardment causes sputtering and a roughening of the surface in a random way. Figure 1.4 shows how the energy of single ion impacts (assumed Gaussian shape) is spread out in space (orange ellipses). Ions that enter a crest at $A_r$, $B_r$, or P deposit their energy at the same depth beneath these locations. Impacting ions in a trough (points $A_v$, $B_v$, and Q) provide the same amount of deposited energy as ions hitting in P and Q. However, the contribution to the sputter yield near the surface varies between troughs (point Q) and crests (point P) if we consider the energy deposition by the secondary ions hitting at $A_r$, $B_r$ and Av, Bv. By comparing the mean energy deposited at P and Q, one can see that ion strikes in $A_V$ and $B_V$ have a bigger effect on sputtering in Q than strikes in $A_r$ and $B_r$ have on sputtering in P. This is because the distance from the mean energy deposition depth to Q is shorter in the trough than in the ridges (at P). Because of this, an ion-induced surface instability, also called surface micro-roughening, makes the surface rougher.

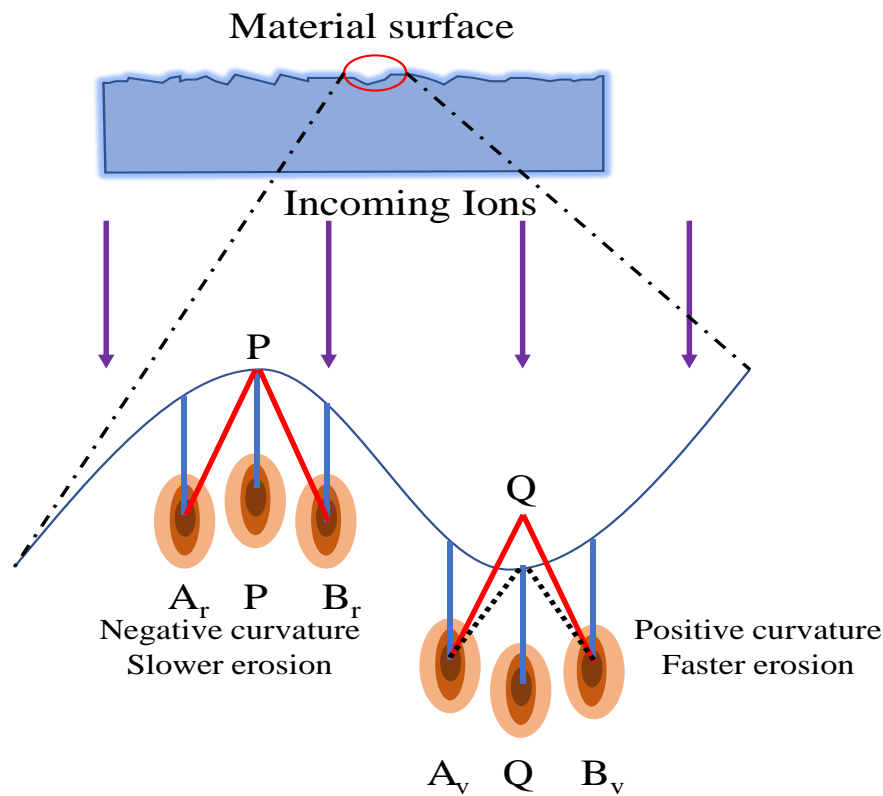

**Figure 1.4:** Schematic drawing of the energy distribution (elliptical shaped color coded) in rough surfaces. The red line and black dashed line represents the corresponding distances from energy deposition to the ridge (P) and valley (Q) region [48][49].

In particular, the BH theory predicts that for low angles of ion incidence, a ripple pattern perpendicular to the ion beam should form, while for grazing incidence, a ripple pattern should become parallel to the ion beam. The BH model mimics some of the most important empirically observed properties of pattern generation and early morphology development, such as pattern orientation relative to the ion beam and exponential rise of ripple amplitude. However, certain experimental data, such as the saturation of the ripple amplitude, cannot be explained within the context of the linear model over extended sputtering durations. This difference was due to the increasing dominance of nonlinear terms over morphology in subsequent periods. For this reason, the ion-induced development of periodic surface structures was described in 1995 by Cuerno and Barabasi [50] using a Kuramoto-Sivashinsky (KS) [48][49] type nonlinear continuum equation. This equation behaves similarly to the linear BH equation in the early time domain. However, the nonlinear terms begin to influence the surface's development at a certain point. Since then, numerous further improvements have been made to the theory. Thus, new physical processes, such as the nonlinear dependency of the erosion rate on the local surface slope, surface viscous flow, ion-induced mass redistribution, ion implantation, stress effects, etc. [47][49] [51][52], have been added.

### 1.5.3 IBE: a tool for tuning Magnetic anisotropy

The application of low-energy ion beam erosion diversified to a wide range of fields in material, physical, chemical and medical sciences. The IBE-induced self-organized structures induce anisotropic optical properties of silver nanoparticle array [53], tailor resistivity anisotropy [54], modify adhesion of neural stem cells [55], used in high-performance $NO_2$ gas sensor [56], tune plasma resonance of Ag nanoparticle arrays [57]. IBE is also used for creating water repellent super hydrophobic surface [58], modification of optical and structural properties of nanowires [59], tuning wetting dynamics [60], nano smoothening and conductivity enhancement of metal thin film [61], improve of anti-bacterial properties of metallic biomaterials [62], polishing of Si in W/Si soft X-ray multilayer structures [63], modulation of structure, morphology and wettability of polytetrafluoroethylene [64], changes Debye temperature and sheet resistance [65], enhances solar-blind photodetection [66], etc.

Magnetism is a collective phenomenon that is sensitive to surface/interfacial morphology, the shape of materials, and the lattice symmetry breaking effect. In ultrathin films, where the role of interface atoms has become relevant or even dominant, nano-structuring opens the possibility of tuning the magnetic anisotropy. Ion beam-induced self-organized nanostructures such as periodic ripple patterns, ordered arrays of the dot, pits, triangular and sawtooth features provide a great option to pattern and tailor the morphology of substrate and thin film. These artificially induced morphological features have dimensions ranging from μm to a few tens of nm and are comparable to the magnetic length scales such as exchange length, domain wall width, equilibrium particle size for coherent rotation, superparamagnetic blocking radius, etc. Thereby affecting magnetic properties strongly. Magnetic anisotropy in the thin film can be induced in two ways: either by deposition of thin film over the patterned substrate or direct ion beam patterning of the thin film surface. In the patterned polycrystalline thin film, shape anisotropy is the main source of MA. However, an intrinsic magneto-crystalline anisotropy is present in the epitaxial thin film. Therefore, an interplay between shape anisotropy, MCA and modified spin-orbit coupling in the epitaxial thin film can give rise to a special kind of magnetic switching behaviour.

## 1.6 Aim and scope of the present thesis.

As discussed in the earlier sections, magnetic thin film and nanostructured materials have huge potential in thin film-based electronic devices. It is also previously mentioned that oblique incidence low energy ion beam erosion (IBE) is a handy, versatile and cost-effective tool compared to the lithographic-based techniques for self-organized nanopatterning of substrate as well as various 2-dimensional polycrystalline and epitaxial thin film surfaces for tuning a large number of physical properties. In this regard, the present thesis aims to investigate the effect of low energy IBE on the growth, structure and morphology of epitaxial and polycrystalline thin film as well as its artificial engineering to induce and flexible tailoring of magnetic anisotropy. The key challenge in this field is to precisely quantify the influence of ions beam on surface structure and morphology and to unequivocally connect the same with magnetic characteristics. Additionally, in ex-situ studies, the thin film surfaces get oxidized and contaminated as soon as they are exposed to the environment. As a result, it is impossible to acquire the ion beam's true influence on the surface structure. Using traditional methods like X-ray diffraction, it is quite challenging to retrieve structural information from a thin, ion beam-modified layer that is just a few angstroms thick. Therefore, in-situ real-time observation of the ion beam modified surface texture, morphology and magnetic anisotropy has been another major objective to correlate and separate them unambiguously.

Furthermore, surface crystallinity, such as texture and structure modification by optimization of ion beam parameters and to test whether it can induce magneto crystalline anisotropy, is another aim of this thesis. Development of understanding of different kinds of magnetic reversal behaviour from the microscopic point of view, such as atomic arrangements, external shape, buried morphology, domain configurations, etc. In addition to this removal of the pre-existing limitations in IBE-induced UMA, for instance, a saturation of magnetic anisotropy with an increment of film thickness or sources of UMA limited to the film surface region due to limited penetration depth of ions, by designing a novel and unconventional experiment. In this regard, the possibility of combining anisotropic morphology and crystallographic texture would be one promising approach. As epitaxial thin film exhibits intrinsic MCA, the study of the interplay or competition between induced and crystalline anisotropy would be another scope of the present thesis. Furthermore, attempts would be made to combine separate methods of inducing magnetic anisotropy, such as oblique angle deposition and thin film deposition on the rippled substrate, to enhance the strength of UMA as compared to each of the individual methods.

# Chapter 2

# Experimental Techniques

This chapter focuses on the diverse experimental techniques used for thin film deposition and characterization throughout the completion of this thesis work. In addition, a thorough description of the ultra-high vacuum chamber utilized for in-situ deposition and characterization is also given.

# Chapter 2: Experimental techniques

## 2.1. Introduction
## 2.2. Thin film deposition techniques
   2.2.1. E-beam evaporation

   2.2.2. Ion beam sputtering

## 2.3. Characterization techniques
   2.3.1. Reflection high energy electron diffraction (RHEED)

   2.3.2. X-ray reflectivity (XRR)

   2.3.3. X-ray Diffuse Scattering (XDS)

   2.3.4. Magneto-optical Kerr effect (MOKE)

   2.3.5. Grazing Incidence Small Angle X-ray scattering (GISAXS)

   2.3.6. Grazing Incidence Wide Angle X-ray scattering (GIWAXS)

   2.3.7. Nuclear Forward Scattering (NFS)

   2.3.8. Atomic Force Microscopy (AFM)

## 2.4. A versatile UHV chamber for in-situ study
   2.4.1. Technical details

   2.4.2. Benefits of in-situ UHV chamber

   2.4.3. Electron beam evaporation set up

   2.4.4. In-situ MOKE set up

   2.4.5. In-situ XRR set-up

   2.4.6. In-situ RHEED set-up

   2.4.7. In-situ ion beam erosion set-up

## 2.1 Introduction

Several thin films and multilayers have been prepared in this thesis to investigate growth, structure, morphology, magnetic properties and to tailor its physical properties by ion sculpting. These samples were prepared and characterized using a variety of experimental procedures, which are summarized in this chapter. Section 2.2 explores thin film deposition procedures in detail, whereas structural, morphological and magnetic characterization methods are covered in Section 2.3. Since most of the current thesis works are carried out in-situ under ultra-high vacuum (UHV) conditions, section 2.4 also includes information on the UHV apparatus and its characterization and deposition methods. The advantages of in-situ investigations over ex-situ studies are also highlighted in the same section.

## 2.2 Thin film deposition techniques

The deposition of thin film and multilayer is critical in the production of micro and nanodevices. The thickness of a thin film is commonly measured to be less than 1000 nanometers. The emission of particles from a source initiates the deposition process. The particles are then transported toward the substrate. At last, the particles accumulate on the surface of the substrate, where they are condensed. The three primary physical vapor deposition (PVD) methods used for thin film deposition are (a) Sputtering, (b) Thermal evaporation and (c) Electron-beam evaporation (e-beam evaporation).

### 2.2.2 E-beam evaporation

E-beam evaporation is a powerful physical vapor deposition process that allows the user to deposit thin film and multilayers with high melting points that are difficult or even impossible to process using standard resistive thermal evaporation [67][68]. E-beam evaporation allows the evaporation of a wider range of metals with higher melting points, such as gold, platinum, tungsten, tantalum, and silicon dioxide. A schematic sketch of the e- beam evaporation set up is shown in Fig. 2.1. The electron beam is generated by an electron gun containing a filament through thermionic emission. Emitted electrons are then accelerated by a high potential difference. A permanent magnetic field is often used to control the trajectory of electrons and focus them on the target material. After the collision, the electrons transfer their kinetic energy rapidly to the target material, which is then converted into thermal energy. Once the temperature of the surface of the target material reaches the sublimation/evaporation temperature, the material evaporates and deposits onto the substrate in the form of a thin film.

The e-beam has a high electrical density, allowing for the efficient evaporation of high-melting-point materials that don't sublimate during thermal evaporation [69]. Electron-beam evaporation requires a high vacuum chamber for evaporation of the target material and to minimize the scattering of evaporated target atoms with background materials [70]. The target material is placed inside a crucible made of copper or tungsten. During e-beam evaporation, chilled water is circulated below the crucible, preventing it from heating through thermal energy. The evaporation occurs very confined region near the electron bombardment site on the surface of the source. As a result, little to no contamination from the crucible occurs. E-beam evaporation is also compatible with using an ion assist source to enhance the desired characteristics of a thin film.

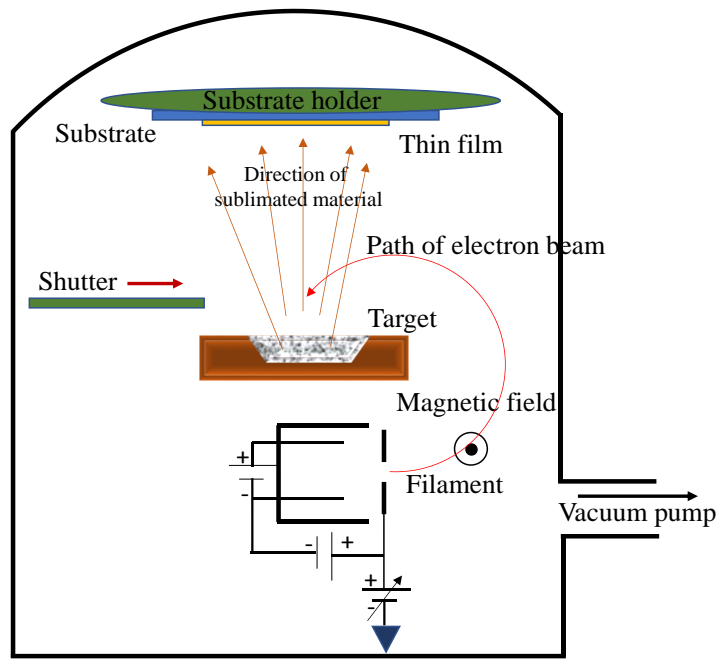

**Figure 2.1:** Schematic sketch of e- beam evaporation set-up.

## 2.2.2 Ion beam sputtering

Ion beam sputtering (IBS) is another physical vapor thin film deposition technique. It is the primary alternative to the evaporation technique. In this technique, an ion source is used to deposit a target material onto a substrate. The ion beam is usually generated by an ion gun which is focused on the sputtering target, and the sputtered target material finally deposits onto the substrate to create a film. The target and substrate are placed inside a high vacuum chamber. A high voltage of 1-10kV is applied to the anode of the ion source that confines stray electrons by creating an electrostatic field. Usually, an inert gas such as Argon is used as the sputtering gas. For efficient momentum transfer, the atomic weight of the sputtering gas should be close to the atomic weight of the target [68][70].

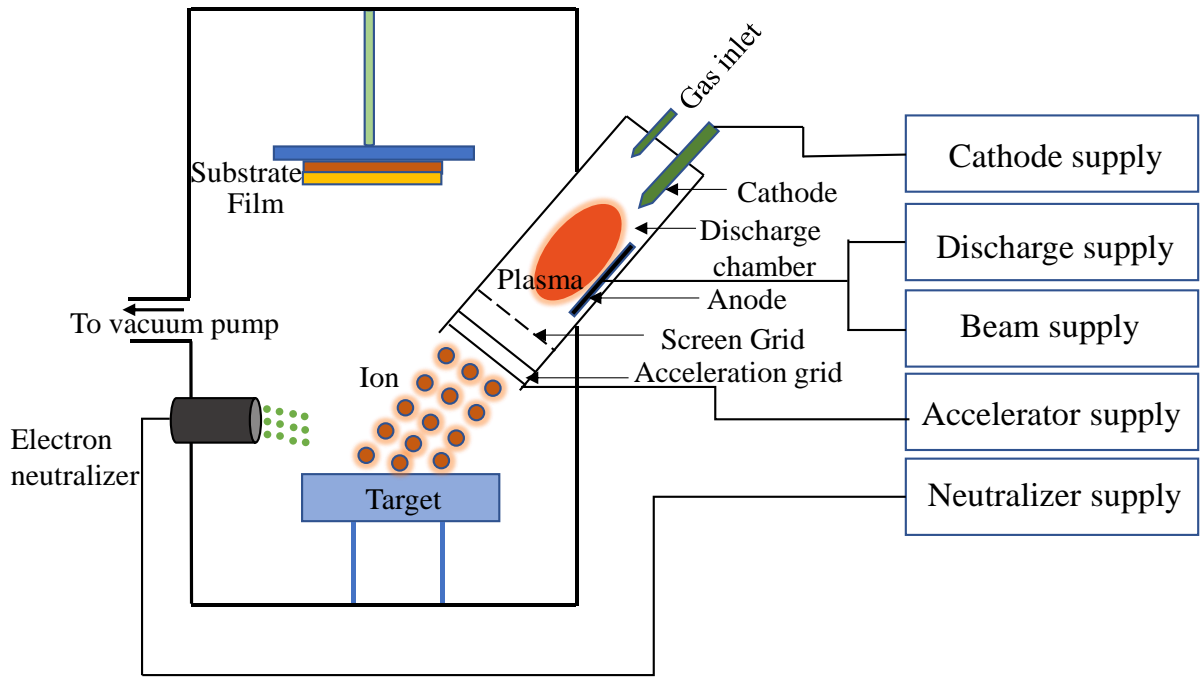

**Figure 2.2:** Schematic diagram of IBS system.

When inert gas is introduced inside the ion source, the gas is ionized by these stray electrons and plasma is formed. These ions are then extracted by the cathode electrode. The resulting ion beam strikes the target surface and transfers its momentum to sputter off the material that gets deposited in the form of thin film on the substrate. A schematic diagram of the IBS set up is shown in figure 2.2. Ion beam deposition processes are advantageous because the sputtering rate, the ionic energy and density can be controlled easily.  Furthermore, control over stoichiometry is better for multi-component systems [69]. It is also suitable for the deposition of heat-sensitive material due to low thermal radiation. Since the target material is directly transformed into the vapor phase through the mechanical process rather than the chemical or thermal process, ideally, all kinds of material can be deposited by this process.

## 2.3 Characterization techniques

### 2.3.1 Magneto-Optical Kerr Effect (MOKE)

Magneto-optical Kerr effect magnetometry is a widely used easy to handle, non-destructive technique for determining the magnetic and magneto-optic properties of a wide range of magnetic thin film, multilayers and magnetic nanostructure systems. The interaction of an electromagnetic wave with the material is sensitive to its magnetic state and gives rise to a magneto-optic effect. The working principle of the MOKE magnetometer is based on the so-called Kerr effect, which describes the change in intensity/plane of polarization of a linearly polarized light after reflection from a magnetized sample [71].

**2.3.1.1 Classical theory of MOKE**

A linearly polarized light can be decomposed into left circularly polarized (LCP) and right circularly polarized (RCP) light of equal magnitude.

$$E_\pm(r,t) = \frac{1}{\sqrt{2}} E_0 (e_x \pm e_y) e^{(kz-iwt)} \qquad (2.1)$$

From the classical viewpoint, the electric field of LCP light will force the electrons to rotate left-circularly, while the electric field of RCP light will induce a right-handed motion around the fixed positive centre. The pair electron-positive centre forms a rotating electric dipole. In the absence of an external magnetic field, the radius of both circular motions will be equal as given by:

$$qE_{L,R} + kr_{L,R} = m\omega^2 r_{L,R} \gg r_{L,R} = \frac{eE/2m}{(\omega^2 - \omega_0^2)} \qquad (2.2)$$

In this scenario, the dielectric constant can be written as [72]:

$$\epsilon_{L,R} = \varepsilon_0 \left(1 + \frac{Ne^2/2m\varepsilon_0}{\omega^2 - \omega_0^2}\right) \qquad (2.3)$$

However, with the application of an external magnetic field, an additional Lorentz force will act on each electron's motion. This force will alter the magnitude of left and right circular motion differently. In this case, the radius for left and right circular motion takes the form [72]:

$$qE_{L,R} + kr_{L,R} + e\omega r_{L,R} B = m\omega^2 r_{L,R} \tag{2.4}$$

$$r_{L,R} = \frac{eE/2m}{(\omega^2 - \omega_0^2 \mp \omega Be/m)} \tag{2.5}$$

It gives rise to a change in the dielectric constant for two oppositely oriented motions.

$$n_{L,R}^2 = \left(1 + \frac{Ne^2/2m\varepsilon_0}{\omega^2 - \omega_0^2 \mp \omega Be/m}\right) \tag{2.6}$$

Since the dielectric constant is directly related to the refractive index of the material, propagation velocity will be different for each component. So, a phase shift will occur and rotation of the plane of polarization. Therefore, the linearly polarized light, after reflection from the magnetic surface, becomes elliptically polarized [71][72]. The former is due to the unequal phases, and the latter one is due to the unequal magnitude of reflected left and right circularly polarized right. The rotation of this polarization axis is known as Kerr rotation, and the amount of rotation is denoted by θ$_{Kerr}$. The Kerr rotation is proportional to the magnetization of the sample.

### 2.3.1.2 Macroscopic electromagnetic/dielectric theory

The macroscopic theory of the Kerr effect is based on the dielectric properties of a medium with a suitable description of Maxwell equations of electromagnetic waves propagating in finite magnetic media.

Normally, when a linearly polarized light is reflected off from a magnetized surface, its polarization changes. The effect of magnetization on the reflected light is solely given by the change of the permittivity tensor with magnetization. The changes in polarization of the reflected light can be shown using the general dielectric tensor, given by [73][74]:

$$\varepsilon = \varepsilon_{xx} \begin{pmatrix} 1 & -iQm_z & iQm_y \\ iQm_z & 1 & -iQm_x \\ -iQm_y & iQm_x & 1 \end{pmatrix} \tag{2.7}$$

where Q is the magneto-optical coupling factor of the material known as Voigt Constant, $\varepsilon_{xx}$ is scaler relative permittivity and $m_i$ are components of magnetization vector M.

The off-diagonal elements of the dielectric tensor are proportional to the magnetization. Therefore, the polarization state of the light at the surface can be correlated with the magnetization of the sample. This kind of correlation is called the magneto-optical response, which the MOKE device measures. The magneto-optical response is the rotation of the polarization of the in-phase component of incoming light. This response also changes the out-of-phase part of the incident light, causing an ellipticity. This change is known as Kerr ellipticity. It is feasible to observe the influence of an external magnetic field on the sample's magnetization by using the correlation between the polarised light that is reflected and the sample's magnetization. The MOKE measurement gives a hysteresis loop whose slope is equal to the material's permeability ($\mu = B/H$).

With the help of this dielectric tensor, a reflection matrix is constructed by solving the Maxwell equations, which is known as the Fresnel reflection matrix. The incident and the reflected component of the electric field are related through Fresnel reflection matrix R as given below [75][76]:

$$\begin{pmatrix} E_s \\ E_p \end{pmatrix}^r = \begin{pmatrix} r_{pp} & r_{ps} \\ r_{sp} & r_{ss} \end{pmatrix} \begin{pmatrix} E_s \\ E_p \end{pmatrix}^i = R \begin{pmatrix} E_s \\ E_p \end{pmatrix}^i \qquad (2.8)$$

Here, Ep and Es are the electric field vector amplitudes for p and s polarized components of electromagnetic waves. Here i and r denote incident and reflected wave, respectively. $r_{mn}$ is the reflection coefficient and is defined as the complex ratios between the reflected (*r*) m polarized and the incident (*i*) n polarized electric field amplitudes: $r_{mn} = E_{rm}/E_{in}$. Therefore, the reflected signal intensity measured by the detector is given by: $I = (E_r)^2$. Note that the reflection coefficients, $r_{mn}$, depend on the refractive index as well as the permittivity of the material under study.

The Kerr rotation($\theta_K$) and Kerr ellipticity($\varepsilon_K$) of s and p polarized light are related to the ratio of these coefficients as follows [71][74][76] [77]:

$$\Theta_K^p = \theta_K^p + i\,\varepsilon_K^p = \frac{r_{sp}}{r_{pp}} \qquad (2.9)$$

$$\Theta_K^s = \theta_K^s + i\,\varepsilon_K^s = \frac{r_{ps}}{r_{ss}} \qquad (2.10)$$

The expression for complex Kerr effect for arbitrary magnetization direction stands as [74]:

$$\Theta_K^{p,s} = \frac{\cos\theta_0 (m_y \tan\theta_1 \pm m_z)}{\cos(\theta_0 \mp \theta_1)} \cdot \frac{in_0 n_1 Q}{(n_1^2 - n_0^2)} \tag{2.11}$$

Where, $\theta_0$, $\theta_1$, $n_0$, $n_1$ are the angle of incidence, angle of refraction, the refractive index of the nonmagnetic medium 0, and the magnetic medium 1, respectively.

### 2.3.1.3 MOKE experimental set-up

The MOKE apparatus is constructed on a vibration-isolated optical table. The set-up consists of a 1mW red He-Ne laser ($\lambda \approx 633$nm), one photo-elastic modulator, one polarizer, one analyzer, one photo diode detector, lock-in amplifier and computer interfacing. A schematic of the MOKE magnetometry set-up is presented in Figure 2.3. The sample is placed at the centre between the two pole pieces of an electromagnet. A polarized beam of He-Ne laser is directed towards the sample surface that passes through a polarizer and PEM. The PEM act as an optical chopper of a given frequency. The operating frequency of PEM must be set to an unique frequency other than the electrical frequencies present in the lab in order to maintain the integrity of the data. The reflected light passes through an analyzer whose polarization axis is kept at an angle $\phi$ with extinction angle from the polarizer and is detected by a photo-diode. The detected signal is amplified by a lock-in amplifier, which is locked to the same frequency as the PEM, reducing the effect of ambient light on the measurement.

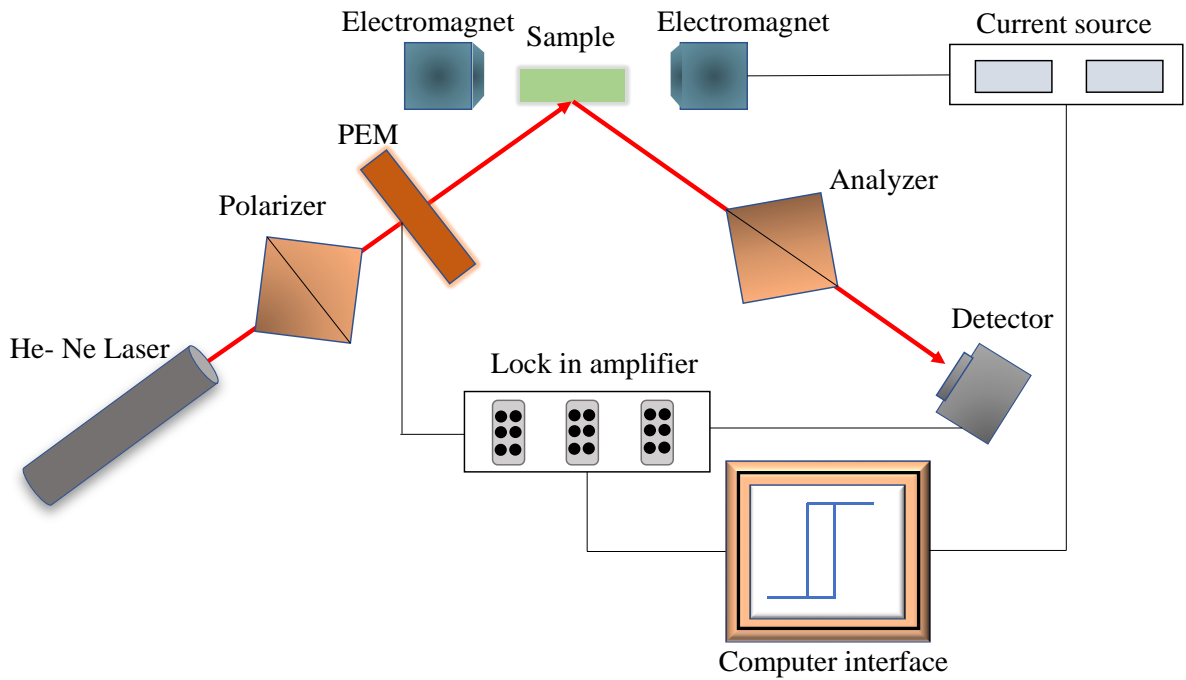

**Figure 2.3:** An aerial perspective of the MOKE set-up is shown in this diagram.

**2.3.1.4 Different geometries of MOKE**

Depending upon the relative orientation of the magnetization vector and plane of incidence of the laser beam, three basis geometries are categorized. These are described below:

1. Longitudinal MOKE: In longitudinal geometry, the magnetization vector is oriented parallel to the plane of incidence of the laser beam and in the film plane. This geometry is sensitive to the determination of the in-plane component of magnetization.

2. Transverse MOKE: In this geometry, the magnetization vector is aligned perpendicular to the plane of incidence of the laser beam and in the film plane. This kind of MOKE measurement is used to determine the perpendicular component of magnetization with respect to the plane of incidence of the laser beam and in the film plane.

3. Polar MOKE: Polar MOKE is used to extract out of plane component of magnetization. In this geometry, the magnetization vector is oriented perpendicular to the film plane and parallel to the plane of incidence of the laser beam.

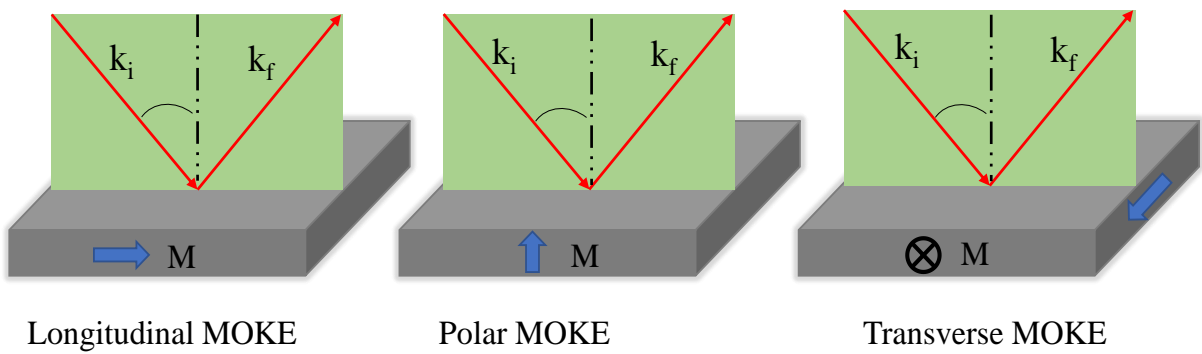

**Figure 2.4:** Different geometries of MOKE measurement.

In recent years, the MOKE has gained a great deal of attention because of its extensive usage in magneto-optical recording systems. MOKE techniques are widely used for the study of magnetic ordering, magnetic anisotropy, coupling between the layers of the multilayer film, and dynamic magnetic studies of film growth.

## 2.3.2 Reflection High Energy Electron Diffraction (RHEED)

Reflection high energy electron diffraction (RHEED) is a surface-sensitive, easy-to-handle, an in-situ compatible diffraction-based technique [78][79]. The instrumental part of RHEED consists of an electron gun as a source of the high energetic electron, a fluorescent screen where diffracted electron from the sample creates a diffraction pattern, and a CCD camera placed outside the vacuum chamber to capture the resulting diffracted image. A schematic of the RHEED geometry is presented in Fig. 2.5. The typical energy of electron used for RHEED measurement varies from 15 to 30 KeV. In an RHEED experiment, the electron is incident at an angle from 1º to 4º from the sample surface. The grazing incidence geometry makes the RHEED technique more surface-sensitive than other diffraction methods and in-situ compatible [80]. The wavelength of the electron can be calculated from the de-Broglie hypothesis: $\lambda = \frac{h}{\sqrt{2mE}} = \sqrt{\frac{150.4}{E(in\ ev)}}$ Å. For an electron of energy 20KeV, the wavelength corresponds to 0.085855Å. Ultra-high vacuum condition is required to perform RHEED measurement. It is due to the elimination of strong scattering of an electron by gas molecules and sensitivity of diffraction pattern by adsorbed impurities. Therefore, the application of this technique is limited to ultra-high vacuum-based systems.

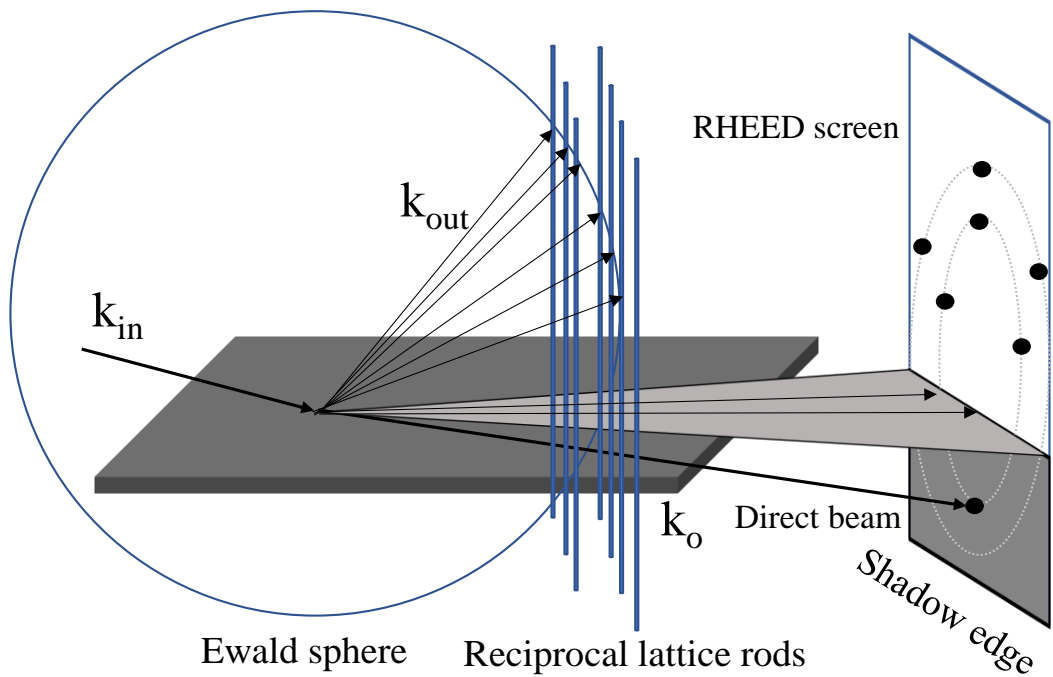

**Figure 2.5**: Schematic of RHEED geometry.

RHEED provides a great opportunity to monitor the real-time growth process of the thin film. In particular, the intensity oscillation arising from periodic surface roughness variation can be used to precisely monitor and control film thickness. RHEED pattern can be used to extract symmetry and surface periodicity of material, texture and crystallinity [81][82][83].

**Different Diffraction Geometries of RHEED**

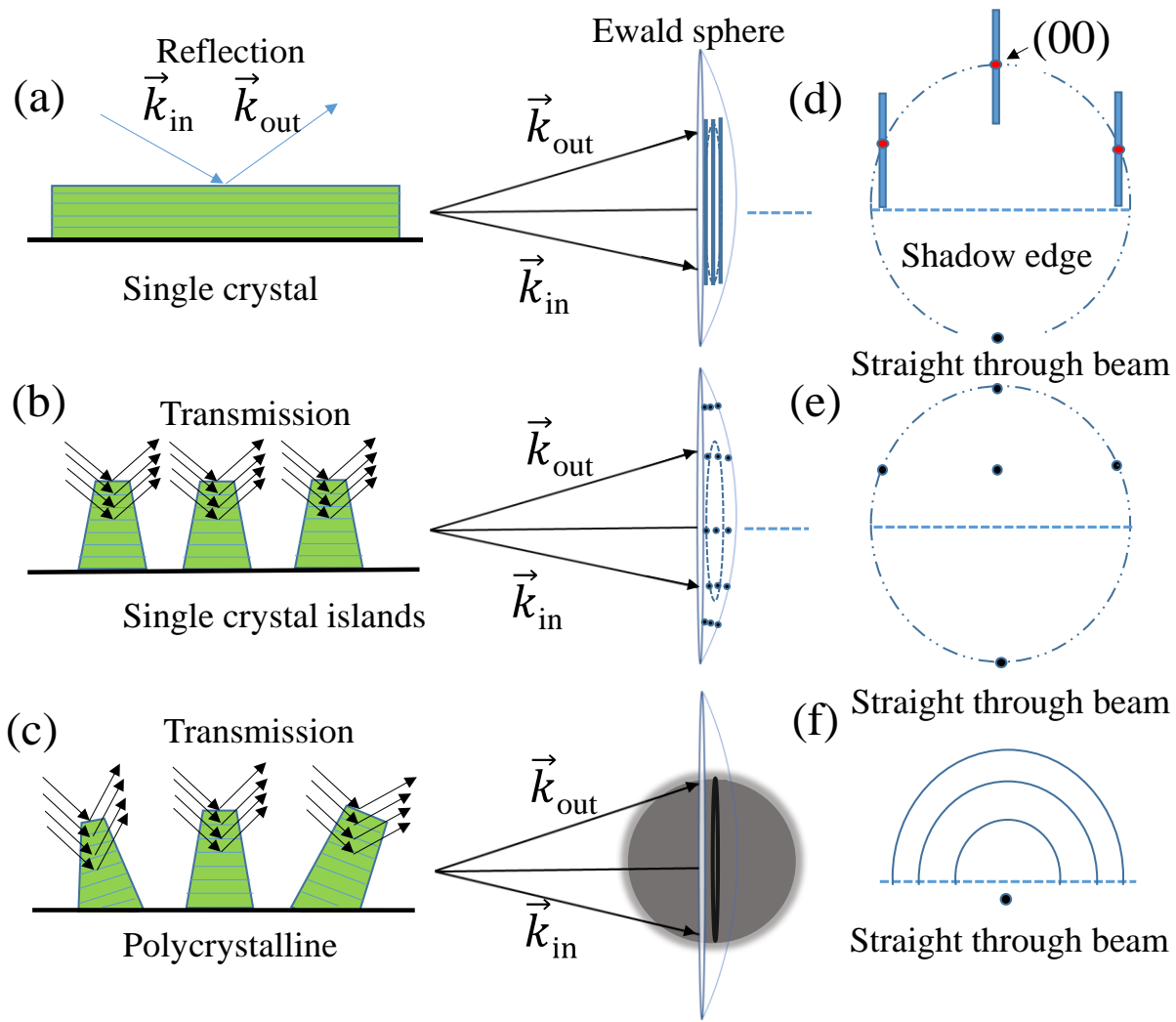

**Figure 2.6:** Schematics of different electron scattering geometries, film morphologies, crystalline structures and their RHEED pattern are shown here: single-crystalline thin film having (a) smooth surface and (b) islands, (c) polycrystalline thin film. The Ewald sphere constructions corresponding to these three structures are also presented alongside.

It is possible to separate RHEED patterns into two types: reflection and transmission, depending on the electron scattering geometry [84]. Reflection patterns are observed from smooth 2-d like film surfaces, while the presence of 3-d like surface roughness gives rise to transmission RHEED pattern. The electron scattering from three crystalline materials—a smooth single-crystal surface, a rough single-crystal film with islands, and a rough polycrystalline film — is shown in Fig. 2.6a-c. The word island generally represents surface roughness. Electrons are reflected from the top layers of a smooth single-crystal surface, but they pass through and scatter from the adjacent surface planes of crystalline and polycrystalline islands. Streaks are observed for a smooth single-crystal surface where the large radius Ewald sphere intersects with a series of reciprocal lattice rods that are perpendicular to the surface. For single crystalline rough surface electron enters from one face of the island and leaves from another face, giving rise to a sharp spot in the diffraction pattern. In the case of the polycrystalline thin film where randomly oriented grains are present, the reciprocal structure consists of the concentric spherical cell. Its interaction with the Ewald sphere creates continuous concentric rings in a diffraction pattern [85].

### 2.3.3. X-ray Reflectivity (XRR)

X-ray reflectivity is a commonly used technique for analyzing surfaces and interfaces of thin film and multilayer systems. The electron density variation with Angstrom resolution can be probed by X-ray reflectometry [86]. Therefore, it perfectly satisfies the criteria for cutting-edge nano samples. X-ray reflectivity measurements are performed in specular conditions, i.e., the incident and scattered angle are the same. Therefore, the resulting momentum transfer vector q= $\frac{4\pi}{\lambda} \sin \alpha_i$ always points perpendicular to the sample surface [87]. The refractive index n of a matter for an x-ray of wavelength ≈ 1Å can be written as [88–90]:

$$n = 1-\delta-i\beta. \tag{2.12}$$

*Here* δ is a wavelength-dependent scattering term: $\delta = \frac{r_e \lambda^2 \rho_e}{2\pi}$ and

*β*: wavelength-dependent absorption part: $\beta = \frac{\lambda}{4\pi} \mu$

Typical orders of magnitude are $\delta=10^{-5}$ and $\beta=10^{-6}$. $r_e$ is the classical radius of electron, $\rho_e$ is the electron density, $\mu$ is the mass density. With $\delta, \beta > 0$, the value of n is slightly less than 1. Therefore, after refraction, the transmitted beam bends toward the surface. When the incident angle $\alpha_i$ is smaller than a critical value $\alpha_c$, the beam experiences total external reflection. Thus, only an evanescent wave is present below the surface, which decays exponentially perpendicular to the surface, and propagates parallel to the surface. The critical angle expression based on total external reflection condition and Snell's law stand as [87][88][91]:

$$\alpha_c^2 = 2\delta = \frac{r_e \lambda^2 \rho_e}{\pi} \tag{2.13}$$

Typical orders of magnitude of $\alpha_c \approx \sqrt{2\delta}$ with $\delta = 10^{-5}$ varies $\approx$ from 0.1° to 0.5°. When $\alpha_i$ is greater than $\alpha_c$, the penetration depth of the transmitted wave sharply increases and propagates into the bulk. At every interface where the X-ray beam experience a change in electron density, a part of the X-ray beam is reflected. The reflected X-ray from different interfaces interferes constructively and destructively as a function of incoming angle that creates the reflectometry pattern known as Kiessig fringes.

The reflected intensity of X-ray can be obtained by considering the continuity condition of the electromagnetic wave at the interface and the Fresnel coefficient of the reflection equation. The specular reflectivity R(θ) at an angle θ for incident radiation of intensity $I_0$ is given by [87][91]:

$$R(\theta) = \frac{I(\theta)}{I_0}$$

$$R(\theta) = \left| \frac{\theta - \sqrt{\theta^2 - \theta_c^2 - 2i\beta}}{\theta + \sqrt{\theta^2 - \theta_c^2 - 2i\beta}} \right|^2 \tag{2.14}$$

Since the specular reflection condition is maintained here, the intensity of the reflected radiation, the so-called Fresnel reflectivity R(θ) can be expressed in terms of the wave vector $\vec{q} = \{0, 0, q_z = 4\pi \sin(\theta)/\lambda\}$ as [91][92]:

$$R(q_z) = \left| \frac{q_z - \sqrt{q_z^2 - q_c^2 - \frac{32i\pi^2\beta}{\lambda^2}}}{q_z + \sqrt{q_z^2 - q_c^2 - \frac{32i\pi^2\beta}{\lambda^2}}} \right|^2 \approx \left( \frac{q_c}{2q_z} \right)^4 \; for \; q_z \gg q_c.$$ Where $q_z = 2k\sin\alpha_i$ is the z component of the wavevector transfer q and $q_c = 2k\sin\alpha_c$ is the critical wavevector transfer

Deviation from the ideal interface or presence of roughness decreases specular reflection. The roughness effect causes damping in the reflectivity pattern and is described by the Debye-Waller-like factor [92]:

$$R_{rough} = R_{ideal} \cdot e^{-q_z^2 \sigma^2} \tag{2.15}$$

The reflection coefficient of the electric field of a single layer using matrix formalism and dynamical theory in terms of reflection coefficients of each interface is written to be [91][92]

$$r = \left| \frac{r_1 + r_2 e^{-2ik_z t}}{1 + r_1 r_2 e^{-2ik_z t}} \right| \tag{2.16}$$

In situations where absorption does not have to be taken into account, the reflected intensity is then,

$$R = \frac{r_1^2 + r_2^2 + 2r_1 r_2 \cos 2k_{z,1} d}{1 + r_1^2 r_2^2 + 2r_1 r_2 \cos 2k_{z,1} d} \tag{2.17}$$

The presence of cosine term in the equation 2.14 implies that the reflectivity curve will oscillate in reciprocal space whose period will be defined by the equality $2k_{z,1}d \approx q_{z,1}d = 2p\pi$, or $q_{z,1} = \frac{2p\pi}{d}$

For a layered system with N layers, the reflection coefficient r equals [91][92]

$$r = r_1 + r_2 e^{iq_{z,1}d_1} + r_3 e^{i(q_{z,1}d_1 + q_{z,2}d_2)} + \ldots\ldots\ldots + r_{j+1} e^{i\sum_{k=1}^{j} q_{z,k}d_k} + \ldots \tag{2.18}$$

In Eq. (2.18), the ratio $r_{j+1}$ of the amplitudes of the reflected to the incident waves at interface j, j+1 is:

$r_{j,j+1} = \frac{q_{z,j} - q_{z,j+1}}{q_{z,j} + q_{z,j+1}}$ with the wave-vector transfer in medium j: $q_{z,j} = (4\pi/\lambda)\sin\theta_j = \sqrt{q_z^2 - q_{c,j}^2}$.

A schematic sketch of the XRR set-up and XRR pattern of 18nm thick Co thin film deposited on Si substrate is presented in Fig. 2.7 and Fig. 2.8, respectively.

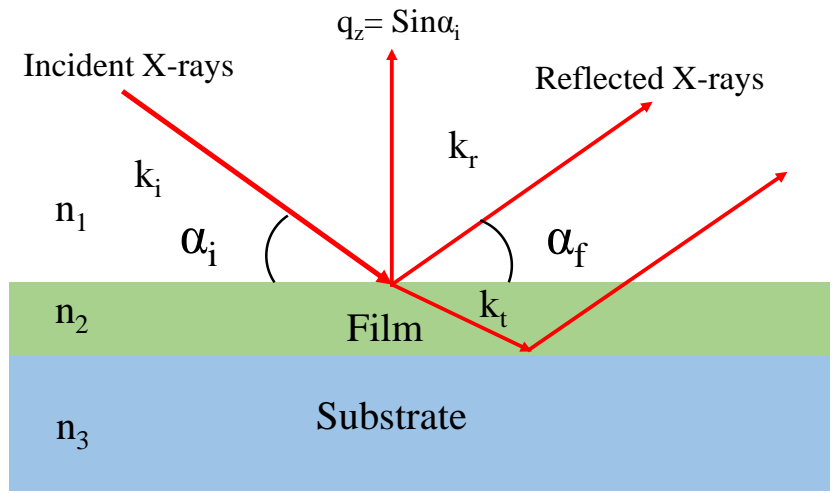

**Figure 2.7:** Schematic of XRR measurement geometry.

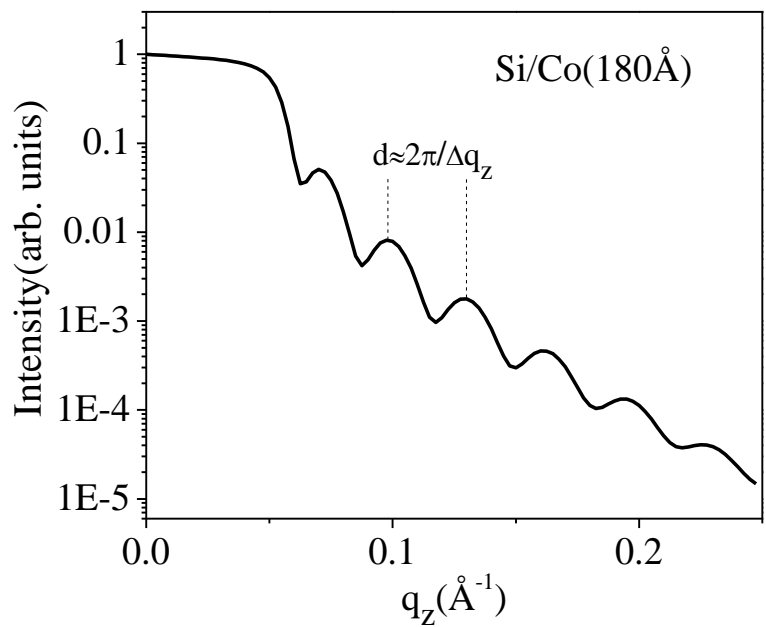

**Figure 2.8:** XRR pattern of Co thin film deposited on Si substrate.

These reflectivity curves provide information about layer parameters such as the individual layer thickness, density profile perpendicular to the surface and the r.m.s. surface/interfacial roughness, regardless of the crystallinity of each layer.

## 2.3.4. X-ray Diffuse scattering (XDS)

Surface morphology greatly impacts physical properties such as magnetism, plasmonic and transport properties of the thin film. X-ray scattering-based techniques are becoming more popular for characterizing surface and interfacial morphology. It is because X-ray scattering is sensitive to electron density differences and always probes along the direction of the scattering vector. When the morphology is anisotropic, it is not sufficient to describe its characteristics via the root mean square (r.m.s.) roughness because different surface profiles with varying height distributions may have the same r.m.s. values. Fig. 2.7 shows the schematic of x-ray diffuse scattering geometry. A monochromatic X-ray beam is incident on the sample surface at an angle of incidence $\theta_i$ and gets scattered at an angle $\theta_f$. In specular condition, $q_i=q_f$ and the resulting momentum transfer vector q coincide with $q_z$ axis and point along the sample's normal direction. We define the momentum-transfer vectors $q_x$ and $q_z$ in parallel and perpendicular directions to the film plane, respectively. However, the rocking or transverse scans are performed at a fixed scattering angle that nearly scans $q_x$ with $q_z$=const [93]. For an ideal interface, X-rays are reflected specularly. However, the surface and interfaces are never ideal or perfectly smooth. Therefore, a diffuse or non-specular component is always present in the scattered X-rays. The XDS reproduces the details of the lateral ordering of the interfacial morphology and the vertical replications of roughness from one interface to another.

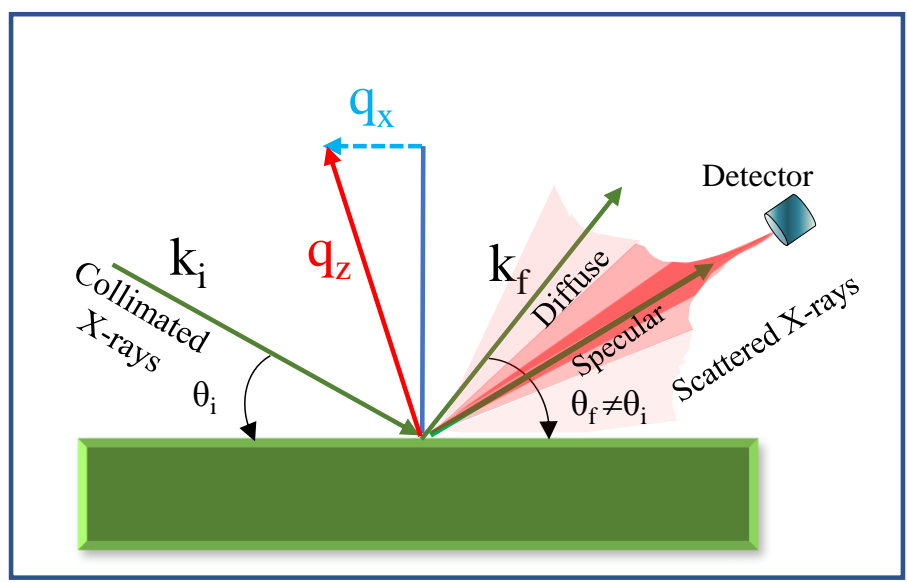

**Figure 2.9:** Schematic of X-ray diffuse scattering measurement.

The statistical roughness of an interface is best described by height-height correlation function: C(x,y) [94][95][96] = <δz(0)δz(x,y)> = $2\sigma^2 \left\{ 1 - \exp\left[ -\left(\frac{\sqrt{x^2+y^2}}{\xi}\right)^{2h} \right] \right\}$. $0 < h < 1$ is referred to as the "roughness" exponent. h ~ 0 jagged or irregular surfaces, h ~ 1 smooth hill-valley structures, σ is the root mean square roughness, which is height fluctuation around average height. ξ is the lateral correlation length defining a typical size of the roughness pattern/ corresponds to the length scale at which a point on the surface follows the memory of its initial value. Diffuse scattering allows us to obtain the details of the height-height correlation function at the interface and model the interface accordingly [97].

## 2.3.5. Grazing Incidence Small Angle X-ray Scattering (GISAXS)

Nanostructured thin films and surfaces have been attracting much attention in recent years as a template for creating anisotropic magnetic, transport, and plasmonic properties. GISAXS is a relatively new technique to analyze the morphology, size, shape, and distribution of nanostructured objects present at the surface and buried interface [98][99]. Such measurements often use high-brightness synchrotron radiation to boost the scattering intensity. Fig. 2.10 shows a schematic of GISAXS geometry. The incident angle $\alpha_i$ is generally kept constant below 1°. It is generally kept close to the critical angle $\alpha_c$ of the material to increase near-surface sensitivity by total external reflection and to minimize unwanted scattering from the substrate as well as the bulk of the material. The scattered beam is detected by a two-dimensional position-sensitive detector in a direction forming an in-plane scattering angle of $2\theta_f$ with respect to $k_i$ and $\alpha_f$ from the sample surface. The momentum transfer vector, defined as **q = k$_f$-k$_i$**, is a function of $\alpha_i$ and $\alpha_f$. X-ray with hard energy regime (~10 keV), corresponding wavelength λ ~ 1 Å, the magnitude of q is of the order of a few nm$^{-1}$. It covers real space distances in the nanoscale range, i.e., 10–100 nm. It makes the GISAXS technique suitable for investigating the nanostructure's size, shape, lateral ordering and arrangement [100].

In GISAXS, the angular coordinates are related to the momentum transfer coordinates according to the following relations [101]:

$q_z = k_0[ \sin(\alpha_i) + \sin(\alpha_f)]$

$q_y = k_0[\text{Sin}(2\theta_f)\text{Cos}(\alpha_f)]$

$q_x = k_0[\text{Cos}(2\theta_f)\text{Cos}(\alpha_f) - \text{Cos}(\alpha_i)]$

$k_0 = 2\pi/\lambda$

For an isotropic surface, GISAXS measurement along any azimuthal direction is sufficient to extract morphological parameters. However, in the case of an anisotropic surface, GISAXS measurement must be performed along the different azimuthal directions to extract complete information about the nanostructured object [102].

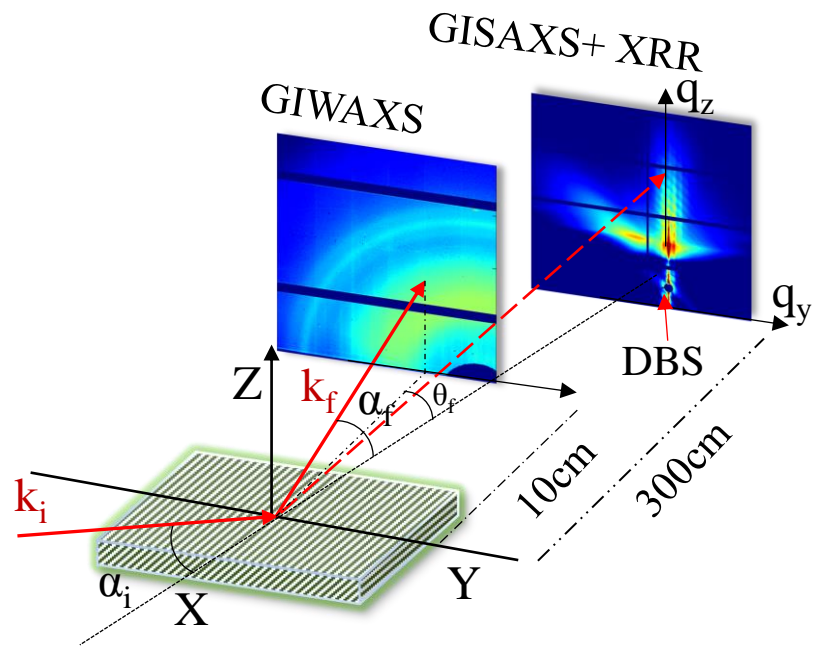

**Figure 2.10:** Schematic of GISAXS and GIWAXS set-up.

The GISAXS method has the following unique advantages:

(i) GISAXS is a fully non-destructive technique provided that the sample can bear hard X-ray exposure.

(ii) The scattered X-ray from a few millimeters square sample area provides statistically averaged information compared to microscopy-based AFM, FESEM, and TEM techniques.

(iii) By varying the incident angle, the probed depth changed. Therefore, it offers the ability to probe from surface to buried interfaces and the bulk of the sample.

(iv) The sample does not require any special environment or preparation condition. Thus, the technique is applicable to various types of environments, from ultra-high vacuum to gas atmospheres, even during chemical reactions, during growth or a catalytic reaction, at ambient to nonambient temperature.

As with most techniques, X-ray scattering also has some drawbacks:

(i) The X-ray signal coming out from sample scales with the sample volume. Therefore, the use of synchrotron radiation is nearly mandatory.

(ii) The extracted information is obtained in reciprocal space. Thus, it has to be ''transformed'' into real space.

(iii) X-ray beam size, divergence, and limitations of optics make it challenging to characterize a single nanoobject.

## 2.3.6. Grazing Incidence Wide Angle X-ray Scattering (GIWAXS)

The wide-angle x-ray scattering technique determines the degree of crystallinity, texture, interplanar spacing, crystal orientation, presence of strain, crystallite size [103] etc. of materials. This technique is analogous to powder diffraction. Similar to the SAXS technique, the X-ray is incident on the sample surface at very grazing incidence. Here, the area detector is placed close to the sample goniometer stage, roughly within 10-15cm. Therefore, it captures a relatively large scattering angle compared to GISAXS. Under this geometrical condition, it can probe length scale down to the interatomic scale of an unit cell [104]. The scattering pattern of WAXS represents the arrangement and orientation of atoms and molecules. Thin film with no crystallographic texture exhibits a sharp ring, film with preferentially ordered crystallographic orientation results in a broken ring or arc-like peaks, and highly ordered crystallographic orientation displays an ellipse-like scattering pattern. The crystallite size (d) is calculated from the position (θ) and full width at half maximum (F.W.H.M.) of a particular peak using the Scherrer formula: $d = \frac{K.\lambda}{FWHM.Cos\theta}$. λ is the wavelength of X-ray, K is a dimensionless shape factor, normally with a value of 0.9, θ is Bragg's angle (scattering angle).

## 2.3.7. Nuclear forward scattering (NFS)

Nuclear forward scattering (NFS) of synchrotron radiation is currently a popular methodology for studying the electronic and magnetic properties of materials [105]. NFS is the temporal counterpart of Mössbauer spectroscopy and provides information analogous to Mössbauer spectroscopy [106][107]. Synchrotron radiation from third-generation sources has a number of advantageous characteristics that make it an excellent option for research in harsh settings where classical Mossbauer spectroscopy is difficult or ineffective. Synchrotron-based NFS is especially beneficial when it is hard to get the radioactive sources ready for Mossbauer spectroscopy, the radioactive source's lifetime is limited, or the experimental set-up demands a collimated or small-size beam.

The short lifetime of an excited nuclear state and the pulse structure of synchrotron radiation is essential to observe NFS. The sample is irradiated with synchrotron radiation that matches the energy difference between the nuclear ground state and an excited state, resulting in nuclear excitation. As the excited state has a limited lifetime, the resonantly scattered photons lag behind in comparison to the non-resonantly scattered (prompt) photons. The usual lifetime of excited states is within the range of 0.2 to 200 ns and corresponds to the bunch structure of modern synchrotron radiation facilities. In theory, this means that the nuclear forward scattering can be measured for any Mossbauer isotope [108][109]. A schematic of the energy levels of the $Fe^{57}$ isotope and a typical NFS spectrum is shown in Fig. 2.11. Fast avalanche photodiode detectors [110] with a time resolution of 1 ns were used to measure the temporal distributions (time spectra) of time-delayed reemitted photons. The moment when prompt electron scattering was found, considered a zero-time reference. The observed time spectra exhibit quantum beats caused by interference of electromagnetic waves emitted by various hyperfine components of the nuclear transition with identical resonance frequencies. The pattern of spectra is governed by the magnetic state of iron nuclei and the magnetic structure of the materials [111]. The spectra were analyzed by fitting the quantum beats using the REFTIM software [112] and taking into account the experimental geometry, the polarization of the synchrotron radiation, and the direction of the external magnetic field.

The NFS investigation was conducted using linearly polarized synchrotron radiation at nuclear resonance scattering beamline P01, PETRA-III, Hamburg, Germany. The synchrotron was operated in the 40-bunch mode with a bunch separation of 192 ns for these measurements.

A time frame of this size is useful for studying the time spectra of nuclear scattering for the $^{57}$Fe Mössbauer isotope, which has a natural lifetime of 141 ns. A beam of linearly polarized x-rays with an energy of 14.4 keV of the first excited state of the $^{57}$Fe nucleus was used to excite the iron nuclei in the samples.

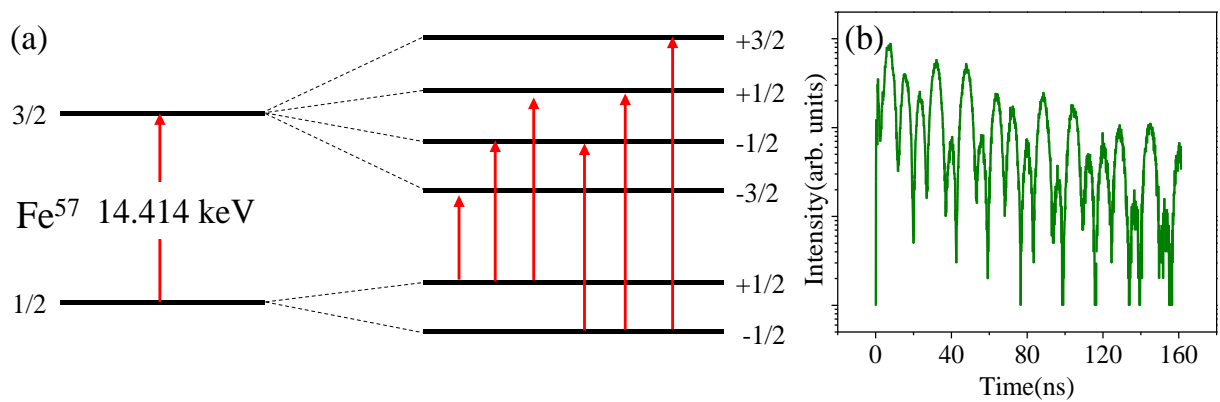

**Figure 2.11:** (a) Energy levels of Fe$^{57}$ isotope. (b) Nuclear forward scattering time spectra.

## 2.3.8. Atomic Force Microscopy (AFM)

Atomic force microscopy is a powerful scanning probe microscope for studying three-dimensional topographical features of samples at the nanoscale. A fundamental shortcoming of the scanning tunnelling microscope (STM) that it can only map surfaces of either conducting or semiconducting material was addressed by the development of atomic force microscopy (AFM) [113]. With the AFM, it is possible to image practically any surface, from polymers, composites and ceramics to glass and biological materials [114].

A standard AFM is made up of a cantilever with a tiny tip (probe), a laser, a position-sensitive photodiode, and an electronic controller. The rear side of the cantilever is coated with a film that reflects light similarly to a mirror. An AFM scans a sample surface using a cantilever with a sharp tip. The average radius of curvature at the end of commercial tips composed of silicon or silicon nitride is 5-10 nm. The cantilever compress or stretches depending on the force between the tip and the sample surface. A laser is pointed at the back of the cantilever, and a position-sensitive photodiode captures the reflected light. The deflection of the cantilever towards or away from the surface is influenced by the raised and lowered features on the sample surface. By using a feedback loop to regulate the height of the tip above the surface, the AFM can build an accurate topographic map of the surface's characteristics [113]. The sample, which is on a translation stage beneath the microscope, may also be moved. Fig 2.12 depicts a typical AFM configuration.

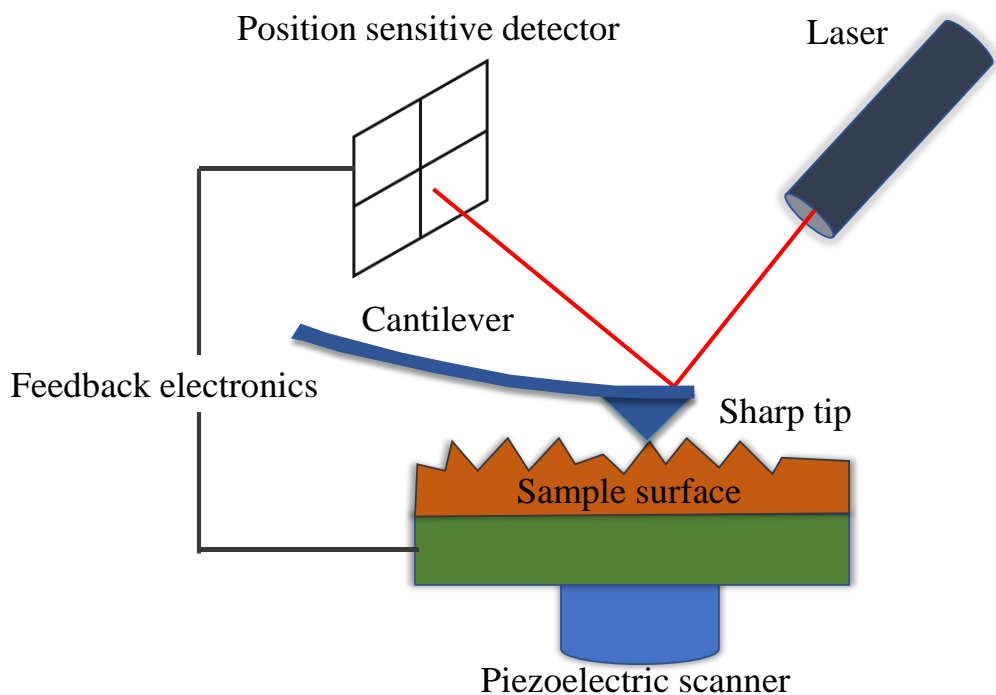

**Figure 2.12:** Schematic of AFM set-up.

## 2.4. A versatile UHV chamber for in-situ study

### 2.4.1. Technical details

During the thesis work, I made extensive use of a versatile in-situ chamber available at the in-situ thin film lab at UGC-DAE CSR, Indore (India), to conduct in-situ studies on thin film characterization. There are two parts to the UHV chamber: the main chamber and the load lock chamber. The sample is loaded and transferred to the manipulator in the main chamber using the load lock chamber. The main chamber has an e-beam evaporation set up for the deposition, MOKE, RHEED, XRR, and four-probe resistivity techniques for characterization and an ion gun for surface patterning/cleaning. The operating principles of these techniques have been thoroughly discussed in the preceding parts of the chapter. A combination of rotary, turbo molecular, and ion pumps achieves a base pressure of $2\times10^{-10}$ mbar in the main chamber and $1\times10^{-8}$ mbar in the load lock.

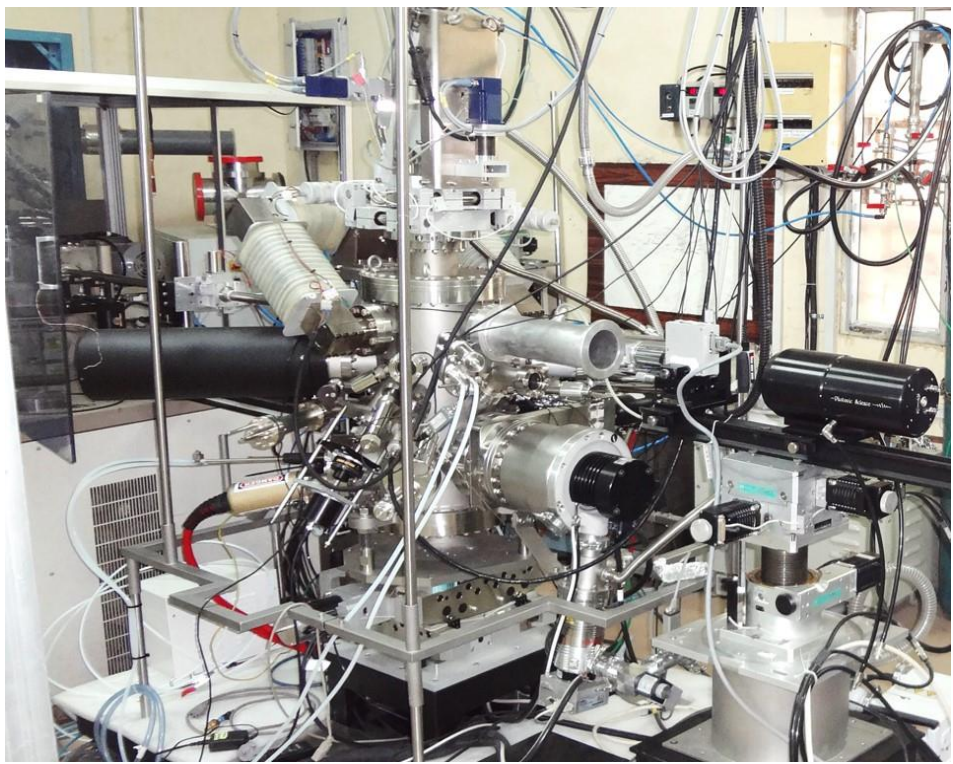

**Figure 2.13:** Actual picture of in-situ UHV set-up.

The real snapshot of the UHV chamber, comprising the deposition and characterization techniques, can be seen in Fig. 2.13, while a cross-sectional view of the chamber can be found in Fig. 2.14. (AutoCAD design). This demonstrates the sophisticated design and operation of various characterization methods, including RHEED, MOKE, ion gun, and others, in a convoluted fashion to carry out an in-situ experiment.

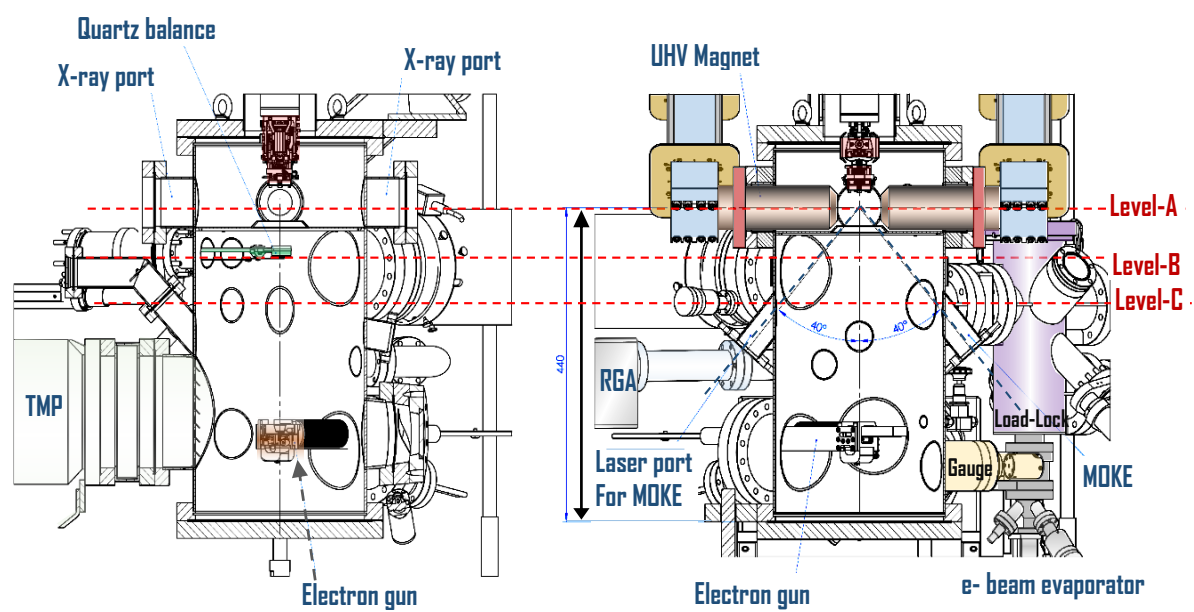

**Figure 2.14:** Cross-sectional view of the chamber.

The technical architecture of the chamber may be understood by splitting it into four levels A, B, C, and D, as illustrated in Fig. 2.14. At Level A, there are four CF 63 flanges that are cantered in a circle. Two of these are used to mount magnetic poles for MOKE measurements, and the other two are used to mount beryllium windows for XRR measurements. There are also two angle ports of CF 35 that are focused on level A and can be used to mount the He-Ne laser and other optical components needed for MOKE measurements. Level B features two connectors for attaching a RHEED source (40 CF) and detector (100 CF). In addition, one 50-degree angle port for the ion gun is directed to this level. MOKE and XRR measurements may be done simultaneously at Level-A, but RHEED measurements and sample cleaning are performed at Level B to minimize electron and ion beam deflection caused by the magnetic field created by the electromagnet during MOKE experiments.

At Level C, a CF 63 port connects the sample transfer chamber to the main chamber. The electron cannon is connected to the main chamber through a CF 100 flange at Level D. On the top 100 CF flange, a 5-axis sample manipulator and a transferrable, specifically constructed sample holder are placed with the sample in a horizontal position. It also has sample heating/cooling (50 K to 1200 K) capabilities, as well as four electrical feedthrough connections to the sample for resistivity and magneto-resistance (MR) studies. An electromagnet with a field capacity of 2500 Oe is connected to magnetic poles for MOKE measurement. The MOKE setup is put together so that MOKE measurements can be made in longitudinal geometry. A fixed lab-based Cu K-α x-ray source with a focused beam is attached to the chamber for the XRR experiment. The whole chamber is positioned on a HUBER stage with tilt motion for scans relative to the source, while the scintillation detector is situated on a combined linear and tilt stage for 2θ scans. The relative positions of these components belonging to distinct measurements are fixed, allowing for the simultaneous execution of several approaches during sample growth.

### 2.4.2. The Benefits of an in-situ Ultra High-Vacuum Chamber

Real-time magnetic, structural, and transport characteristics of films and growth may all be measured in the same chamber without exposing the film to the environment.

2. All measurements may be taken concurrently while and after the film is deposited. This means that the magnetic, transport, and structural properties can be studied as they change with parameters like film thickness, annealing temperature, etc.

3. A sample manipulator with angular rotation capability is appropriate for azimuthal angle-dependent magnetic characteristics (magnetic anisotropy study).

4. Unlike earlier systems in the literature, where measurements are done with different techniques separately, and the results are figured out by putting all the data together, the present set-up is promising for surface and interface analysis of magnetic thin film nanostructures by combining all the techniques as an in-situ measurement.

### 2.4.3. Electron beam evaporation set-up

The electron gun is connected to the bottom side of the main chamber via a CF100 flange, as previously mentioned. A 35 CF port is utilized for electrical high voltage feedthrough of the electron gun on the same port. A 3kW e-gun (TELEMARK Model: No.-528) with four crucibles was utilized to deposit thin films. Figure 2.15 is an illustration of an e-gun with four crucibles. This e-gun is vacuum compatible and has a 1.5 cc material capacity per crucible. The port's neck is shaped in such a way that the active crucible is virtually in the middle of the chamber. For the e-gun, the mechanical shutter is mounted just above the crucibles using push-pull feedthrough. The distance between the e-gun and the substrate holder is around 50 cm, resulting in a homogeneous film thickness. This distance may be adjusted according to our needs by adjusting the manipulator (sample holder) up and down (around 10 cm).

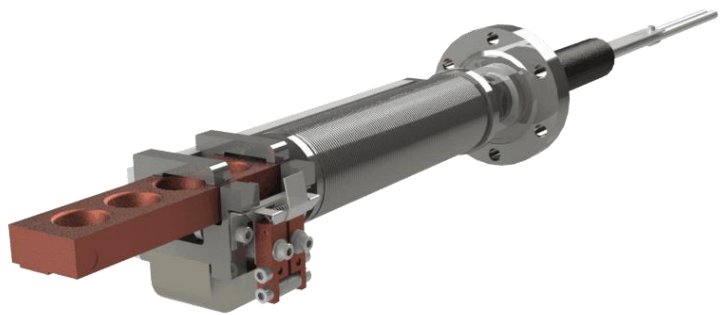

**Figure 2.15:** Image of a Telemark made e-gun set-up installed in the UHV setup used for e-beam evaporation.

## 2.4.4. In-situ MOKE set-up

Figure 2.16 depicts a schematic of the MOKE apparatus associated with the UHV chamber. Its fundamental principle is the same as described in section 2.3.1. An intensity stabilized He-Ne laser ($\lambda$= 632.8 nm), polarizers, PEM, lock in-amplifier, analyzer, a Si photo diode detector, and a computer for interface control make up the MOKE system, which has been adjusted for good sensitivity. The polarized laser beam enters the UHV chamber via a CF 35 port after passing through a PEM. This port is slanted at 45 degrees, so the laser falls at a 45-degree angle on the sample. Inside the chamber, the laser reflects from the magnetized sample placed between the poles of the electromagnet poles and exits through another CF 35, which is similarly oriented at 45 degrees. The electromagnet can generate a field of up to 2500 Oe at the location of the sample in the film plane, i.e., MOKE measurement can only be performed in longitudinal geometry. The electromagnet is supplied with electric current by a Kepco, 6 Amp power supply (MODEL NO: BOP 72-6ML, 48862420). The reflected light is then passed through an analyzer with a polarization axis that is approximately (2° from maximum extinction) crossed with the polarizer axis. As a polarizer and analyzer, two Glan-Taylor prisms with antireflection coating are used. Finally, the photo diode detector detects the reflected light from the sample.

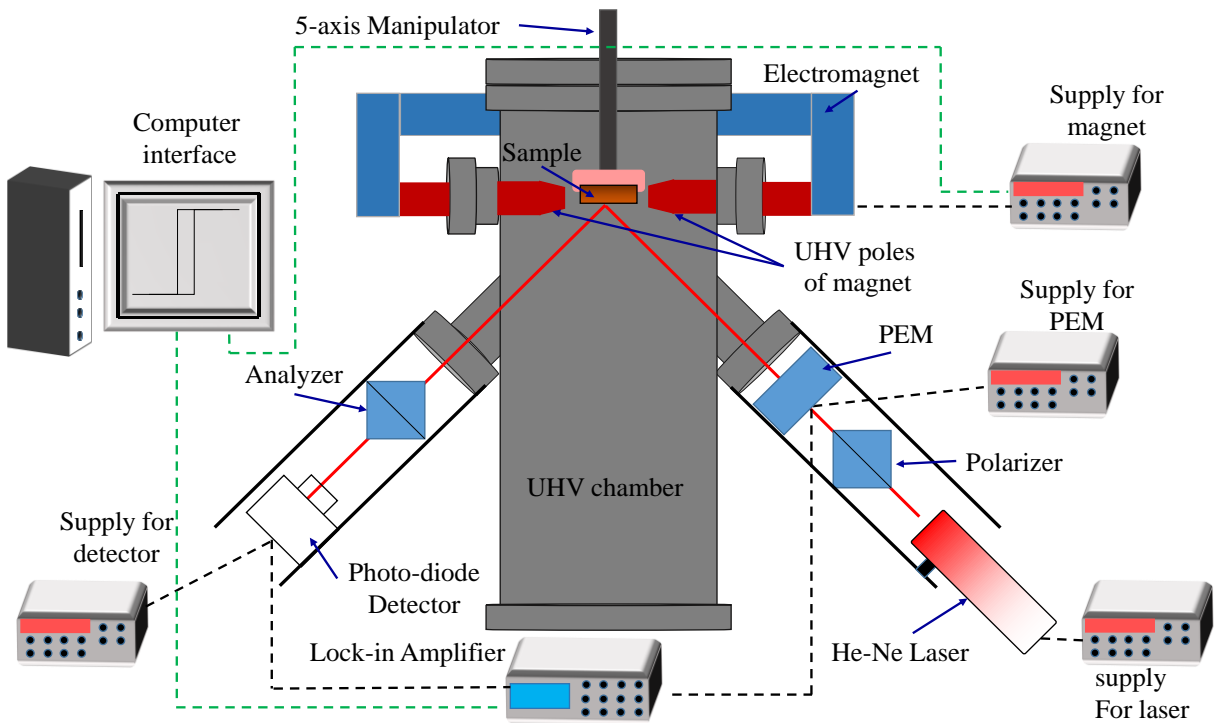

**Figure 2.16:** Schematic of in-situ MOKE set-up.

## 2.4.5. In-situ XRR set-up

XRR components are connected to two 100CF ports at level A in order to conduct XRR measurements concurrently with MOKE (see Fig. 2.14). Figure 2.17(a) shows a schematic of an XRR setup connected directly to the UHV chamber. The Bragg-Brentano geometry is used to do XRR measurements, in which the source is stationary while the sample and detector are in motion to obtain θ-2θ geometry. A collimated x-ray beam which is directed at the sample surface enters into the UHV chamber via a beryllium window. After reflection from the sample surface, it exists through another beryllium window and reaches the detector.

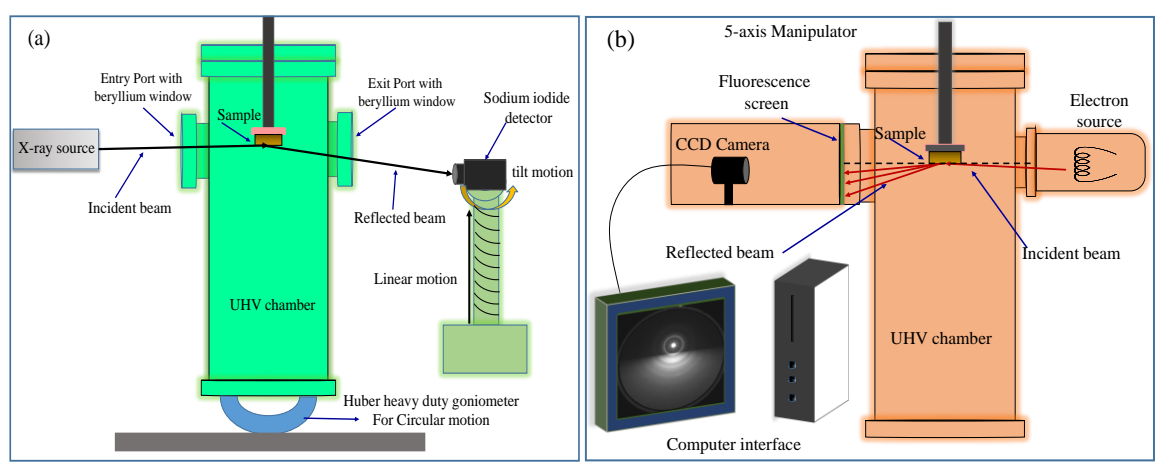

**Figure 2.17:** Schematic of (a) XRR and (b) RHEED set-up attached to the UHV chamber.

## 2.4.6. In-situ RHEED set-up

RHEED components such as electron gun and detector are connected to the UHV chamber at level B along the diametrically opposite end. In order to prevent the deflection of electrons by the magnetic field of the electromagnet positioned at level A, the RHEED components are intentionally attached to the UHV chamber at level B. The electron gun is enclosed in a vacuum case, and a minimum pressure of $1\times 10^{-5}$ mbar is needed to operate the electron gun. STAIB instruments provide the electro gun source with an energy range of 500 eV-30 keV and a maximum operating current of 100 mA. Beam location and incidence angle may be accurately changed by controlling electronic grid voltages. The reflected beam from the sample creates a diffraction pattern on the fluorescence screen, which is captured by the CCD camera and seen on the computer screen in real-time. Figure 2.17(b) is a sketch of how RHEED is set up and connected to the UHV chamber.

## 2.4.7. In-situ ion beam erosion set-up

The ion source IS40C1 attached to the in-situ UHV chamber is a compact, easy-to-use extractor type ion gun for broad area UHV sample surface cleaning and patterning. It is mounted on a DN 40CF flange. The ion source generates an ion current of density >70μA/cm$^2$ (Argon) with a Gaussian beam profile. A IS40-PS power supply with fine adjustment of both beam energy and ion density drives the IS40C1 ion source. Yttrium-oxide coated Iridium filament is used here as a cathode. The working pressure of this ion gun ranges between 10$^{-5}$ to 10$^{-6}$ mbar. The ion source has an option to vary ion energy from 0.12keV to 5keV and electron ionization current ($I_{emis}$) from 0.01mA to 10mA. Furthermore, the source to sample distance can also be varied between ±5cm. The extracted ions are focused onto the sample surface at an angle of 50° from the sample surface normal. However, the sample manipulator stage can change the ion incidence angle on the sample to the extent of ±10°. The beam size varies from 3-20mm depending upon the ion energy and working distance. An inert gas such as Ar is mainly used for operation. However, reactive gases ($O_2$, $H_2$ and hydrocarbon) can be used with a reduced lifetime [115].

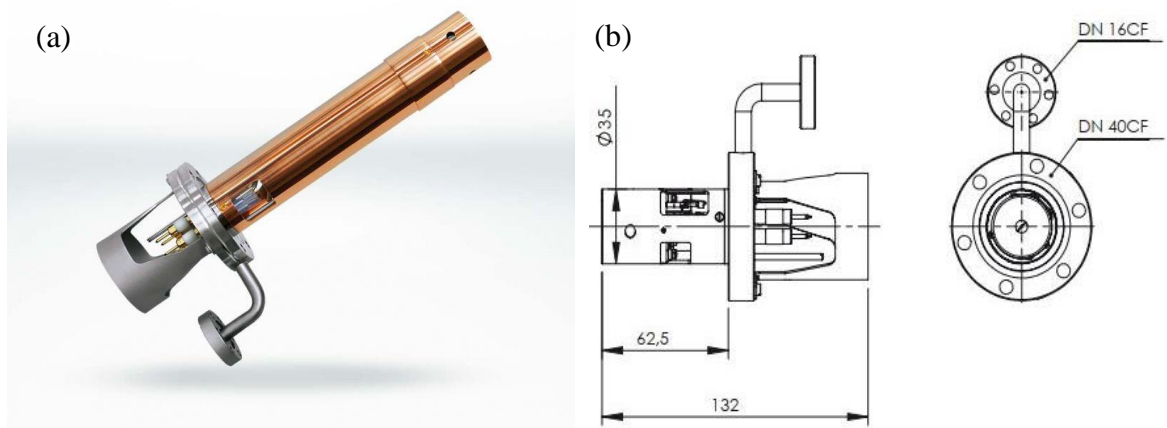

**Figure 2.18:** (a) Digital and (b) AutoCAD image of ion source attached with the UHV chamber.

# Chapter 3:

# Columnar nanostructure on ion beam patterned templates

**Understanding the origin of large magnetic anisotropy in obliquely grown columnar thin films**

The artificial tailoring of magnetic anisotropy by manipulating surface and interface morphology is attracting widespread interest for its application in spintronic and magnetic memory devices. In this chapter, oblique angle deposition on a nanopatterned rippled substrate is presented as a novel route of inducing large in-plane uniaxial magnetic anisotropy (UMA) in magnetic thin films. For this purpose, Cobalt films and rippled $SiO_2$ substrates have been taken as a model system for the present study. Here, nanopatterned substrates are prepared by low energy ion beam erosion (IBE), above which films are deposited obliquely along and normal to the ripple directions. A clear anisotropy in the growth behavior has been observed due to the inhomogeneous in-plane organization of adatoms in the form of columns. The increased shadowing effect in the films deposited obliquely normal to the direction of the ripple patterns causes preferential coalescence of the columns along the substrate ripples, resulting in stronger in-plane UMA in the film. This peculiarity in magnetic behavior is addressed by considering the morphological anisotropy governed by the enhanced shadowing effect, the shape anisotropy and the dipolar interactions among the magneto-statically coupled surface ripple and buried nanocolumnar structure.

# Chapter 3: Columnar nanostructure on ion beam patterned templates.

**Understanding the origin of large magnetic anisotropy in obliquely grown columnar thin films**

**3.1 Introduction**

**3.2 Growth and characterization of Co thin film deposited obliquely on nanopatterned ripple substrate:**

    3.2.1 Preparation and characterization of rippled substrate.

    3.2.2 Oblique angle deposition of Co thin film on the rippled substrate.

    3.2.3 Detailed characterization of film morphology and structure

        3.2.3.1 GISAXS study of film morphology and buried structure.

        3.2.3.2 Surface morphology characterization by FESEM.

        3.2.3.3 Characterization of film crystallographic structure by GIWAXS.

        3.2.3.4 Magnetic anisotropy and domain dynamics studied using MOKE based Kerr microscopy.

**3.3 Oblique and normal deposition of Co film on flat and patterned substrate to compare the strength of magnetic anisotropy**

**3.4 Origin of large UMA in light of shape anisotropy and dipolar interaction.**

**3.5 Conclusions**

## 3.1 Introduction

Interfacial morphology and symmetry breaking effect at the nanoscale strongly influence the magnetic response of a material such as magnetic anisotropy [24], coercivity [116], magnetization reversal process [117], hysteresis [118], magneto-resistance [119]. In particular, magnetic anisotropy that relies on the preferred orientation of the spin alignment [19] of a thin film structure has extensively found its application in the spintronics industry [3], perpendicular spin valves [120], magnetic tunnel junction devices [121] as well as other applications in magnetoelectronics [122], high-density storage media [123], and magnetic sensors [2]. Therefore, tailoring magnetic anisotropy (MA) by modifying the surface and interface morphology is a fascinating field of research. In past decades, mainly multistep lithography-based techniques like optical [124] and electron beam nanolithography [125], X-ray [126] and ion beam lithography [127] have been utilized to fabricate such well-defined nanostructures. However, these methods often encounter numerous shortcomings [13], e.g., low yield, high cost, the need for high temperatures, limitations of fabricating the large-scale surface areas or a restriction to isotropic patterns, which hinders practical applications [128]. Therefore, cheaper, high throughput and time-saving approaches for synthesizing nano-dimensional morphological features with well-controlled size and shape distribution are thus extensively searched. Within this context, one practical, promising approach to overcome these limitations consists of using a self-assembly procedure that can be achieved by oblique angle deposition (OAD) [129] and ion beam erosion (IBE) [14],[130] methods. Both methods have the capability of patterning surface morphology and modifying physical properties, which has emerged as a promising bottom-up approach in recent decades.

In oblique angle deposition, columnar structures or elongated grains are formed due to the shadowing effect and limited diffusion of adatoms [131][132]. This morphological symmetry breaking phenomenon induces uniaxial magnetic anisotropy (UMA) [7],[133]. However, it lags from creating lateral periodic ordered arrays on bare flat substrates owing to random nucleation of the deposited material [134]. This must be avoided in some applications like three-dimensional photonic crystals or surface-enhanced spectroscopy. Similarly, the induced UMA is of lower strength due to the absence of long-range ordering in morphology. Moreover, its strength decreases with the increment of film thickness due to exchange interaction and merging of columns. On the other hand, IBE has proved its applicability in the formation of well-defined periodic surface nanoripples to a wide range of crystalline or amorphous materials. The ripple pattern formation is well understood in terms of a balance between two competitive counteracting processes induced by curvature-dependent preferential IBE that tries to roughen the surface and thermally activated smoothening mechanisms like surface diffusion or viscous flow of material [135][136]. The film deposited on rippled substrates exhibits strong magnetic anisotropy [137],[138][139], which is driven by morphological anisotropy. Direct nanoscale ripple patterning can also induce in-plane UMA in epitaxial [140] and polycrystalline [119],[19] thin films. For the film deposited over the rippled substrate, the saturation in the strength of UMA with higher film coverage due to the merging of ripples with their nearest neighbours is still an open issue [23,24,137,141]. In addition, when IBE is applied to a magnetic film to create correlated surface morphology, the initial film thickness must be high as it only modifies the top surface. Therefore, new and novel routes with the in-depth realization of the process involved in nanoscale control of magnetic properties are highly desired. One promising approach would be the combination of nanostructuring with the self-assembled process, so-called templated self-assembly. As the topographic condition strongly influences thin film's growth, oblique incidence defocused broad ion beam surface patterning potentially offers an easy and inexpensive solution to prepattern the surface, which can be used as a template for the overgrowth structures [102,142,143]. Several discrete studies on morphology and magnetic anisotropy of the two distinguished systems, i.e., obliquely grown film on bare substrate and film deposited on the rippled patterned substrate, have been investigated thoroughly for different systems [2],[119],[7,19,138,140]. Few works on coupled oblique-ripple systems have been done in the recent past, emphasizing morphology or optical property [53,102,144,145]. Despite the interest in the above-mentioned studies in many systems, similar studies on magnetic systems have been neglected mostly. This may be owing to the complicated nature of the problem because the surface/interface morphology, film thickness and crystallinity, buried nanocolumnar structure, magnetic domain configuration, the interplay of surface and magnetostatic interaction etc., actively participates in deciding the magnetic energy of the system. Furthermore, in earlier works [23],[116],[138],[141], on the normal deposition of magnetic systems, such work was performed on the rippled substrate having an amplitude of hardly 1-2 nm and wavelength less than 50 nm, so could not take advantage of shadowing, like done in the current work with a higher ripple amplitude. To our knowledge, no reports have been published on a combined oblique-rippled magnetic system where microscopic domain dynamics and detailed microstructural characterization by utilizing surface and buried interface sensitive X-ray scattering techniques are involved.

In this study, both the methods have been combined in an intriguing way such that a stronger in-plane magnetic anisotropy, as compared to either of the individual methods, has been obtained in polycrystalline Co film. Samples have been prepared by subsequent deposition of Co film at oblique incidence on pre-patterned $SiO_2$ substrate fabricated by IBE. Under this condition, inhomogeneous nucleation due to self-shadowing and self-masking effect induced by the height of wrinkled surface topography is used to control the morphology and magnetic anisotropy in the film. In contrast to the conventional characterization techniques where patterned film structures were probed by local direct imaging techniques like atomic force microscope (AFM), field emission scanning electron microscope (FESEM) restricted over $1\mu m^2$ area or even less; the present study emphasizes nano structural analysis by synchrotron-based grazing-incidence small-angle X-ray scattering technique which is capable of extracting information of the surface, sub-surface region and even from the buried interface with good spatial resolution, high angular resolution and statistical accuracy. In addition, we demonstrate that the films exhibit UMA of different strengths interpreted as the consequence of three contributions: in-plane self-organization of Co over layers controlled by shadowing effect, shape anisotropy and dipolar interaction between the rippled surface topography and buried nanocolumnar structure.

## 3.2 Growth and characterization of Co thin film deposited obliquely on nanopatterned ripple substrate

### 3.2.1 Preparation and characterization of the rippled substrate.

A commercially available amorphous $SiO_2$ wafer has been used as a template for the formation of ripple patterns. Before ion irradiation, an arrow was marked on the substrate surface to be considered for reference direction of ion beam irradiation. Irradiation on the surface of the substrate was performed with a defocused 700 eV Ar+ ions having flux ~$1.3\times 10^{14}$ ions/$cm^2$.s by an electron cyclotron resonance-based broad-beam ion-source (GEN-II, Tectra GmbH, Germany) at an incidence angle of 65º with respect to the surface normal for 20 minutes. AFM measurements were performed in the tapping mode in a commercial AFM instrument from Agilent Technologies 5600LS to study the topography of the substrate and to obtain the orderliness of the nanopatterns on the surface. Before the magnetic thin film deposition, the topography of the patterned substrate is characterized in detail using AFM and GISAXS measurements. The AFM image of the surface topography of the substrate is presented in Fig. 3.1a. Clear traces of periodic ripple pattern-oriented perpendicular to the projection of ion beam on the substrate surface is observed. As shown in the inset of Figure 3.1a, the autocorrelation image has been extracted from the AFM image to measure the morphological ordering quantitatively [146]. This image is periodic and exhibits a remarkable degree of coherence. The line profile of the autocorrelation function and AFM image along the direction of the ripple wave vector is presented in Figure 3.1b. The ripple wavelength (λ) determined from this image is 67±3 nm. The average height of the rippled pattern substrate is 3.3 nm which will support and enhance the shadowing effect on the deposited overlayer.

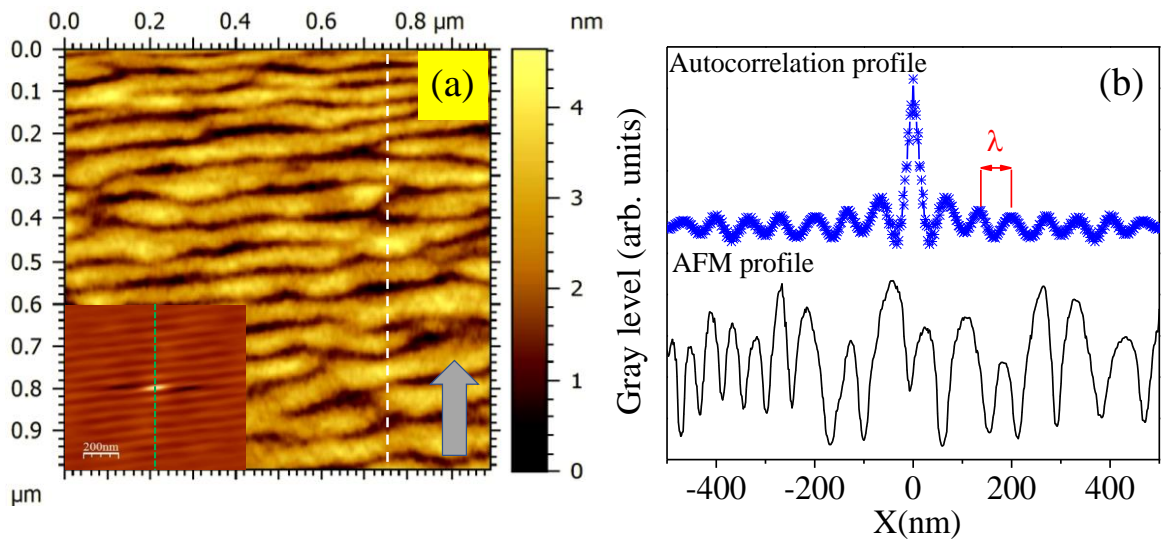

**Figure 3.1:** (a) AFM image of SiO$_2$ substrate after IBE. The arrow indicates the projection of the ion beam on the substrate surface. The inset of figure (a) is the corresponding two-dimensional auto correlation function. (b) Line profile of autocorrelation function and AFM image taken parallel to the ripple wave vector direction as pointed out by dashed lines.

Figure 3.2a shows the GISAXS spectra of the rippled SiO$_2$ substrate with X-ray propagation direction parallel to the length of the ripple. The image shows that streaks extending in the q$_z$ direction at different q$_y$ values are present. Here q$_z$ and q$_y$ is the scattering vector along the z and y-direction, respectively. Such GISAXS image is mainly due to the periodic lateral variation of the electron density; therefore, it provides information on the lateral height correlation of the surface. Figure 3.2b gives the corresponding intensity profile extracted from the GISAXS image in Fig. 3.2a by taking a stripe along the horizontal direction for a fixed q$_z$ value corresponding to the Yoneda region [147]. The high intense specular rod at q$_y$=0 nm$^{-1}$ and two side satellite peaks at q$_y$=±0.092 nm$^{-1}$ in Fig. 3.2b arise due to the periodicity of the ripples. Ripple wavelength ($\lambda$)= $2\pi/q_y$; is found to be 68nm. This value of the ripple wavelength is very close to the value obtained from AFM measurements.

It may be noted that the left satellite peak has a higher intensity than the right side of the satellite peak. This asymmetry in the satellite peaks originates due to the asymmetric shape of the ripples caused by different local sputtering rates on the front and rare side of ripples [144],[148],[149]. The lower intensity satellite peak corresponds to the steeper slope of ripples [144]. The slopes of the ripple pattern have been calculated using the methodology developed by Babonneau *et al.* [144] and found to be 14.5° and 17°.

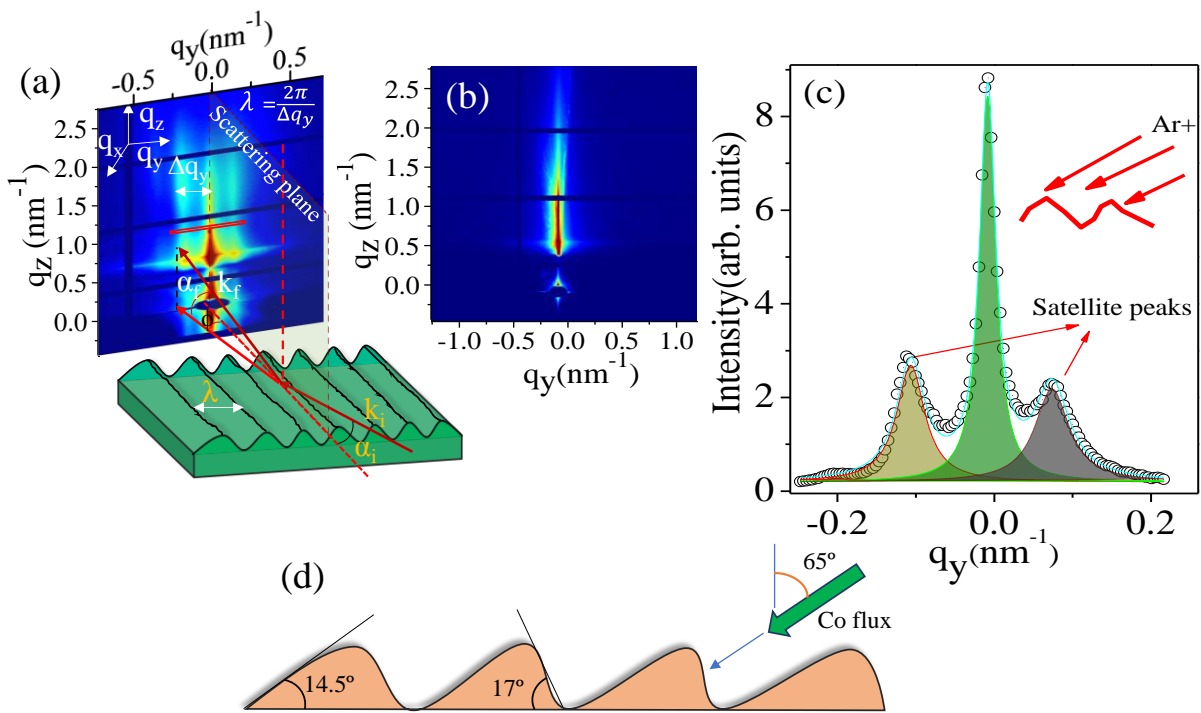

**Figure 3.2:** The GISAXS spectra of the rippled SiO$_2$ substrate with X-ray beam aligned (a) parallel and (b) perpendicular to the ripple direction. The schematic illustration of the scattering geometry used in GISAXS measurements is also shown simultaneously. (c) The corresponding in-plane 1D line profile extracted from the red-marked rectangular portion in (a). (d) Schematic representation of asymmetric ripple profile used in this study.

### 3.2.2 Oblique angle deposition of Co thin film on the rippled substrate.

These rippled substrates were used as a template to deposit Co thin film. Three portions of the same rippled substrates were mounted onto a sample manipulator inside an UHV chamber having a base pressure of ~10$^{-10}$ mbar. Among them, two substrates are positioned in such a way that their surface normal makes an angle 65º from Co atoms' flux arrival direction, in one case parallel (along $\phi=0°$, denoted as $OAD_{0°}$) and in another case perpendicular (along $\phi=90°$, denoted as $OAD_{90°}$) to the ripple direction. $\phi$ is the angle in the azimuthal direction with respect to the path of the ripple length. The third rippled substrate was placed with its surface normal parallel to Co atoms' flux arrival direction ($RIP_{norm}$). Furthermore, Co film was deposited separately at the same oblique angle as that of rippled substrates on a bare Si substrate (without surface cleaning treatment) in order to draw a comparison on the magnetic properties ($OAD_{flat}$). Co film of nominal thickness 50nm as monitored in situ by calibrated quartz crystal monitor was deposited onto the substrates simultaneously using e-beam evaporation at room temperature (RT) with a deposition rate of $\approx 0.2$Å/s. A schematic of deposition geometry is shown in Fig. 3.3.

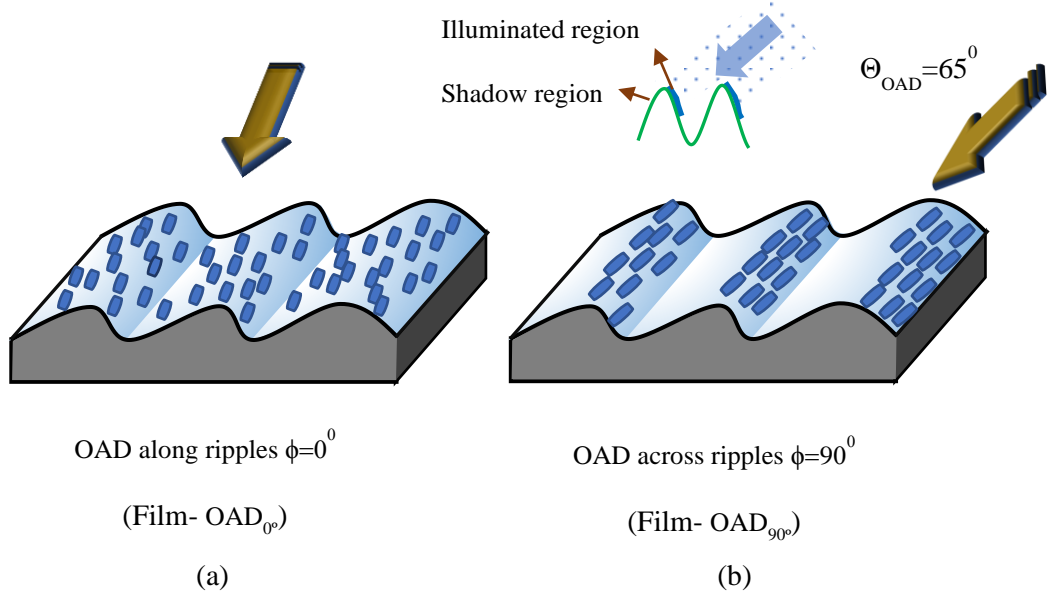

**Figure 3.3:** Schematic diagram of the Co deposition direction with respect to the surface normal for the film deposited (a) along and (b) across the ripples. Samples (a) and (b) are denoted as $OAD_{0°}$ and $OAD_{90°}$, respectively.

### 3.2.3 Detailed characterization of film morphology and structure.

For film thickness, surface morphology, buried film/substrate interface and structural characterization, X-ray reflectivity (XRR), GISAXS and GIWAXS measurements were performed simultaneously. The schematic of the X-ray scattering setup is shown in Fig. 3.4a. Here $k_i$ and $k_f$ denote the incident and reflected wave vectors, $\alpha_i$ is the X-ray incidence angle while 2θ and $α_f$ is in-plane and out-of-plane scattering angle. The non-specular intensity away from the specular rod contains information of the film structure. Horizontal cuts taken at the Yoneda peak region along $q_y$ direction provide information of the lateral structures of the material parallel to the film surface, while the off specular vertical cut along $q_z$ offers information about the structure perpendicular to the substrate [150].

### 3.2.3.1 XRR and GISAXS study of film morphology and buried structure.

XRR measurements are done using 2D PILATUS 1M detector. The XRR data was obtained by changing the incidence angle from 0.038° to 1.87° while integrating one single scattering pattern. This procedure was done at two different detector positions to avoid shadowing from any intermodular detector gaps [150]. Figure 3.4(b,c) shows the 2D XRR images of the OAD$_{0°}$ and OAD$_{90°}$ film, respectively. The 1D line profile, which gives exit angle *versus* specularly reflected X-ray intensity, is extracted from the respective central vertical streak present at $q_y = 0$ nm$^{-1}$ and plotted in Fig. 3.4d. It exhibits well-defined Kiessig fringes due to the interference between the X-ray waves scattered by the film/substrate interface and the film surface. The film thickness is extracted by fitting XRR data using Parratt's recursive equations [151] and plotted as a continuous line in the same graph. The observed thickness ($d^{OAD}_{\phi_0}$ and $d^{OAD}_{\phi_{90}}$) and rms roughness ($\sigma^{OAD}_{\phi_0}$ and $\sigma^{OAD}_{\phi_{90}}$) of OAD$_{0°}$ and OAD$_{90°}$ films are found $d^{OAD}_{\phi_{90}} = 36$nm, $d^{OAD}_{\phi_0} = 33.2$nm, $\sigma^{OAD}_{\phi_{90}} = 2$nm, and $\sigma^{OAD}_{\phi_0} = 1.5$nm, respectively. From the XRR fitting, it is found that the electron density decrease from the value ≈2.244Å$^{-3}$ for OAD$_{0°}$ sample to the value ≈2.067Å$^{-3}$ for OAD$_{90°}$ sample, which corresponds to nearly 8% reduction in electron density of the layer. This indicates the formation of more porous film in OAD$_{90°}$ sample due to enhanced shadowing by the ripples. It may be noted that, although the same effective thickness was grown identically on both the substrates using pre-calibrated thickness monitor but 8% increase in the thickness for OAD$_{90°}$ sample is mainly due to the increase of porosity in the film associated with the enhanced shadowing deposition of OAD$_{90°}$ sample.

GISAXS measurements are performed for samples OAD$_{0°}$ and OAD$_{90°}$ by keeping the X-ray beam at a grazing-incidence angle of 0.4° along (in-plane azimuthal angle ϕ=0°) and perpendicular (in-plane azimuthal angle ϕ=90°) to the length of ripple. GISAXS 2D spectra corresponding to OAD$_{90°}$ samples along both the ϕ directions are shown in Fig. 3.5(a,b).

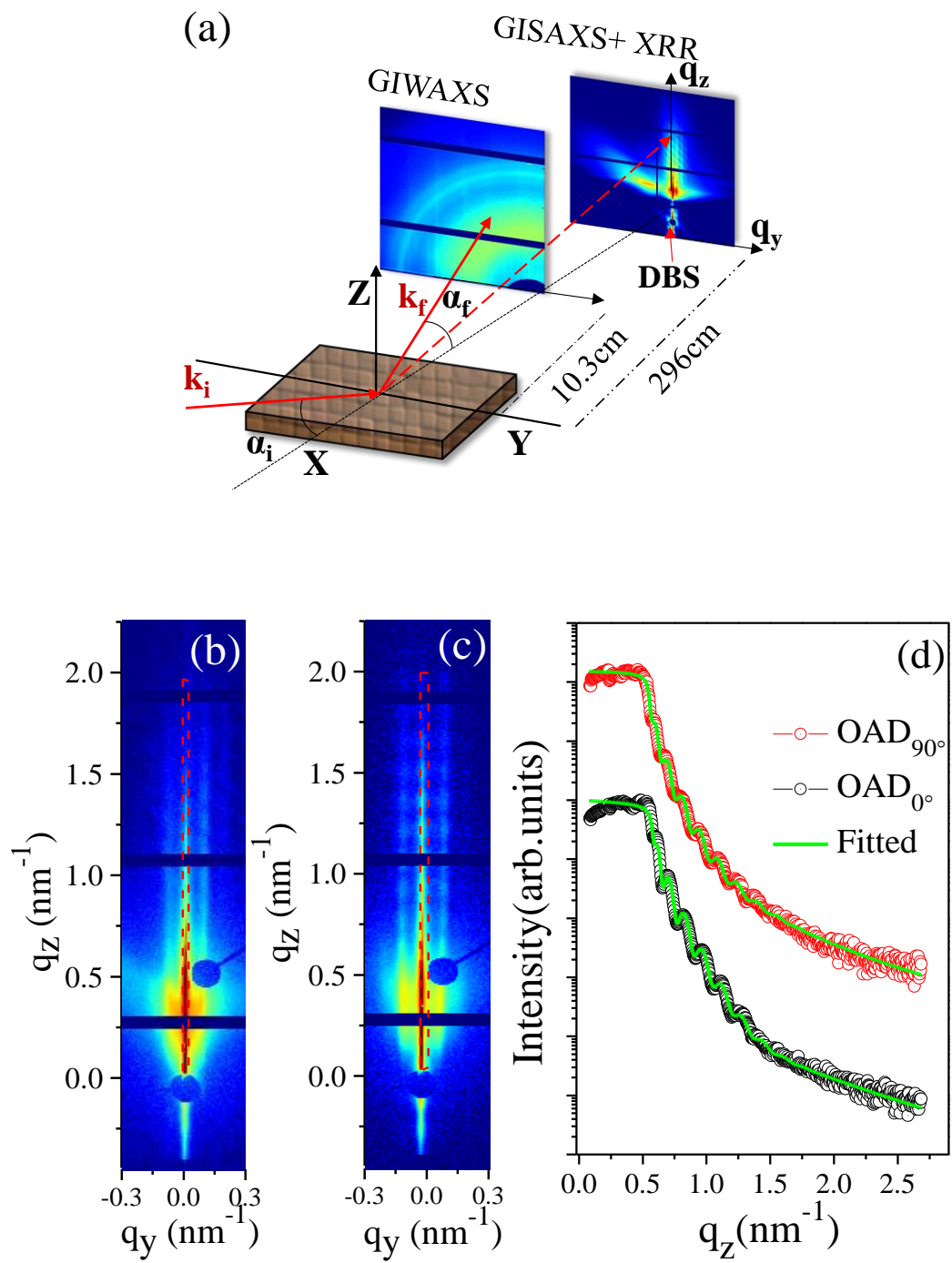

**Figure 3.4:** (a) Schematic of the X-ray scattering set-up for XRR, GISAXS and GIWAXS measurements. 2D XRR patterns of (b) $OAD_{90°}$ and (c) $OAD_{0°}$ film are presented, respectively. (d) 1D intensity line profile extracted along $q_z$ direction as guided by the dashed red line. The hollow circles represent the experimental data, and green lines represent the fitted data using Parratt formalism.

Drastic differences in both images, as listed below, are mainly due to the anisotropic morphology of the sample along two azimuthal directions.

i) Along $\phi=0°$, asymmetric distribution of the scattered intensity on the left side of the specular rod ($q_y= 0$ nm$^{-1}$) is observed. This asymmetry is absent in the case of measurements across ($\phi=90°$) ripples.

ii) Compared to measurements along $\phi=90°$ (Fig. 3.5b), several satellite peaks are visible on two sides of the specular rod in Fig. 3.5a.

iii) Several inclined fringes at a certain angle appear in the specular rod and satellite peaks in the Fig. (3.5a), whereas these fringes are horizontal in the specular rod only in Fig. 3.5b.

To understand the data more clearly and correlate the same with sample morphology, intensity (I) versus in-plane momentum transfer vector ($q_y$) profiles are extracted around the Yoneda wings from both images (as marked in Figure 3.5a) and plotted in Fig. 3.5c. Scattered broad diffused peak at around $q_y= -0.42$nm$^{-1}$ indicating the presence of inclined laterally ordered nanostructure arrangement in Co film. The lateral spacing of the nanostructure, as calculated, is found to be 14.8 nm. Observation of the broad peak only along -$q_y$ side confirms the formation of inclined columnar nanostructures [100,152–155]. The angle ($\beta$) of tilted columnar morphology with respect to substrate normal is obtained based on the slope of intensity elongation I (Figure 3.5c) with respect to the vertical plane. It is found to be $\beta=58°$ for the deposition angle $\alpha=65°$ from the surface normal. A dashed inclined red line guided to the eye for the inclined fringes in Fig. 3.5a. It resembles the inclined scattering rods typical from the inclined faceted surface of nanocolumns [154]. Estimation of the inclination of the nanocolumns, as obtained based on intensity distribution inclination, is 52° from the surface normal. It closely matches the earlier reported $\beta$ value (58°) and gives a more accurate value than obtained using the broad diffused intensity feature present along the negative $q_y$ side. Satellite peaks at around $q_y= +0.0772$ and $-0.0733$ nm$^{-1}$, indicating the presence of laterally ordered nanostructure due to the periodic ripple arrangement, which corresponds to the lateral spacing $\approx 84$nm. It is important to note that the relative intensities of the satellite peak along -$q_y$ and +$q_y$ side are reversed after Co deposition (compared with substrate GISAXS in Fig. 3.2a). i.e., the peak intensity on the positive $q_y$ side becomes higher than the one on the negative $q_y$ side. It indicates that the Co atoms nucleate preferentially due to geometrical shadowing on the facets of the rippled surface illuminated during the Co deposition, resulting in a reduced slope on the side of deposition [102],[128].

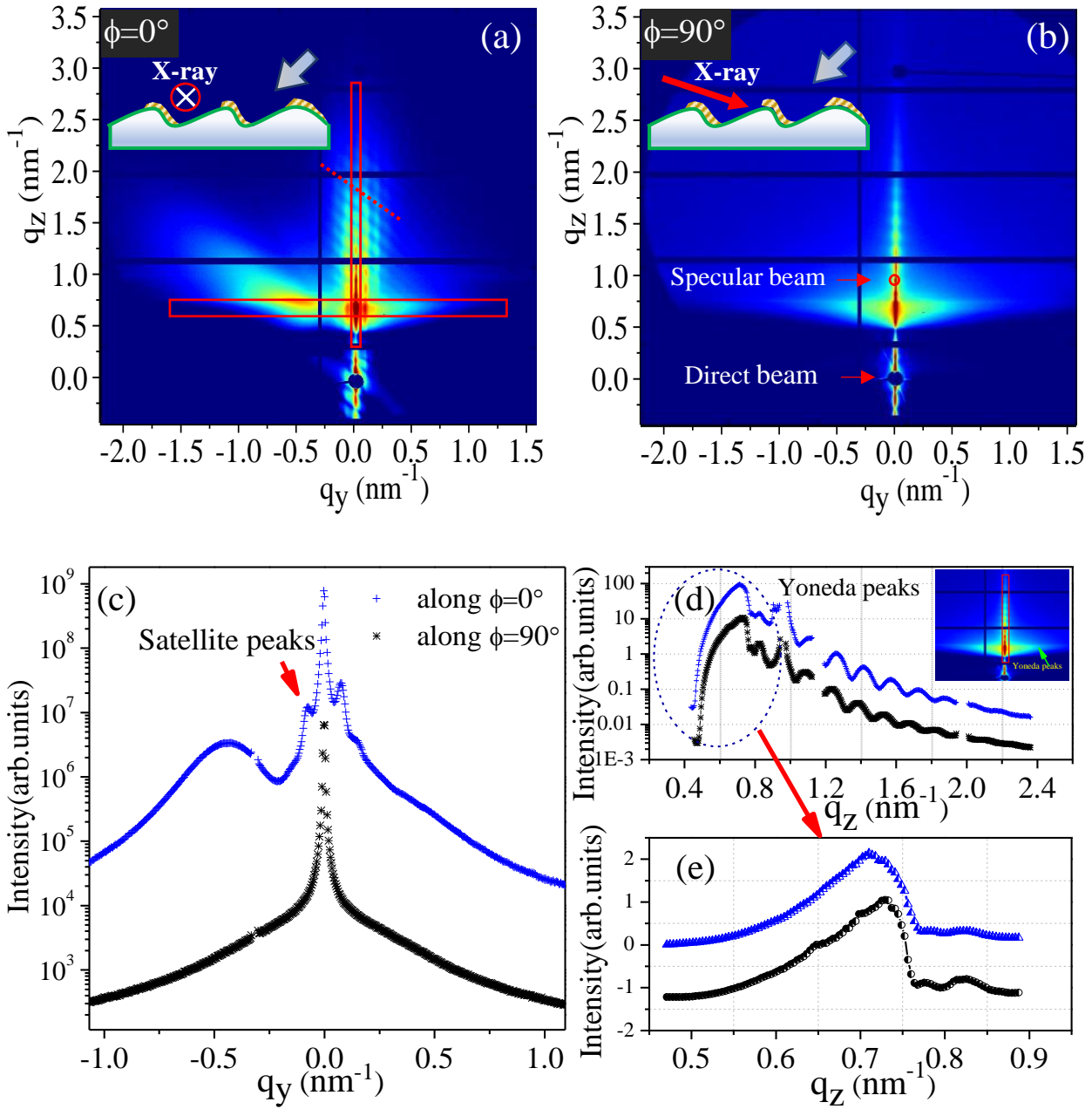

**Figure 3.5:** GISAXS spectra of the OAD$_{90°}$ sample taken (a) along ($\phi=0°$) and (b) across ($\phi=90°$) to the ripple pattern. (c)1-D in-plane intensity profile extracted around Yoneda peak from (a) blue line and from (b) black line. (d) The corresponding $q_z$ line scan at $q_y=0$ nm$^{-1}$. (e) The zoomed portion of (d) around the Yoneda region as marked by an oval dashed line.

In Fig. 3.5d, 1-D line cut taken at $q_y=0$nm$^{-1}$ along $q_z$ direction is plotted for both the images. Intensity oscillations similar to Kiessig-like fringes [23],[144] are arises in both the plots due to constructive interferences of X-ray scattered from film and film-substrate interface. As expected, the oscillation periods, which are linked with the film thickness ($d_{Co}= 2\pi/\Delta q_z$, where $\Delta q_z$ is the average period of oscillation), are almost the same in both directions. On the other hand, the position of the Yoneda peak, which is the measure of electron density, appears slightly at a higher angle ($\alpha_f=0.23°$, $q_z=0.73$ nm$^{-1}$) along $\phi=90°$ as compared to its orthogonal direction ($\phi=0°$, $\alpha_f=0.21°$ $q_z=0.71$ nm$^{-1}$). It means the formation of a higher electron density distribution of nanocolumnar structure takes place along the Co flux arrival direction. Moreover, as per the earlier studies for OAD on planner templates, a significant shadowing effect is responsible for this asymmetry in the electron density along the azimuthal direction [100].

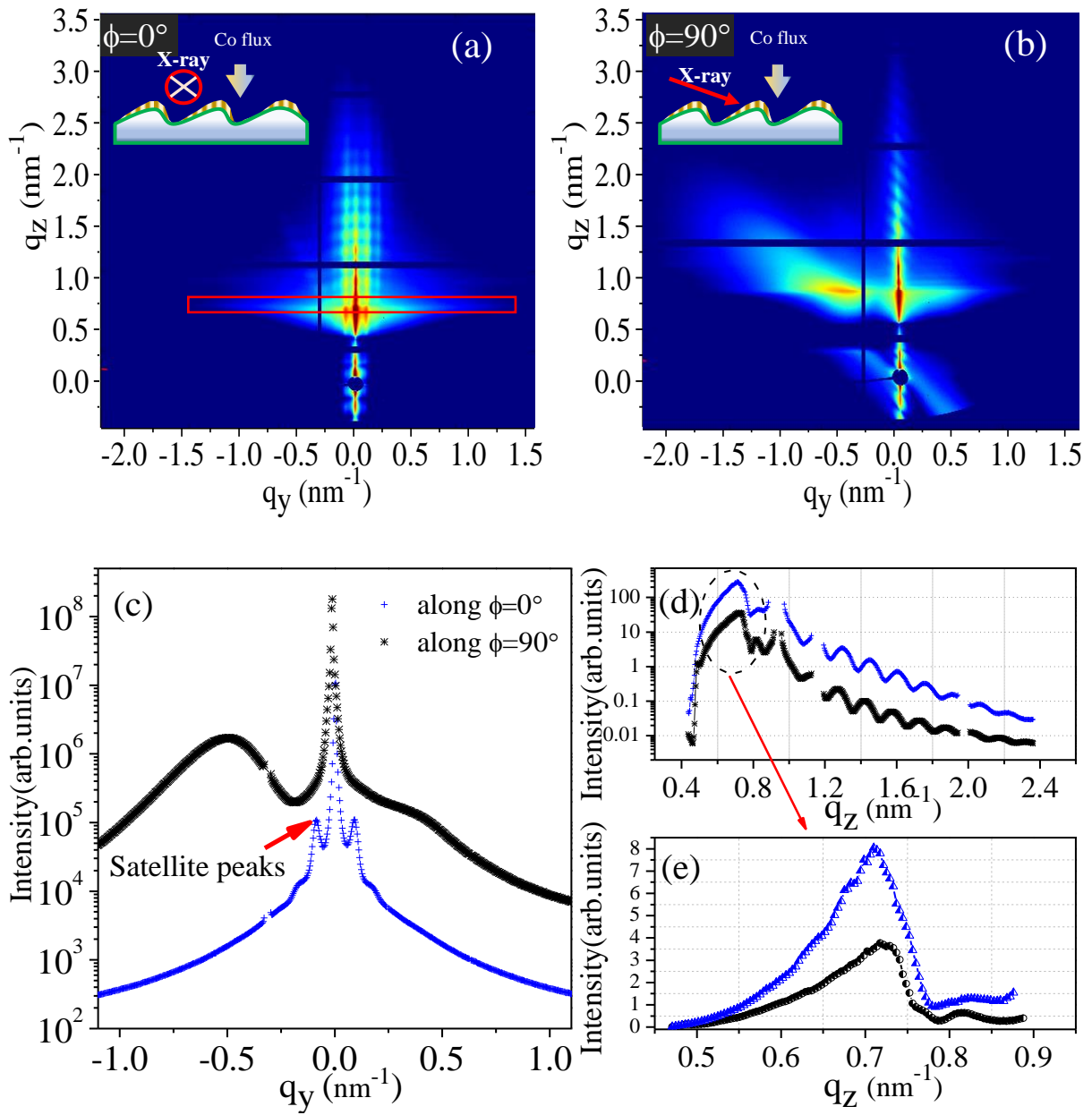

**Figure 3.6:** GISAXS spectra of the sample OAD$_{0°}$ taken (a) along ($\phi=0°$) and (b) across ($\phi=90°$) to the ripple pattern. (c) 1-D in-plane intensity profile extracted around Yoneda peak from (a) and (b)

for both orientations, $\phi = 0°$ (blue line) and $\phi = 90°$ (black line). (d) The corresponding $q_z$ line scan at $q_y=0$ nm$^{-1}$. (e) The zoomed portion of the marked region of (d) around Yoneda peak.

Figure 3.6 (a,b) gives GISAXS 2D images collected for OAD$_{0°}$ sample (Co deposition along the length of the ripples) by keeping the X-ray beam along the length of the ripple ($\phi=0°$) and perpendicular ($\phi=90°$) to it, respectively. A completely different scenario compared to OAD$_{90°}$ sample appears for this case. Here the scattered intensity distribution is symmetric along the ripple direction. In contrast, inclined broad peak intensity is observed at 90° to ripple direction. It suggests that the formation of the nanocolumnar structure takes along the length of the ripples. The angle of inclination of the columns, as obtained from Fig. 3.6b, is ~43°, whereas the average separation between nanocolumns is found to be 12.7nm. I *versus* $q_y$ profiles also extracted from images 3.6(a,b) are plotted in Fig. 3.6c. In contrast, to sample OAD$_{90°}$, the equal intensity of satellite peaks is mainly due to a relatively less shadowing effect. Similar to the OAD$_{90°}$ sample, the slight difference in the Yoneda peak, as shown in Fig. 3.6e, reveals the electron densities differences along $\phi=0°$ and $\phi=90°$ directions. It is interesting to note that the average ripple wavelength value is 71nm, which is significantly lower than 84nm in the case of the OAD$_{90°}$ sample. The origin of the same is discussed later in this chapter

**3.2.3.2 Characterization of film crystallographic structure by GIWAXS.**

Figure 3.7(a,b) gives GIWAXS 2D diffraction images, collected for both samples by keeping X-ray beam at grazing-incident angle ~0.4° along the length of the ripple ($\phi=0°$). It may be noted that due to the geometric constraints and also shadowed by the substrate, only one-fourth parts of the diffraction pattern are visible. Three uniform arc patterns which correspond to (100), (002), and (101) diffraction planes of hcp Co-phase are identified. Arcs with almost uniform intensity distribution suggest that the crystallites are randomly oriented, i.e., polycrystalline nature of the film. The broad halo arc between $q_r \approx 12$ nm$^{-1}$ to $\approx 24$ nm$^{-1}$ is related to the presence of diffusely scattered radiation due to the amorphous substrate [104]. Similar features have been observed for OAD$_{0°}$ sample. Overall integrated intensities of the concentric arcs from both images ranging from $q_r=26.55$nm$^{-1}$ to $q_r=36.15$nm$^{-1}$ are obtained using DPDAK software and plotted in Fig. 3.7(c,d), respectively. Except sharpening of the rings in the case of the OAD$_{0°}$ sample, no significant change in the ring pattern has been observed. The sharpness of the peaks may be related to the elongated columnar microstructure, which is created by the oblique angle deposition. The strongest intensity of the (002) peak indicates that it is favourably grown compared to the other orientated crystallites. The hcp

(100), (002) and (101) peaks of Co were well fitted with the help of Gaussian shape and lattice constant, crystalline grain size corresponding to the peaks were obtained using the Scherrer equation [156].

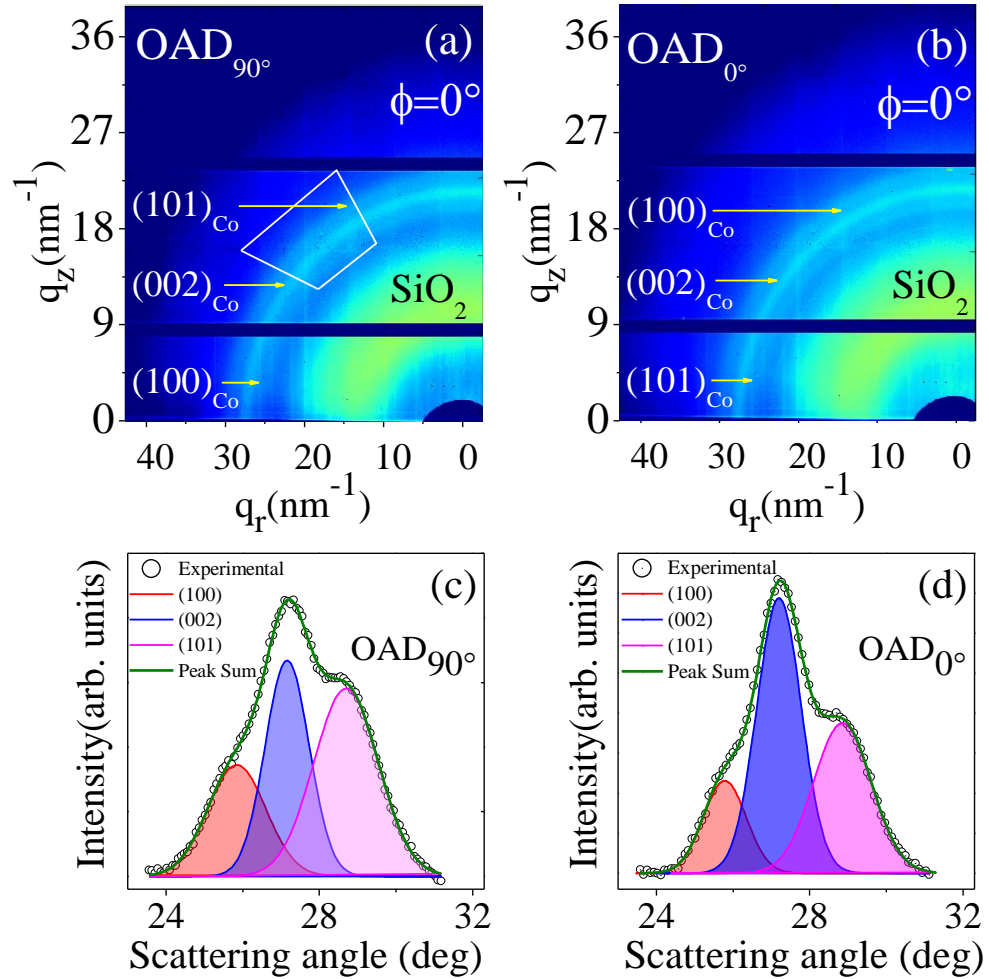

**Figure 3.7:** 2D GIWAXS images of the sample (a) $OAD_{90°}$ and (b) $OAD_{0°}$ taken along ripple direction. Crystal peaks corresponding to (100), (002) and (101) spacings are evident. (c) and (d) represent the corresponding radially averaged integrated intensity distribution of the marked white section of the sample $OAD_{90°}$ and $OAD_{0°}$, respectively.

The lattice constant deduced from the obtained Bragg peak positions is a=2.52Å and c=4.06Å for $OAD_{90°}$ geometry which is close to the standard bulk values $a_{bulk}$ =2.51Å and $c_{bulk}$ =4.07Å, respectively. However, 0.24% elongation in a value and 0.45% contraction in the c value indicates some tetragonal distortion present in the crystal structure, which may be due to random stress developed in the film. OAD films are generally found to show stress lattices due to the porous nature of the films [157], and hence, minor stress in Co could be correlated to OAD. The average crystallite size of (100), (002) and (101) oriented Co crystallites are found to be 30.52Å, 40.4Å, 27.54Å for OAD$_{90°}$ and 41.32Å, 40.04Å and 30.81Å for OAD$_{0°}$ sample, respectively. This asymmetric crystallite size may be due to the preferred growth of crystallites in distinct crystalline directions.

**3.2.3.3 Surface morphology characterization by FESEM**

The film surface morphology was imaged by FESEM (FEI Nova Nano SEM) measurement, and image processing was performed using Gwyddion software [158]. Figure 3.8a and b give FESEM images of the surface topography of OAD$_{90°}$ and OAD$_{0°}$ samples, respectively. The images display the formation of periodic ripple-like structures in Co film with grooves running parallel to the substrate ripple direction. The Co atoms' flux arrival direction is marked by the white arrow in the images. A quantitative measure of the surface modulation coherence is obtained from the 2D-Fourier transformation calculated by Gwyddion software [158] and shown at the bottom inset in this figure. It clearly displays two spots in k space along the ripple direction whose intermediate separation is equivalent to an average surface undulation of ~ 160nm and 77nm for samples OAD$_{90°}$ and OAD$_{0°}$, respectively. In the top view micrograph in Fig. 3.8(a,b), it is possible to appreciate a certain lateral association of nanocolumns. This feature is common in this type of OAD system (lines guided to the eyes are added to highlight this nanocolumns association) [100],[159]. The surface morphology characteristics extracted from the line profile perpendicular to the ripple direction are plotted in Fig. 3.8(c,d). The observed difference in the wavelength of the morphology for the two samples can be understood as follows. In an ion beam patterned surface, fluctuations exist both in wavelength and height of the morphology. Besides this, when materials are deposited perpendicular to the ripple direction, it faces an enhanced shadowing effect induced by the height of the ripples. Therefore, materials are preferably deposited on top of the hilly regions of ripple that are tall enough to serve as nucleation sites for the subsequent columnar development, whereas the shorter ones are masked from receiving material. As the growth proceeds, the aggregation of columns grown over taller mounds suppress the growth of the shorter neighbouring ones. This introduces a large tilted structure having a wavelength very different from the substrate. However, when materials get deposited parallel to the ripple, this shadowing effect is missing. Therefore, the grown structure follows the same substrate morphology. Thus, we have observed that in the OAD$_{90°}$ sample, the periodicity of the surface morphology is about twice that of the OAD$_{0°}$ sample as well as the substrate.

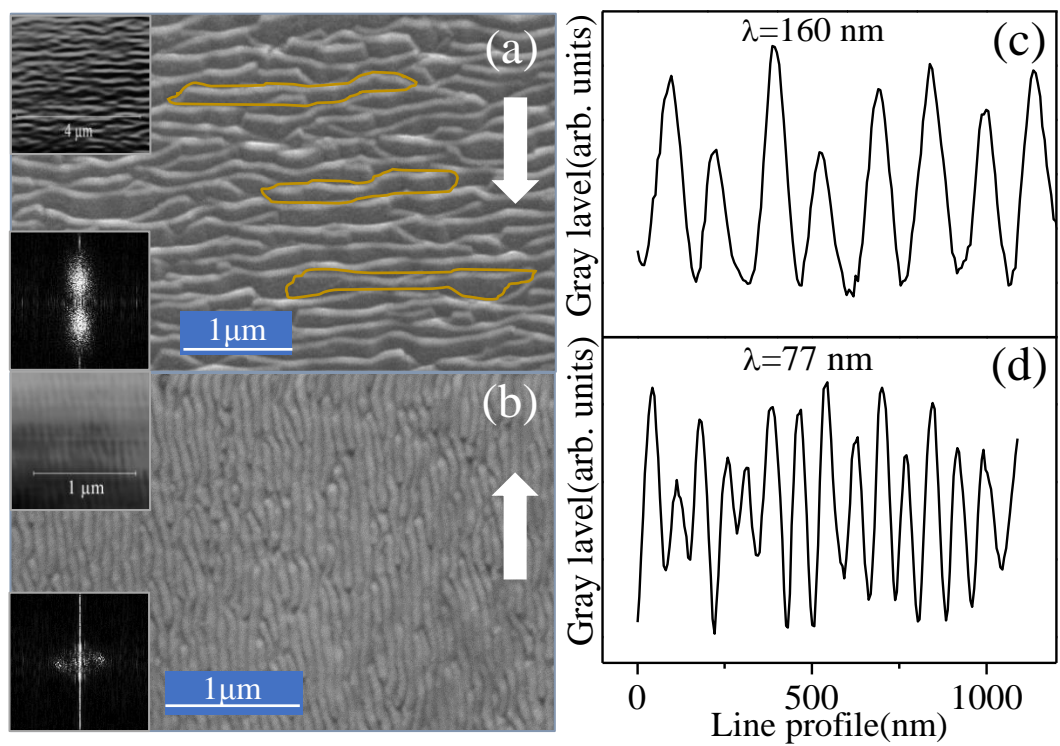

**Figure 3.8:** FESEM images of (a) $OAD_{90°}$ (b) $OAD_{0°}$ sample showing the formation of periodic ripple-like structures on the Co film surface. The lateral association of nanocolumns in the direction perpendicular to the growing direction has been highlighted with yellow lines. The white arrow indicates the projection of Co atoms' flux arrival direction. The inset shows 2D Fourier transformation and autocorrelation function images. (c-d) Line profile extracted perpendicular to ripple direction from (a) and (b), respectively.

### 3.2.3.4 Magnetic anisotropy and domain dynamics studied using MOKE-based Kerr Microscopy.

Magnetic anisotropy, switching behaviour and domain dynamics were analyzed by Kerr microscopy-based magneto-optical Kerr effect (MOKE) measurement manufactured by Evico Magnetics Ltd. Germany in longitudinal geometry. The magnetic characterization of both samples was carried out by azimuthal angle-dependent MOKE measurements in longitudinal geometry. The analysis was focused on the determination of strength and direction of magnetic anisotropy with respect to the direction of ripples and OAD. Fig. 3.9(a,b) shows the hysteresis loops along easy ($\phi = 0°$) and hard ($\phi = 90°$) directions of $OAD_{90°}$ and $OAD_{0°}$ samples, respectively. There is a strong variation in the shape of the hysteresis loop with $\phi$. For $\phi = 0°$, which is along the ripple direction, the hysteresis loop is almost square, whereas, along $\phi = 90°$ direction, the hysteresis loop rounds off.

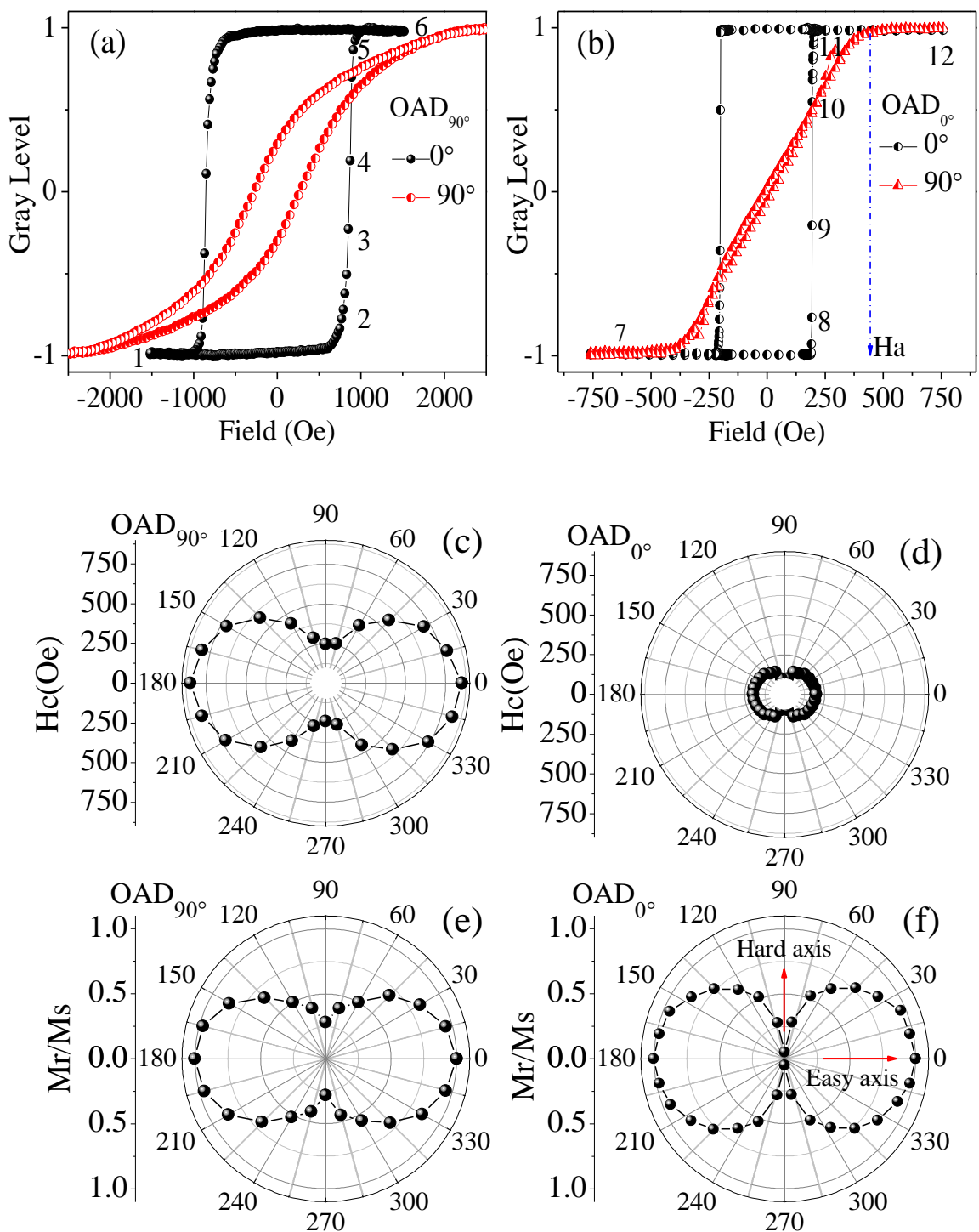

**Figure 3.9:** Longitudinal MOKE hysteresis loop taken along (ϕ=0º, black circle) and across (ϕ=90º, red circle) to the ripple pattern of the sample (a) OAD$_{90°}$ and (b) OAD$_{0°}$, respectively. Ha represents the anisotropy field calculated at the value of saturation magnetization along the hard axis. Polar plot of coercivity (c, d) and normalized magnetization (e, f) of the samples OAD$_{90°}$ and OAD$_{0°}$, respectively.

It is a clear indication of the existence of in-plane magnetic anisotropy in both samples. The observed angular dependences of the coercivity (Hc) and normalized magnetization (Mr/Ms) are also shown in polar plots in Fig. 3.9(c-f), which further confirms that the film possesses in-plane magnetic anisotropy of uniaxial in nature (UMA). The direction of the easy axis of magnetization (EAM) in both samples is along $\phi = 0º$ (along the length of the ripple).

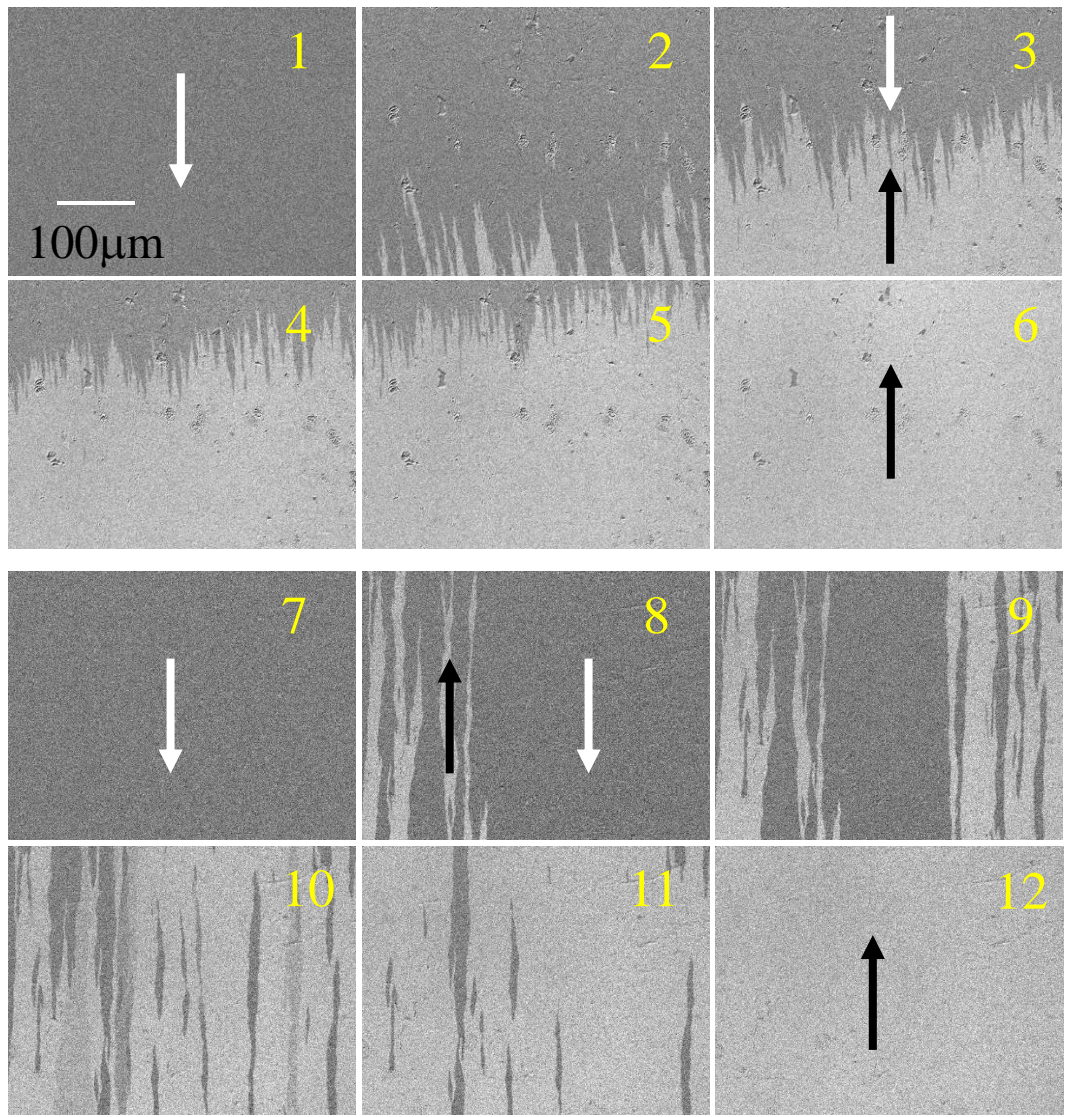

**Figure 3.10:** Domain images captured through Kerr microscopy measurement for sample $OAD_{90º}$ (1-6) and $OAD_{0º}$ (7-12) at different switching field values as marked in the hysteresis loops in figure 3.9(a-b).

For further insight into the origin of strong magnetic anisotropy, the domain dynamics were simultaneously captured through Kerr microscopy. The domain images at different field values corresponding to the hysteresis loops captured along the easy magnetic axis are reported in Fig. 3.10. The single most striking observation that can be drawn from the data comparison is two distinct types of domain configuration are present between the two films. We find that in the OAD$_{90°}$ sample, the domain walls appearing have zigzag or 'saw-tooth' like shapes forming needle-like structures at the boundary that separate two magnetization regimes. On the other hand, in the OAD$_{0°}$ sample usual straight domain wall, i.e., two anti-parallelly saturated magnetized regions separated by 180° walls oriented along the easy axis of magnetization, is present. The aforementioned zigzag arrangement of the domain is observed in magnetic thin films with in-plane uniaxial magnetic anisotropy having charged walls separating two areas of opposite head-on magnetization direction, and its appearance is associated with minimization of the magnetostatic energy [160],[161]. Moreover, the appearance of the zigzag domain is accompanied by an increase in the anisotropy energy [160],[162].

For quantitative estimation of the strength of UMA, the anisotropy constant $K_u$ is extracted using the formula $K_u = H_a M_s/2$ [19],[24],[163],[164], where $M_s = 1422$ emu/cm$^3$ is the saturation magnetization of Co, $H_a$ is the experimental anisotropy field determined along the hard axis direction at the value of the saturation magnetization. The $K_u$ values are $1.68 \times 10^6$ and $3.66 \times 10^5$ erg/cm$^3$ for OAD$_{90°}$ and OAD$_{0°}$ samples, respectively. Surprisingly, the strength of UMA in the OAD$_{90°}$ film is approximately five times higher in magnitude compared to the other one.

## 3.3 Oblique and normal deposition of Co film on flat and patterned substrate to compare the strength of magnetic anisotropy

Unlike epitaxial thin films, polycrystalline films lack long-range structural order [6]. Therefore, getting the required UMA in the polycrystalline magnetic thin film is always challenging. Extensive work has been carried out in the literature, where such anisotropy is induced in most of the polycrystalline and amorphous films by oblique deposition on the flat substrate [7],[129],[165] and normal deposition on ripple patterned substrates [138],[166]. However, it provides limited UMA at higher thickness due to masking of ripples in the case of patterned substrate [23],[137],[141]. On the other hand, columns collapse with each other in the case of OAD [167]. In order to compare the strength of UMA in OAD$_{90°}$ and OAD$_{0°}$ samples with conventional studies in literature, separately, a set of two films of similar thickness were deposited normally on the same rippled substrate (RIP$_{norm}$) and obliquely (at 65°) on a flat substrate (OAD$_{Si}$). All the deposition parameters were kept identical to the previous samples used in the chapter. MOKE hysteresis loops for both samples (RIP$_{norm}$ and OAD$_{Si}$), taken parallel and perpendicular to ripple and OAD direction, are shown in Fig. 3.11a,b, respectively.

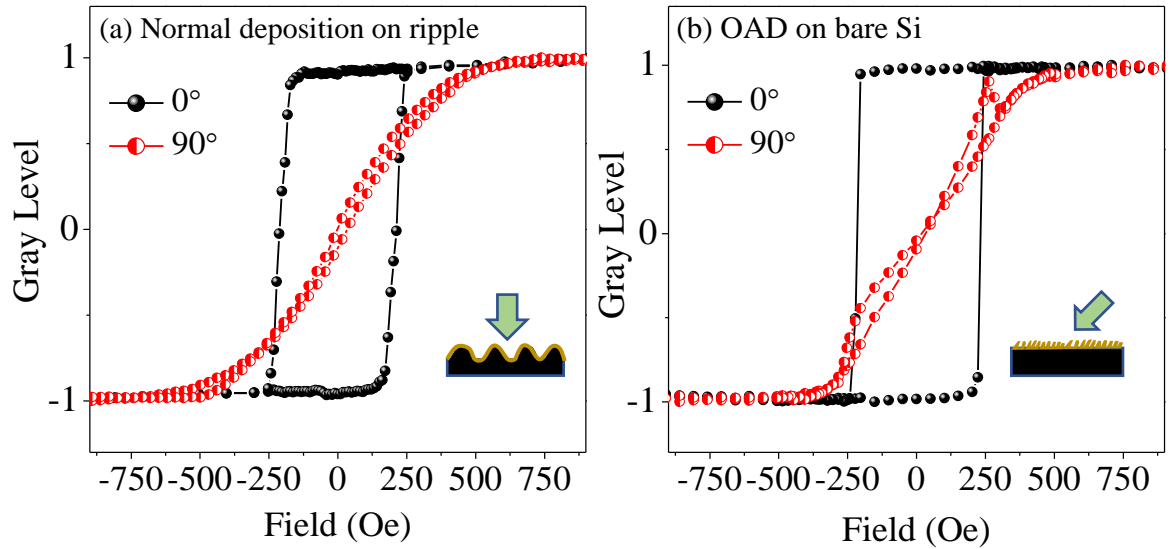

**Figure 3.11:** MOKE hysteresis loop of films deposited (a) normally on the rippled substrate and (b) obliquely on bare Si substrate.

Both the films exhibit UMA of the strength of $4.82\times10^5$ and $3.05\times10^5$ erg/cm$^3$, respectively. Therefore, it is clear that both these films (RIP$_{norm}$ and OAD$_{Si}$), including the OAD$_{0°}$ film from the previous set of films, exhibit UMA of almost the same strength. However, the OAD$_{90°}$ film displays the highest strength of UMA, nearly five times higher than the other samples, as discussed. Furthermore, the strength of UMA is even higher as compared to any polycrystalline Fe, Co thin film in the literature utilizing any method in the literature [7],[19],[24],[137],[141],[168].

## 3.4 Origin of UMA in light of shape anisotropy and dipolar interaction.

By the combined analysis of the data, it is clear that the origin of the strong UMA is sensitively dependent on the growth conditions, where the film and the substrate morphology actively take part in enhancing UMA significantly. The main source of UMA in ferromagnetic materials is the spin-orbit interaction (SOI) in single crystals. Therefore it could be often realized in the epitaxial ferromagnetic films and is known as magneto-crystalline anisotropy (MCA) [169]. It may be noted that present films are polycrystalline in nature, and therefore, MCA is absent due to the randomness of the grains. Moreover, the growth of the film is on amorphous substrates, therefore, it averages out the possibility of step, strain and substrate structure-induced contribution to UMA in the films. The origin of UMA in polycrystalline and amorphous films is often related to the shape anisotropy in the thin films [19],[116], generated through slanted columnar growth in the film at oblique angle deposition and by growing film on patterned substrates [137],[141],[166], where the direction of the easy axis is always aligned along the length of the structure due to the minimization of the magnetostatic energy along this direction.

In the present case, the origin of UMA in films is mainly attributed to the shape anisotropies, i) due to the overall shape of the ripple, which would try to align the easy axis along their length [24],[141], and ii) due to the tilted nanocolumns which try to align the easy axis along the projection of columns in the film plane [170]. The observed direction of the easy axis is the resultant of these two shape anisotropies. It may be noted that the underneath structure of surface topography of the films consists of columnar structure as revealed by GISAXS measurement, which is oriented parallel and perpendicular to ripple direction in $OAD_{0°}$ and $OAD_{90°}$ film, respectively. Therefore, in the case of $OAD_{0°}$, the overall shape of the ripple and tilted nanocolumns within the ripple would try to align the easy axis along their length. On the contrary, for $OAD_{90°}$ film, the directions of the easy axis due to ripple and columns are normal to each other. In this case, the alignment of the easy axis along the length of the ripples can be understood in terms of the relative qualitative strength of shape anisotropy in ripples and columns. If we approximate the shape of the structures (ripples or columns) as a prolate spheroid [7], the shape anisotropy is expressed as

$$E_s = \frac{(N_a - N_c)M_s^2 \sin^2\psi}{2} \qquad (1),$$

where $N_c$ and $N_a$ are the demagnetization factors along and normal to the prolate shape. $\psi$ is the angle between the direction of saturation magnetization ($M_s$) and prolate structure. The strength of $N_c$ and $N_a$ is related to the aspect ratio of length over width. It is clear from FESEM and GISAXS measurements that the aspect ratio of length over the width for ripple structure is much higher as compared to columnar structures; accordingly, $N_a/N_c$ ratio for ripples will be higher as compared to tilted columns. Specially, in the case of the $OAD_{90°}$ sample, $N_a/N_c$ will be much higher due to self-shadowing induced longer ripples. As a consequence, the contribution of shape anisotropy owing to the ripple structure will be greater, resulting in easy axis alignment along the length of the ripples. It is interesting to note that from the above discussion, it is expected that UMA in sample $OAD_{0°}$ would be greater than $OAD_{90°}$ sample.

However, the opposite scenario was observed from the experimental results, where the OAD$_{90°}$ sample has much higher anisotropy as compared to the OAD$_{0°}$ film. Contradiction in the obtained results is an indication that the UMA in the present case cannot only be understood on the basis of the simple alignment of ripples and columns in the film.

The origin of strong UMA in this film is related to the minimization of dipolar interactions between the ripples and columns and, accordingly, explained by considering the dipolar interaction between the rippled structure. Assuming each ripple as a magnetic dipole and magnetostatic interactions among ripples, which can be approximated as dipole-dipole interactions, the dipolar energy term within an assembly of interacting ripples is given by [171]

$$\sum_{i>j} \frac{\vec{m_i}.\vec{m_j} - 3(\vec{m_i}.\hat{r_{ij}})(\vec{m_j}.\hat{r_{ij}})}{r_{ij}^3} \qquad (2)$$

The vector **r**$_{ij}$ connects the sites i with j having magnetic moment **m**$_i$ and **m**$_j$, respectively. Therefore, as the separation between the interacting ripple structures decreases, they start to interact via their dipolar stray fields. The dipolar interaction is inversely proportional to the separation between the ripples. This magnetostatic coupling between two dipoles, therefore, causes the pair to have a uniaxial anisotropy with an easy axis along the line joining the dipoles and reduces the effective uniaxial anisotropy field. It has been shown that with the decrease in the separation of the ripples, the anisotropy strength decreases along the axis of the ripple [172–174]. In the present case, since the ripple structure in the OAD$_{90°}$ sample has large porosity and separation due to the substantial masking effect, the weak dipolar interaction among them would result in a significant increase in the shape anisotropy as compared to that of other samples. It gives rise to strong anisotropy along the length of the ripples that lead OAD$_{90°}$ sample to exhibit enhanced UMA. Therefore, the utility of the enhanced shadowing effect on a pre-patterned substrate is underlined here for uplifting the strength of UMA. We believe that the present approach for inducing large UMA is not limited to Co thin film, but related effects may be expected for all kinds of amorphous and polycrystalline magnetic thin films and multilayer structures. Our technique clearly has an advantage over conventional oblique angle deposition on a flat substrate and normal deposition on the rippled substrate.

## 3.5 Conclusions

Detailed morphological and magnetic characterization of Co films deposited obliquely on a nanorippled SiO$_2$ substrate is reported here. Low energy broad ion beam erosion is used to form a self-assembled periodic array of rippled morphology on $SiO_2$ substrate above which films are deposited obliquely along and perpendicular to ripple direction. The films exhibit distinct characteristics in growth behavior and morphology, having different correlation lengths and anisotropic electron density distribution. The strong asymmetric scattering pattern in GISAXS measurements reveals the formation of tilted columnar nanostructure and preferential film growth at the buried interface due to the shadowing effect induced by the height of ripples. Furthermore, it is found that both films exhibit UMA. However, their domain configuration has a striking difference, and the strength of UMA is almost one order of magnitude higher for the film deposited utilizing the geometrical shadowing effect induced by the height of the ripples. The anomaly in magnetic behavior is attributed to the shape anisotropy and the influence of dipolar interactions among the buried nanocolumnar and surface ripple structures. The present observation explores the crucial role of initial surface topography together with the shadowing effect on the formation of self-organized nanostructured thin film and hence, UMA. In our view, these results provide a new, simple and cost-effective avenue of fine tailoring the strength of morphology-induced UMA in the polycrystalline thin film for possible application in different fields of next-generation high-frequency devices such as thin-film inductors, microwave filters and microwave absorbers, where large magnetic anisotropy is required to ensure an operating ferromagnetic resonance frequency in the gigahertz range.

# Chapter 4:

# Ion beam erosion induced surface texturing and anisotropic morphology

Ion beam, a tool to prepare polycrystalline films with tunable uniaxial magnetic anisotropy

A detailed in-situ investigation to understand the origin of IBE-induced magnetic anisotropy in polycrystalline thin films is presented in this chapter. A combined role of structure and morphology of polycrystalline Fe thin film on the evolution of uniaxial magnetic anisotropy (UMA) subject to several cycles of ion beam erosion (IBE) process is investigated through a set of self-consistent in-situ experiments. This is an unique study as film deposition, IBE and all characterization are done in-situ. Compared to the earlier studies, where IBE was applied to modify surface morphology, the present study demonstrated the use of low-energy ions to modify surface structure. In the initial stages of IBE, surface texturing of Fe film imprints a recognizable magneto-crystalline anisotropy (MCA) with the easy axis of magnetization along the IBE direction. Further erosion results in the development of well-defined correlated morphology that incorporates reinforcement of stray dipolar fields and switches the easy axis of magnetization by 90°. The observed unusual crossover is ascribed to be originating from an interplay between the relative weight of MCA and shape anisotropy, which is successfully explained by considering modified surface structure, morphology and the relative strength of the dipolar stray fields. The present phenomenological understanding provides a promising option for the fine tailoring of the strength of UMA by controlling the crystallinity of the surface structure with anisotropic morphology.

# Chapter 4: Ion beam erosion induced surface texturing and anisotropic morphology

Ion beam, a tool to prepare polycrystalline films with tunable uniaxial magnetic anisotropy

**4.1 Introduction**

**4.2 In-situ characterization of Fe thin film deposited on Si substrate**

**4.3 In-situ study of the evolution of morphology, surface structure and related induced magnetic anisotropy of Fe thin film subject to several cycles of IBE**

    4.3.1 IBE of Fe thin film

    4.3.2 Evolution of Magnetic anisotropy studied using MOKE.

    4.3.3 Detailed characterization of film morphology by XDS.

    4.3.4 Characterization of film crystallographic structure by RHEED.

**4.4 Discussions**

**4.5 Conclusions**

## 4.1 Introduction

The field of thin film magnetism has increased rapidly over the past few decades, mainly due to the presence of peculiar properties such as magnetic anisotropy, exchange bias, interlayer coupling etc. In particular, magnetic anisotropy (MA), which confers a preferred direction on the spin of a thin film structure, has attracted a great deal of interest as it determines the magnetic behavior of a system in response to an external magnetic field. Recently, artificially created thin-film nanostructures that exhibit magnetic anisotropy rapidly actuated extensive research efforts because of the current interest in the spintronics industry [3],[175], perpendicular spin valves [120], magnetic tunnel junction devices [121],[176] as well as other applications in magneto-electronics [122], high-density storage media [177],[178] and magnetic sensors [1],[22]. In addition, an electromagnetic device with a data transfer rate in the GHz region requires soft magnetic material with comparatively large in-plane uniaxial magnetic anisotropy to achieve high ferromagnetic resonance frequency. Therefore, soft magnetic thin films and multilayers with flexible and tailorable magneto anisotropic properties are currently a major field of research in condensed matter physics, where surface and interface science are at the forefront of this development. It holds the key to developing a better understanding of thin film nanostructures with desired strength and functionality through controlled modification of the surface interface morphology and structure.

In the case of epitaxial or crystalline thin films, the primary source of MA is spin-orbit interaction [169],[179], which gives rise to the preferred magnetization direction with respect to the crystallographic structure of materials and is responsible for magneto-crystalline anisotropy (MCA). Unlike epitaxial films, random crystallographically oriented grains are present in polycrystalline thin films, and long-range atomic ordering is absent in amorphous thin films [6]; hence it is difficult to induce MCA in such polycrystalline and amorphous thin films. However, such films can still manifest MA, where the key source of MA is surface morphology-induced shape anisotropy [15,18,20,180,181] produced through the artificial morphological engineering of surfaces and interfaces of thin films. As a matter of fact, the study of MA in such thin films is a striking topic, and in particular, in-plane uniaxial magnetic anisotropy (UMA) induced by nanoscale patterned morphology is a topic of current interest in spintronics. In this regard, the oblique incidence ion beam erosion (IBE) process has been demonstrated to be a handy and powerful tool to induce self-organized nanoscale morphological structures on the substrates [22],[128],[182] as well as 2D thin film surfaces [20],[183],[184]. In the literature, the oblique incidence IBE process has been used extensively to form nanometer-scale ripples patterns on the surfaces. Depending on surface temperature and ion incidence angles, these ripples are aligned parallel or perpendicular to the IBE direction [185],[186]. Magnetic thin films with such microstructure reveal a pronounced UMA with an easy axis of magnetization along the direction of ripples [20],[187] on the surface. Even when a magnetic film deposited on ion beam induced nano-rippled substrates, UMA was found due to the formation of replicated correlated ripple structure in the film [18],[138]. Considerable research has already been conducted in this area. In general, the origin of UMA is being understood in terms of the model proposed by Schlömann [21], where the stray dipolar fields generated by rippled morphology are responsible for UMA in the magnetic films.

Besides a good agreement with the proposed model in most of the earlier works [137], strong deviation from the theory has also been observed in various studies. For example, Miguel et al. [180] have grown 100 nm thick cobalt film on a flat Si substrate, and the surface of the film has been irradiated with ion beam to form the pattern on the surface. Anisotropy energy variation with ripple magnitude has been further compared with the theory, and deviation of the observed anisotropy energy variation from the theory is attributed to the possible discontinuity in the film as a result of IBE [180]. Chen et al. [188] studied ripple-patterned ultra-thin Co films on MgO (001) substrate, and the origin was understood by considering the role of surface and volume contribution, whereas the observed surface contribution in thickness dependence magnetic anisotropy was attributed to unknown reasons. The main problem in this area is to accurately determine the effect of ions beam on surface structure and morphology and to correlate the same with magnetic properties unambiguously. Despite knowing the effect of the ion beam on surface structure modification of epitaxial [189],[190] as well as polycrystalline films [191–194], the role of ion-beam induced structure modification was not taken into account along with surface morphology modification in the literature while studying the origin of UMA in such thin films. The main difficulty faced in doing the same is the extraction of unambiguous information extracting precise information about the ion beam-induced structural and surface morphological changes simultaneously in conjunction with UMA. So far, most of the studies in the literature were performed ex-situ by preparing a series of ion beam sputtered thin film samples [15,18,20,180,185]. Therefore, it is challenging to keep deposition and ion beam parameters identical in different samples to compare UMA. Moreover, as soon as the thin films are exposed to the ambient, their surface gets oxidized. Therefore, the genuine effect of the ion beam on the surface structure cannot be obtained. It is also difficult to extract structural information of a few angstroms thick ion beam modified thin layer on the surface using conventional techniques such as x-ray diffraction (XRD).

In the present work, the effect of oblique incidence low energy $Ar^+$ IBE on the surface of evaporated Fe film has been studied in-situ using magneto-optical Kerr effect (MOKE), X-ray diffuse scattering (XDS) and surface sensitive reflection high energy electron diffraction (RHEED) measurements. In contrast to the earlier works, where experiments were performed ex-situ on a series of samples separately, and the results were interpreted by combining the data, in the present case, a combined role of structure and morphology of polycrystalline Fe thin film on the evolution of uniaxial magnetic anisotropy (UMA) with the increasing cycle of IBE process is investigated in-situ. The present findings have been used to better understand the origin of UMA and its unusual 90° crossover originating from two competing anisotropy in Fe film. The present understanding provides the capability to modulate the strength and the direction of UMA according to the desired functionality in different applications.

## 4.2 In-situ characterization of Fe thin film deposited on Si substrate

Fe film of nominal thickness ~10 nm was deposited on $SiO_2/Si(001)$ substrate using electron beam evaporation technique with a deposition rate of ≈ 0.02nm/s at room temperature inside an UHV chamber having base pressure of ~ $10^{-10}$ mbar. Film thickness and deposition rate were monitored in-situ using a water-cooled calibrated quartz crystal monitor. The amorphous $SiO_2$ layer on the surface of Si prevents Fe from coming into direct contact with Si and forming a silicide layer, hence preventing the creation of a magnetic dead layer at the Si/Fe interface [25]. It also excludes possible epitaxy-induced MCA in Fe film due to the Si (100) substrate crystallinity [137].

The sample was also characterized inside the same UHV chamber equipped with the facilities for *in-situ* characterizations using MOKE, RHEED, XRR and XDS techniques [195], along with an ion gun focusing on the sample surface at an angle ~50º from the surface normal. A polarized He-Ne laser light (λ = 632.8 nm) and other optical components are attached to the chamber and aligned to the sample surface. Hysteresis loops were recorded in longitudinal geometry by applying a variable magnetic field at the sample location using an UHV compatible electromagnet, which can provide a maximum field of about 3000 Oe with a minimum step of the field 1 Oe. The penetration depth of MOKE He-Ne laser for Fe thin film is about 15 to 20 nm. Therefore, a Fe film thickness of about 10nm was selected to get the MOKE signal from the total depth of the film. The sample holder is attached to a rotatable manipulator so as to get an azimuthal angle (magnetic field in the different in-plane direction of the sample) dependent on magnetic properties. MOKE and X-ray scattering measurements are performed simultaneously at the same sample height location, whereas RHEED measurements and ion beam erosion process are performed at different locations in the same chamber to avoid deflection of electrons and ions due to the magnetic field generated by electromagnet during MOKE measurements. The XRR, MOKE and RHEED measurement of the as-deposited film is presented in the next section while discussing the IBE effect on the film. In short, the film grows in a polycrystalline state and exhibits soft magnetic properties without any magnetic anisotropy.

## 4.3 In-situ study of the evolution of morphology, surface structure and related induced magnetic anisotropy of Fe thin film subject to several cycles of IBE

### 4.3.1 IBE of Fe thin film

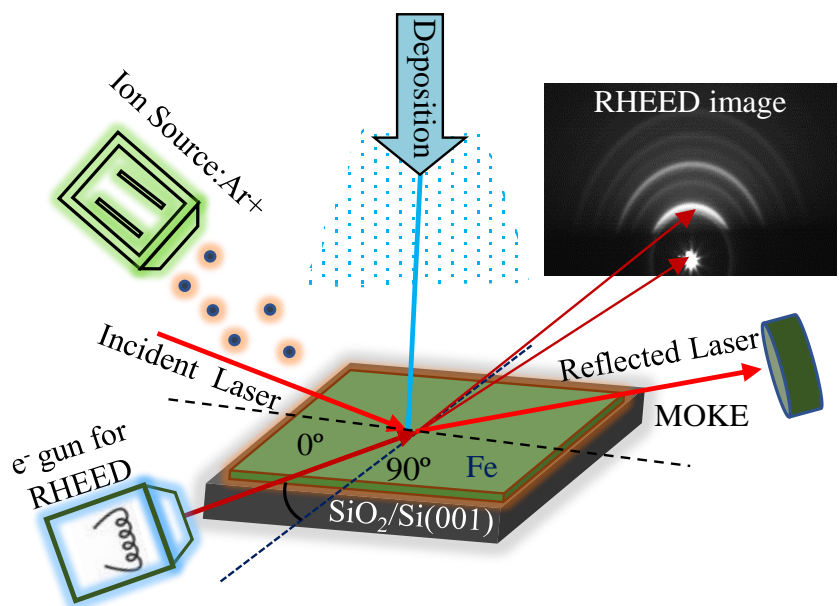

**Figure 4.1:** A schematic illustration of the deposition (e- beam evaporation), characterization (MOKE & RHEED) and ion beam erosion geometry used in this study. Fe thin films are deposited normally on the substrate surface. $Ar^+$ ion sculpts the Fe film surface at an angle of 50º w.r.t. to the surface normal. θ=0º and 90º indicate the respective measurement directions; 0º is along the direction of ion beam erosion.

The Fe film surface was bombarded in repeated cycles of 10 minutes with Ar+ ions having 2 keV energy and ion current density 9 µA/cm$^2$, which corresponds to a mean flux of 5.61×10$^{13}$ ions/cm$^2$/sec. In order to get magnetic information in relation to the direction of IBE, MOKE measurements were done by rotating the sample with respect to the IBE direction. In-situ XRR measurements were performed to extract the actual thickness of the film after each step of IBE, while XDS measurements were carried out in non-specular conditions so as to get information about the in-plane correlation of the morphology and the values of correlation length (ξ).

### 4.3.2 Evolution of film thickness with IBE

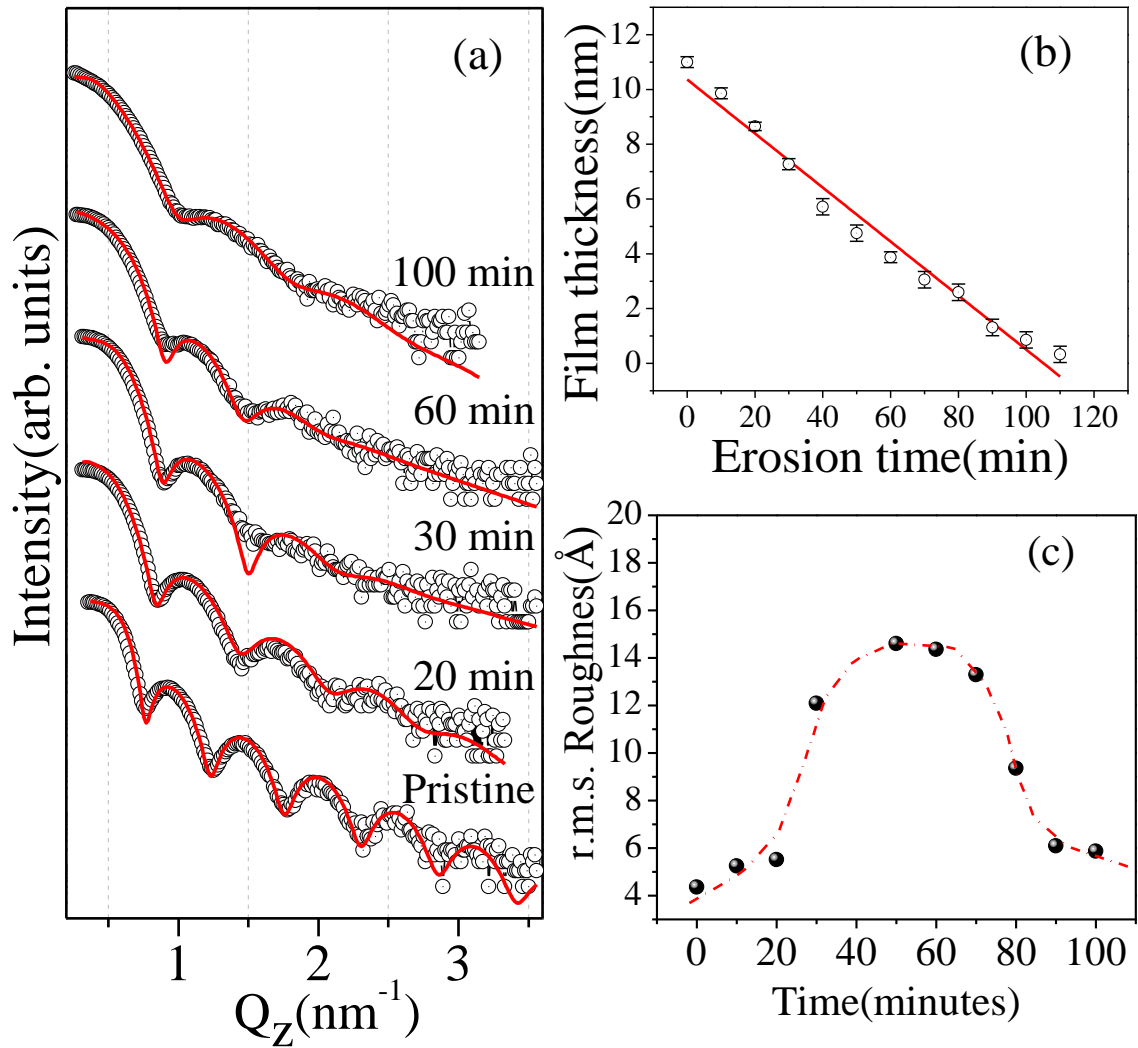

**Figure 4.2:** (a) XRR plots of Fe film in the as-prepared state as well as after 20, 30, 60 and 100 minutes of erosion. The red lines represent the fitting to the experimental data (hollow circles) using

Parratt formalism [151]. (b) Film thickness variation with erosion time. (c) Variation of r.m.s. roughness with IBE time.

Figure 4.2(a) gives representative XRR patterns collected after several cycles (10 min each) of Fe film erosion. The presence of periodic multiple peaks (Kiessig fringes) indicates good structural order and morphological uniformity within the film. Increasing Kiessig fringes spacing with the erosion cycle confirms that the erosion process effectively decreases Fe layer thickness, whereas the increase in damping of Kiessig fringes and slope of the pattern is mainly due to the variation in surface roughness [196]. As extracted after fitting the experimental data, film thicknesses are plotted with erosion time in Fig. 4.2(b). The average erosion rate obtained from the thickness slope versus erosion time plot is ~0.1 nm/min. The variation of r.m.s. roughness with erosion time is plotted in Fig. 4.2(c).

### 4.3.3 Evolution of Magnetic anisotropy studied using MOKE.

Hysteresis loops obtained in-situ along ($\theta=0°$) and across ($\theta=90°$) with respect to the IBE direction after each step of erosion are plotted in Fig. 4.3(a & b). A systematic decrease in the height of the hysteresis loop (Kerr Signal) with the erosion cycle clearly indicates a decrease in Fe layer thickness due to the erosion process. Variation in the Kerr signal can be readily interpreted by keeping in mind that for thin film, the Kerr signal is proportional to film thickness [140] in the thin film region. Film thickness extracted based on Kerr signal analysis is found to be consistent with the film thickness obtained using XRR measurement.

A systematic variation in coercivity (Hc) along and across the IBE direction with the increasing number of erosion cycles is plotted in Fig. 4.4(a). Throughout the evolution of $H_C$, several regimes may be distinguished. In the early stages of ion bombardment, after initial 10 minutes of erosion, decrement in Hc from 37.5Oe to 22.5Oe could be due to the annihilation of some random stresses which might have been generated in the as-deposited film during the deposition process [20],[197],[198]. However, up to 30 min of erosion, the magnitude of Hc varies almost isotropically with slight increment along the aforementioned directions. Whereas, beyond 30 min, we evidence significant distinction in Hc between $\theta= 0°$ and $90°$ direction, which indicates the onset of UMA with easy axis lying along IBE direction ($Hc_{\theta=0°}>Hc_{\theta=90°}$). Surprisingly, after 60 minutes of erosion, Hc along $\theta=90°$ surpassed the rise in Hc along $\theta=0°$.

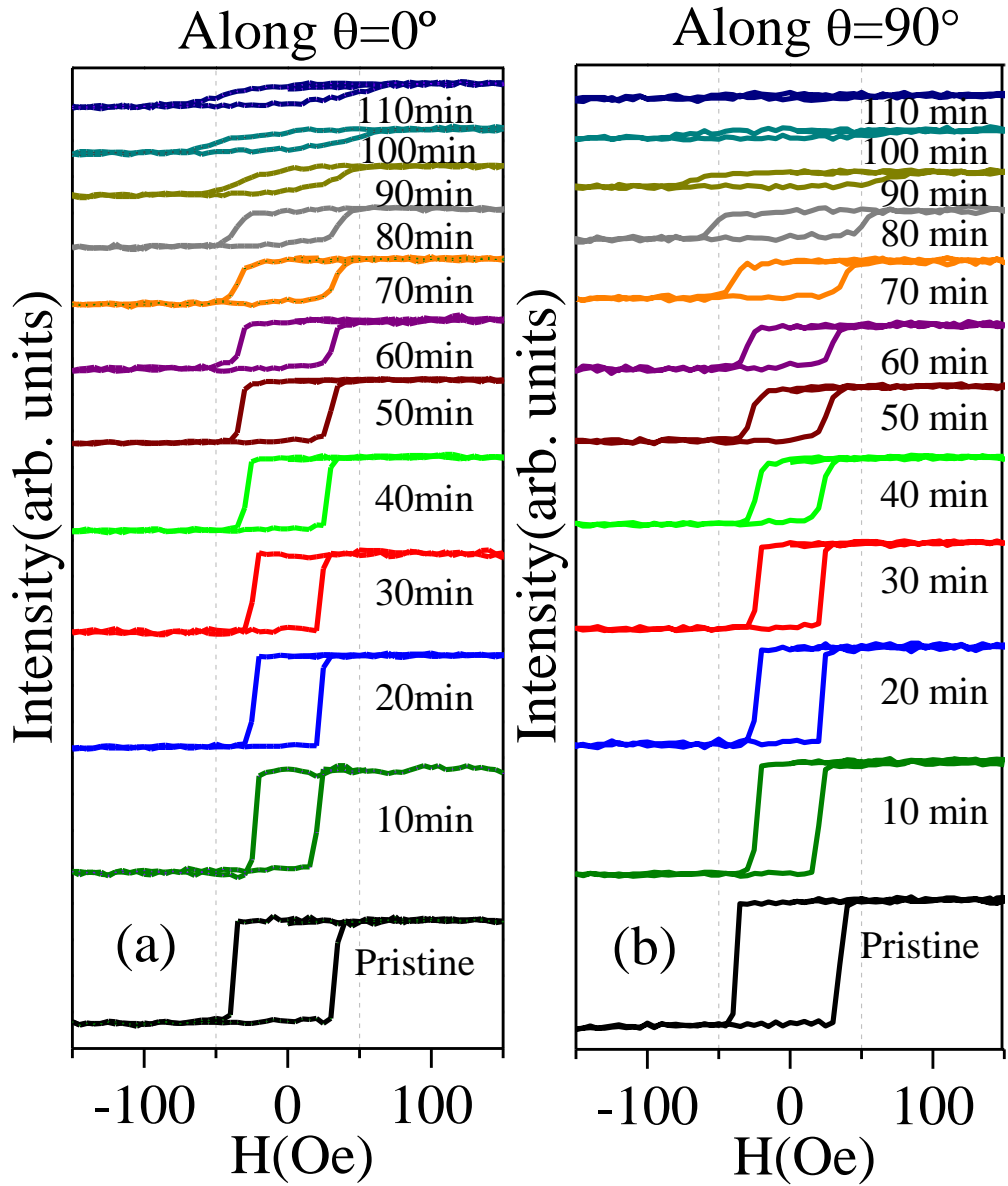

**Figure 4.3:** Longitudinal MOKE Hysteresis loop taken (a) along (θ=0°) and (b) across (θ=90°) to the ion beam erosion direction after each step of the erosion cycle.

In order to facilitate the above observation, we classified Fig. 4.4a into three different regions; I ($Hc_{\theta=0°} \approx Hc_{\theta=90°}$), II ($Hc_{\theta=0°} > Hc_{\theta=90°}$) and III ($Hc_{\theta=0°} < Hc_{\theta=90°}$). In particular, to gain further insight into the switching behavior of UMA corresponding to these three regions, the azimuthal angle ($\theta_{AZI}$) dependence of Hc is plotted for thicknesses corresponding to points A, B and C, as marked in the polar plots (b, c and d) of Fig. 4.4a. It is clear from the polar plot that $H_C$ in I-region is almost independent of $\theta_{AZI}$, which confirms the absence of any UMA in the film. However, in II-region, after 30 min of erosion, dumbly shaped variation in Hc with $\theta_{AZI}$ confirms the presence of (UMA) with an easy axis along the direction of IBE ($\theta_{0°}$).

Interestingly, after 60 min of erosion (III-region), the UMA rotates 90° with easy axis now along θ$_{90°}$ [normal to ion beam sputtering (IBS) direction] can be noticed.

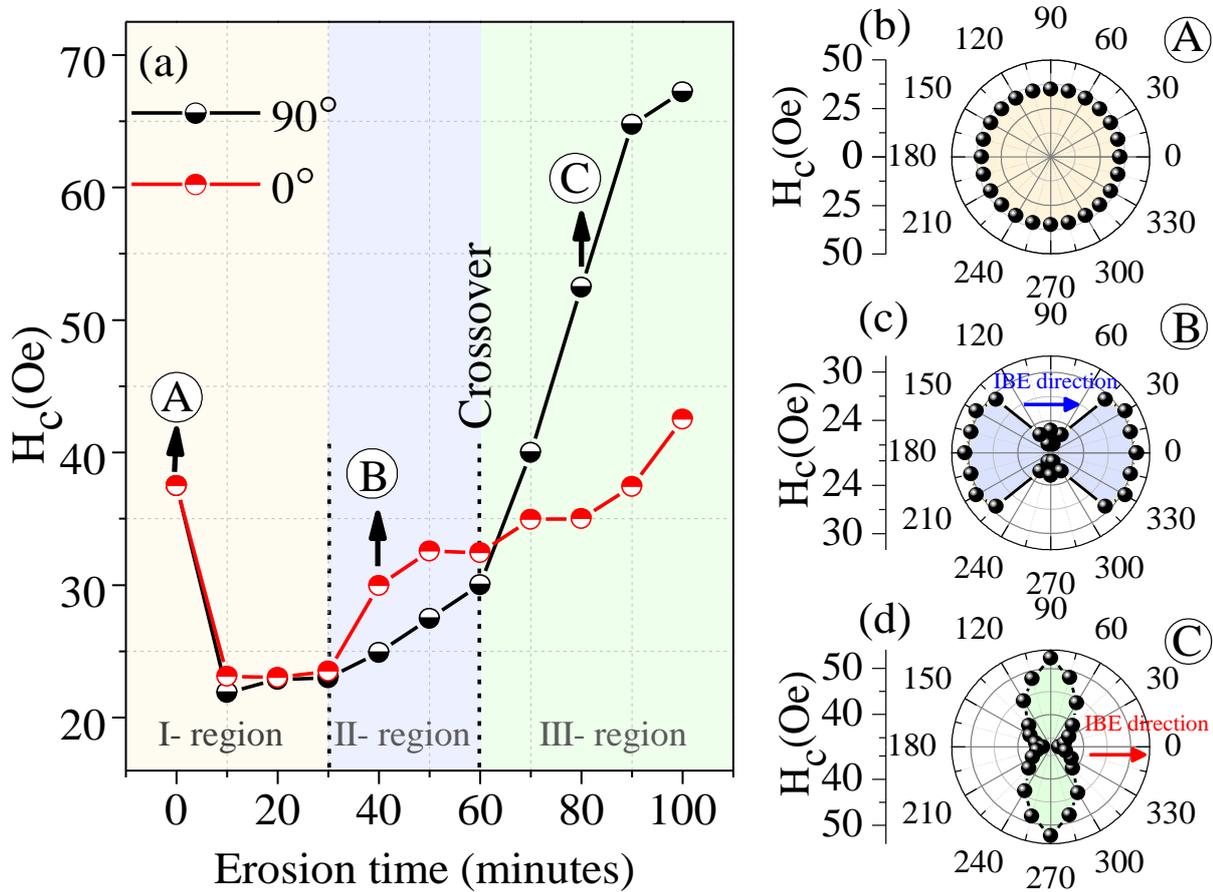

**Figure 4.4:** (a) Variation of coercivity as a function of erosion time along (θ=0°) and across (θ=90°) to IBE directions. Polar plots of the Hc versus θ of (b) pristine (A in region-I), (c) after 40 minutes of erosion (B in region-II) and (d) 80 minutes (C in region-III) of erosion.

To understand the origin of the UMA and its crossover, the effective anisotropy constant is extracted using $K_{eff}$ = H$_a$M$_s$/2 [180], where H$_a$ is the anisotropy field calculated from the hard axis configuration where the hysteresis curve reaches saturation [17,199,200] and M$_S$ is the saturation magnetization of Fe (for bulk Fe ~1714 emu/cm$^3$). Products of $K_{eff}$ and the nominal Fe layer thickness are plotted as a function of Fe layer thickness in Fig. 4.5. It may be noted that $K_{eff}$ contribution can be divided into the combined effect of volume and surface anisotropies and expressed as:

$$K_{eff} = K_v + \frac{K_s}{d} \qquad (1)$$

Where, $K_v$ is volume contribution, $K_s$ is surface contribution and $d$ is the film thickness.

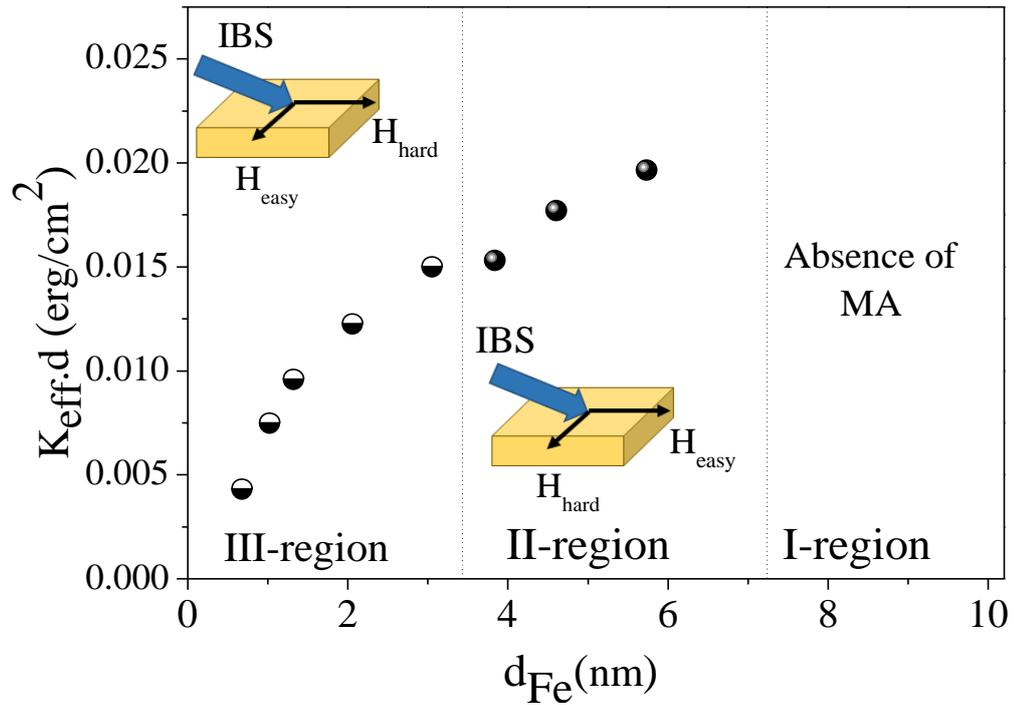

**Figure 4.5:** Variation of effective magnetic anisotropy energy ($K_{eff}$) vs Fe film thickness.

It is clear that a plot of ($K_{eff} \times d$) *versus d* will give a straight line with a slope as $K_v$ and intercept being the surface term $K_s$ of UMA. In Fig. 4.5, the $K_{eff} \times d$ curve does not follow a straight line but deviates significantly. Present observation differs from earlier observation by Kai Chen et al. [188], where $K_{eff} \times d$ varies linearly with film thickness for films deposited on the patterned substrate. The current observations indicate the presence of an additional contribution of an unknown factor, which plays a significant role in the origin of UMA in both regions as well as is responsible for UMA crossover.

Further, RHEED and XDS measurements were also performed after different steps of IBE and analyzed in detail to develop a clear understanding of the observed results.

### 4.3.4 Evolution of film surface structure studied using RHEED.

RHEED patterns in the as-prepared state (I region) as well as after erosion are obtained along and normal to the IBE direction, whereas some representative RHEED images for 20 min, 40 min (II region) and 70min (III region) of IBE along θ=90° direction (normal to IBE direction) are shown in Fig. 4.6(a)–4.6(d), respectively.

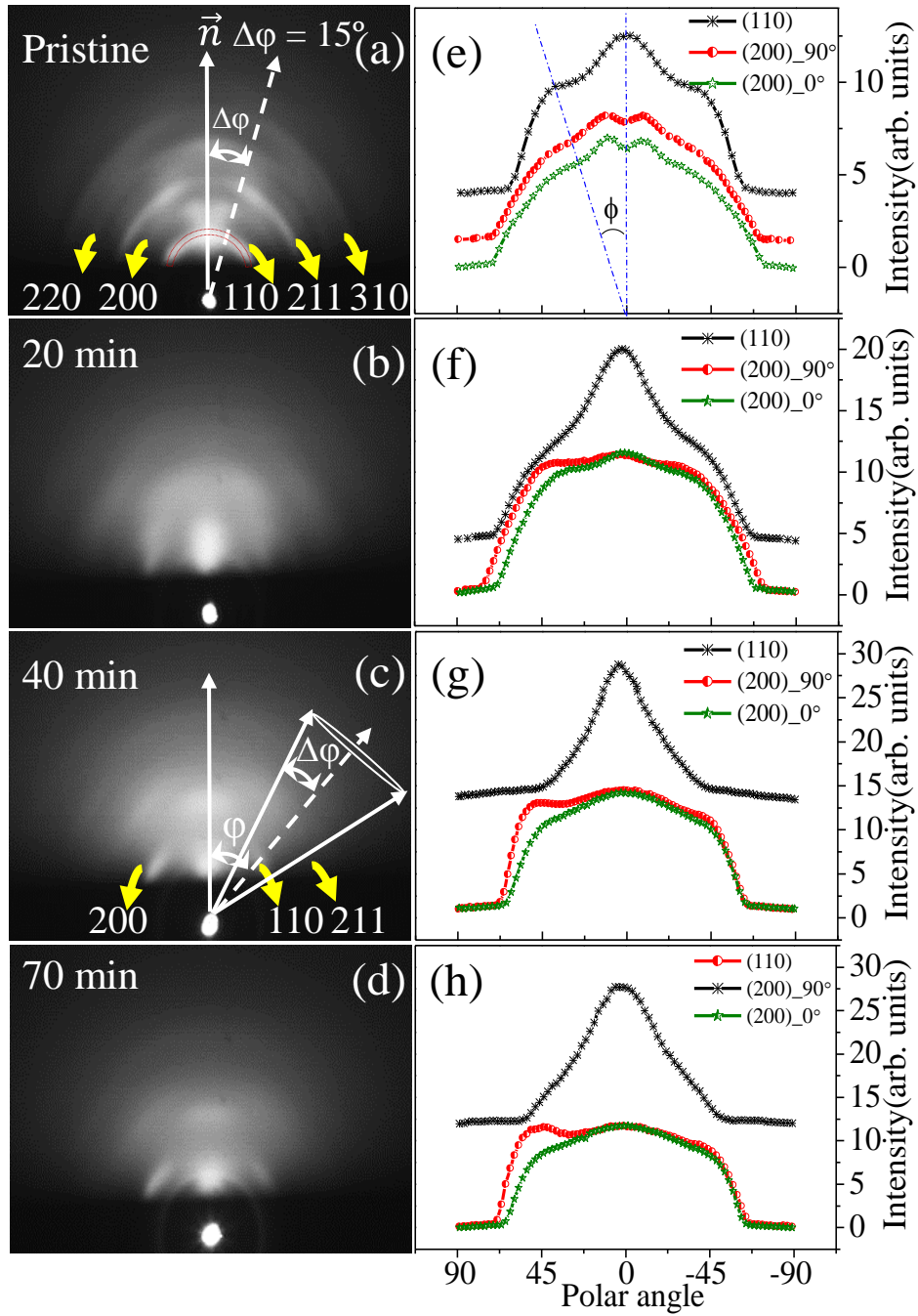

**Figure 4.6:** Representative RHEED images in (a) as prepared as well as after (b) 20 minutes (c) 40 minutes (d) 70 minutes of IBE. The corresponding polar-angle dependent (0° to ±90°) intensity distribution of the (110) and (200) rings are also presented in (e)-(h), respectively.

Sets of concentric rings (Debye rings) corresponding to different diffraction planes (110), (200), and (211) of Fe film are identified [201]. Non-uniform intensities along the arcs of the rings reveal the preferred orientation (texturing) of the Fe planes in both the as-prepared and sputtered states [78,82,202]. It is in line with the observation of other researchers [192], where films with bcc structure, such as Fe, exhibit fiber texturing with <110> direction being normal to the surface such that closed packed planes orient parallel to the surface of the substrate to minimize the surface energy. Hazy RHEED images after IBE are dominant due to the background electron scattering, which increases with increasing surface roughness [203]. In order to understand a substantial rearrangement of intensities in the rings (along the arcs) after IBE, (110) and (200) rings (Fig. 4.6a-4.6d) are integrated radially and plotted with polar angle $\phi$ in Fig. 4.6(e) to 4.6(h) [204], where $\phi$ is defined as the angle measured from the vertical axis perpendicular to the substrate ($\phi =0°$ is at the top of the ring). Variation of $\phi$ from 0° to 90° gives information about out of plane ($\phi =0°$) to in-plane ($\phi =90°$) texture [80]. The sharp decreasing intensities observed at the higher angle ($55 \pm 3^0$) are due to the geometrical constraints or in-plane diffraction masking by the sample [204]. The background rectification is applied to the distinct rings by deducting background intensity variation slightly to the outside of the ring [204].

Extracted line profile of the (110) and (200) diffraction rings in Fig. 4.6(e) to 4.6(h) suggests that, in sputtered film, the intensity of the (110) ring is substantially higher at $\phi=0°$ as compared to pristine (as prepared) film. This behavior is the characteristic of an increment in preferential alignment (texturing) of the <110> direction (pointed normal to the film surface) after IBE. It may be noted that <200> direction makes an angle of 45º with <110> direction in bcc structure. Hence increased texturing of (110) peaks normal to the film surface will be responsible for the relative increase in textured (200) planes at 45° direction normal to the film surface (and 45° inclination towards in-plane direction). The same has also been confirmed by increased (200) peak intensity about $\phi= \pm 45°$ in Fig. 4.6(e) and 6(h). Here we would like to emphasize that ion beam-induced modifications are expected only on the film surface due to the low energy (~2 keV) IBE process used in the present experiments. Moreover, a closer look reveals that the intensity distribution of (200) ring along and across the IBE direction modifies differently due to differences in sputtering yield along different azimuthal directions of Fe surface grains.

**4.3.5 Evolution of film surface morphology studied using X-ray diffuse Scattering.**

X-ray diffuse scattering (XDS) measurement is a complementary technique of atomic force microscopy (AFM) that can provide morphological information over macroscopic dimensions in a non-destructive way with good statistical accuracy. In addition, X-ray scattering-based methods can be performed in-situ under UHV conditions. In the present study, in-situ XDS measurements in non-specular scattering (NSC) conditions are performed along the direction of erosion after 30 min (I region), 50 min (II region) and 90 min (III region) of IBE to reveal the evolving in-plane correlation in surface morphology. The geometry for NSC is shown in Fig. 4.7(a-b), where the z-axis is perpendicular to the film surface, and x-z plane is the plane of scattering. Here, $k_i$ and $k_f$ denote the incident and scattered wave vector, respectively. Specular condition is given in terms of the momentum transfer vectors $q_x=q_y=0$, $q_z>0$ and the non-specular intensity is measured with a parallel momentum transfer component $q_x$, (along erosion).

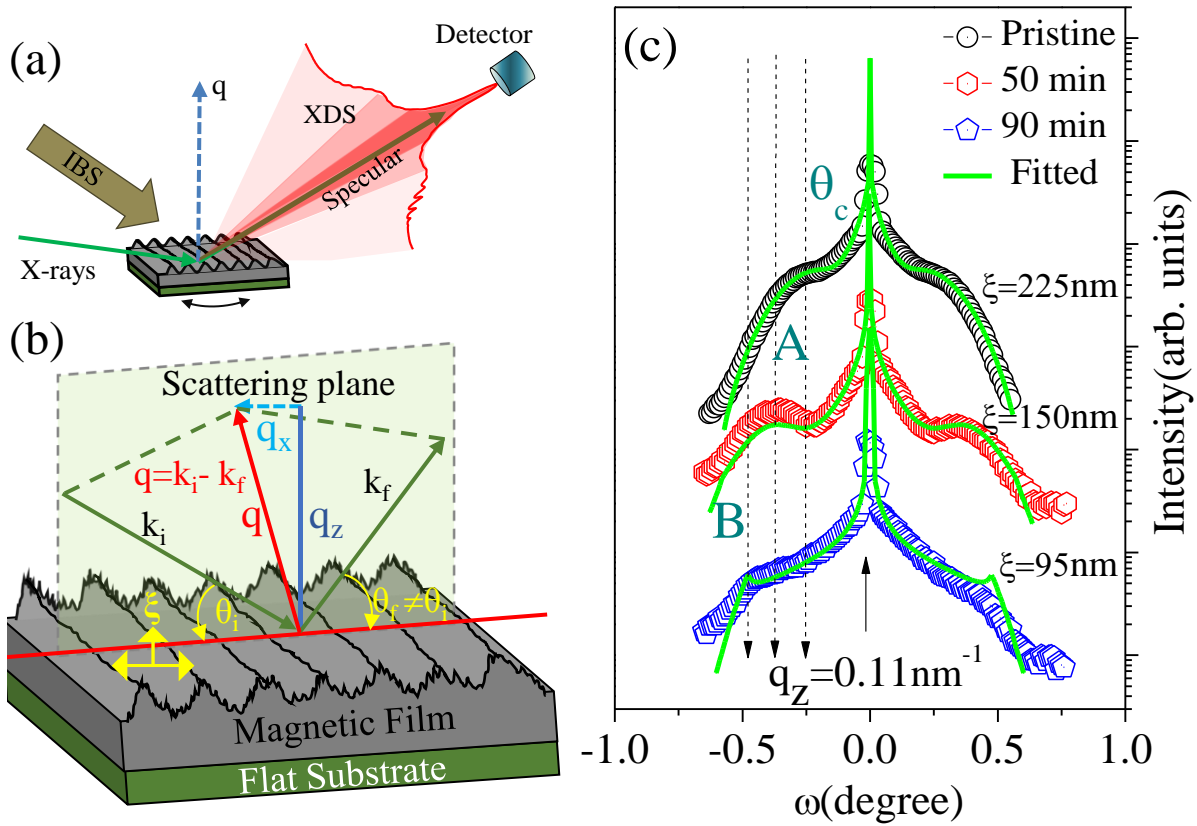

**Figure 4.7:** (a) Schematic illustration of XDS geometry with respect to the ion beam direction and surface morphology (b) Representation of in-plane ($q_x$) and out-of-plane ($q_z$) wave vector in non-specular experiments. (c) Fitted XDS spectra as a function of the rocking angle $\omega$ ($=\theta_i-\theta_{sc}$) at fixed $q_z$ value 0.11nm$^{-1}$. The spectra have been shifted vertically for clarity purposes, and the scattered intensity is shown on a logarithmic scale. The vertical lines are "guides to the eyes" to show the correlation peak shift, which is representative of the respective in-plane correlation lengths in the sample.

The XDS measurements are carried out using X-ray of wavelength λ=0.154 nm by keeping the scattering angle θ$_{sc}$ fixed at 0.65º (above the critical angle) and varying the angle of incidence θ$_{sc}$ from 0º to 2θ$_{sc}$. XDS patterns are presented in Fig. 4.7(c). The broad peaks appearing on each side of the specular peak indicate the formation of correlated structure [205],[206] (along $q_x$) on the surface produced by the ion beam induced self-organization process [14],[188]. Progressively shifting of the peak away from the specular peak with increasing erosion time is related to decrement of the lateral separation between correlated structures which are separated in real space by a distance equivalent to ($\frac{2\pi}{q_x}$). To get quantitative information, XDS data is fitted using the TRDS_sl program [207] in the framework of distorted wave Born approximation (DWBA) formalism [94], where the distribution of height and its correlation in the *x, y* plane is expressed by the height-height correlation function C(x,y) [208–210] which can be approximated by an exponential correlation function-

$$C(x,y) = \sigma^2 \exp(-|R|/\xi)^{2h} \qquad (2)$$

where $R = (x^2 + y^2)^{1/2}$, σ is r.m.s. roughness, ξ is the lateral correlation length and exponent *h*(0<*h*<1) is the jaggedness parameter. ξ describes the lateral ordering of the morphology on the surface and measures the length scale of the correlation (average distance between peaks and valleys on the surface morphology), *h* is a measure of the degree of surface irregularities. The variables σ, ξ and *h* are taken as a fitting parameter to fit experimental data and are reported in *Table I*. It may be noted that both *h* and ξ parameters decrease up to 0.12 and 95 nm, respectively, after 90 minutes of erosion. Whereas σ$_{rms}$ increases maximum up to 1.44 nm after 50 minutes of erosion and subsequently decreases to 0.61 nm after 90 minutes of erosion. As the values of ξ and *h* decrease with erosion, it indicates that the in-plane correlation length of the correlated part of the roughness is getting shorter and turning the film surface more jagged.

In order to provide visual evidence of the formation of correlated morphology on the film surface, a separate Fe film of thickness 15nm was deposited on SiO$_2$/Si substrate in the same UHV chamber under identical conditions (as mentioned above). The film surface was eroded with the Ar+ ion for 70 minutes in the identical condition as that of 10nm Fe film. The film surface morphology is imaged by NT-MDT S.I. AFM in contact mode, as shown in Fig. 4.8 and image processing was performed using WSxM software [211]. We observe clear traces of quasi-periodic ripple pattern with grooves oriented perpendicular to the projection of IBE direction on the film surface. The r.m.s. roughness and correlation length of the pattern is found to be 1.25nm and 208nm, respectively. These values are comparable to that obtained for 10 nm Fe film from XDS measurement. Therefore, the direct imaging technique AFM supports our earlier conclusions obtained from XDS in favor of the formation of correlated rippled morphology.

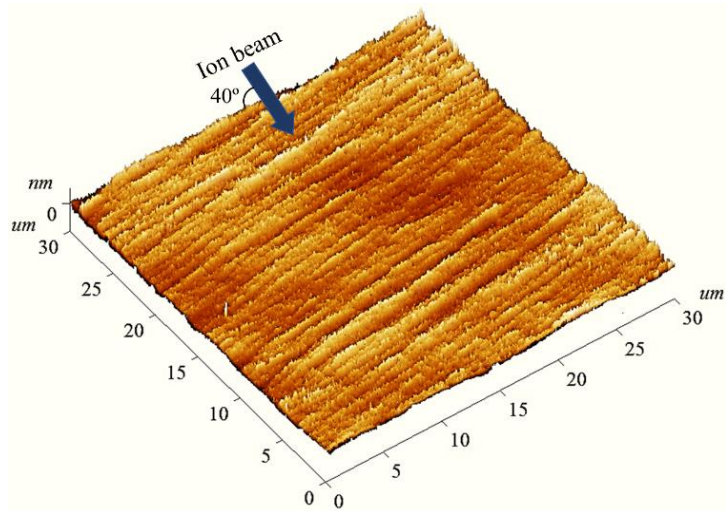

**Figure 4.8:** AFM image of Fe film surface after 70 minutes of erosion time. The pattern has correlation length ξ= 208nm and r.m.s. roughness σ= 1.25 nm. The arrow indicates the projection of ion beam on the film surface.

***Table 1:*** Simulated parameters obtained by fitting XDS data. $H_a^{cal}$ and $H_a^{moke}$ are anisotropy field obtained from Schlömann model and MOKE data, respectively. The easy axis (EA) direction of UMA with respect to the ion beam erosion (IBE) direction is shown in the last column.

| IBS time (minutes) | Regions in Fig.4.3 | $H_a^{moke}$ (Oe) | d(nm) | ξ(nm) | h | $\sigma_{rms}$ (nm) | $H_a^{cal}$ (Oe) | UMA direction |
|---|---|---|---|---|---|---|---|---|
| Pristine | I | none | 11 | 225 | 0.2 | 0.44 | 5 | **No UMA** |
| 30 | II | 34.97 | 7.37 | 170 | 0.17 | 1.21 | 77 | **EA ∥ IBS** |
| 50 |  | 44.85 | 4.60 | 150 | 0.15 | 1.44 | 192 |  |
| 90 | III | 84.55 | 1.32 | 95 | 0.12 | 0.61 | 256 | **EA ⊥ IBS** |

## 4.4 Discussions

The observed magnetic modifications can be explained by considering contributions to the UMA coming from magnetostatic and magnetocrystalline effects. It was already suggested by XDS results that the surface gradually evolves from a flat morphology to a well-defined correlated morphology which in turn can give rise to stray dipolar field generated across the anisotropic surface. In order to compare erosion and thickness-dependent magnetic anisotropic field $H_a^{moke}$ as obtained through MOKE, we begin our analysis by attempting calculation of anisotropy field $H_a^{cal}$ on the basis of the theory of demagnetization field due to correlated anisotropic roughness [21], which can be expressed as [15];

$$H_a^{cal} = 4\pi M_s \frac{\pi \sigma_{rms}^2}{\xi \times d} \tag{3}$$

where $M_s$ is the saturation magnetization, $\sigma_{rms}$ is the r.m.s. roughness of the periodically patterned profile with the wavelength ξ and d is the film thickness.

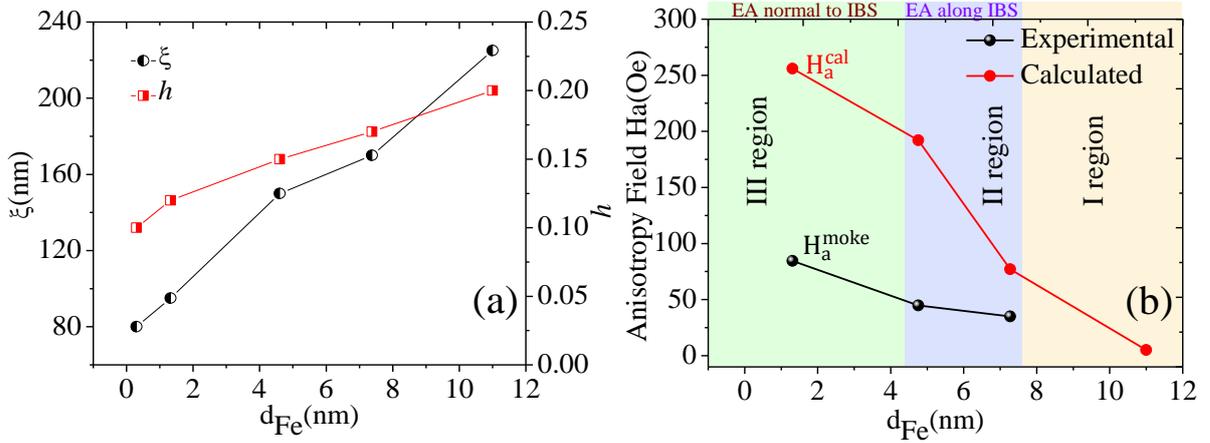

**Figure 4.9:** At selected thickness, (a) variation of ξ and h parameter of Fe film surface and (b) variation of anisotropy field values calculated (based on demagnetization field generated through the patterned surface) and experimentally obtained with film thickness.

$H_a^{cal}$ clearly shows a strong dependence on the product ξ×d, which indicates higher magnetic anisotropy in the film corresponds to the short in-plane correlation and reduced film thickness. The values of $H_a^{cal}$ are calculated for the pristine as well as films after 30, 50, and 90 minutes of IBE and listed in table 1. The values of ξ, h and $H_a^{moke}$, $H_a^{cal}$ are also plotted in Fig. 4.9(a) and 9(b) with $d_{Fe}$, respectively. Up to 30 minutes of IBE, UMA is absent due to large values of ξ =170-225nm and $d_{Fe} \approx 7.4$nm. The calculated value of $H_a^{cal}$ is also negligible for this region. It may be noted that $H_a^{cal}$ for region II, indicates weak magnetic anisotropy with easy axis normal to IBE direction, which is in contradiction to the experimentally observed $H_a^{moke}$, where the easy axis of UMA is found along the direction of IBE (please see Fig. 4.4). As ions are bombarded on film at an angle of 50º, therefore, it can induce self-organized ripple pattern formation normal to the IBE direction [14]. Hence, the dominating origin of UMA would be something else rather than stray fields induced UMA due to periodic roughness in Fe.

The origin of UMA in this region can be rationally understood in terms of ion beam-induced surface structure modification. As observed through RHEED measurement, ion beam could be able to induce texturing at the surface of the Fe film, which has been started much earlier to the proper formation of patterned surface morphology. Here, we would like to emphasize that Fe possesses inherent uniaxial magnetocrystalline anisotropy (MCA), with the <100> axis being the easy direction. Therefore, texturing could be the main factor responsible for magnetic anisotropy in this region.

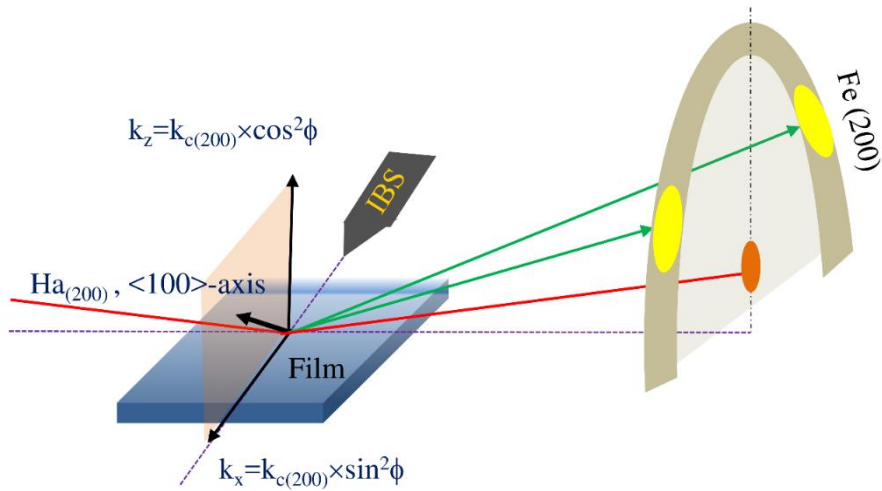

**Figure 4.10:** Schematic to understand the Fe (200) planes orientation with respect to IBE direction.

This picture is schematically represented in Fig. 4.10, where we have shown that the <100> direction of Fe aligns preferably at 45° off-normal direction perpendicular to IBE. If $k_c$ is the MCA constant along <100>, then $k_c\cos^2\phi$ and $k_c\sin^2\phi$ are the z and x component of the MCA along perpendicular and in film plane (along IBE direction), respectively. It is interesting to note that the z component ($k_{c200}\cos^2\phi$) is not expected strong enough to overcome a large demagnetizing field of ~$4\pi M_s$ (where $M_s$ is saturation magnetization) of the film. The perpendicular alignment of spins is "screened" by the demagnetizing field, causing an absence of the perpendicular magnetic anisotropy component in the film. Therefore, the projection of <100> ($k_{c200}\cos^2\phi$) along the beam direction is responsible for inducing uniaxial magnetic anisotropy (UMA) with an easy axis along the IBE direction. In the earlier stages of erosion UMA induced by correlated surface, morphology is relatively weak due to both large values of ξ and film thickness. Therefore, the stray field-induced UMA normal to the IBE is weak, and the MCA decides the direction of UMA in this region. With increasing IBE time, well-defined correlated ordered patterns with the shorter length scale develop normal to the IBE direction, which is responsible for imprinting well-defined UMA through stray field-induced shape anisotropy in region III. In particular, we notice that the value of the anisotropy field that we have extracted from our experimental data shows a systematic tendency to rise with the increasing erosion cycle. Thus, increased strength of UMA due to shape anisotropy overcomes MCA and is responsible for the direction of easy axis rotation normal to the IBE (90° rotation) direction. Therefore, the delicate interplay between the relative weight of MCA and shape anisotropy is attributed here as the origin of crossover in UMA. In this sense, from our present study, it appears that low-energy IBE causes substantial morphology modification and reordering of surface structure. Consequently, our phenomenological consideration of MCA arising from the surface structure modification along with the relative strength of the dipolar stray field originating from surface morphology successfully explain the experimental findings.

## 4.5 Conclusions

In conclusion, the evolution of morphology and structure as well as magnetic properties of Fe film after various erosion stages are observed and comprehensively studied in detail by exploiting MOKE, RHEED, XRR and XDS techniques in-situ. In contrast to the studies in the literature, the present work unfolds the intermediate states of the modified structure and morphology of Fe thin film and selectively deals with their combined role in the creation and interpretation of UMA. Present results reveal that, in the initial stages of IBE, a surface-type preferential crystallographic orientation of <100> axis along the direction of IBE imprints a recognizable MCA with easy axis of magnetization along the IBE direction. Further erosion of Fe film develops a well-defined correlation within the morphology. It enhances shape anisotropy through the generation of stray dipolar fields, resulting in the crossover from MCA to shape anisotropy along with the switching of the easy axis by 90° at a specific value of film thickness or erosion cycle. Controlled surface crystallinity and morphology modification, along with precise characterization by self-consistent technique under in-situ conditions, makes it possible to understand the unusual UMA crossover. The present study provides possibilities to modulate UMA's strength and direction according to its application in designing next-generation magnetic sensors and electromagnetic devices for operation in gigahertz band region where their superior efficiency and thermal stability demand soft magnetic materials with tunable in-plane UMA.

# Chapter 5

# Ion beam modification during film growth

Method of enhancing the limit of uniaxial magnetic anisotropy in thin films

This chapter describes an unique approach to engineering and enhancing the strength of oblique incidence ion beam erosion (IBE) induced in-plane uniaxial magnetic anisotropy (UMA) by simultaneous modification of film morphology and texture. For the present study, a cobalt thin-film system is selected as a model system due to its advantage of easy growth and magnetocrystalline easy anisotropy axis along its "c" axis. We have seen in the previous chapter that IBE on the film surface can lead to textured surface topography that can induce magnetic anisotropy. However, limited penetration depth restricts ion beam-induced modification to the surface and sub-surface region. Therefore, we focus our efforts on sequential deposition and subsequent IBE of the film rather than the traditional post-growth IBE of film. Detailed in-situ investigation insights that the film grows in a biaxially textured polycrystalline state with the formation of nanometric surface ripples. The film also exhibits pronounced UMA with an easy axis oriented parallel to the surface ripple direction. Remarkably, the induced UMA is about ten times larger than the reported similar kind of earlier studies. The capability of imposing in-plane crystallographic texture throughout the whole layer of the film giving rise to magneto-crystalline anisotropy, along with the shape anisotropy of nanometric surface ripple, enhances the strength of the UMA, hence, illustrating the unibersal applicability of the present method.

# Chapter 5: Ion beam modification during film growth

Method of enhancing the limit of uniaxial magnetic anisotropy in thin films

**5.1   Introduction**

**5.2   Preparation of thin film by sequential deposition and ion beam erosion process.**

**5.3   In-situ characterization of film surface morphology, structure, and induced uniaxial magnetic anisotropy of Co thin film.**

   5.3.1 Characterization of film crystallographic structure by RHEED.

   5.3.2 Detailed characterization of film morphology by XDS.

   5.3.3 Magnetic anisotropy studied using in-situ MOKE.

   5.3.4 Discussion in light of Schlömann's theory.

**5.4   Temperature-dependent in-situ study of film structure, morphology and magnetic anisotropy.**

**5.5    Comparative study with a post-growth ion beam eroded thin film.**

**5.6    Conclusions**

## 5.1 Introduction

Magnetism in low-dimensional systems such as magnetic nano-structures, ultrathin film and multilayers is of great interest from both fundamental and applied points of view. Therefore, it has become an active field of research. In particular, uniaxial magnetic anisotropy (UMA) and magnetization reversal are the key properties that find application in high-density magnetic memories [212], magnetic sensors [213], spintronics and wireless communication devices [3]. Therefore, researchers have been developing several strategies such as oblique angle deposition [214], magnetic field and stress annealing [215], engineering of the interface [216] etc., for flexible and fine tailoring of UMA. In this respect, ion beam erosion (IBE) has been demonstrated as a versatile and handy tool to induce UMA by engineering surface and interface morphology through the self-assembled formation of nanometric patterns on the surface [15–20]. Ferromagnetic films deposited on the pre-patterned nanostructured substrate [15–17] (bottom-up approach) or post-growth IBE of the film surface [18–20] (top-down approach) deposited on planner substrate are found to imprint a UMA. However, these studies suffer from their limitations. For example, in the top-down approach, ion beam-induced modifications happen only on the top surface due to the limited penetration depth of ions [20].

Furthermore, the initial film thickness must be sufficient for the development of a homogenous pattern as well as for maintaining film continuity [22]. On the other hand, in the bottom-up approach, the strength of UMA falls off after a critical film thickness due to the merging of ripples crests with their nearest neighbors [15],[23],[25]. Moreover, due to the presence of randomly oriented grains in the polycrystalline film, it is always challenging to induce magneto crystalline anisotropy (MCA) by modifying film texture through preferred grain alignment. However, earlier it has been observed that due to minimization of magneto-elastic energy and coupling of MCA with shape anisotropy, the c-axis gets oriented either perpendicular to ripple or parallel to the long axis of column and nanowire [16],[217]. Therefore, one promising approach would be inducing crystallographic texture aligned parallel to the ripple direction for intensifying UMA. Despite being aware that the interaction between the ion beam and thin film can alter the structure or crystallinity of material [26], no studies in the literature have considered this contribution to enhancing UMA in thin films. This may be owing to the ex-situ characterization where surface oxidation and contamination destroys film surface structure. Therefore, in order to extract genuine and unambiguous information about surface structure, film morphology and its correlation with magnetic anisotropy, the use of surface sensitive techniques and more importantly in-situ characterization is highly required.

Within this context, we have followed a sequential deposition and IBE procedure for simultaneously tuning the crystallographic texture and morphology to achieve strong UMA. In-situ reflection, high energy electron diffraction (RHEED) and X-ray diffuse scattering (XDS) provided information about film structure and morphology. At the same time, the magneto-optical Kerr effect (MOKE) gave information about magnetic anisotropy, thus making it possible to correlate the film structure and morphology with that of UMA in the film. Furthermore, unlike the conventional "top-down" approach, the present method permits textured growth throughout the whole layer of the film. Thus, the film exhibits strong UMA originating from a combined effect of MCA and shape anisotropy.

## 5.2 Preparation of thin film by sequential deposition and ion beam erosion method.

The sample preparation and characterization are done inside an ultra-high vacuum (UHV) chamber having a base pressure of $\approx 5 \times 10^{-10}$ mbar or better. It is equipped with facilities for thin film growth using the electron beam evaporation technique and in-situ characterization using MOKE, RHEED, X-ray reflectivity (XRR) and XDS measurements.

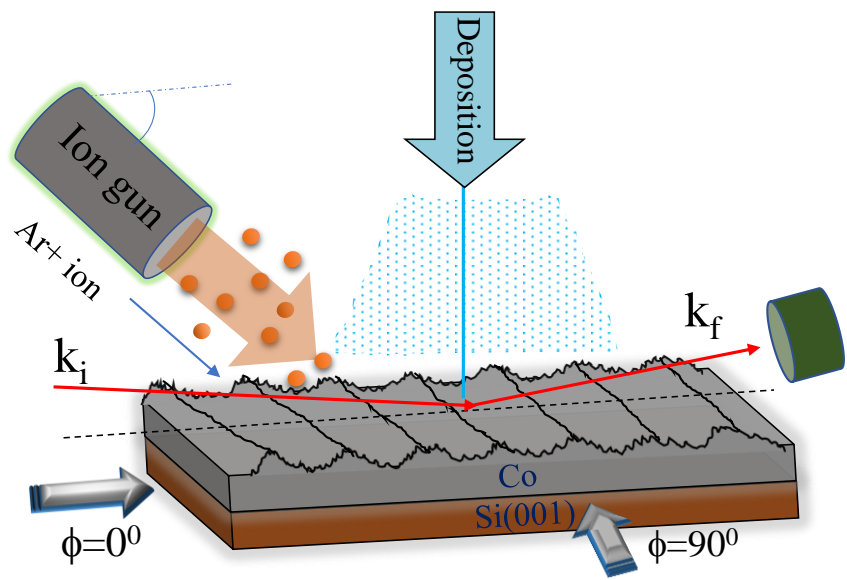

**Figure 5.1:** Schematic of experimental geometry.

In the present study, a Si (001) wafer covered with a native oxide layer is used as a substrate. Co film of thickness ≈20nm is deposited in 20 steps on this substrate with a deposition rate of ≈0.2 Å/s as readout by a water-cooled calibrated quartz crystal monitor. Each step of the sequential deposition-erosion process consists of 1nm thick Co film deposition followed by 2 minutes of film surface erosion using a 2keV Ar+ ion beam at an angle ~50º from the surface normal. The ion flux on the sample was approximately $10^{12}$ ions/cm$^2$ as separately measured by a Faraday cup. A schematic of experimental geometries with all relevant equipment and directions is presented in Fig. 5.1.

## 5.3 In-situ characterization of film surface morphology, structure and related induced magnetic anisotropy of Co thin film.

MOKE, RHEED, XRR and XDS were performed at a few intermediate steps as well as after complete deposition. In order to draw intercorrelation among structural, morphological and magnetic anisotropy in conjugation with the direction of IBE, all the measurements are performed by rotating the sample with respect to the IBE direction.

### 5.3.1 Characterization of film crystallographic structure by RHEED.

Figure 5.2(a) represents the RHEED diffraction pattern of the film taken after deposition of 1nm thick Co film. The diffraction pattern consists of symmetric, concentric and continuous Debye rings with uniform intensity distribution. This is a signature of random nucleation of crystallites which confirms the polycrystalline nature of the film. The rings corresponding to different planes are identified from their position and marked in Fig. 5.2(a). It may be noted that the first ring from the center is quite broad compared to the other rings due to the convolution of (101), (002) and (102) rings [218]. The corresponding RHEED pattern after 2min of IBE is given in Fig. 5.2(b). A drastic change in the ring pattern can be seen in the figure. It exhibits asymmetric and nonuniform intensity distribution with the presence of broken arcs along the Debye rings. It confirms the existence of the preferred orientation (texturing) of the different crystallites. So, a similar process is further followed for 20 steps which include sequential film deposition and successive IBE. The RHEED image taken after the 5$^{th}$ cycle and the whole process along ϕ=0° and ϕ=90° are shown in Fig. 5.2 (c, d) and 5.2(e, f), respectively. It is clear from the figure that the film displays pronounced texturing even after 20 steps of the process.

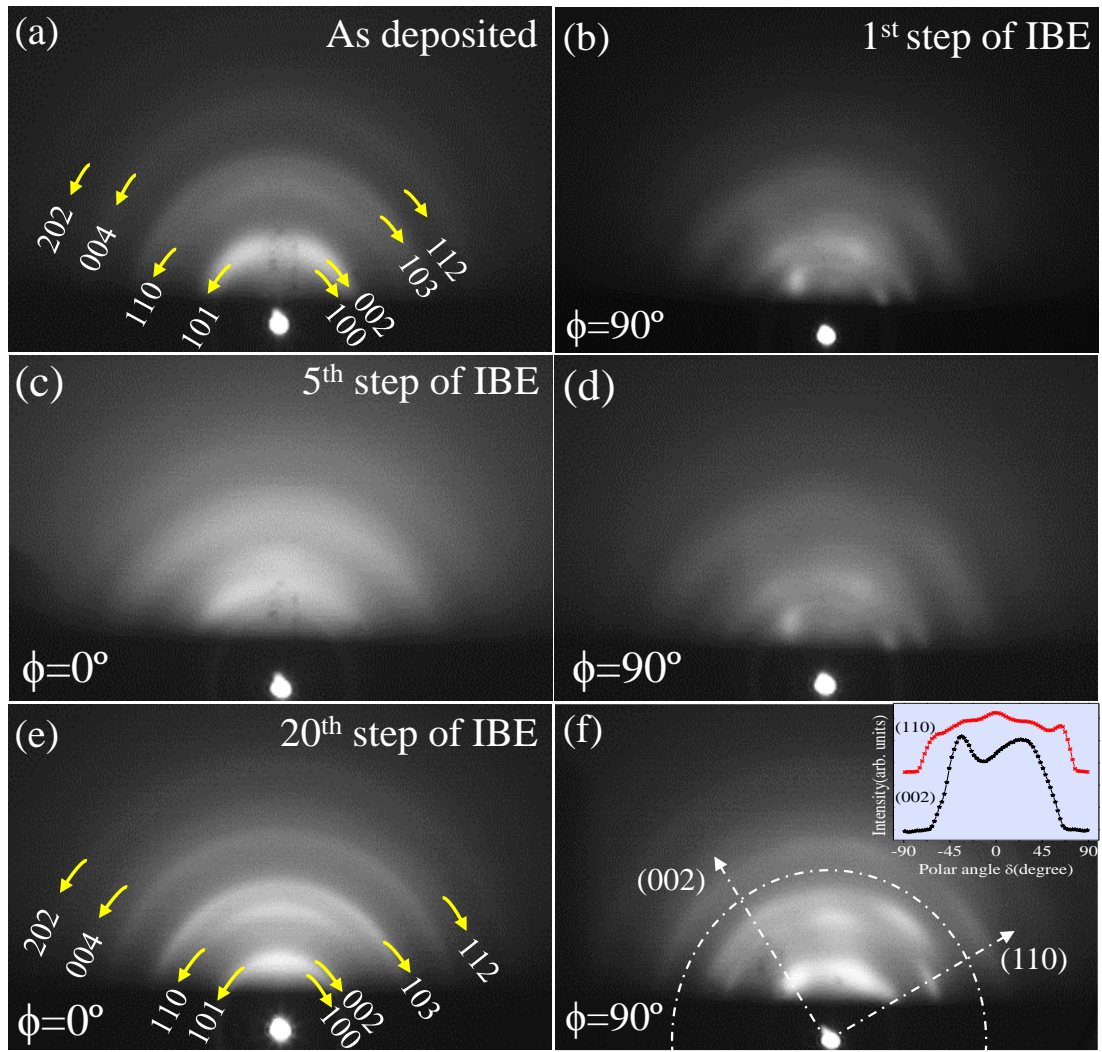

**Figure 5.2:** (a) In-situ RHEED images taken after 1nm film deposition. (b) The corresponding RHEED images after 2min of IBE. RHEED images taken (c,e) along and (d,f) across to the IBE direction after the 5th and 20th steps of deposition-erosion. The dotted line with an arrowhead in (f) indicates the surface normal of planes.

In order to give insight into the substantial arrangement of the distribution of ring intensity and texture axis, the radially integrated intensity distributions taken at the position of the (002) ring are plotted with polar angle δ in the inset of Fig. 5.2(f) where δ is defined as the angle measured from the vertical axis perpendicular to the substrate (δ=0º is at the top of the ring). This selection is driven by the fact that <002> direction is the easy axis of MCA. We found that the [110] and [002] texture axis is tilted from the surface normal by ~ 60º and ~30º, respectively, which matches the standard angle between the two directions. In the case of hcp structure, <0002> texture is thermodynamically preferred for a physical vapor deposited film because films with planes of highest atomic density oriented parallel to substrate are thermodynamically preferred due to minimization of surface free energy. Whereas, in this sequential deposition and etching method, the origin of the texture formation can be understood by the difference in the degree of the ion channeling effect or ion-induced anisotropic radiation damage in the crystal plane resulting from the oblique angle energetic Ar ion bombardment [219] and epitaxial layer overgrowth favoring the nucleation of grains with lower sputter yield orientation. For hcp structure $<11\bar{2}0>$ is the easiest channeling direction. Therefore, due to easy channeling, the probability of surviving the $(11\bar{2}0)$ grains with a crystallographic axis oriented along the IBE direction are more compared to other-oriented grains. Thus, with the help of the present method, crystallographic texture can be induced throughout the whole layer of film, and the texture axis can be tilted away from the normal of the film surface to the ion flux direction [83].

## 5.3.2 Detailed characterization of film thickness and morphology by XRR and XDS.

In-situ X-ray scattering techniques, XRR and XDS, have been used to study both the specular and diffuse scattering to get the in-plane and out plan surface morphology of the film.

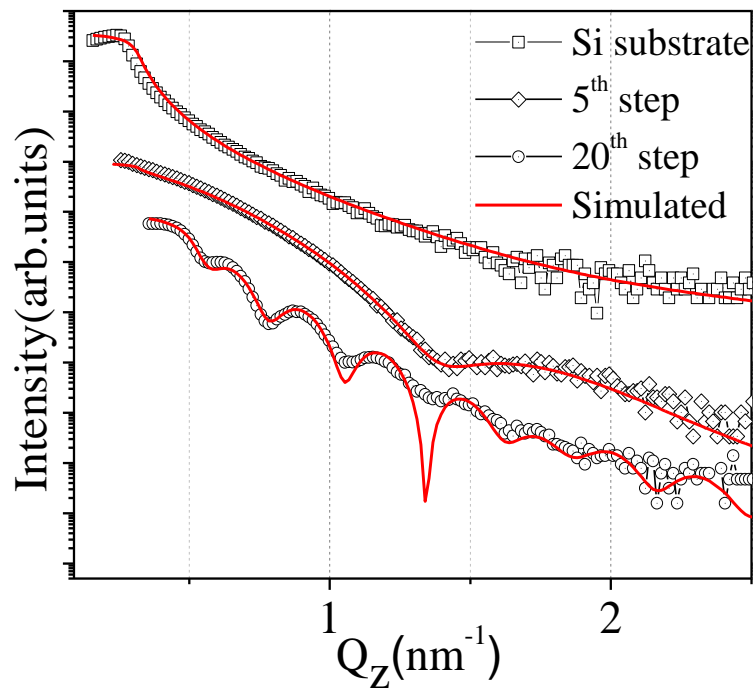

**Figure 5.3:** In-situ XRR plots of Si substrate, Co thin film at the end of 5$^{th}$ and 20$^{th}$ cycle. The hollow lines represent the experimental data, and the red lines represent the best fit to experimental data using Parratt formalism.

Figure 5.3 gives the XRR pattern of the bare Si substrate and Co thin film after the 5[th] and the final 20[th] step of the deposition-erosion process. The presence of periodic multiple peaks (Kiessig fringes) indicates good structural order and morphological uniformity within the film. XRR patterns have been fitted using Parratt's formalism [151]. The extracted root mean square (r.m.s.), roughness ($\sigma$), and film thickness (d) are as follows: $\sigma$ =0.5nm for substrate, d= 4.6nm and $\sigma$ = 0.6nm for the film after the 5[th] cycle, d=18nm, $\sigma$=0.9nm for the film after the 20[th] cycle.

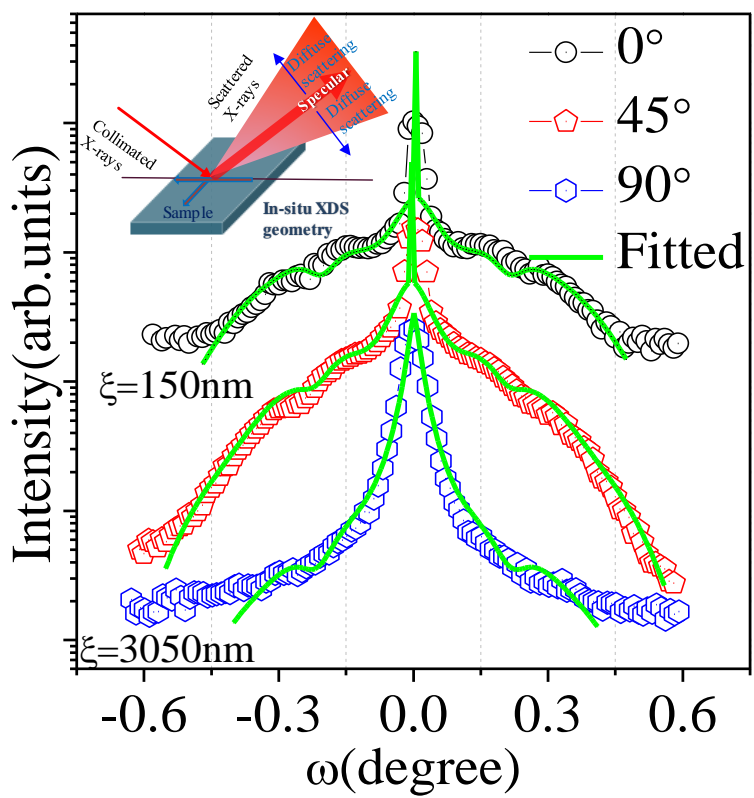

**Figure 5.4:** Plots of in-situ XDS measurement as a function of the rocking angle $\omega$ (=$\theta_i$−$\theta_{sc}$) at the fixed $q_z$ value 0.896 nm$^{-1}$ along different azimuthal angles with respect to the IBE direction. The spectra have been shifted vertically for clarity purposes. Inset shows the schematic of XDS geometry.

In order to extract statistically averaged lateral correlations of surface morphology, XDS measurements are performed along different azimuthal directions ($\phi$=0°,45°,90°) and shown in Figure 5.4. The experiment was carried out by keeping the scattering angle $\theta_{sc}$ fixed at 0.62° (corresponding $Q_z$ value = 0.896 nm$^{-1}$) and varying the angle of incidence $\theta_i$ from 0° to 2$\theta_{sc}$ [19]. A significant difference in the scattered intensity pattern suggests morphological anisotropy on the surface relative to the IBE direction. Along ϕ=90º, the XDS curve exhibits a distribution of scattered intensity only near-specular region (at ω≈ 0). In contrast, small subsidiary peaks or maxima are present along ϕ=0º on both sides of specular reflection. The intensity oscillations correspond to a characteristic length scale of periodic surface morphology. XDS data is fitted using TRDS_sl program in the framework of distorted wave-born approximation formalism to get quantitative information of surface morphology, where the distribution of height and its correlation in the *x,y* plane is expressed by the height-height correlation function [208–210] $C(x,y) = \sigma^2 \exp(-|R|/\xi)^{2h}$, where $R = (x^2 + y^2)^{1/2}$, σ is r.m.s. roughness, ξ describes length scale of correlated lateral ordering on the surface, exponent *h* is the jaggedness parameter. The fitted values of the parameters ξ and h are found to be 150nm and 0.25 for ϕ=0º, 3050nm and 0.42 for ϕ=90º, respectively. Since ξ corresponds to the length scale at which a point on the surface follows the memory of its initial value, the obtained values suggest that short and long axis of the anisotropic surface morphology is oriented along ϕ=0º and ϕ=90º respectively. It also confirms formation of periodic ripple like pattern with ridges running perpendicular to the IBE direction.

### 5.3.3 Magnetic anisotropy studied using In-situ MOKE.

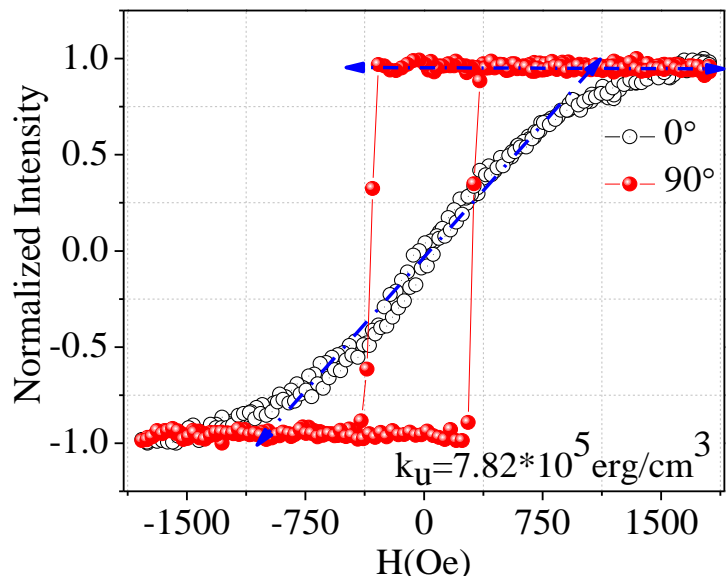

**Figure 5.5:** In-situ longitudinal MOKE hysteresis loops along ϕ=0° and ϕ=90º.

The magnetic characterization of the sample was carried out in-situ by MOKE measurements in longitudinal geometry, and analysis was focused on the determination of strength as well as the orientation of the in-plane anisotropy axis. Figure 5.5 displays the representative MOKE hysteresis loops measured with external field H applied along $\phi=0°$ and $\phi=90°$ directions. The hysteresis curve along the IBE direction displays no opening in the middle of the loop, suggesting magnetization switching in this direction is reversible and proceeds via coherent rotation in the applied field direction. However, along the $\phi=90°$ direction, the hysteresis curve exhibits very high squareness having a coercive field of 400 Oe with almost 100% normalized remanence magnetization (Mr/Ms). Mr/Ms for angular range 0° to 360°, is plotted in Fig. 5.6(a). It exhibits a regular eight-like shape which indicates a strong UMA with an easy axis of magnetization oriented 90° to the IBE direction. As the magnetization reversal displays two-fold symmetry, it has been modeled with the following cosine-like function: $\frac{M_r}{M_s} = a|\cos(\phi - \phi_0)| + b$ to extract the exact orientation and degree of anisotropy. Here, $a$ defines the degree of anisotropy, b arises due to isotropic contribution, $\phi_0$ is the angular offset between the easy axis and the zero setting of the rotational sample stage. From fitting, the obtained value of $a$ is 94% and $\phi_0$ is 3°.

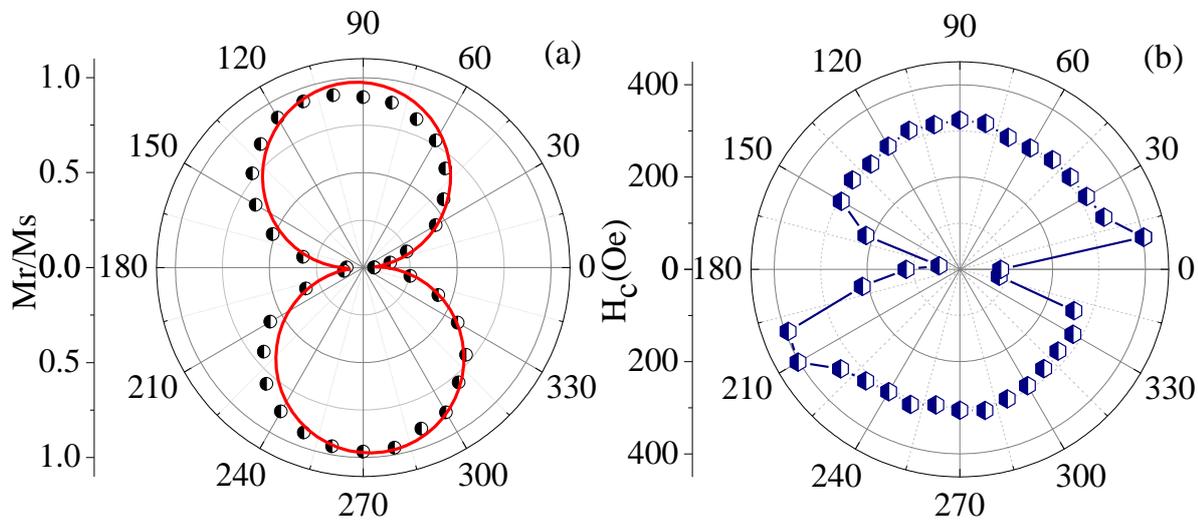

**Figure 5.6:** A polar plot of (a) the normalized magnetization and (b) coercivity.

Thus, a strong UMA is present with an easy axis of magnetization oriented almost 90° to the IBE direction. To estimate the strength of UMA, the experimental value of UMA energy, $K_U$, has been calculated using the relation $K_U = M_s H_a/2$. Here, Ms is the bulk saturation magnetization value (1422 emu/cm$^3$) of Co. $H_a$ is the anisotropy field, which has been determined by the extrapolation of the linear magnetization regime obtained along the hard axis direction at the saturation magnetisation value. $H_a =1100$ Oe is also marked by the blue line in Fig. 5.5. The value of $Ku \approx 7.82 \times 10^5$ erg/cm$^3$ is obtained, which is one order of magnitude greater than similar previous studies [15–18], [23]. For instance, Sarathlal et al.[16] have obtained $Ha \approx$ 120 Oe of 18nm thick Co film, Bukharia et al. [137] have obtained $Ha$ ≈100-150 Oe for Co with thickness ranging from 5-60nm, Liedke et al. [15] have obtained anisotropy field in the range 100-250 Oe for 2-40nm Co, Fe and Py film deposited over the rippled substrate. J. Berendt [220] obtained $Ha \approx$ 10 Oe of 20nm thick permalloy film deposited on the holographic grating. Zhang et al. [20] have achieved $Ha \approx$ 75 Oe only by IBE of Fe film. Compared to these studies, almost one order higher $Ha \approx$ 1100 Oe has been achieved. Until now, in this field, except in the present study, Miguel et al. [221] have achieved the highest value of Ha≈800Oe by directly modifying the surface morphology of Co thin film with large ripple amplitude up to 20 nm as the leading variable. The film surface is modified in all the films by single erosion step. Therefore, only the top surface is expected to be modified. No evidence of crystallographic texture could be due to the top surface contamination in ex-situ studies. Thus, it is evident that sequential film deposition and IBE of film imprint a stronger UMA. Furthermore, it also provides an opportunity to grow textured thin film on a non-textured substrate.

### 5.3.4 Discussion in light of Schlömann's theory.

It is noteworthy that the magnetostatic contribution to the total strength of induced UMA for a magnetic film with anisotropic surface roughness is evaluated from morphological parameters using Schlömann's formula [21] $Ku(shape) = 2\pi M_S^2 \frac{\pi\sigma^2}{\lambda d} = 2\pi M_S^2 (Na - Nc)$ and a good agreement between theoretical and experimental values are obtained in several previous studies [17],[23],[166]. Here, Na and Nc are the demagnetization factor along the short and long axis. $Ku = 1.4 \times 10^4$ erg/cm$^3$ is obtained using the simulated values extracted from XDS measurement. This significant difference between the experimental value obtained from the hysteresis loop and the theoretical value calculated from Schlömann's formula indicates that the induced UMA does not solely originate from shape anisotropy due to the rippled topography. Earlier it has also been observed that c-axis of Co gets aligned along the long axis of nanowire or column due to coupling of magneto-crystalline and shape anisotropies that favors a coherent rotation. Co has an intrinsic magneto-crystalline easy axis along <002> direction. As revealed from RHEED measurement, this axis is textured, in the film plane along ripple direction. Therefore, both shape anisotropy and MCA gets coupled together in the same direction and enhances the strength of UMA.

In order to provide visual evidence of the formation of ordered rippled morphology, a separate film has been deposited in identical conditions. The corresponding AFM image of the film surface is shown in Fig. 5.7. It exhibits periodic ripples with grooves running perpendicular to the projection of IBE direction on the film surface and confirms the formation of unidirectional correlated rippled morphology in the film that also complements the earlier in-situ XDS measurement.

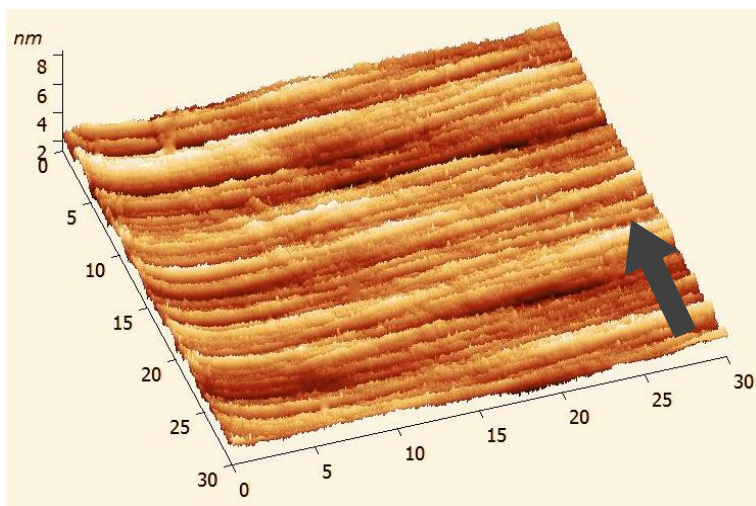

**Figure 5.7:** The AFM image of the Co film surface over an area of $30\times30\mu m^2$. The grey arrow indicates the projection of the ion beam on the film plane.

## 5.4 Temperature-dependent in-situ study

We further extended our work on the temperature-dependent study of the film crystallinity, morphology and related UMA to understand the origin of this large induced UMA. We annealed the sample from room temperature (RT) to 350ºC at an interval of 50ºC. At each temperature, the sample was heated for 30 minutes, whereas all the measurements were performed by cooling down the sample to RT.

The RHEED images taken at 200°C and 350˚C along $\phi=0º$ and $\phi=90º$ to the IBE direction are shown in Fig. 5.8(a-d). The RHEED pattern at 350°C consists of symmetric Debye rings of uniform intensity distribution characteristics of the isotropic polycrystalline nature of the film. Thus, it is revealed that annealing removes the modulation of the rings' intensity as well as annihilates the texturing developed through this sequential etching and deposition method.

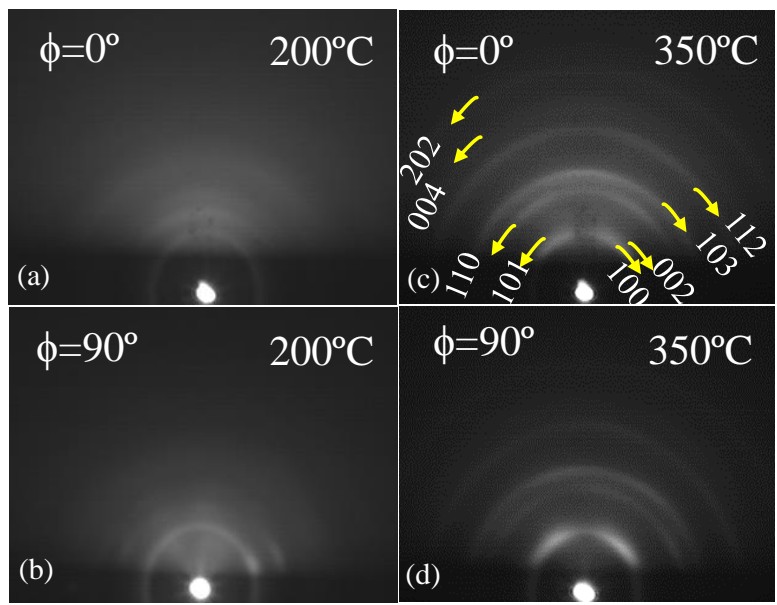

**Figure 5.8:** RHEED images taken at (a,b) 200°C and (c,d) 350°C along φ=0° and φ=90° directions.

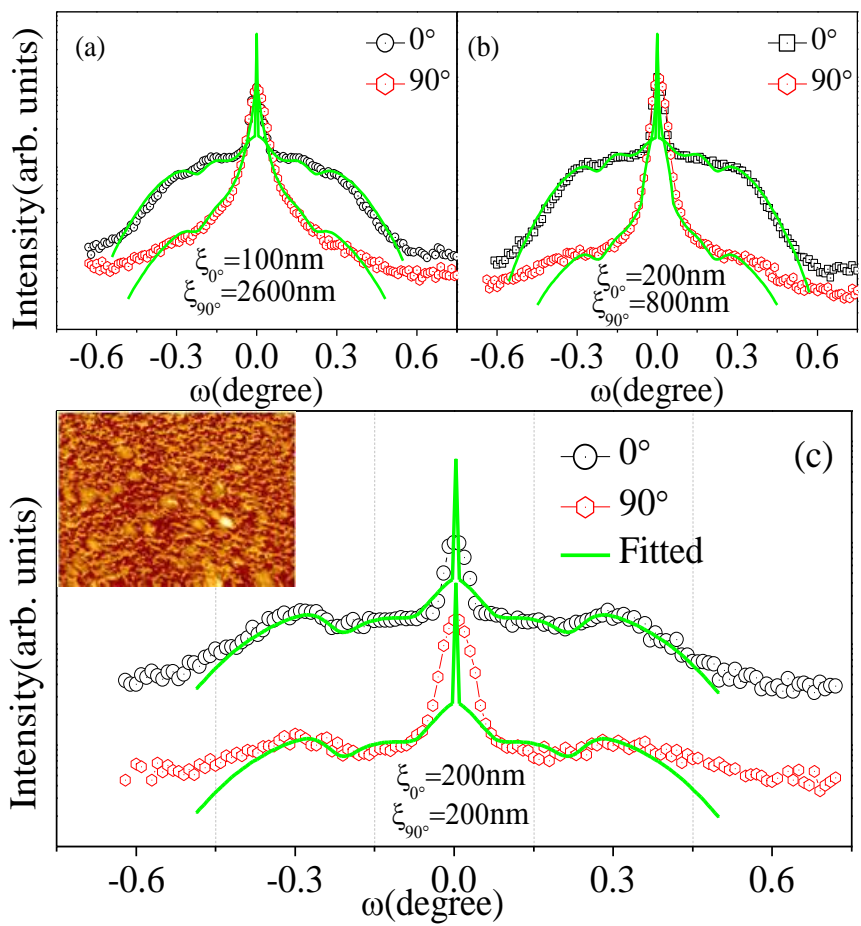

**Figure 5.9:** XDS curves taken at (a) 200°C, (b) 300°C, (c) 350°C. The inset of (c) represents the AFM image of the film surface taken after annealing at 350°C.

The XDS measurement performed at 200°C, 300°C and 350°C is presented in Fig. 5.9. From the graph, one can observe that the asymmetry of the XDS curve between ϕ=0° and ϕ=90° direction gradually decreases with an increase in temperature. The XDS measurement at 350°C, as shown in Fig. 5.9c, exhibits a symmetrically developed identical intensity pattern around the central peak. This behavior is characteristic of relaxation of the unidirectional nanopattern present on the surface due to migration of the Co atom and corresponds to isotropic surface morphology. The corresponding surface topography of the film as imaged by AFM is presented in the inset of Fig. 5.9c. No directional morphology is observed. From these results, we conclude that both the shape asymmetry and the preferential grain distribution are eliminated from the film by annealing. More importantly and consequently, it also confirms that the observed behavior of UMA is caused by the film texture and morphology.

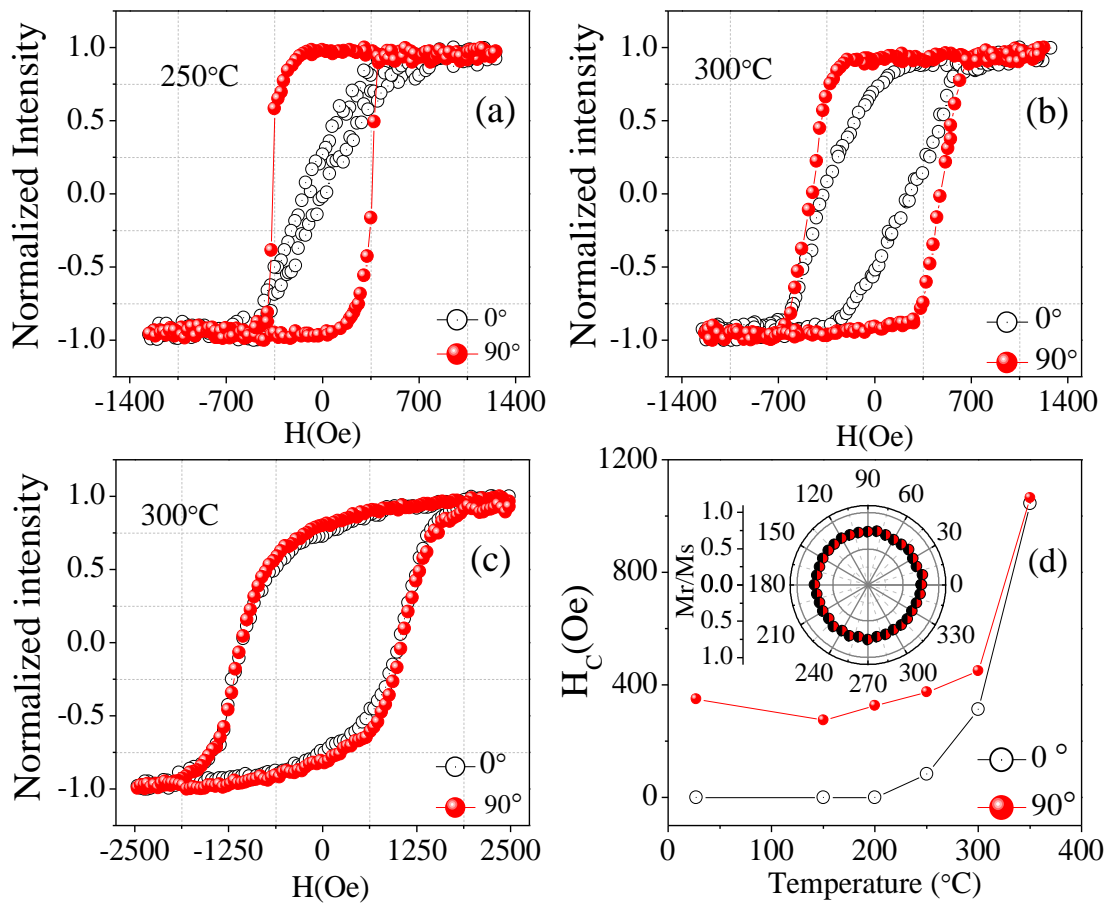

**Figure 5.10:** MOKE hysteresis loop taken along ϕ=0° and ϕ=90° after annealing for 30 minutes at (a) 250°C, (b) 300°C and (c) 350°C. (d) Variation of coercivity with annealing temperature. The inset shows the corresponding polar plot of normalized remanence at 350°C.

In order to study the evolution of UMA, MOKE measurement was performed at different annealing temperatures along with RHEED and XDS measurements. The MOKE hysteresis loop taken at 250°C, 300°C and 350°C is presented in Fig. 5.10(a-c). The variation of coercivity at each annealing temperature as extracted from the MOKE hysteresis loop is also presented in Fig. 5.10(d). We observe that at temperatures around 250ºC, the difference between coercivity values starts to decrease, and at 350ºC, it disappears. The corresponding MOKE hysteresis loop in Fig. 5.10(c) also displays identical hysteresis behavior. Furthermore, the polar plot of normalized remanence at 350ºC, as presented in the inset of Fig. 5.10(d), has converted from a dumble shape into a circular shape. This further confirms the isotropic magnetic nature of the film.

## 5.5 Comparative study with a post-growth ion beam eroded thin film.

Another Co thin film of comparable thickness was deposited on the Si substrate to compare the strength of induced UMA. Its surface was eroded by identical ion beam parameters and all the characterizations are done in-situ.

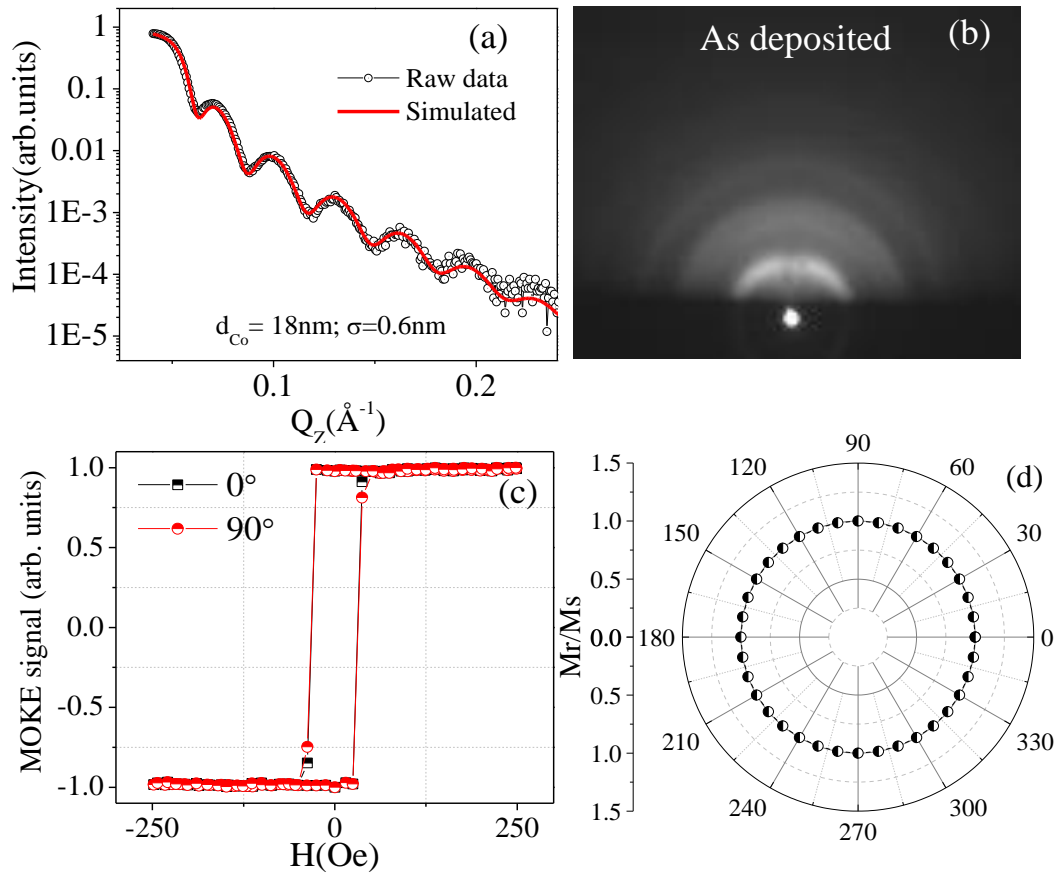

**Figure 5.11:** (a) In-situ XRR pattern of Co thin film deposited on Si substrate. In-situ (b) RHEED

images and (c) MOKE hysteresis loop of the as-deposited Co thin film. (d) Polar plot of normalized remanence.

The in-situ XRR pattern and RHEED image of the film is shown in Fig. 5.11(a) and 5.11(b), respectively. The film thickness and r.m.s. roughness, as obtained by fitting the experimental data, is found to be 18nm and 0.6nm, respectively. The RHEED pattern exhibits uniform intensity distribution along the Debye rings. Thus, it indicates that the film grows in the polycrystalline state without any crystallographic texture. The corresponding MOKE hysteresis loop was taken along and across to the IBE direction as presented in Fig. 5.11(c), and it displays identical magnetization switching behavior. Furthermore, the polar plot of normalized remanence, as shown in Fig. 5.11(d), confirms that magnetic anisotropy is absent in the film.

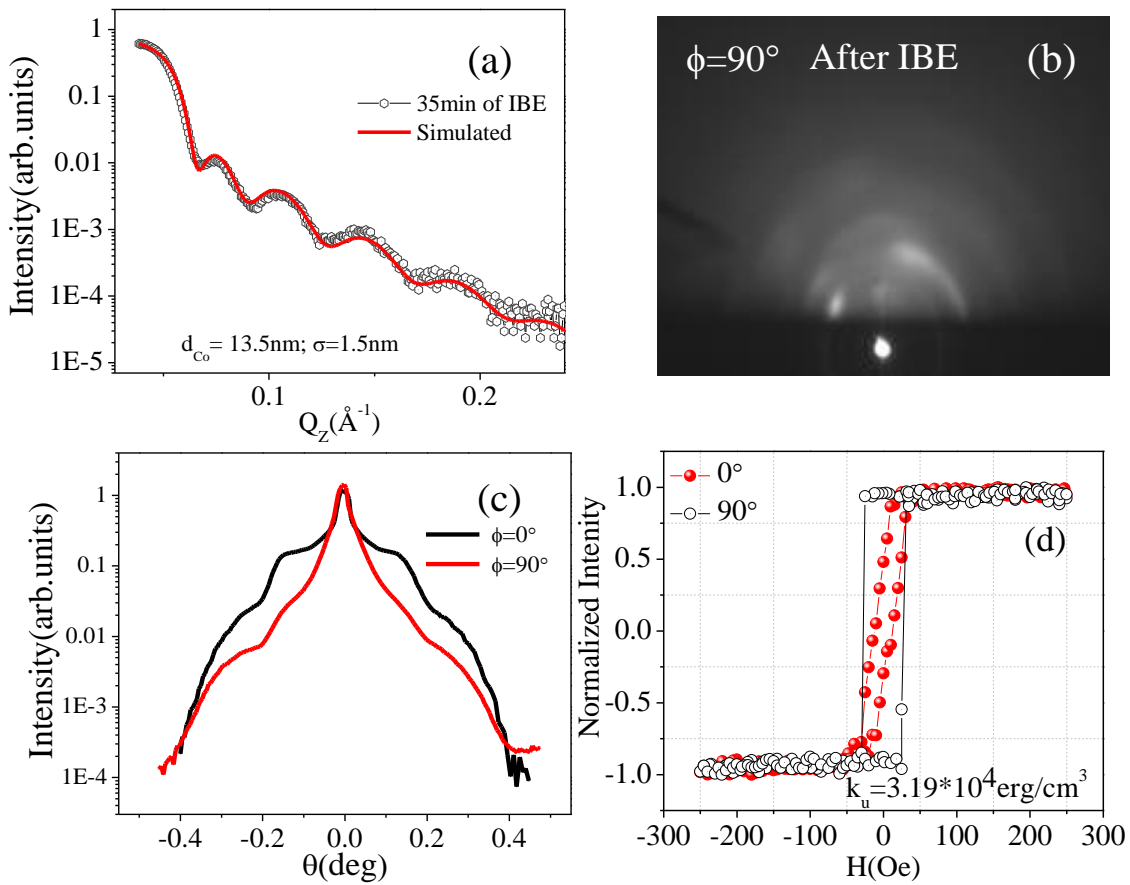

**Figure 5.12:** In-situ (a) XRR pattern and (b) RHEED images taken along $\phi=90°$ after IBE for 35 minutes. (c) X-ray diffuse scattering taken along $\phi=0°$ and $\phi=90°$ after IBE. (d) In-situ MOKE hysteresis loop taken along and across to the IBE direction.

Figure 5.12a shows the reflectivity pattern of the thin film after 35minute of IBE. The residual film thickness and r.m.s. roughness is found to be 13.5nm and 1.5nm, respectively. The corresponding diffuse scattering pattern taken along and across the IBE direction is shown in Fig. 5.12b. The asymmetric distribution of scattered intensity indicates the formation of asymmetric morphologic patterns on the film surface. The corresponding RHEED pattern, as shown in Fig. 5.12c, also displays clear development of surface texturing after IBE. The MOKE hysteresis loop taken along and across to the IBE direction is presented in Fig. 5.12d. A clear anisotropy behaviour with an easy axis of magnetization oriented perpendicular to the IBE direction is observed. The strength of UMA has also been calculated and found to be $3.19 \times 10^4$ erg/cm$^3$. Compared to this, the strength UMA induced by the sequential deposition-erosion process is one order magnitude higher.

## 5.6 Conclusions

Co film has been prepared by sequential deposition and IBE process, and its morphology, surface texture and magnetic anisotropy have been characterized in-situ. The film grows in a highly biaxially textured polycrystalline state and exhibits strong UMA, roughly one order larger than that similar kind of the previous studies. Detailed analysis reveals that the induced UMA not only originates from shape anisotropy but MCA due to crystallographic texture also contributes to enhancing the strength of UMA. Furthermore, a temperature-dependent study indicates that the UMA disappears due to the relaxation of crystallographic texture and anisotropic surface morphology. Therefore, the novelty of the present approach is the ability to imprint film texture throughout the whole film layer along with surface morphology modification, consequently not restricting the source of UMA limited to the thin surface region of a polycrystalline ferromagnetic film for flexible tuning and enhancement of the strength of UMA.

# Chapter 6

# Ion beam induced modification on Epitaxial thin film

## Understanding and controlling magnetic properties

In the previous chapters, we have demonstrated that thin film deposited over rippled substrate displays UMA due to shape anisotropy, and direct ion beam erosion of thin films exhibits additional UMA due to both shape and induced crystalline anisotropy. This chapter aims to extend this notion by combining induced anisotropy with inherent crystalline anisotropy of epitaxial thin film. So, on the one hand, we can study more about how these two different types of anisotropy interplay and affect each other. On the other hand, a successful combination of the two will give us a new opportunity to modify the magnetic properties of a material. A proper material system that satisfies a number of characteristics is required to explore these two features. First, it must be simple to deposit the desired magnetic material on top of substrates and must show crystalline anisotropy in the same plane. Symmetry in the intrinsic anisotropy must vary from the induced uniaxial anisotropy so that the two may be clearly distinguished. The iron thin film matches these requirements, which is easy to grow and displays an intrinsic four-fold magneto-crystalline anisotropy due to its cubic structure. Compared to the previous chapters, this chapter deals with an epitaxial thin film, bringing some exciting and significant results. This chapter aims to study the different magnetic switching behavior arising from the interplay of intrinsic crystalline anisotropy and induced uniaxial magnetic anisotropy. This chapter is divided into two parts. The first part describes the growth and magnetic anisotropy of 20nm thick Fe film grown on ion eroded Ag(001) substrate that has been studied in-situ. On the other hand, in the second part, IBE was directly applied in different cycles on epitaxial Fe thin film grown on a MgO(001) substrate to study the evolution of its surface structure and magnetic anisotropy.

**Part I:**

    **Magnetic anisotropy of epitaxial thin film on ion beam eroded substrate**

**Part II:**

**Magnetic anisotropy and surface reconstruction of epitaxial thin film induced by IBE**

# Chapter 6 I

# Magnetic anisotropy of epitaxial thin film on ion beam eroded substrate

**Magnetic anisotropy of epitaxial Fe film on ion-sculpted Ag(001) substrate"**

*An In-situ Investigation*

This chapter describes the growth and magnetic anisotropy of Fe thin film grown on ion-sculpted Ag(001) substrate, which has been studied in-situ using RHEED and MOKE. Before this, the temperature and temporal evolution of the rough-to-smooth transition of the ion beam-eroded Ag (100) surface were studied using RHEED by the passage from spotty to streaky RHEED patterns at two different temperatures. Fe film is found to grow epitaxially on the Ag surface and exhibits epitaxial island formation up to ~1nm film thickness. Ion beam sculpting of the substrate surface prior to the deposition of the Fe film induces an UMA in Fe film, which couples with the cubic four-fold anisotropy of bcc-Fe, leading to the multiple-step jump in magnetic hysteresis loops. Magnetic switching and its correlation with the domain dynamics of the film, as studied by the Kerr microscopy technique, revealed a striking difference in domain structure depending upon the direction of the applied magnetic field with respect to crystallographic direction. It is found that the magnetization reversal takes place via a "single-step jump" mechanism through the sweeping of 180° domain walls, whereas the combination of coherent rotation and the sweeping of 90° domain walls is responsible for a "double step jump" magnetization reversal.

# Chapter 6 Part I:

# Magnetic anisotropy of epitaxial thin film on ion beam eroded substrate

**Magnetic anisotropy of epitaxial Fe film on ion-sculpted Ag(001) substrate**

**An In-situ Investigation**

## 6.I.1  Introduction

## 6.I.2  Evolution of surface morphology studied using RHEED and substrate patterning.

## 6.I.3  Growth and characterization of Fe thin film.

6.I.3.1 Real-time epitaxial Fe thin film growth monitored through RHEED.

6.I.3.2 In-situ study of magnetic anisotropy using MOKE.

6.I.3.3 Ex-situ study of domain dynamics by KERR microscopy.

## 6.I.4  Discussions

## 6.I.5  Conclusions

## 6.I.1 Introduction

The interest in the growth and magnetic properties of single-crystalline thin films on metal, semiconductor and insulating substrates has attracted much attention in recent years for their technical application and fundamental research in magnetism. This is mediated by the availability of high-quality single crystalline substrate and advancements in growth and probing technology. Furthermore, such systems can provide unique magnetic properties, e.g., giant-magneto resistance [222], enhanced size of the magnetic moments [223], magnetic anisotropy [224] and magnetic hyperfine fields [225], which are otherwise not possible in their bulk counterpart. Furthermore, the peculiar properties of these 2D materials make them essential for the next generation spintronics devices such as hybrid semiconductor-magnetic memories [226], magnetoelectric coupling devices [227], magnetic field sensors [228], spintronic terahertz emitters [229], tunnel magnetoresistance [230], giant-magneto resistance [231], magnetic tunnel junction devices [232], spin valve [233] etc.

In view of the above facts, Fe thin films on different single crystal substrates such as MgO [234], GaAs [235], Au [236], $BaTiO_3$ [237] and Cu [238] have been studied extensively. Among all Fe film on Ag, single crystal is of particular importance since they form a well-defined sharp interface as they are entirely immiscible. In addition, the in-plane lattice mismatch is only 0.8%, allowing the growth of high-quality epitaxial superlattice films [239] and exhibiting well-defined cubic four-fold magnetic anisotropy (MA) due to its crystallographic structure. However, this system often presents an extra Uniaxial magnetic anisotropy (UMA) due to film-substrate lattice mismatch, substrate miscut, and film-substrate bonding [240]. This extra UMA, along with cubic four-fold MA in the film, gives rise to several exciting phenomena such as the split of the hysteresis loop [187],[241], field-induced spin reorientation transition [24] etc. As the presence of MA in thin films dictates the switching characteristics of a material in the presence of an external magnetic field, therefore, it is crucial to understand the nature of MA to control the magnetization reversal process according to the requirement. Furthermore, the strength and symmetry of MA must be tailored and appropriately optimised for its proper realisation in thin film-based devices.

It is found that MA is strongly sensitive to interfacial morphology and symmetry [241][22,169,242]. Therefore, artificial tailoring of interfacial morphology has dimensions comparable to that of magnetic length scales. e.g., exchange length, domain wall width etc., significantly impacts the magnetization reversal process. In recent past years, oblique incidence ion beam erosion (IBE) is emerging to be a handy and powerful tool to modify the morphology of the substrate [138,186,243,244] as well as the film [20,136,245] surface. Magnetic film in epitaxial or polycrystalline state grown on such templates or post-growth IBE of the film is found to display an additional UMA [138],[185],[241],[22] and is responsible for the complex magnetization reversal process. Here the switching process of magnetization is found to be sensitive to different crystallographic and anisotropic axes. For example, Liedke et al. [24] have found thickness-dependent competing volume and surface contribution to the induced UMA in Fe, Co and $Ni_{81}Fe_{19}$ films deposited on rippled Si substrates. Chen et al. [188] have studied morphology-induced UMA in Co film due to ion erosion of MgO substrate. Bisio et al. [241] have isolated step contribution to the UMA in nanostructured Fe/Ag film. Moroni et al. [187] have studied the evolution of MA in nanostructured Co/Cu system with ion dose. Sekiba et al. [246] have tuned MA by the formation of nanoscale ripple through ion sculpting of Co/Cu thin films. However, in most of these studies, the preparation of substrate or film template and its characterization is done ex-situ. Therefore, interface oxidation and adsorption of contamination is an issue. Besides this, IBE is predominantly applied on semiconducting and insulating substrates, mainly Si and MgO is used, which gets amorphized after ion irradiation, hence restricting the growth of the epitaxial film. On the contrary, a single-crystalline metal substrate does not get amorphized after the ion pact and opens up the possibility of epitaxial magnetic film growth on such a template. Therefore, this type of system can be used as a potential candidate where competition or interplay of magneto crystalline anisotropy (MCA) and shape anisotropy can be studied. Moreover, till date magnetic characterization is only centered on MOKE measurement, thereby magnetic domain structures of these systems remain quite unexplored. Therefore, a clear understanding on the magnetic reversal process of single layer film is prerequisite for understanding magnetization reversal behaviour and separation of inter layer coupling in case of multilayer film.

In the present work, epitaxial Fe film is grown on an Ar+ ion sculpted Ag(001) substrate to induce additional UMA in Fe film by modifying substrate surface morphology. Cleaning of the substrate and film growth is monitored through reflection high energy electron diffraction (RHEED), whereas magnetic anisotropy and magnetization reversal mechanism of the Fe film have been studied in detail using in-situ magneto-optical Kerr effect (MOKE) and ex-situ Kerr microscopy. Present work provides details of Fe growth on ion eroded Ag crystal and made it possible to correlate the magnetic properties with that of morphology and structure of the film. In contrast to the earlier similar studies, where only macroscopic magnetization switching processes are discussed, current work shows that a combination of macroscopic MOKE and microscopic Kerr microscopy provides insightful enlightenment of the often-complex reversal mechanisms rising in epitaxial thin films.

## 6.I.2. Evolution of surface morphology studied using RHEED

In the present study, Ag(001) substrate purchased from Bayville Chemical Supply Co., Inc. having dimensions of 15 mm diameter × 2 mm height with two faces polished to 1 nm, is used as a substrate. Present experiments have been performed inside an ultra-high vacuum (UHV) chamber with a base pressure of ~$5\times10^{-10}$ mbar or better to avoid oxidation and absorption of contaminations on the substrate and film surface. The substrate is mounted on a sample holder attached with heater assembly on a high-precision rotatable UHV-compatible manipulator.

Figure 6.I.1 gives the electron diffraction patterns (RHEED images) of the Ag crystal surface before (a) and after (b) surface the cleaning process. Depending on the energy of the electron and target material, the electron's mean free path is a few nanometers to tens of nanometers in solids because of the large scattering cross section [247]. This makes the RHEED technique surface sensitive, where the electrons probe even smaller depths (only a few topmost atomic layers) because of the small incident angle. Therefore, the absence of an electron diffraction pattern before the surface cleaning process may correspond to the contaminated or oxidized thin layer at the crystal surface. The crystal surface was cleaned by several cycles of ion beam erosion process using ultra-high-purity 99.999% Ar+ gas and subsequent annealing at 600°C for 30 minutes. This process was continued until a sharp RHEED pattern was observed. The diffraction pattern in Fig. 6.I.1(b) is dominated by five relatively narrow and intense streaks corresponding to (02), (01), (00), (0$\bar{1}$) and (0$\bar{2}$) Bragg reflections of the Ag surface. After ion beam erosion, the well-defined streaks pattern in Fig. 6.I.1(b) indicates that the cleaning process has efficiently removed the adsorbed contaminated surface layer. In the reflection geometry, the bottom part of the diffraction image is shadowed by the substrate; therefore, all RHEED images are seen with only the upper part of the diffraction pattern. The straight-through beam in the shadow part is the incident electron beam without hitting the substrate [84]. The lattice constant, calculated from the horizontal separation of the streaks, is found to be (4.075 ± 0.04) Å, which corresponds to the fcc Ag structure. Here,

a minor deviation in our measured lattice constant from the standard value of 4.085Å [248] may be due to the uncertainty in the measured distance of the RHEED screen from the sample.

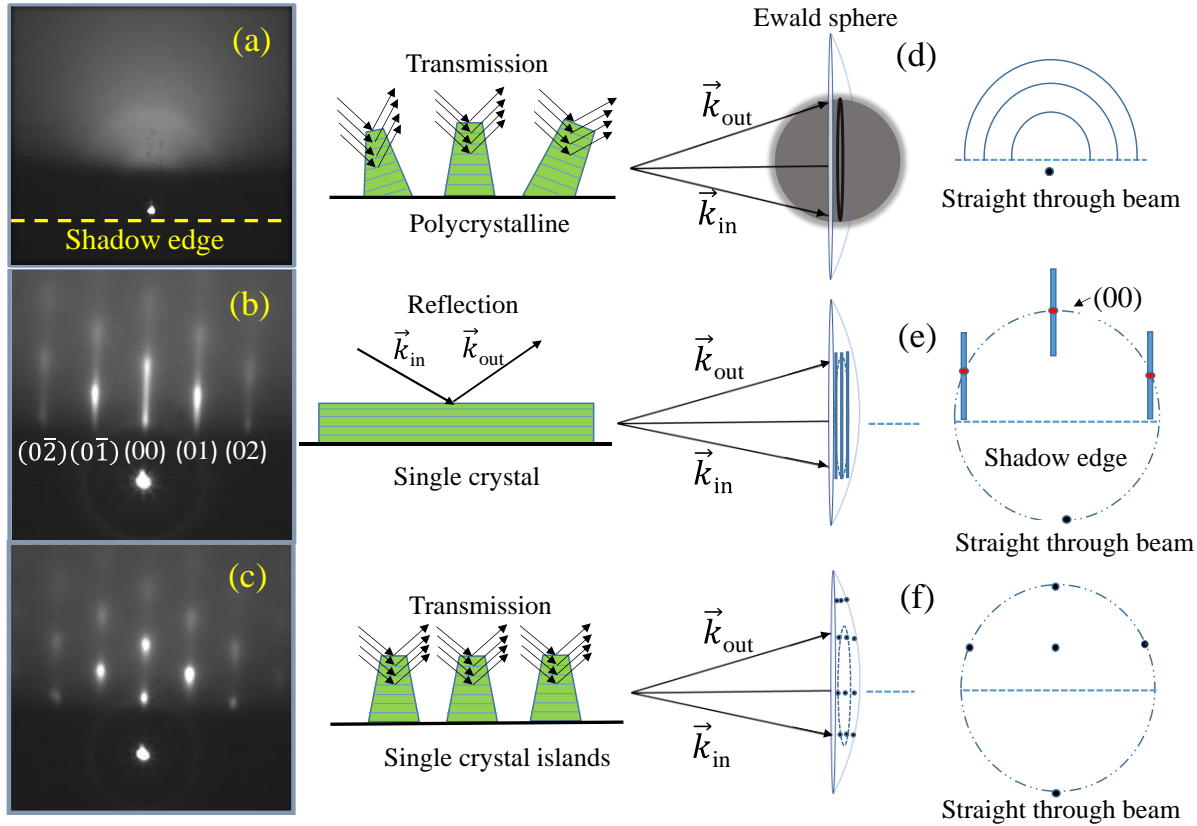

**Figure 6.I.1:** RHEED images of Ag (100) substrate (a) before and (b) after the cleaning process as mentioned in the text. The schematics of electron-scattering geometries of (d) polycrystalline film, (e) single-crystal with a smooth surface, and (f) single-crystal with islands. $k_{in}$ and $k_{out}$ are the wave vectors of the incident and scattered electron beams, respectively. The Ewald sphere constructions of electron scattering and the possible RHEED patterns are shown in front of the respective figure [84],[202].

It may be noted that if the ion beam erosion process is continued without subsequent annealing, the streaks get broken and have converted into spotty patterns. A RHEED image comprised of spots, as shown in Fig. 6.I.1(c) clearly indicates a rough surface [202]. As shown schematically in Fig. 6.I.1(f), the RHEED patterns are formed by the transmission of electrons through three-dimensional atomic arrangements of these small 3D structures on the surface. It is also clear that the surface is monocrystalline, as the spotty RHEED patterns are arranged in a well-defined manner on the vertical streaks, which correspond to the parallel atomic planes on the crystal surface (in the case of a rough surface, the electrons can enter into ''individual monocrystalline islands'' through their side surface). Therefore, we have studied the rough-to-smooth transition of the ion beam eroded Ag (100) surface using reflection high-energy electron diffraction (RHEED) at two different temperatures. Crystal surface morphology self-evolution with time, after ion beam sputtering at an ion energy of 2 keV for 30 minutes, is continuously monitored by collecting RHEED images. Figure 6.I.2 shows RHEED patterns of Ag surface taken at various times viz. 5 to 140 minutes (along the [110] directions). It is clear from the images that initially, the diffraction consists of spotty patterns, which become sharper and streakier with increasing time.

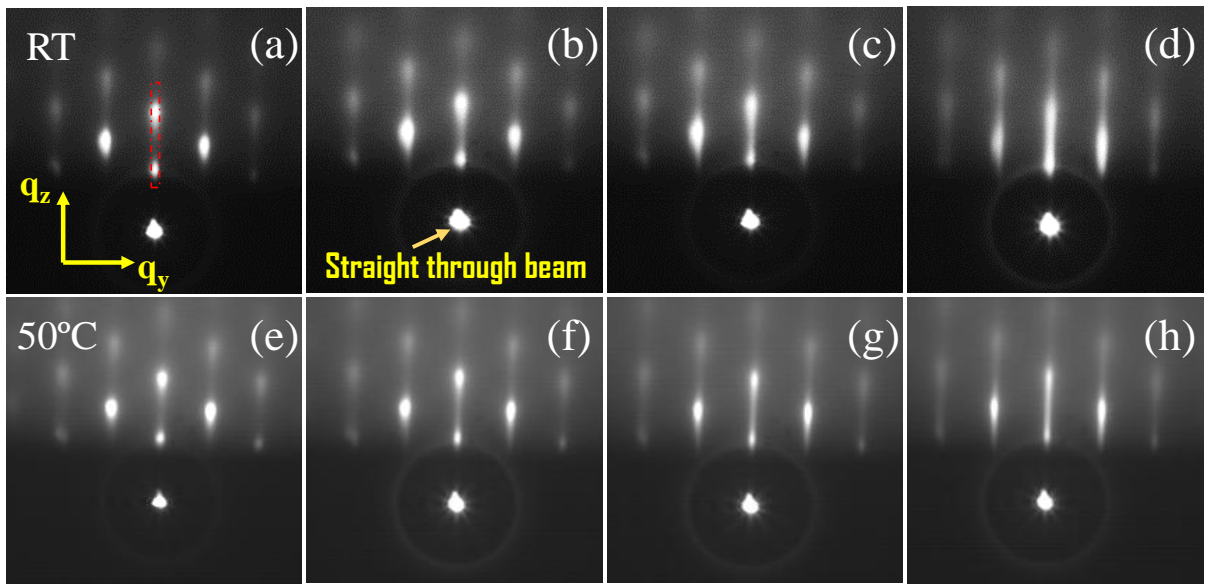

**Figure 6.I.2:** Representative RHEED images of an Ag single crystal (along [110] direction) taken at different times after 30 min ion beam erosion, keeping the substrate temperature fixed at (a through d) RT and (e through h) 50°C.

To understand it more clearly, as marked in one of the images in Fig. 6.I.2, $q_z$ vs intensity plots were extracted from one of the streaks for all RHEED images and are plotted in Fig. 6.I.3. The two peaks in the plot correspond to the intensities of the spots crossing the marked area, where the momentum transfer along the z direction is levelled as $q_z$. The increasing relative intensity in the intermediate portion between two spots, where the spotty and wide diffuse streaks change into sharper streaks, indicates the change in surface morphology from the 3D rough pattern to a 2D-like smooth surface at RT. Notably, for a sample temperature of 50°C, spotty areas and streaks in the RHEED patterns are sharper than in the sample at RT because of the stress relaxation and improvement of the crystal structure [249]. Notably, the morphology change from the 3D rough pattern to 2D-like smooth surface morphology for the crystal at 50°C takes place earlier than the sample at RT.

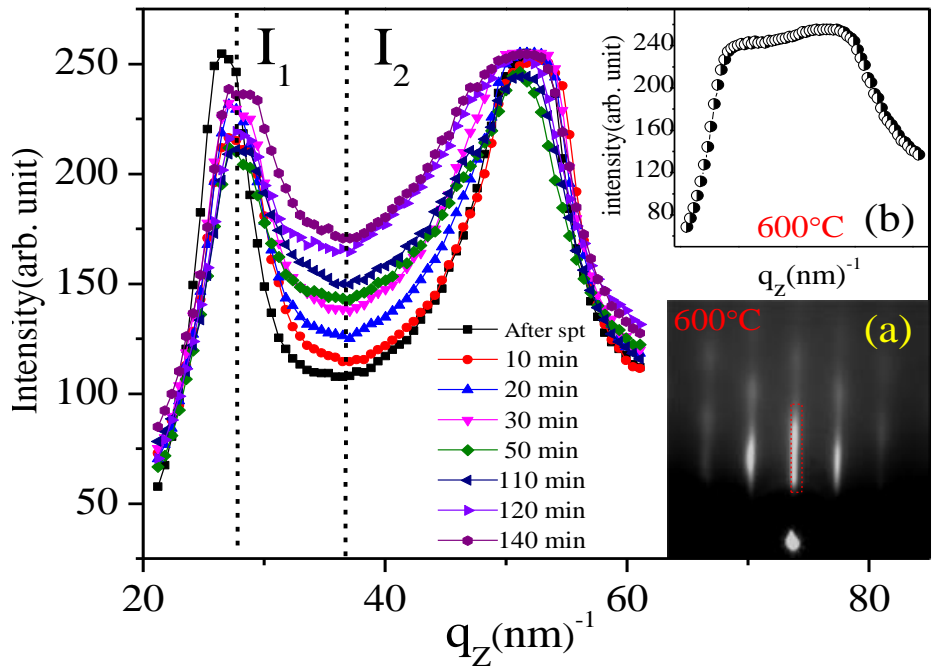

**Figure 6.I.3:** Representative $q_z$ vs intensity plots extracted from one of the streaks in vertical direction for all RHEED images. Insets in (a) and (b) correspond to the measurements after annealing at 600°C for 1 h.

To obtain the self-evolution of the morphology or smoothness with increasing time, $f_r$, defined as $(I_1/I_2)$, was calculated from each RHEED image. As shown in Fig. 6.I.3, $I_1$ is the intensity of the brightest spot in the central streak, and $I_2$ is the reflected electron beam intensity in the intermediate area between the two nearest spots in the central streak. The inset in Fig. 6.I.3 corresponds to the measurements after annealing at 600°C for 1 hour. Here the value of $(I_1/I_2)$ is close to 1, which confirms the smoothest surface. In the case of a perfectly smooth surface (almost uniform intensity distribution in the streak), $f_r$ should be close to unity; therefore, the quantity $(1-f_r)$ is a measure of the relative smoothness of the surface. Values of $(1-f_r)$ obtained from Fig. 6.I.3 are plotted in Fig. 6.I.4. As shown by the fitted data, one may note that the $(1-f_r)$ decreases almost exponentially with increasing time. With a further increase in time, no significant changes were observed after 120 minutes for an Ag (100) crystal kept at RT. For a temperature of about 50°C, the observed trend of a decrease in $(1-f_r)$ was found to be similar to that at RT, but the achieved smoothness value was about 12% more. The interaction of a low-energy ion beam with the solid surface leads to various physical and chemical phenomena responsible for various defects [250][251] on a crystal surface, including misorientation, steps, partial surface disorder and twinning. Therefore, the evolution of the solid surface morphology during ion beam interaction is governed by the interplay between the dynamics of sputtering and smoothing due to material transport during surface diffusion [252][253]. When a surface is rough after sputtering [254][255], it is difficult to determine the self-material transport and the process involved in surface evolution at room temperature on long-term scale. Nevertheless, the qualitative study of smoothing kinetics and time evolution using RHEED under UHV condition is important for understanding not only the single-crystal surface but also epitaxial thin films where the kinetics of the surface processes controls the surface reactions and improves the material properties [256][257]. In the present case, surface self-evolution after sputtering, when the temperature was low for the surface to achieve thermal equilibrium, diffusion and relaxation of randomly distributed defects and vacancies, determined the surface morphology and roughness [258]. With an increase in temperature, the increased surface diffusion led to a smoother surface. This approach allowed us to follow the qualitative time evolution of the surface morphology of the Ag (100) surface, which underwent a self-transition from a rough-to-smooth surface (3D rough surface to a 2D-like smooth surface). Once the saturation in smoothness was reached, no significant changes in surface morphology with time were observed.

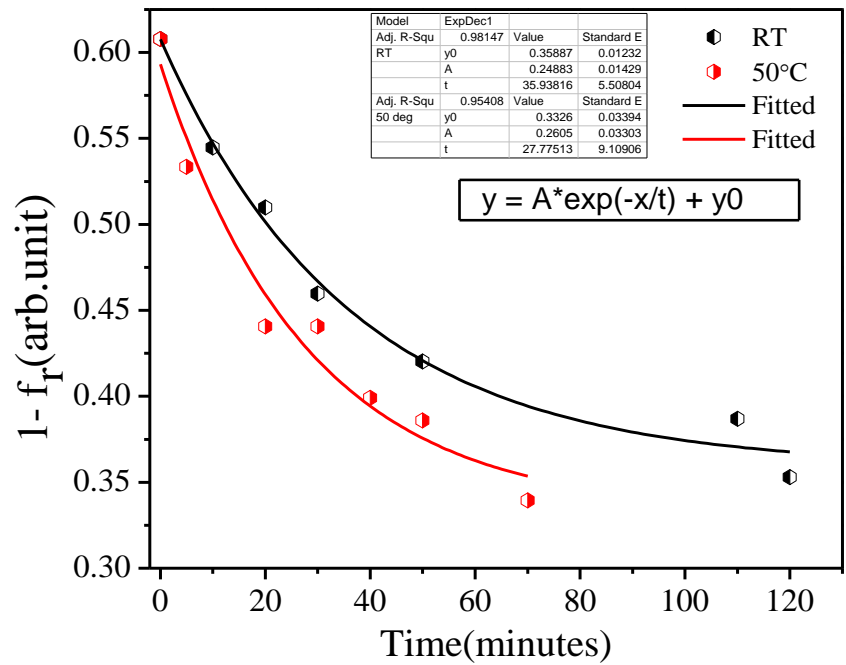

**Figure 6.I.4:** A measure of smoothness (1-$f_r$) plotted with time for RT and 50°C.

To provide microscopic evidence of the formation of correlated morphology, the Ag(001) substrate was ion eroded along [100]$_{Ag}$ direction with Ar+ ion of energy 2keV for 20 minutes at an angle of incidence 50º from the surface normal. After IBE, the substrate was removed from the UHV chamber for AFM measurements. The surface morphology of the sputter-etched surface was imaged by Bruker BioScope Resolve atomic force microscope in contact mode, and image processing was performed using WSxM software [259]. The AFM images of Ag single crystal substrate taken before and after the IBE process is shown in Fig. 6.I.5. The differences in their morphology between both the images are visible; step or anisotropy in the surface morphology (ripple-like features) oriented perpendicular to the ion beam projection is observed. The morphology has r.m.s. roughness ≈1.89 nm and average periodicity (average correlation length of the morphology) ≈101.2 nm.

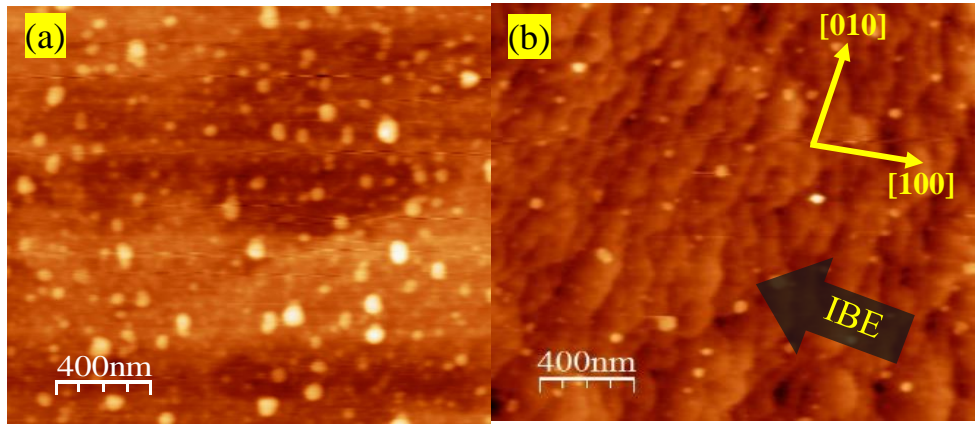

**Figure 6.I.5:** AFM image of Ag(001) substrate after IBE with 2KeV Ar+ ion for 20 minutes at an angle of incidence 50º from the surface normal. The black arrow indicates the projection of ion beam direction on the substrate.

An Ag(001) substrate of the same specifications as described earlier is used to grow Fe thin film and study its magnetic properties. In the previous section, we have seen that impurity or contaminated layers remain present on the substrate surface. Therefore, substrate surface cleaning is an essential step to achieve lattice matching induced epitaxy. Figure 6.I.6(a) represents a schematic of IBE and orientation of correlated surface morphology with respect to the different Ag crystallographic directions. Figure 6.I.6(b-f) represents the electron diffraction pattern (RHEED images) of Ag(001) single crystal surface taken at different stages of surface cleaning and IBE process. Fig. 6.I.6(c) and 6(d) depict RHEED images captured along [100]$_{Ag}$ and [110]$_{Ag}$ directions, respectively, after cleaning by several cycles of 0.8keV Ar+ ion sputtering and subsequent annealing at 600ºC. The well-defined streaks and a low background intensity in the RHEED pattern in Fig. 6.I.6(c-d) indicate that the cleaning process has removed the adsorbed contaminated surface layer and corresponds to the ordered 2-dimensional (2D) single-crystal surface.

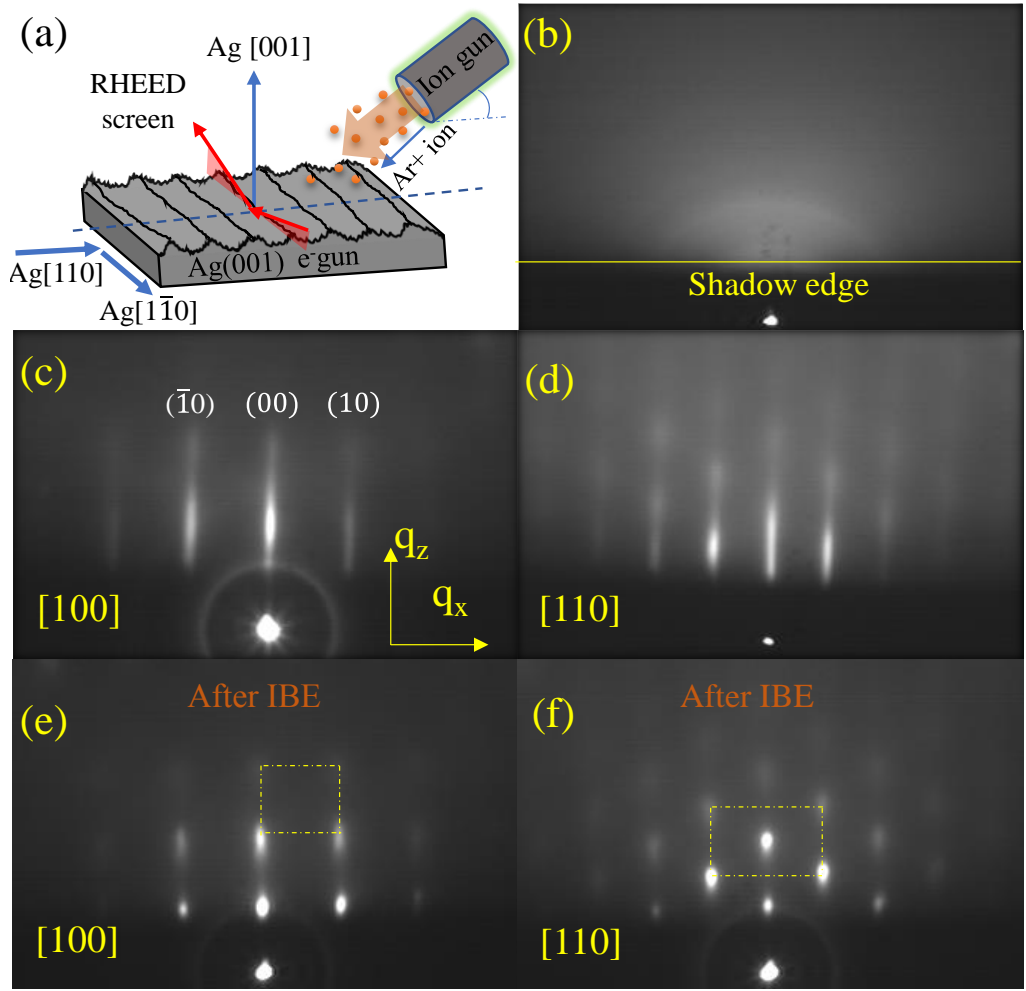

**Figure 6.I.6:** (a) Schematic illustration of RHEED measurement and the direction of IBE with crystallographic orientations in Ag substrate. In-situ RHEED patterns from Ag(001) surface (b) before and (c-d) after the UHV cleaning process along $[100]_{Ag}$ and $[110]_{Ag}$ direction, respectively. (e-f) RHEED image along $[100]_{Ag}$ and $[110]_{Ag}$ direction after IBE.

Before film deposition, with an object of the modification of the substrate morphology, its surface is bombarded with Ar+ ion of energy 2keV having an average flux of $3.74 \times 10^{13}$ ion/cm$^2$/sec for 20 minutes at an angle of incidence 50º from the surface normal. The ion beam projection on the substrate surface was close to $[110]_{Ag}$ direction, which corresponds to the thermodynamically preferred step orientation direction of the fcc Ag structure [260]. The RHEED image of the substrate surface after the ion irradiation process is presented in Fig.

6.I.6(e-f). The major change observed from this image is that the long streaks present in Fig. 6.I.6(c-d) are broken and have been converted into dotted or spotty patterns. A close inspection reveals that the spots in the RHEED pattern of Fig. 6.I.6(e) exhibit a square shape and in 6(f) displays a centered rectangular shape arrangement. This type of RHEED patterns are formed by the transmission of electrons through three-dimensional atomic arrangements, which may be small islands or oriented steps present on the surface [261–264]. Here, it is informative to mention that, in several earlier studies, it was found that oblique incidence ion erosion of metallic single crystal surfaces such as Ag(001), Ag(110), Cu(110) creates an unbalance between the number of preferred steps oriented along and across to the ion beam direction and clear traces of ripple is formed [186],[187],[243] [265].

### 6.I.3. Growth and characterization of epitaxial Fe thin film

#### 6.I.3.1 Real-time growth of epitaxial Fe thin film monitored through RHEED

In order to prevent any growth-induced anisotropy in the Fe film, deposition is done normal to the substrate surface. Fe film thickness is monitored in-situ during deposition by a water-cooled calibrated quartz crystal monitor. The real-time growth process is recorded using RHEED, which provides information about the growth mode of Fe.

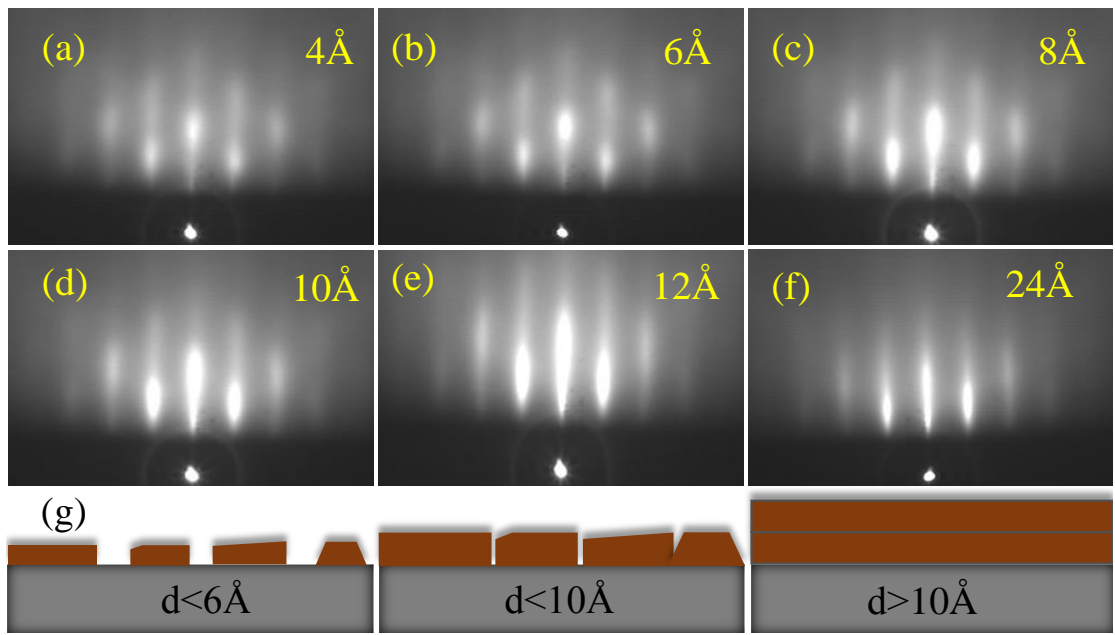

**Figure 6.I.7:** In situ RHEED patterns of Fe film surfaces along [110]$_{Ag}$ direction, taken at film thicknesses of (a) ~4Å (b) ~6 Å, (c) ~8 Å, (d) ~10 Å, (e) ~12 Å and (f) ~24Å respectively. (g) Schematic of the growth process of Fe on Ag(001) substrate.

Figure 6.I.7(a)–7(f) represents in-situ RHEED patterns collected during the growth of Fe on ion irradiated Ag (001) substrate. Here images (a) to (f) correspond to 4Å, 6Å, 10Å, 12Å and 24Å thick Fe film, respectively. The diffraction patterns were recorded using the e⁻ beam incidence along [110] $_{Ag}$ direction with an incident angle of 2.4° as estimated from the position of the shadow edge, the specular spot, and the direct beam. We observe that the intensity of RHEED images decreases just after Fe starts wetting the Ag substrate due to the increment of diffuse scattering originating from Fe adatoms on the substrate surface. The patterns show that the Fe adatoms follow a similar mesh of Ag(001) and grow in a single-crystalline form.

Moreover, diffraction from Fe initially consists of spotty patterns, which become sharper and streakier with increasing Fe thickness. This indicates that Fe grows in a 3-dimensional (island) monocrystalline form in the early stages of growth before complete wetting of the Ag(001) surface. This island form growth is mediated by the difference in surface free energy between Ag and Fe surfaces and the presence of surface steps (or directional roughness) on the Ag surface. Besides this, the sharpness of RHEED streaks with film thickness indicates an improvement in the growing film crystalline quality as a reduced in-plane crystalline coherence length leads to the broadening of diffraction spots [266]. The lattice constant calculated from the streaks spacing is found to be (2.89±0.04) Å which is consistent with the bcc Fe crystal which is 2.86 Å. The deviation in our measured lattice constant is due to the uncertainty in the measured distance of the phosphor screen from the sample.

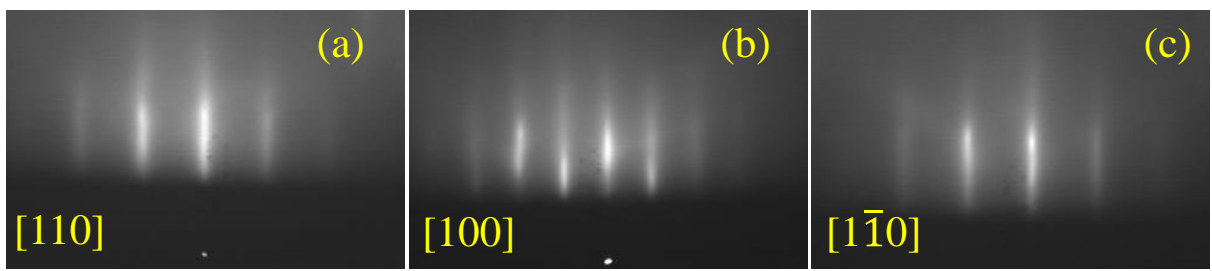

**Figure 6.I.8:** In-situ RHEED images of the 20nm thick Fe film along (a) [110], (b) [100] and (c) [1$\bar{1}$0] direction respectively.

Figure 6.I.8(a-c) represent RHEED patterns corresponding to [110], [100] and [1$\bar{1}$0] directions respectively of the final 20nm thick Fe film. The symmetry of the diffraction pattern captured by rotating the sample indicates the epitaxy of Fe with Ag having well-known crystallographic orientations (001)[110]$_{Fe}$|| (001)[100]$_{Ag}$ [239],[267]. The RHEED pattern is still characterized by a predominantly two-dimensional (2D) behaviour.

**6.I.3.2 In-situ study of magnetic anisotropy by MOKE**

MOKE in longitudinal geometry investigated the magnetic switching behaviour of the 20nm thick epitaxial film in situ. The azimuthal angle-dependent MOKE hysteresis loop is recorded using a polarized He-Ne laser light ($\lambda$ = 632.8 nm) and a UHV compatible electromagnet in longitudinal geometry over the full 0° to 180° range by rotating the sample about the substrate normal using high precision manipulator manufactured by PREVAC, Poland. The in-situ MOKE measurement was performed after deposition of the final 20nm thick Fe film to avoid deflection of RHEED electron due to the magnetic field generated by the electromagnet during MOKE measurements. It may be noted that to find the accurate orientation of the anisotropy axis, hysteresis loops close to the cubic hard axis are taken with a precision of ≈1º. Furthermore, the reproducibility of the loops was checked by repeating each loop two to three times. To get information about magnetization reversal and the origin of magnetic anisotropy, MOKE hysteresis loops were collected by rotating the sample with respect to the magnetic field along different in-plane directions $\varphi$. Figure 6.I.9 displays representative hysteresis loops taken along different $\varphi$ directions measured from $[100]_{Fe}$ direction. It is observed that along $[100]_{Fe}$ direction ($\varphi=0º$) direction, the hysteresis loop exhibits a single-step rectangular loop where magnetic moments saturate immediately after the single Barkhausen jump. Therefore, the magnetization reversal occurs through nucleation and propagation of the reversed domain. Here, one can see that the normalized magnetization ($M_r/M_s$) value is close to 1, therefore, considered the easy magnetization axis.

Moreover, a small coercive field of ≈ 3mT indicates the low number of domain wall pinning centers due to the good crystalline quality of deposited Fe film. However, as the sample is rotated by an amount of 45º i.e., along $[110]_{Fe}$ direction, the shape of the loop changes. Here, as the magnetic field is gradually reduced from a positive saturation state, the normalized magnetization decreases close to 0.7 (MsCos45º) and exhibits a sudden irreversible jump to a negative magnetization state. After that, as the field increases, the magnetization continuously rotates in the field direction until it gets saturated in the negative magnetization state. It is characteristic of fourfold in-plane magnetic anisotropy that is present in the film as expected for epitaxial bcc Fe film. However, for intermediate directions, magnetization switches through two clear Barkhausen jumps. Two two-step loops corresponding to $\varphi=41º$ and 311º are presented in Fig. 6.I.9. A clear plateau separates these steps to a stable intermediate state. Two-step switching behaviour is observed around the $[110]_{Fe}$ axis within an angular range of ≈ ±15°. Therefore, it appears that when the external field is applied adjacent to the cubic hard axis, it becomes energetically more favourable for magnetization to switch through the two-jump process. A closer inspection reveals that three-step loops are also present in the magnetic switching process close to the cubic hard axis. Two representative three-step loops at φ= 313° and 43° are included in Fig. 6.I.9. In addition to this, a distinct feature is observed by comparing hysteresis loops along equivalent crystallographic directions. A close look-over reveals that the loops are asymmetric for equivalent symmetric directions such as 45° and 135°, 43° and 313°, 41° and 311°, as reported in Fig. 6.I.9. This behaviour deviates from the ideal characteristics of biaxial MA and indicates the presence of an additional component of MA.

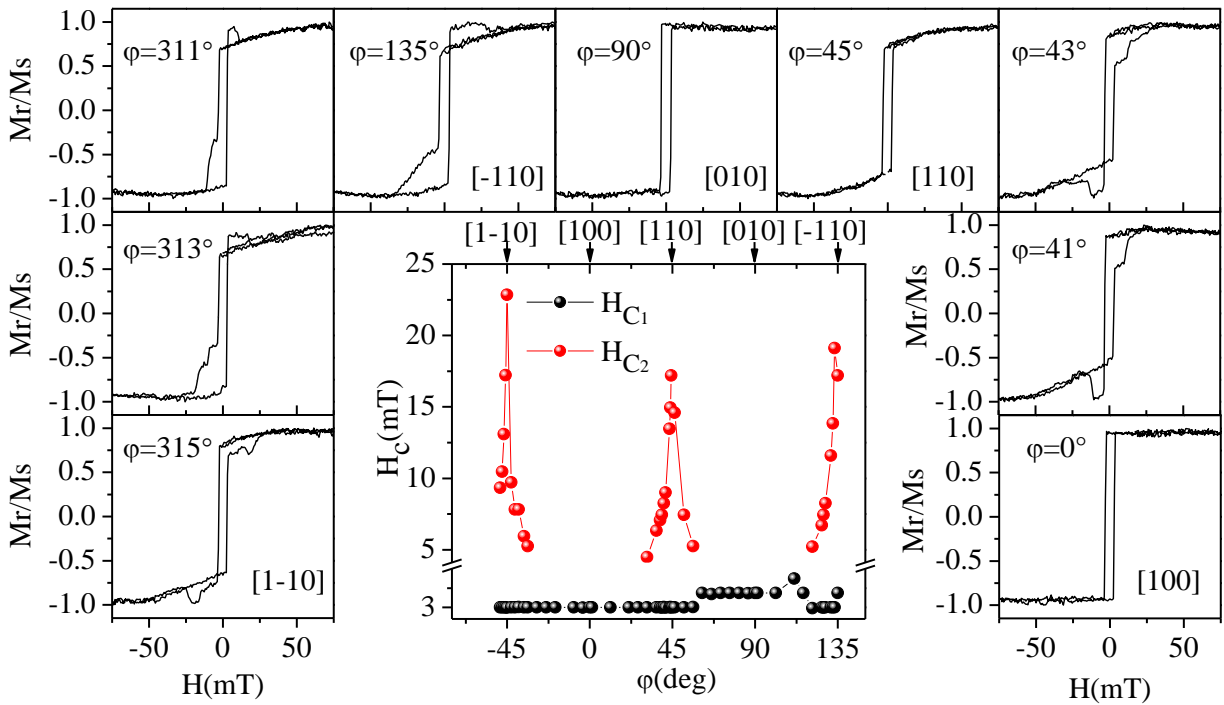

**Figure 6.I.9:** Longitudinal MOKE hysteresis loops for different in-plane field orientations with respect to [100]$_{Fe}$ direction. The experimental one jump switching fields H$_{c1}$(black bubbles) and two jump switching fields H$_{c2}$(red bubbles) as a function of the field orientation angle φ.

The switching fields extracted from the hysteresis loop for the one-step (we refer to this switching field as Hc1) and for the two-step (we refer to the lower switching field as Hc1 and to the higher switching field as Hc2, respectively) loops as a function of the field orientation is presented at the center of Fig. 6.I.9. The Hc1 displays an abrupt jump at φ≈ ±15° (±2.5° uncertainty) from the cubic hard axis in the range 45°<φ<135°. Furthermore, the values of Hc1 in the range 45°<φ<135° are greater compared to -45°<φ<45°. As per the fact that the switching field reflects the symmetry of the anisotropy axis, it can be concluded from the switching behaviour of magnetization reversal that the extra UMA is oriented in the vicinity of the cubic easy axis. Therefore, the magnetic easy axis is not equivalent to the present case. The $[100]_{Fe}$ direction is the intermediate "hard-easy" in-plane magnetic direction, whereas the $[010]_{Fe}$ direction is the easiest "easy- easy" direction due to the presence of UMA.

### 6.I.3.3. Ex-situ study of domain dynamics by KERR microscopy

After in-situ characterization, the sample was taken out of the UHV chamber, and its domain dynamics were studied under a Kerr microscope to investigate the local process of magnetization reversal, which cannot be extracted from the MOKE hysteresis loop and also to give further insight into the microscopic origin of the different magnetization reversal process. The MOKE hysteresis loop recorded through KERR microscopy along φ ≈ 0º, 40º, 45º and 135º are presented in Fig. 6.I.10, 11 and 12, respectively. The domain images corresponding to different hysteresis loops at different transition fields marked by the numerical numbers are enclosed beside the hysteresis loops. The color contrast in the domain images represents the relative orientation of spins with respect to the field direction.

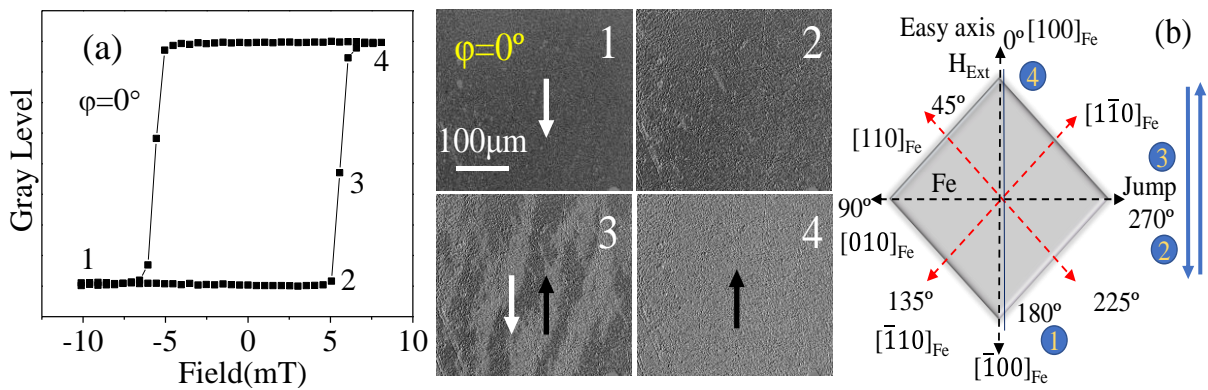

**Figure 6.I.10:** (a) MOKE hysteresis loop as obtained from Kerr microscopy and corresponding domain images at different field values. (b) Schematic of 1-step switching process.

As reported in Fig. 6.I.10, it is observed that along the $[100]_{Fe}$ direction, magnetization reversal takes place through nucleation and propagation of the 180° domain wall that gives rise to the classic rectangular hysteresis loop. At point 1, in the saturation state, all domains are oriented along $[\bar{1}00]_{Fe}$ direction and the resulting domain image is dark in contrast. At point 2, brighter contrast domains start to nucleate, characteristic of $[100]_{Fe}$ oriented domain. These domains develop with increasing field and simultaneously present with the dark contrast domains at point 3. Further increment in the field causes the growth of bright contrast domains all over the sample. Finally, at point 4, all spins are oriented in a positive saturation state, i.e., along $[100]_{Fe}$ direction. For better visualisation, a schematic sketch of this one-jump reversal process is shown in Fig. 6.I.10(b).

Figure 6.I.11 represents the hysteresis loop and the corresponding domain images taken close to $[110]_{Fe}$ direction at φ=40º. In contrast to the magnetization reversal along the φ=0º direction, we found that magnetization switching occurs through two separate 90° domain wall nucleation at two switching fields, giving rise to two-step hysteresis loops. At point 5, all the spins align along the field direction for a sufficiently large applied negative field. Therefore, the corresponding domain is dark in contrast. As we gradually reduce the magnetic field, the magnetization tries to rotate towards the nearest easy axis, i.e., along $[\bar{1}00]_{Fe}$ at point 6. No domain walls were observed in a remanent state, indicating that the Fe film was still in a single-domain state. Further field increment in the positive direction breaks the single domain state. It causes appearances of $[010]_{Fe}$ oriented domains (that sweep via 90º domain wall motion, 1$^{st}$ jump) to grow through Barkhausen jump at point 7 and get saturated close to point 8. On further successive increments of field, 90º domains are nucleated in which magnetization is oriented along $[100]_{Fe}$ direction at point 9(2$^{nd}$ jump). Beyond this applied field strength, the magnetization coherently rotates away from $[100]_{Fe}$ direction to lie towards the applied field orientation at point 10. Fig. 6.I.11(b) shows a schematic sketch of this two-jump reversal process. One can note that these two jumps occur when the magnetization passes over one of the two cubic hard-axis directions in a sample.

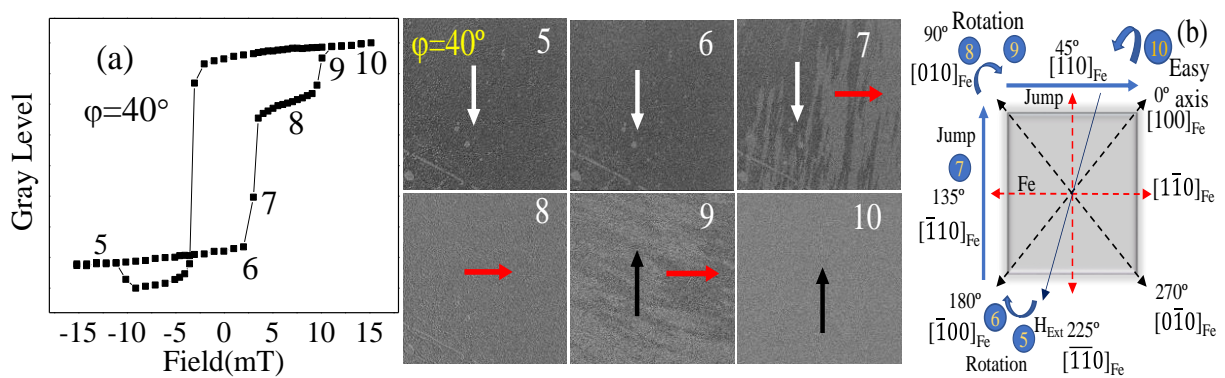

**Figure 6.I.11:** (a) MOKE hysteresis loop close to the hard axis at φ=40º as obtained from Kerr microscopy and corresponding domain images at different field values. (b) Schematic of the 2-step switching process.

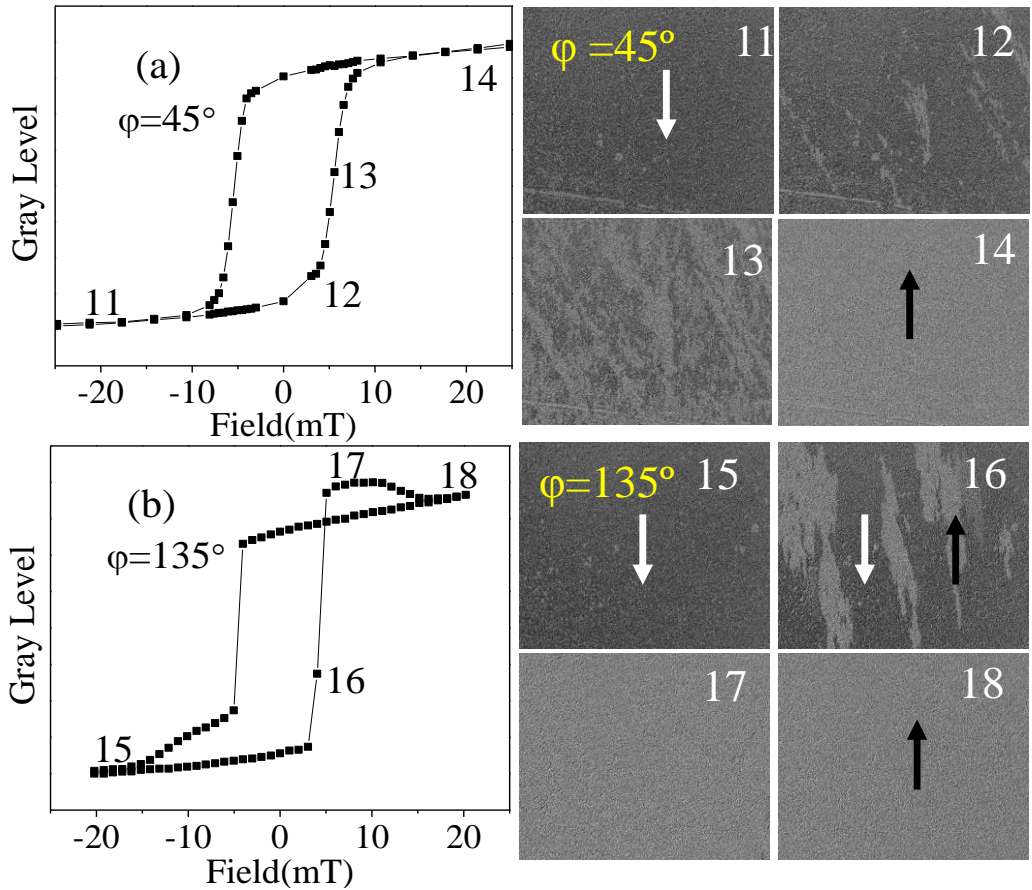

**Figure 6.I.12:** MOKE hysteresis loop along the cubic hard axis at (a) φ=45º and (b) φ=135º as obtained from Kerr microscopy and corresponding domain images at different field values.

The hysteresis loop and domain dynamics along $[110]_{Fe}$ direction are presented in Fig. 6.I.12(a). One can observe that magnetization reversal along this direction occurs via a combination of rotation and nucleation of domain walls. At point 11, all the moments are saturated in a negative magnetization state. Reduction of magnetic field causes rotation of moments along the nearest easy axis. The field increment in the positive direction causes domains oriented along $[110]_{Fe}$ direction to grow via points 12 and 13 that gets saturated at point 14. The domain dynamics along another equivalent cubic hard axis at φ=135º is reported in Fig. 6.I.12(b). Compared to the cubic hard axis at φ=45º, the magnetization switching along this direction occurs through a combination of irreversible jump and rotation of domain walls. The deviation of the hysteresis loop and domain dynamics for the equivalent crystallographic axis indicates that extra UMA strongly affects the switching behaviour. Note that a similar kind of hysteresis behaviour is also observed via in-situ MOKE measurement. Therefore, Kerr microscopy measurement complements the earlier observations in in-situ MOKE measurements with an in-depth understanding of the microscopic origin of different magnetic switching behaviour.

## 6.I.4. Discussions

The above observations indicate that anisotropy of domain nucleation and the growth processes significantly affect the magnetization reversal behavior of the film. It may be noted that Ag(001) surface is intrinsically isotropic; therefore, the observed magnetic anisotropy can be directed related to IBE-induced modified surface morphology. In epitaxial thin films, the sources of extra UMA are mostly referring to interface-induced contributions such as atomic steps due to miscut of the substrate or strain arising from film-substrate lattice mismatch [41], film substrate bonding [268], dangling bonds [269], directions of the reconstructed surface of the underlying substrate [270], artificially created uniaxial surface symmetry of substrate [271] or by oblique angle deposition [272]. Due to the lattice symmetry breaking at the interface, film grown on vicinal or stepped surface displays UMA originating from the unbalance between the density of steps and low symmetry environment. Here, the main source of UMA is the roughness-induced demagnetizing field or shape anisotropy [273],[274] and modified spin-orbit interaction [169][275]. Application of oblique incidence IBE of single-crystalline metal surface leads to the self-assembled formation of wrinkled rippled topography with side wall consisting of parallel steps oriented along ripple direction [187],[260],[275],[276]. In the present case, as the substrate surface is eroded with an ion beam at an angle of 50º, the formation of ripple-like oriented morphology on the surface is highly probable [14]. The formation of oriented step or ripple morphology has also been confirmed through AFM measurement provided earlier for Ag(001) substrate eroded in identical conditions as that of the substrate used for film deposition. This pattern formation breaks the rotational symmetry of the substrate surface, and causes unbalance in the energy of the deposited overlayer, inducing an extra UMA. Therefore, film grown on such a template displays a morphology-induced UMA superimposed with intrinsic crystalline anisotropy. In earlier studies, a similar hysteresis behaviour has also been observed by Milik et al. in an obliquely deposited Fe/MgO system [277] and by Morley et al. in an epitaxial Fe/GaAs system [278]. This asymmetrical hysteresis behaviour is attributed to extra UMA. Millik et al. [54] have deposited Fe magnetic thin films on MgO substrate at an oblique angle, and the origin of UMA is attributed to the shape anisotropy due to columnar growth in the film. Films were prepared with a protective layer on top to prevent the surface oxidation of Fe. In another study by Morley et al. [55], UMA in epitaxial Fe films on GaAs substrates is ascribed to interfacial bond formation. This study showed that a Cr overlayer/capping layer on a 30 ML thick Fe film significantly influences uniaxial contribution to the magnetic anisotropy. In both studies, the origin of this asymmetrical hysteresis behaviour is attributed to the presence of extra UMA and its coupling with already present biaxial magnetic anisotropy in the films. Compared to the previous studies [54,55], the present study performed on the Ag (100) substrate by irradiating it at an incidence angle 50º. It breaks morphological symmetry on the surface and induces UMA with an easy axis normal to the ion beam direction. Most such studies have been performed ex-situ or by covering the magnetic layer with a protective nonmagnetic layer. In comparison, present work done in-situ, where substrate cleaning, correlated surface patterning, film growth and characterization were done using a versatile UHV system [279]. As a result, the observed multi-jump hysteresis loops are understood unambiguously in terms of a combination of the intrinsic cubic anisotropy Kc and the presence of extra UMA Ku. Here interplay between the strengths of the cubic anisotropy, UMA and domain wall pinning energy decides different magnetic switching processes in hysteresis loops. In fact, the orientation and relative strength of Ku forces it to behave unevenly from a magnetic viewpoint for different equivalent crystallographic directions. In an extreme case, it is possible that the strength of UMA overcomes the intrinsic cubic anisotropy, and it converts from four-fold to two-fold symmetry. Therefore, two steps switching occur when $(K_U/K_c)<1$, whereas for $(K_U/K_c)>1$, i.e., for uniaxial nature, only one step loop is found. In the present case the ratio of Ku/Kc is found to be ≈ 0.88% [280][281]. Thus, it can be concluded from the present observations that a considerable amount of interface induced UMA is present in the film. In particular, magnetic systems have been found to be very sensitive to the interfacial morphology affecting the magnetic switching mechanism and the corresponding domain dynamics. Ultimately, the present study on epitaxial thin film superimposed with UMA provides us a window to explore the interplay between crystalline anisotropy and interface induced UMA. It also opens up the possibility of controlling magnetic switching by artificial manipulation of energy levels for potential application in multi-axial magnetic sensors. Therefore, proper optimization of oblique incidence IBE will provide an additional option which must be further explored in detail for precise tuning of morphology induced MA.

## 6.I.5. Conclusions

The growth and magnetic characterization of Fe film on ion-sculpted Ag(001) substrate have been studied using RHEED and MOKE measurements. RHEED provides film morphology and structure during growth, whereas MOKE presents detailed magnetic properties. Thus, it made it possible to correlate the evolution of film structure and morphology with that of UMA in the film. The film grows epitaxially on the Ag surface with crystallographic orientations $(001)[110]_{Fe}$ ∥ $(001)[100]_{Ag}$ via Volmer–Weber mechanism, where islands grow larger to form a continuous film at around 1 nm thickness. Ion beam sculpting of the substrate surface prior to the deposition of the Fe film induces an UMA in Fe film, which couples with the intrinsic cubic four-fold anisotropy of bcc-Fe. It leads to the multiple-step jump in magnetic hysteresis loops and a striking difference in domain structure depending upon the direction of the applied magnetic field with respect to crystallographic direction. It is found that the magnetization reversal takes place via a "single-step jump" mechanism through the sweeping of 180° domain walls, whereas the combination of coherent rotation and the sweeping of 90° domain walls is responsible for a "double step jump" magnetization reversal. The understanding acquired from the present study on the critical role of magnetic anisotropy responsible for different kinds of magnetic switching behaviour can be further extended for multilayer systems where often a separation of interlayer exchange coupling and anisotropy is needed.

# Chapter 6 II

# Magnetic anisotropy and surface reconstruction of epitaxial thin film induced by IBE.

### Reversible surface modification of Epitaxial Fe thin film on MgO(001) substrate

Engineering of thin film surface morphology and structure is one of the essential technological assets for regulating the physical properties and functionalities of thin film-based devices. In this part of the chapter, we study the evolution of surface structure and magnetic anisotropy of epitaxial Fe thin film grown epitaxially on MgO(001) substrate subject to several cycles of IBE. Ultrathin Fe film grows in 3D island mode and exhibits intrinsic fourfold magnetic anisotropy. After a few cyclic processes of IBE, the film displays uniaxial magnetic anisotropy that leads to a split in the hysteresis loop. Furthermore, we present clear and conclusive evidence of IBE mediated (2×2) reconstruction of the Fe surface. We also demonstrate that thermal annealing can tune UMA and surface reconstruction reversibly. Therefore, the feasibility of the IBE technique by proper selection of ion flux and energy for modification of surface structure has been demonstrated apart from conventional tailoring of morphology and tuning of UMA. Thus, the present work paves a way to further explore the IBE-induced self-assembling phenomena.

# Chapter 6: Part-II

# Magnetic anisotropy and surface reconstruction of epitaxial thin film induced by IBE.

**Reversible surface modification of Epitaxial Fe thin film on MgO(001) substrate**

## 6.II.1 Introduction

## 6.II.2 In-situ characterization of epitaxial Fe thin film deposited on MgO (001) substrate.

6.II.2.1 Structural analysis by in-situ RHEED measurement.

6.II.2.2 Magnetic characterization by in-situ MOKE measurement.

## 6.II.3 In-situ study of the evolution of surface structure and induced magnetic anisotropy of Fe thin film subject to several cycles of IBE.

6.II.3.1 IBE of Fe thin film.

6.II.3.2 Characterization of crystallographic film structure by RHEED.

6.II.3.3 Evolution of Magnetic anisotropy studied using in-situ MOKE.

## 6.II.4 Temperature-dependent study of UMA and surface structure.

## 6.II.5 Layer thickness dependent study of magnetic anisotropy and surface reconstruction.

## 6.II.6 Conclusions

## 6.II.7 Overall conclusions

## 6.II.1 Introduction

Magnetic thin films are an area of intense study that has expanded tremendously in the previous decade. The atomic-scale investigation of the link between ultrathin epitaxial film and magnetic properties is a challenging problem due to the presence of several effects such as surface/interface morphology [139], interface diffusion [282] and hybridization [225], film structure [283], stain [284], etc. On the other hand, these dependencies open up the possibilities of the fine tailoring of magnetic properties according to the desired functionalities. Magnetic anisotropy is a crucial characteristic of ferromagnetic materials from a fundamental perspective to study magnetic switching and for numerous practical uses, such as magnetic memory devices [212]. The structure and morphology of ultrathin films directly affect their magnetic anisotropies. The fundamental processes behind magnetic anisotropies include dipolar and spin-orbit interactions [199]. The dimensionality and external shape of the particular ferromagnetic body and the local lattice symmetry are all important factors in magnetic anisotropy. Advances in sample preparation methods, availability of ultra-high vacuum chambers, and the affordability of in-situ compatible atomic scale characterization tools have enabled the growth and characterization of ultrathin films.

Low energy ion beam erosion is a new, versatile and handy tool that has found its application for tailoring various physical properties in diverse fields such as photonics [53], transport [54], sensing [56], plasmonic [57], surface chemistry [58] bio-medical applications [62] etc. It is observed that low energy IBE of thin film and substrate surface create a wide range of morphology, including periodic ripple, nanoholes, arranged arrays of dots and pits [15][119] etc., depending upon the ion energy and angle of incidence. The periodic ripple-like pattern is formed due to competition between curvature-dependent ion beam sputtering that roughens the surface and ion or temperature-dependent viscous flow that smooths the surface [14]. Since the structure and morphology of ultrathin films directly affect their magnetic anisotropies, researchers are utilizing the IBE technique for fine and flexible tailoring of magnetic anisotropy. The thin film deposited over the rippled substrate or direct IBE of thin film was found to induce an additional component of UMA due to shape anisotropy in polycrystalline [128][137] and epitaxial thin film [22][187]. However, due to intrinsic crystalline anisotropy in the epitaxial thin film, a competition between shape and crystalline anisotropy is often observed [139]. This competition gives rise to the multistep switching process, splitting of hysteresis loop [285] etc. Therefore, investigation in this field is ongoing.

It may be noted that the interaction of solid surfaces with low-energy ions can also cause mass redistribution when sputtering is negligible in the near-surface region [286][287][288]. Therefore, surface structure can also be altered with the low-energy ion beam. However, no studies have considered the corresponding ion beam interaction induced structural changes in the polycrystalline and epitaxial thin film. In earlier chapters, we have shown through in-situ investigation that IBE of the polycrystalline thin film changes surface crystallographic texture. This motivates us to further explore the effect of IBE on magnetic anisotropy and structural alternation of epitaxial thin film.

Fe/MgO system has recently received much attention due to its substantial interfacial PMA [289] and applicability in TMR devices [290]. For the present study, we have chosen Fe/MgO system for its easy growth and epitaxial relationship with the MgO substrate. Epitaxial Fe film was grown in the UHV chamber on MgO substrate. The Fe film surface was etched by Ar ion in several steps, and the evolution of surface structure and magnetic anisotropy was studied in situ simultaneously. We observed that the film exhibits an additional UMA that causes a split in the hysteresis loop. Interestingly, a 2×2 surface reconstruction has also been observed for the first time. These results offer a new dimension to the current knowledge of self-assembled surface morphology development during IBE and propose that modelling studies be conducted to capture these events.

## 6.II.2 Fe thin film deposition and its in-situ characterization

For the present study, a commercially available MgO(001) substrate is mounted on a sample holder attached with heater assembly on a high-precision rotatable UHV compatible manipulator. Fig. 6.II.1(a) shows the RHEED images of the MgO substrate taken after loading it inside the UHV chamber. The absence of a diffraction pattern is due to the presence of impurity and contaminated surface layers. A cleaned and ordered substrate surface is prepared by cycles of Ar+ ion beam sputtering using ultra-high-purity 99.999% Ar+ gas and subsequent annealing until a sharp RHEED pattern operated at an electron acceleration voltage of 20 keV and 1.45A emission current is observed. The RHEED images of MgO substrate taken along $[110]_{MgO}$ and $[100]_{MgO}$ directions after several cycles of the cleaning process are shown in Fig. 6.II.1b and 6.II.1c, respectively. The presence of sharp streaks confirms the single-crystalline nature of the substrate. Fe film of thickness 2.5nm is deposited on this cleaned substrate. To prevent any growth-induced anisotropy in the Fe film, deposition is done normal to the substrate surface. Fe film thickness is monitored in-situ during deposition by a water-cooled calibrated quartz crystal monitor. Furthermore, to draw a comparative study of the dependence of induced UMA and surface reconstruction on the film thickness and the direction of IBE, two separate films of relatively higher thickness, 4.2nm and 25nm, were also deposited.

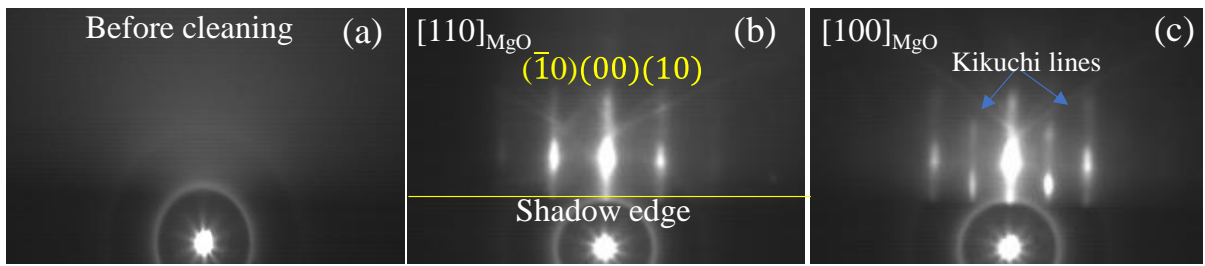

**Figure 6.II.1:** In-situ RHEED images of MgO substrate taken (a) before and (b,c) after surface cleaning along $[110]_{MgO}$ and $[100]_{MgO}$ direction, respectively.

### 6.II.2.1 Structural analysis by in-situ RHEED measurement.

After deposition, the film quality and crystallinity were checked by in-situ RHEED measurement. The corresponding RHEED images of the Fe film surface are presented in Fig. 6.II.2(a-c). The presence of streaks in the images indicates the epitaxial growth of Fe film with its well-established crystallographic orientations $(001)[100]_{Fe} \parallel (001)[110]_{MgO}$. It may be noted that the streaks are not continuous but rather spotty. This type of RHEED pattern is observed due to 3D-like diffraction from island-like structures present on the surface [262]. It indicates that the Fe film surface is not atomically flat, which may be due to the island form growth of Fe on the MgO(001) surface. A close inspection reveals that the spots in the RHEED pattern of Fig.6.II.2a exhibit a centred square shape and in 2(b) displays a rectangular shape arrangement, which reflects the symmetry of the [100] and [110] direction of Fe as it possesses bcc structure.

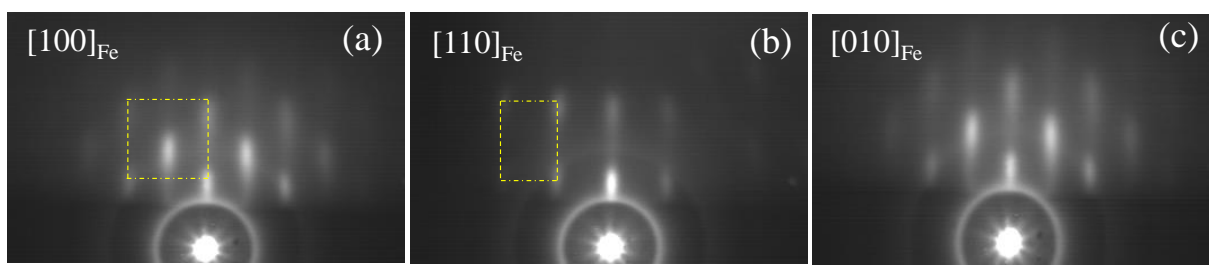

**Figure 6.II.2:** In-situ RHEED images of 25Å thick Fe film epitaxially grown on the MgO substrate along (a) $[100]_{Fe}$, (b) $[110]_{Fe}$, and (c) $[010]_{Fe}$ direction.

**6.II.2.2 Magnetic characterization by in-situ MOKE measurement.**

The magnetic characterization of the sample was carried out in-situ by MOKE measurements in longitudinal geometry, and analysis was focused on the determination of strength as well as the orientation of the in-plane anisotropy axis. The azimuthal angle-dependent MOKE hysteresis loop is recorded using a polarized He-Ne laser light ($\lambda$ = 632.8 nm) and a UHV compatible electromagnet in longitudinal geometry. The MOKE hysteresis loop taken by applying magnetic field H along [100], [010] and [110] directions of Fe is presented in Fig. 6.II.3. It is observed that MOKE loops along [100] and [010] direction exhibits very high squareness with normalized magnetization (Mr/Ms) close to 1, whereas MOKE loop along the [110] direction have normalized magnetization close to 0.7(MsCos45º).

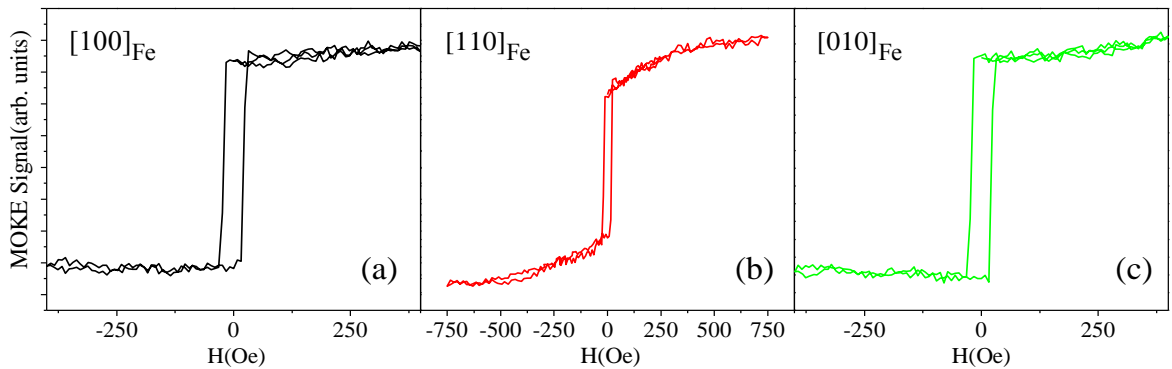

**Figure 6.II.3:** In-situ MOKE hysteresis loop along (a) $[100]_{Fe}$, (b) $[110]_{Fe}$ and (c) $[010]_{Fe}$ direction of Fe.

The polar plot of remanence (Mr) value normalized to its saturation (Ms) value is plotted in Fig. 6.II.4. The shape of the graph indicates that the film exhibits well-defined biaxial magnetic anisotropy which is expected for bcc Fe structure.

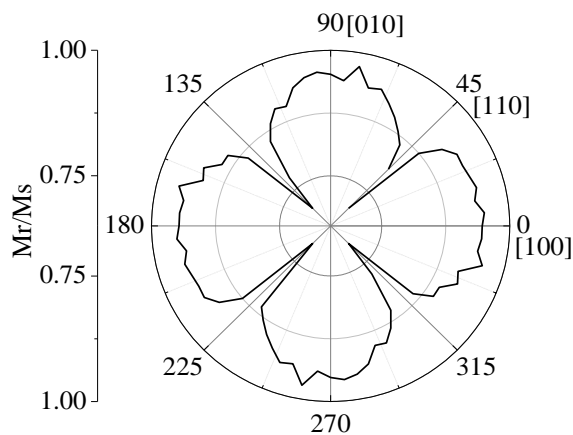

**Figure 6.II.4:** The polar plot of normalized remanence value.

## 6.II.3 In-situ study of the evolution of surface structure and related induced magnetic anisotropy of Fe thin film subject to several cycles of IBE.

### 6.II.3.1 IBE of Fe thin film.

After initial in-situ characterization, the film surface was eroded with Ar ion step by step. Each step consists of 2.5 min IBE by 1keV Ar ion at an angle of incidence 50° from the surface normal. The projection of the ion beam on the film surface was along $[100]_{Fe}$ direction, which corresponds to the thermodynamically preferred step orientation direction. To draw a correlation between surface structure and magnetic anisotropy, their evolution was investigated as a function of IBE time. Furthermore, to draw intercorrelation among structural, morphological and magnetic anisotropy in conjugation with the direction of IBE, MOKE and RHEED measurements were performed by rotating the sample with respect to the IBE direction.

### 6.II.3.2 Evolution of film crystallographic structure studied by RHEED.

The evolution of RHEED images taken along $[100]_{Fe}$ and $[110]_{Fe}$ direction at an interval of 2.5min IBE is presented in Fig. 6.II.5. It is observed that after 7.5 min of IBE, an intermediate line in between integer streaks appears, which becomes more prominent after the 4$^{th}$ cycle (10min) of IBE. In order to extract quantitative information about the position of new streaks, the line profiles have been extracted from the RHEED images as a function of IBE time and have been plotted in Fig. 6.II.5(i). We found that the new streaks are present precisely in the middle between the integer streaks.

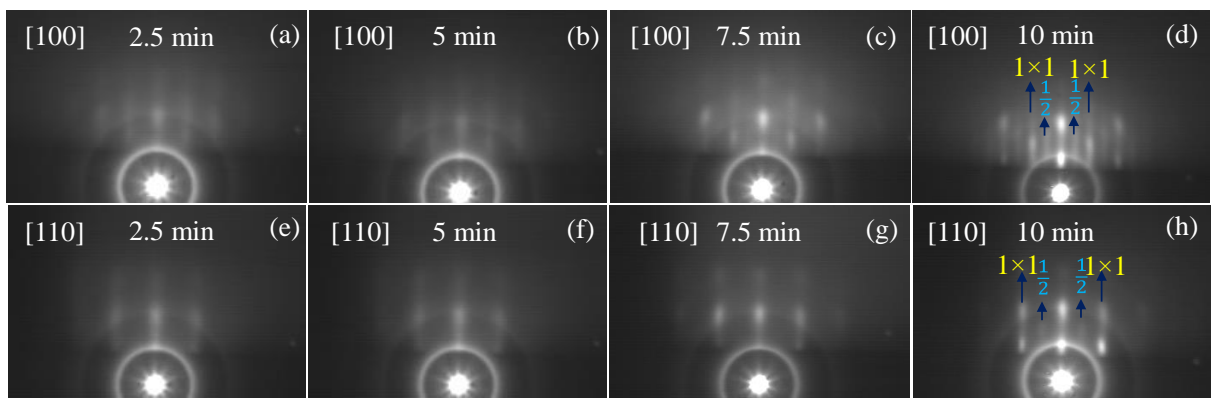

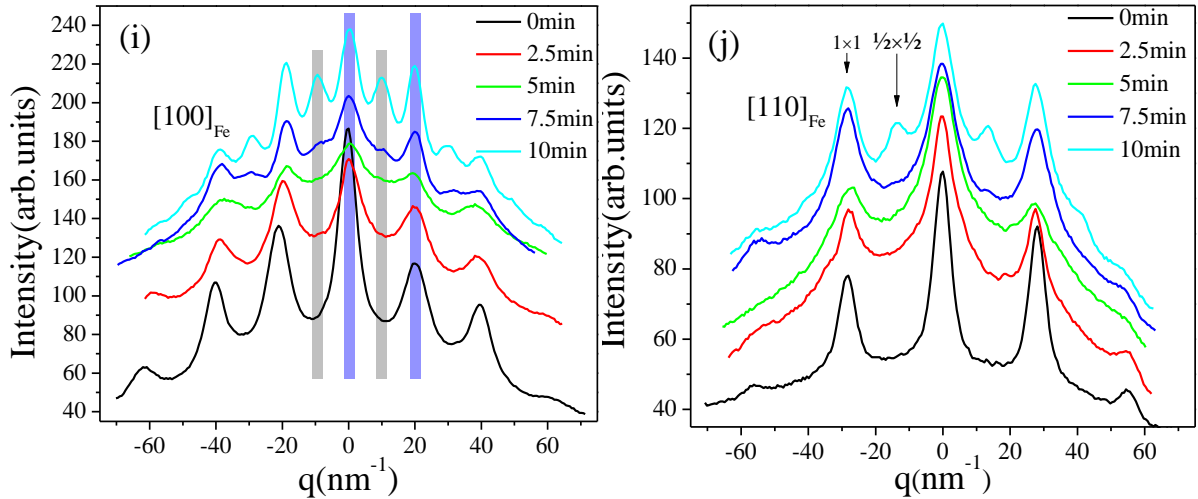

**Figure 6.II.5:** In-situ RHEED images taken along (a-d) $[100]_{Fe}$ direction, (e-h) $[110]_{Fe}$ direction after successive 2.5min ion beam erosion. Line profile of RHEED intensity distribution pattern after successive 2.5 min of IBE extracted from RHEED image taken along (i) $[100]_{Fe}$ and (j) $[110]_{Fe}$ direction.

This is a signature of 2×2 reconstruction of surface structure. A similar diffraction pattern is also recorded along equivalent crystallographic directions, and they are essentially symmetric with respect to the central streak.

### 6.II.3.3 Evolution of Magnetic anisotropy studied using in-situ MOKE.

The evolution of the hysteresis loop along ($\phi=0°$) and across ($\phi=90°$) to the IBE direction, i.e., along [100] and [010] directions, after 2$^{nd}$ and 4$^{th}$ cycles, is reported in Fig. 6.II.6. One can observe that the height of the MOKE signal decreases with the IBE cycle. This is because, in the ultrathin regime, the saturated MOKE signal's height is proportional to the film thickness. Therefore, this change can be rationally explained by considering the decrement of film thickness associated with the sputter removal of the film [19]. A closer inspection reveals that after 10min of IBE, the loop along [100] direction has split into two semi loops. The magnetic remanence also decreases from ≈1 in the as-deposited state to ≈0.15 after the 4$^{th}$ cycle of IBE. On the other hand, hysteresis loops perpendicular to the IBE direction preserves their shape as they were in as prepared condition. It may be noted that both [100] and [010] directions are crystallographically and magnetically equivalent. Therefore, this change in the shape of the loop is associated with the change in the magnetization switching process due to the onset of an extra UMA [291–293]. In the present case, UMA easy axis is oriented perpendicular to the IBE direction, i.e., along the [010] direction. Therefore, the [010] is the easiest direction, whereas the [100] direction is an intermediate direction due to the presence of cubic easy and uniaxial hard axis.

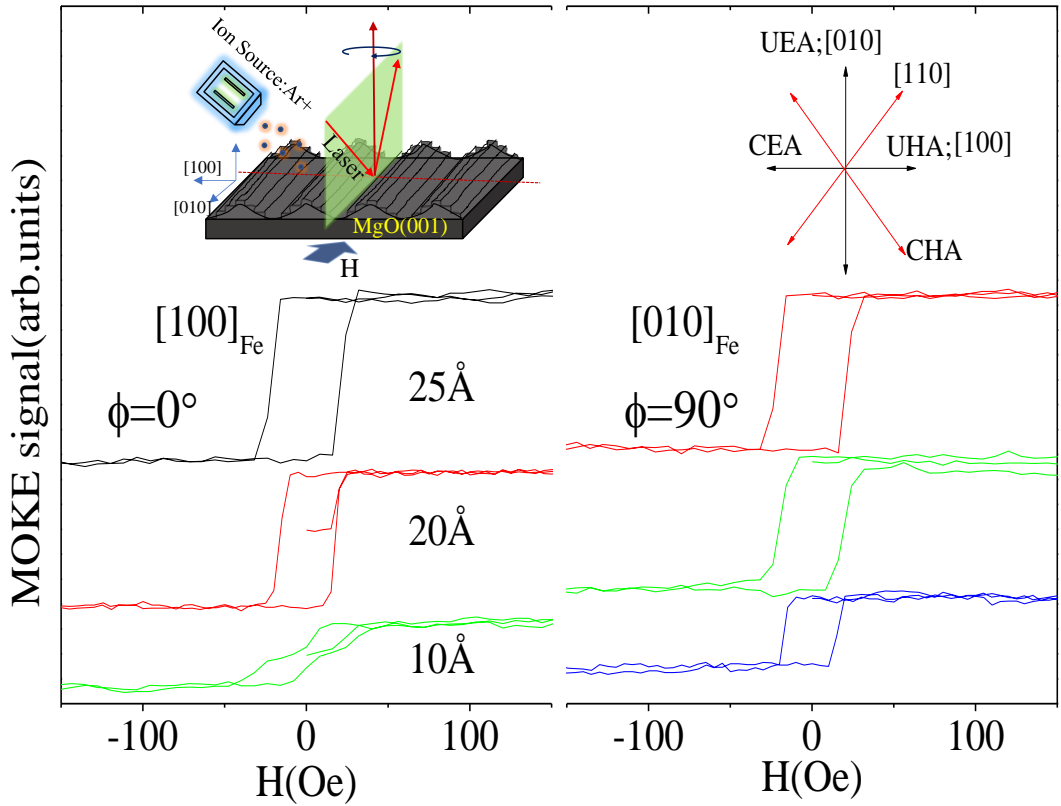

**Figure 6.II.6:** In-situ MOKE hysteresis loop taken along ϕ=0º and ϕ=90º to IBE direction.

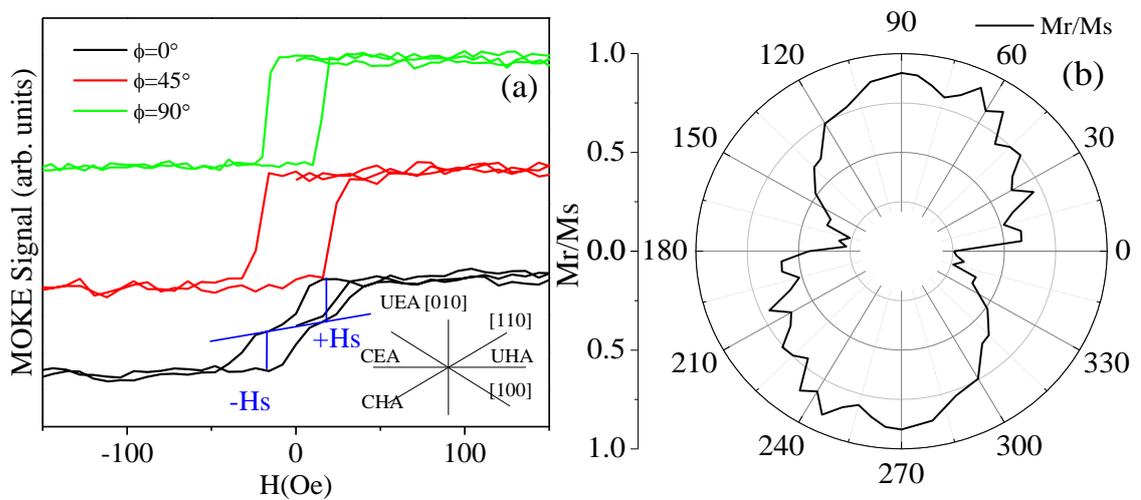

**Figure 6.II.7:** (a) In-situ MOKE hysteresis loop after 4$^{th}$ cycle of IBE taken along 0º,45º and 90º to IBE direction. (b) Polar plot of normalized remanence value after 4$^{th}$ cycle of IBE.

Quantitative value of the strength of induced UMA can be obtained by measuring the shift field Hs which is proportional to the split in the loop. Hs is defined as the difference between zero fields and the center of one shifted loop, i.e., Hs = (Hs2 -Hs1)/2, as pointed out in Fig. 6.II.7(a) and directly connected to UMA constant Ku=MsHs, where Ms=1737 emu/cm$^3$ is the saturation magnetization of Fe. In the present case Ku = 3.13*10$^4$ erg/cm$^3$ has been observed. [187],[294]. The strength of the induced UMA, which is proportional to the shift in the semi-loop, is lower than the intrinsic cubic anisotropy.

## 6.II.4 Temperature-dependent study of surface structure and induced UMA

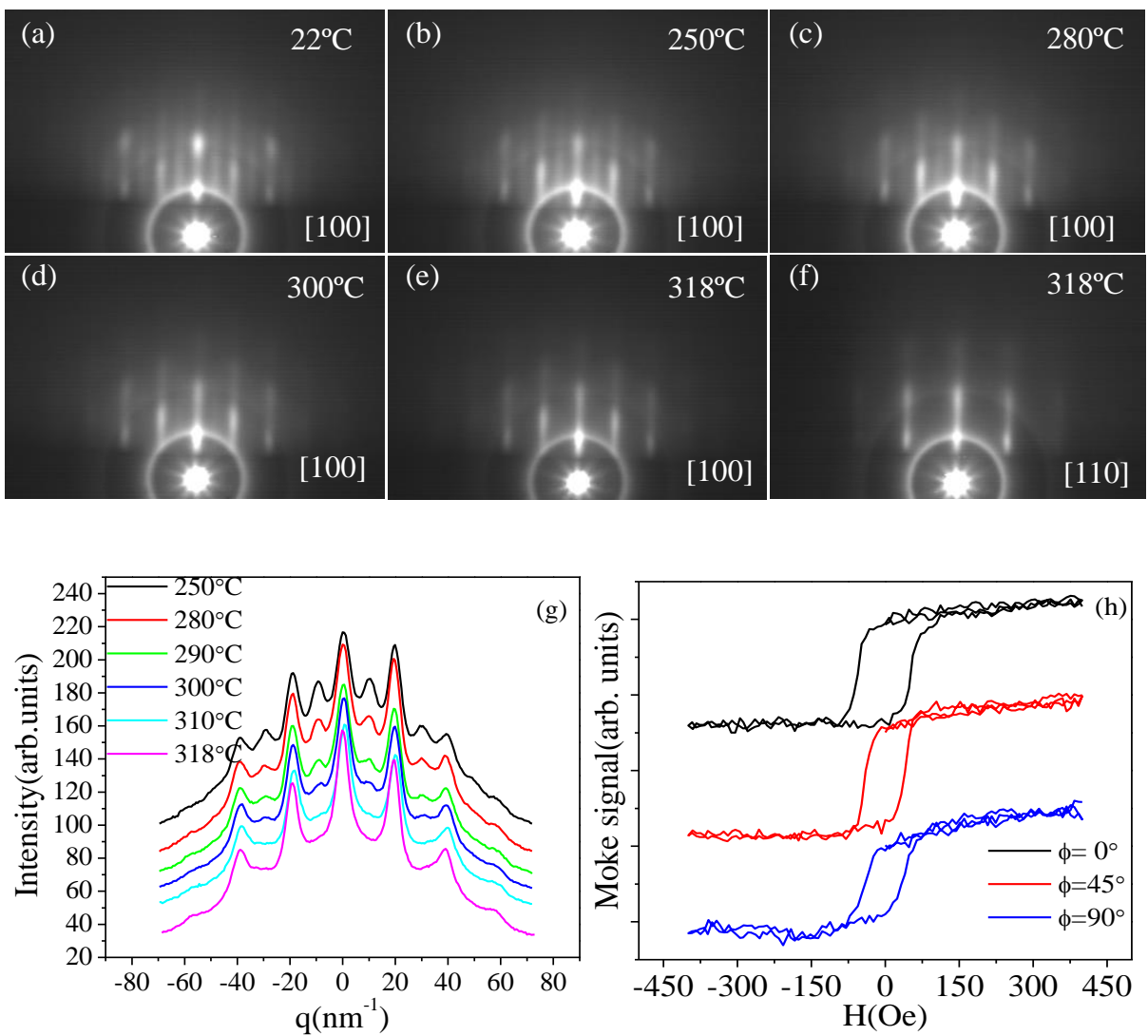

**Figure 6.II.8:** (a-f) Temperature-dependent evolution of 2×2 reconstructed RHEED pattern. (g) Line profile of RHEED intensity distribution pattern at different temperatures extracted from RHEED

image along [100]$_{Fe}$ direction. (h) In-situ MOKE hysteresis loop taken at room temperature after annealing at 320ºC along different orientations with respect to the IBE direction.

The structural evolution was investigated by RHEED as a function of the temperature. The RHEED images taken along [100]$_{Fe}$ direction at different temperature is presented in Fig. 6.II.8(a-e). We observe that the additional streaks start to disappear close to 300ºC and completely disappear at 320ºC. The line profile drawn at different temperatures from RHEED images is reported in Fig. 6.II.8(g). It also confirms that the intensity of the additional streaks gradually decreases with temperature and completely disappears at 320ºC. Therefore, thermal annealing has converted the 2×2 reconstructed structure into its original 1×1 structure.

The MOKE hysteresis loop was taken at 320ºC along ϕ=0º, 45º and 90º directions with respect to the IBE direction, as displayed in Fig. 6.II.8(h). We observe that the hysteresis behavior with two semi loops along [100]$_{Fe}$ has disappeared, and all the hysteresis loops along different crystallographic directions exhibit almost similar magnetization switching behavior. Therefore, it suggests that the induced UMA has disappeared.

## 6.II.5 Layer thickness dependent study of magnetic anisotropy and surface reconstruction.

To study the effect of layer thickness on surface reconstruction and magnetic anisotropy, two other films of relatively higher thickness, 4.2nm and 25nm, were deposited on the same MgO(001) substrate. Each film grows epitaxially on the substrate, as confirmed from RHEED images. The film surface was eroded with identical ion fluence and ion energy with which Fe thin film of 2.5nm thickness was eroded. However, for 4.2 nm thick film, the projection of IBE on the sample surface was along the magneto-crystalline hard axis, i.e., [110] direction of Fe, to explore the dependence of surface reconstruction and induced magnetic anisotropy on the direction of IBE. The corresponding MOKE hysteresis loop taken before and after IBE along ϕ=0° and ϕ=90° to the IBE direction is shown in Fig. 6.II.9. The major change observed is the decrement of the height of the MOKE signal due to the removal of thin film material from the surface. In the present case, no split in the hysteresis loop was observed. However, the RHEED images shown in Fig. 6.II.10 indicate that the surface structure has been converted to a 2×2 reconstructed surface.

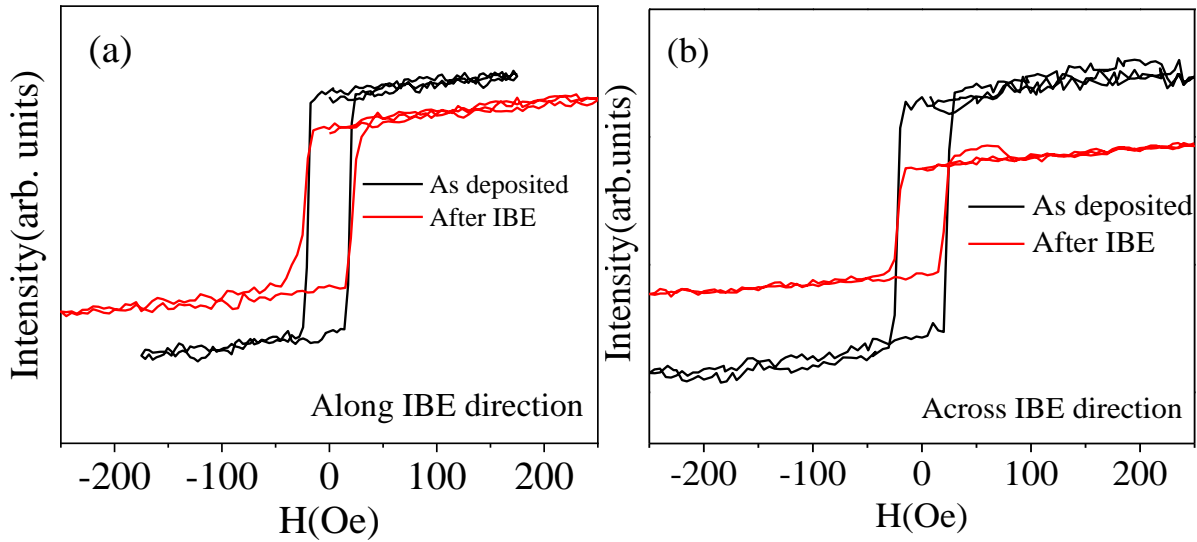

**Figure 6.II.9:** MOKE hysteresis loop of epitaxial Fe thin film having initial film thickness 4.2nm taken along (a) 0° and (b) 90° to the IBE direction in as-deposited state (black line) and after IBE (red line).

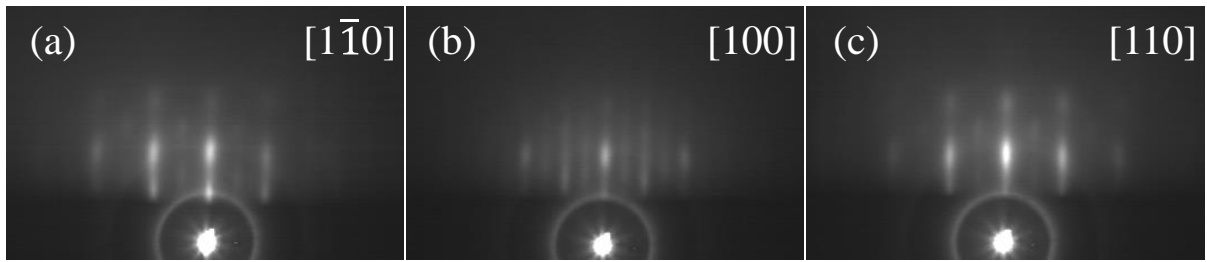

**Figure 6.II.10:** (a-c) RHEED images of 2×2 reconstructed epitaxial Fe thin film having an initial thickness of 4.2nm along different crystallographic directions.

Figure 6.II.11 displays the RHEED images after IBE of the epitaxial Fe thin film having an initial film thickness of 25nm. The RHEED images indicate that the surface structure has converted to a 2×2 reconstructed surface structure in the present case. Therefore, the present observations indicate that surface reconstruction is independent of layer thickness and the direction of IBE. The corresponding MOKE hysteresis loop taken along and across the IBE direction is presented in Fig. 6.II.12. The hysteresis loop before and after IBE exhibits identical magnetic switching behavior. It indicates that the induced UMA has a negligible effect on the overall magnetic switching behaviour. This is because the ion beam-induced morphology and surface structure modification are limited to the surface and near-surface region due to the low energy ion penetration depth. Therefore, the bulk of the film remains undisturbed. Thus, the induced UMA is relatively weaker than the dominant contribution of intrinsic MCA present in the film that decides the overall magnetic switching behavior.

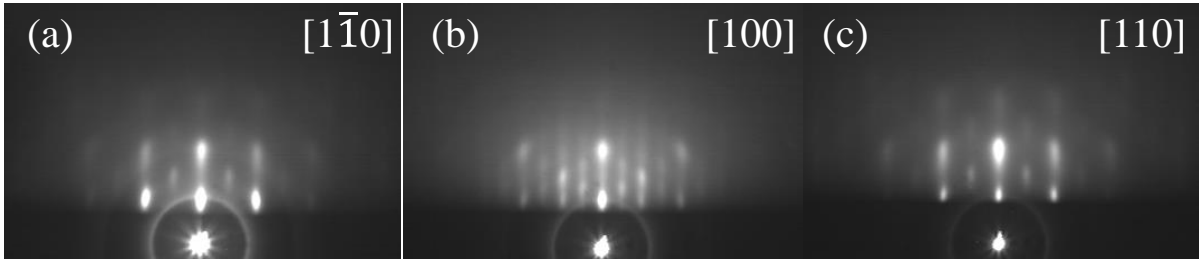

**Figure 6.II.11:** (a-c) RHEED images of 2×2 reconstructed epitaxial Fe thin film having an initial thickness of 25nm along different crystallographic directions.

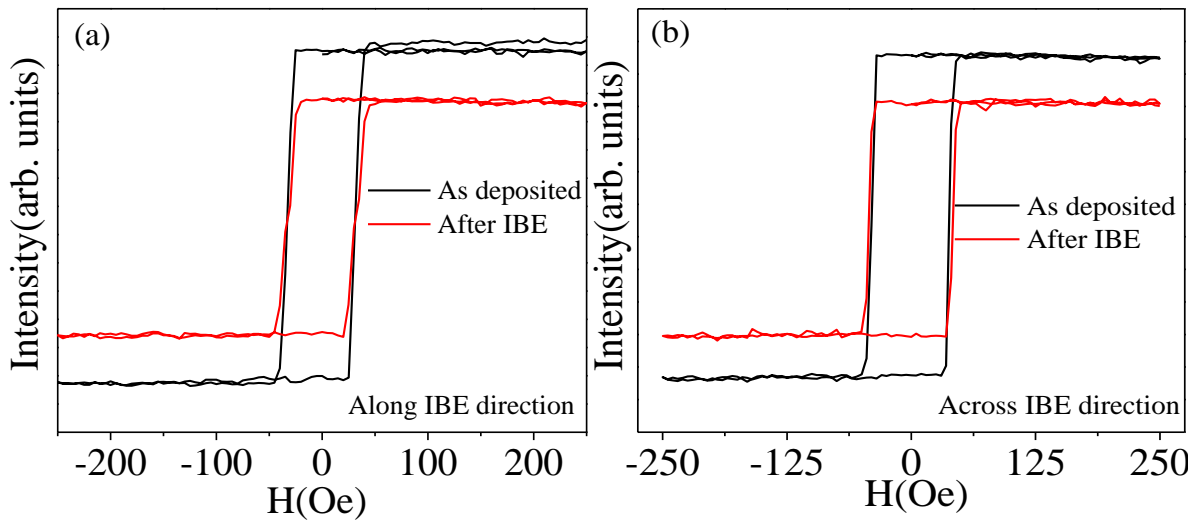

**Figure 6.II.12:** MOKE hysteresis loop of epitaxial Fe thin film having initial film thickness 25nm taken along (a) 0° and (b) 90° to the IBE direction in the as-deposited state (black line) and after IBE (red line).

The energy of anisotropic magnetization related to biaxial (Kc) and uniaxial (Ku) anisotropy constant via the relation: $E = \frac{K_C}{4}\sin^2(2\varphi) + K_U \sin^2(\varphi) - \boldsymbol{M}.\boldsymbol{H}$ [275][280]. Since the film is epitaxial in nature, an intrinsic MCA is present. The origin of induced UMA can be devoted to the MCA due to the unbalance between the number of steps oriented parallel and perpendicular to the IBE direction, low coordinated atom present at step edge or surface defects sites as well as shape anisotropy due to the ripple structure itself [117],[246],[295]. The formation of unidirectional correlated morphology breaks the four-fold symmetry of the Fe surface and induces an UMA oriented parallel to the ripple direction. Since the ion sculpting process is mainly limited to the surface and subsurface layers only, magnetic anisotropy with strength proportional to the inverse of film thickness is expected. Therefore, the appearance of UMA Ku after 10 min of IBE in epitaxial thin film having an initial film thickness 2.5nm results from the gradual build-up of the density of atomic steps along the $[010]_{Fe}$ directions, along with gradually increasing magnetostatic contributions [295].

The physical origin of surface reconstruction involves displacement of the surface atom, increment or decrement of surface density due to the addition or removal of the surface atom that leads to minimization of the number of surface atom dangling bonds, improvement of the compactness of outermost surface layer etc. [296][297][298] The interaction of energetic ions with solid surfaces causes erosion(sputtering) and displacement/ transportation of materials on the surface and near-surface regions. Noble gas ions with energies of about 1 keV have traditionally been used for experiments. In situations when the incoming ions' energies are just a few hundreds of electron volts or less, sputtering is almost non-existent [286][299]. However, the formation of nanostructures in the form of ripples and disordered arrays of spikes have also been seen in this low-energy regime [286][300][301]. In this low-energy domain, mass redistribution (MR) caused by the inelastic displacement of atoms significantly increases due to the transfer of momentum from the incident ions to those atoms residing near the solid surface [287][288][302]. Based on ion energy and target material, hundreds of materials may be transported to another place without sputtering. The ion flux used for the present experiment ranges from $10^{11}$-$10^{12}$ ions/cm$^2$, which is approximately 3-5 orders of magnitude lesser than the conventional ion fluxes used for ordered patterned formation. Therefore, ion-induced mass redistribution by the exposure of material with low energy and low flux causes surface reconstruction.

## 6.II.6 Conclusions

The feasibility of the IBE technique for modification of magnetic anisotropy of Fe/MgO(001) system from a purely biaxial state to a biaxial - uniaxial mixed state has been presented. Furthermore, we have shown that apart from the tailoring of morphology and tuning of UMA, the IBE technique can be further extended to modify film surface structure. From these results, we believe that RHEED is a helpful method to further resolve structural details and questions about self-assembled nanostructures, as demonstrated in this study.

## 6.II.7 Overall Conclusions

Epitaxial Fe films have been grown successfully on Ag(001) and MgO(001) substrates. We observe that magnetic switching behavior is very sensitive to the surface and interfacial morphology. Fe film deposited on ion eroded Ag(001) substrate exhibits multiple-step jump in magnetic hysteresis loops due to coupling between intrinsic cubic anisotropy and induced UMA. Furthermore, magnetic switching and its correlation with the domain dynamics of the film, as studied by the Kerr microscopy technique, revealed a striking difference in domain structure depending upon the direction of the applied magnetic field with respect to crystallographic direction. On the other hand, the IBE of epitaxial Fe film leads to a split in the hysteresis loop. Interestingly, in this case, an ion beam-mediated 2×2 surface reconstruction has been observed for the first time.

# Chapter 7:

# Summary and scope for future work

This chapter presents a quick review of the experimental work addressed in chapters 3, 4, 5 and 6 of this thesis. In addition, potential future work is also provided.

# Chapter 7: Summary and scope for future work

    7.1 Summary of the results

    7.2 Scope for future work

## 7.1 Summary of the results

Magnetic anisotropy is a key parameter of magnetic materials as it decides the response in the presence of an external magnetic field. The artificial tailoring of magnetic anisotropy by manipulating surface and interface morphology is attracting widespread interest for its application in spintronic and magnetic memory devices. In this perspective, the primary focus of this thesis has been inducing and tailoring magnetic anisotropy in thin film-based systems via the engineering of structure and surface/interfacial morphology. For this purpose, instead of multistep, expensive lithographic techniques, we have utilized low-energy IBE as a handy, cost-effective, and useful tool for producing magnetic nanostructures and tailoring its surface structure, morphology and magnetic anisotropy. Furthermore, we have also proposed various new and unconventional methods (oblique angle deposition on the rippled substrate, sequential deposition-erosion) for enhancement of UMA and proved its effectiveness via experiments. To get concurrently excellent magnetic characteristics while preserving the integrity of the surface structure and the morphology proved quite challenging due to the interplay of different contributions, oxidation and contamination of surfaces. Therefore, it was necessary to conduct in-situ investigation to get genuine characteristics and understand their interdependencies. We have performed most of the present thesis work utilizing an in-situ UHV chamber. However, advanced synchrotron-based techniques such as GISAXS and NFS were also utilized to complement the in-situ observations. In this study, we addressed unresolved issues relating to investigating and interpreting low-energy ion-solid interaction-induced surface texturing and structure modification on thin films' magnetic characteristics. This contribution is critical for understanding the complicated magnetization behavior in magnetic multilayers and nanostructures. The overall conclusions of the present thesis, according to the chapters, are summarized below.

Chapter 3 presents oblique angle deposition on a nanopatterned rippled substrate as a novel route of inducing large in-plane uniaxial magnetic anisotropy (UMA) in magnetic thin films. For this purpose, Cobalt films and rippled $SiO_2$ substrates have been taken as a model system. Here, nanopatterned substrates are prepared by low energy ion beam erosion (IBE), above which films are deposited obliquely along and normal to the ripple directions. A clear anisotropy in the growth behavior has been observed due to the inhomogeneous in-plane organization of adatoms in the form of columns. The strong asymmetric scattering pattern in GISAXS measurements reveals the formation of tilted columnar nanostructure and preferential deposition of materials at the buried interface due to the shadowing effect induced by the height of ripples. The increased shadowing effect in the films deposited obliquely normal to the direction of the ripple patterns causes preferential coalescence of the columns along the substrate ripples, resulting in stronger in-plane UMA in the film. Furthermore, their domain configuration has a striking difference. A comparative study drawn on UMA with thin film deposited normally over a rippled substrate and obliquely on a bare Si substrate indicates that the thin film deposited obliquely perpendicular to the ripple direction has UMA almost one order of magnitude higher compared to the others. This peculiarity in magnetic behavior is addressed by considering the morphological anisotropy governed by the enhanced shadowing effect, the shape anisotropy and the influence of dipolar interactions among the magnetostatically coupled self-organized surface ripple and buried nanocolumnar structures. The present observation explores the crucial role of initial surface topography and the shadowing effect on the formation of self-organized nanostructured thin film and hence, UMA.

The influence of ion beam erosion on the structure, morphology, and induced magnetic anisotropy of polycrystalline thin films is discussed in chapter 4. The evolution of surface morphology and structure, as well as magnetic properties of Fe thin film subject to several cycles of ion beam erosion (IBE) process, are observed and comprehensively studied in detail by exploiting MOKE, RHEED, XRR and XDS techniques in-situ. In contrast to the studies in the literature, the present work unfolds the intermediate states of the modified structure and morphology of Fe thin film. In the initial stages of IBE, a surface-type preferential crystallographic orientation of <100> axis along the direction of IBE imprints a recognizable MCA with an easy axis of magnetization along the IBE direction. The preferred ion channelling mechanism in the sub-surface region is responsible for this surface texturing. In earlier studies, IBE-induced surface texturing was not observed due to oxidation and contaminations in ex-situ studies. Therefore, the contribution of surface texturing was not taken into account. Further erosion results in the development of well-defined correlated morphology that incorporates reinforcement of stray dipolar fields and switches the easy axis of magnetization by 90° at a specific value of film thickness or erosion cycle. The observed unusual crossover is ascribed to be originating from an interplay between the relative weight of MCA and shape anisotropy, which is successfully explained by considering modified surface structure, morphology and the relative strength of the dipolar stray fields. Controlled surface crystallinity and morphology modification, along with precise characterization by self-consistent technique under in-situ conditions, makes it possible to understand the unusual UMA crossover. Thus, the present phenomenological understanding provides a promising option for the fine tailoring of the strength of UMA by controlling the crystallinity of the surface along with anisotropic morphology.

Chapter 5 describes an approach to engineering and enhancing the strength of oblique incidence IBE induced in-plane UMA by simultaneous modification of film morphology and film texture. Cobalt film and Si substrate have been taken as a model system to meet this objective. We focus our efforts on sequential deposition and subsequent IBE of the film rather than the traditional post-growth IBE of the film, where IBE-induced modifications are limited to the near-surface regions. Detailed in-situ investigation insights that the film grows in a biaxially textured polycrystalline state with the formation of nanometric surface ripples. The film also exhibits pronounced UMA with an easy axis oriented parallel to the surface ripple direction. Remarkably, the induced UMA is about ten times larger than the reported similar kind of earlier studies. The possibility of imposing in-plane crystallographic texture giving rise to magneto-crystalline anisotropy, along with shape anisotropy of rippled morphology, enhances the strength of the UMA. Therefore, the novelty of the present approach is the ability to imprint film texture throughout the whole film layer along with surface morphology modification, consequently not restricting the source of UMA limited to the thin surface region of a polycrystalline ferromagnetic film for flexible tuning and enhancement of the strength of UMA. The present findings can be further extended to systems characterized by different crystallographic structures and magnetic properties and show the universal applicability of the present approach.

Chapter 6 deals with the effect of IBE on magnetic anisotropy and surface structure in the epitaxial thin film. We observe that magnetic switching behaviour is very much sensitive to the surface and interfacial morphology. Fe film deposited on ion eroded Ag(001) substrate exhibits a multiple-step jump in magnetic hysteresis loops due to coupling between intrinsic cubic anisotropy and induced UMA. Furthermore, magnetic switching and its correlation with the domain dynamics of the film, as studied by the Kerr microscopy technique, revealed a striking difference in domain structure depending upon the direction of the applied magnetic field with respect to crystallographic direction. On the other hand, the IBE of epitaxial Fe film leads to a split in the hysteresis loop. Here, the symmetry of magnetic anisotropy has been converted from a purely biaxial state to an almost uniaxial state. Interestingly, in this case, an ion beam-mediated 2×2 surface reconstruction has been observed for the first time. Therefore, apart from the conventional tailoring of morphology and tuning of UMA, the present observation highlights the IBE technique's capability for the modification of film surface structure and seeks further intensive study on structures of self-assembled nanostructures.

## 7.2 Scope for future work

The current research has shown several novel and intriguing outcomes related to IBE-induced morphology, structure and magnetic anisotropy of thin film. These findings not only revealed the potential of IBE but also pointed the way forward for further research. As a proper ending to the current thesis, the next section discusses a few potential future scopes. Future research opportunities in this domain are enormous, beginning with optimization of ion beam parameters for precise control over ion beam patterned morphology and its applications in diverse fields. It would be interesting to combine various contributions of anisotropy such as exchange bias, magnetic field annealing, and deposition in the presence of a magnetic field for the thin film deposited over the rippled patterned substrate to induce two components of anisotropy simultaneously for fine-tuning the strength and direction of UMA according to the requirement. In this direction, stress can also be another parameter that can be induced artificially by the deposition of thin film on a patterned substrate curved by the application of external mechanical force. An excellent investigation would be to determine the function of IBE in structural distortions, defects, and defect-mediated magnetic characteristics of thin films. Till now, most of the studies have been performed with single-element materials. In the case of binary materials, the difference in sputtering yields between two materials creates a layer enriched with the material having a lower sputtering yield. Hence, the evolution of the morphology and magnetic properties of thin films made of binary compounds would be another domain to explore. The magnetic nanostructures exhibit a special kind of magnetic switching behaviour yet to be understood from microscopic points of view, such as atomic arrangements, external shape or even domain dynamics etc.

Isolated magnetic nanostructures and nanowires can be prepared by proper selection of angle of deposition, the thickness of the layered material using ion beam patterned rippled substrate as a template and can be further studied. The thermal stability of oblique angle deposited columnar nanostructure and corresponding magnetic anisotropy is still an issue. Improvement of thermal stability by incorporating nonmagnetic material can open up plenty of novel studies. Material incorporation during ion beam patterning affects the sputtering yield and can influence the self-organized pattern formation below the critical angle. Therefore, the dynamics of nanopattern formation and magnetic properties during IBE with magnetic material co-deposition deserve further attention. Furthermore, the use of ion beam and ion beam engineered nanostructured surface are not limited to the field of magnetism and extended to plasmonic, nanoelectronics, photovoltaics, biomaterials, and sensing. Therefore, the possibility of growth in future is enormous.

# Appendix

## Morphology and magnetic anisotropy of self-organized Fe nanostructures on Si(001) substrate fabricated by ion beam erosion with simultaneous Fe incorporation.

Low energy ion beam erosion (IBE) is emerging both as a strategy to pattern the surface topography employed as templates for the growth of functional materials and fine tailoring of physical properties. We report structure, morphology and related induced magnetic anisotropy in Fe thin film deposited on Si substrate subject to concurrent IBE during deposition. Small angle and wide-angle x-ray scattering pattern reveals the formation of unidirectional correlated nanometric morphology and textured polycrystalline nature of the film. The surface morphology of the film as imaged by atomic force microscope indicates the presence of elongated grains or aligned rows of islands. The film exhibits an uniaxial magnetic anisotropy (UMA) due to the shape anisotropy of the self-organized structure, whose orientation is found to be dependent on the angle of incidence of the ion beam relative to the sample surface. The present study has an important implication for simultaneous modification of film texture and morphology for fine and flexible tailoring of UMA.

### A.1 Introduction

Ion beam erosion-induced surface nanopatterning is a new and powerful technique. It has proved its effectiveness in producing ordered nanostructures in a wide range of target materials, including metals, insulators and semiconductors, over relatively wide surface areas (up to several tens of $cm^2$) in a short time scale (typically a few minutes) [303]. Furthermore, ion beam fabricated nanopatterns are being used as a template for the growth of magnetic or plasmonic nanostructures, microelectronic device manufacturing, catalysis, nano-magnetism, plasma electronics, and so on. The formation of this nanostructured surface (mainly consists of ripple, dot, pit, cone, and pyramids) [287] is well explained considering surface instability caused by competition between two counteracting processes: curvature-dependent sputtering that tries to roughen the surface and temperature or ion-induced diffusion of material that tries to smoothen the surface. Recently it has been observed that metal incorporation or the presence of impurities during ion beam sputtering (IBS) not only influences the self-organized pattern formation but can modulate nanopatterns too [303][304] [305]. The simultaneous employment of metal atoms affects the sputtering yield during IBS, resulting in various physical and chemical processes (island creation or phase separation). For example, adding metal impurities will make it easier for patterns to form below the critical incidence angle of the ion beam, even though no correlated structure was seen when there were no impurities [306]. It will also make it easier to switch from one shape to another, improve orderliness, change symmetry, and so on [307][308][309]. Even though (usually undesirable) the presence of metal impurities makes things more complicated, these findings add a new aspect to understanding surface morphology evolution during ion sputtering and nanopatterning of multi-component material. It suggests further investigation, exploration, and theoretical modelling of metal impurity-assisted ion beam-induced nanostructures. Because of this, the study and research of metal impurity-assisted ion beam-driven nanostructures are of great importance.

In light of this growing interest, we report the detailed structural, morphological and magnetic anisotropy of self-assembled Fe nanostructures formed on Si substrate by co-deposition-erosion (IBED) method. The structural and morphological characterization was done by synchrotron-based grazing incidence small angle X-ray scattering (GISAXS) and grazing incidence wide X-ray scattering (GIWAXS) measurement. GISAXS reveals the formation of ordered morphology, while GIWAXS indicates that the film is in a textured polycrystalline state. The surface morphology, as imaged by AFM measurement, suggests the formation of aligned rows of nano-islands or elongated grains. The nanostructures also exhibit an UMA with an easy axis of magnetization that is found to be dependent on the angle of incidence of the ion beam with respect to the sample surface.

### A.2 Sample preparation

The sample was prepared inside a thin film deposition chamber having a base pressure of $6\times10^{-7}$ mbar or better. Chamber is attached with a commercial Kaufman-type ion gun for thin film deposition and ion beam irradiation experiments. The ion beam impinged on the Fe target surface at a fixed angle of incidence 45º with respect to the surface normal of the Fe target. Before irradiation with $Ar^+$ ions, Si substrates were mounted onto a custom-built sample holder, as shown in Fig. A.1. The angle of incidence (α) onto the Si substrate can be varied from 0º to 45º with respect to the sample surface using the custom-built sample holder. The sample holder is placed within the ion flux region so that the ion beam can simultaneously irradiate both the Si substrate and the target Fe, thereby, enabling irradiation and concurrent metal deposition onto the Si substrate. As the distance from the Fe target changes, the deposition flux on the Si substrate also changes. Thus, it allows one to study the erosion effect with impurity/foreign atom coverages. Furthermore, a $^{57}$Fe foil was placed on the Fe target to perform nuclear forward scattering experiment to study interface resolved magnetism. For the present study, a sample series has been prepared by varying the ion incidence angle with respect to the sample surface (α) from 0° to 45°, ion irradiation time, and ion energy. However, in this section, we have discussed only two samples deposited at an angle α = 10° and 20° and denoted them as $IBED_{10°}$ and $IBED_{20°}$, respectively.

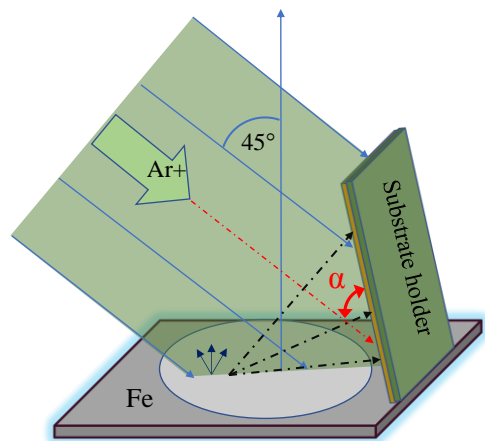

**Figure A.1:** Schematic of experimental geometry.

## A.3 Sample characterization

XRR measurements of the samples are performed using a 2D PILATUS 1M detector. Figure A.2(a) shows a representative 2D XRR image of the $IBED_{10°}$ film. The XRR data was obtained by changing the incidence angle from 0.038° to 1.87° while integrating one scattering pattern. This procedure was done with two different detector positions to avoid shadowing from intermodular detector gaps [150]. The 1D line profile, which gives angle versus reflected x-ray intensity, is extracted from the respective central vertical streak at $q_y$ =0nm$^{-1}$ and plotted in Fig. A.2(b). The film thickness is extracted by fitting XRR data using Parratt's recursive

Appendix

equations [151] and plotted as a continuous line in the same graph. The fitted thickness ($d$) and rms roughness ($\sigma$) is 8.2nm and 1.4nm for the IBED$_{10°}$ sample, 4nm and 1.7nm for the IBED$_{20°}$ sample, respectively.

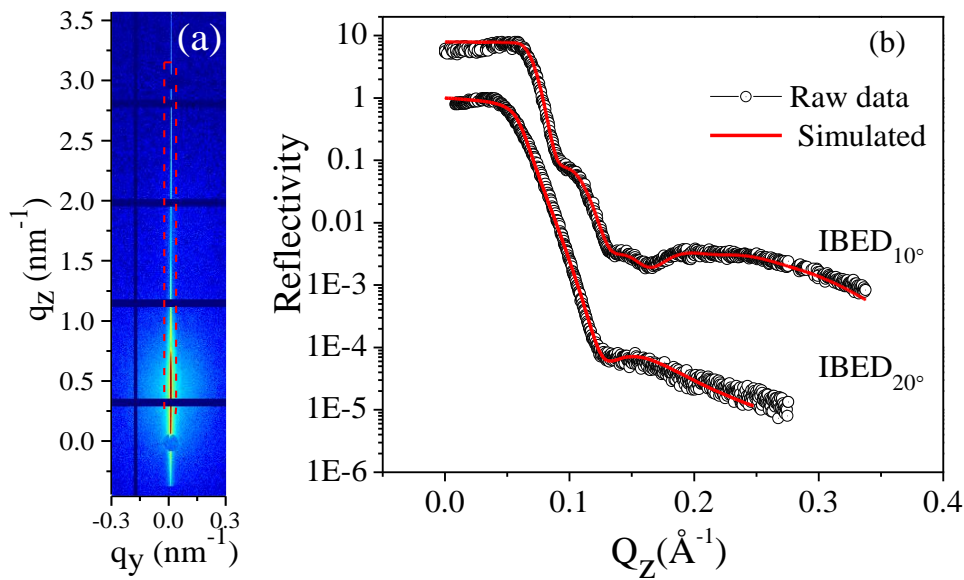

**Figure A.2:** (a) 2d XRR patterns of IBED$_{10°}$ sample. (b)1d intensity line profile of IBED$_{10°}$ and IBED$_{20°}$ sample, extracted along q$_z$ direction as guided by the dashed red line in (a). The red lines represent the fitted experimental data using Parratt formalism, and the hollow circles represent the experimental data.

GISAXS measurements are performed for the sample by keeping the X-ray beam at a grazing incidence angle of 0.4° along (in-plane azimuthal angle $\phi=0°$) and perpendicular (in-plane azimuthal angle $\phi=90°$) to the projection of ion beam on the sample surface. It contains information about the shape, size, and distribution of the self-organized nanostructures and their inter-correlation. GISAXS 2D spectra along both the $\phi$ directions of IBED$_{10°}$ and IBED$_{20°}$ samples are shown in Fig. A.3(a,b) and A.3(c,d), respectively. Drastic differences in both images are mainly due to the anisotropic morphology of the sample along two azimuthal directions. As compared to measurements along $\phi=0°$ (Fig. A.3a), several satellite peaks (fringes) are visible on two sides of the specular rod in the Fig. (A.3b). Several fringes on both sides of the specular rod indicate the formation of correlated morphology. In IBED$_{20°}$ sample, asymmetric distribution of the scattered intensity on the right side of the specular rod (q$_y$= 0 nm$^{-1}$) is noticed along $\phi=90°$ direction in Fig. A.3(d). This asymmetry is absent along $\phi=0°$. The inclined intensity elongation observed along the +q$_y$ side in Fig A.3(d) has an average inclination angle of ≈ 10° to the surface normal. This kind of scattering spectra is observed due to the scattering of x-rays from the facet of nanostructures and is related to the symmetry of the nanostructures [310].

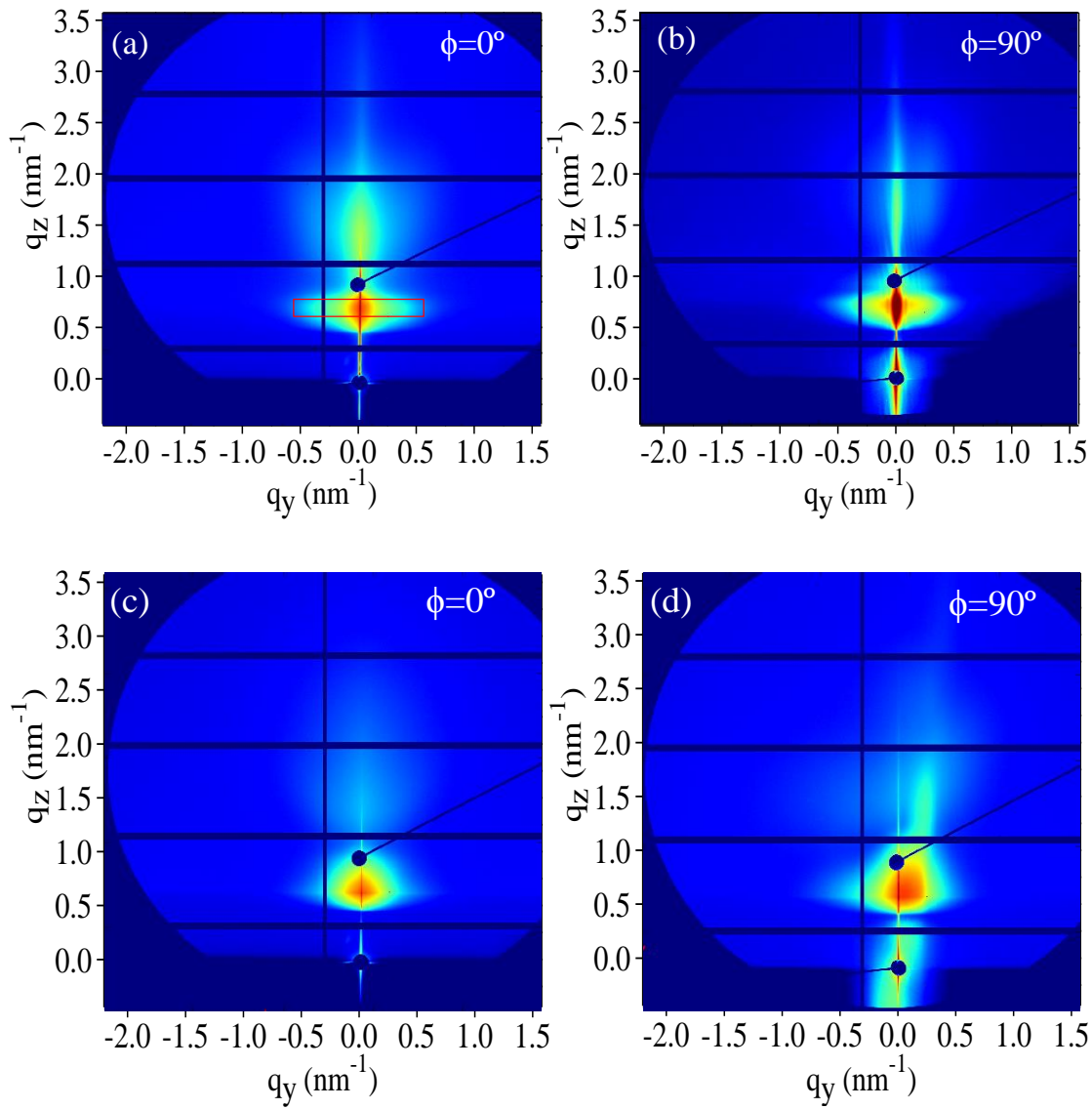

**Figure A.3:** GISAXS spectra of the (a,b) IBED$_{10°}$ and (c,d) IBED$_{20°}$ sample taken along (ϕ=0°) and across (ϕ=90°) to the projection of ion beam on the sample surface direction.

To understand the data more clearly and to correlate the same with sample morphology, intensity (I) vs in-plane momentum transfer vector ($q_y$) profiles are extracted around the Yoneda wings (as marked by the red box in Fig. A.3a) from both the samples and plotted in Fig. A.4. Satellite peaks at around $q_y$= +0.0772 nm$^{-1}$ in Fig A.4a along ϕ=90° indicates the presence of laterally ordered nanostructure in IBED$_{10°}$ sample. The average lateral spacing of this self-assembled nanostructure is ≈ 81nm, as obtained from the line profile. On the other hand, the line profile from IBED$_{20°}$ sample along $\phi=90°$, as plotted in Fig. A.4b, exhibits a shoulder on the right side of the central peak. This asymmetry in scattered intensity arises from scattering from the nanostructured object with an unequal slope. However, a symmetrical line profile along $\phi= 0°$ direction indicates that the x-ray encounters a symmetric surface morphology along this direction. Thus, from these scattering patterns, the symmetry properties of the nanostructured objects can be determined using GISAXS as long as they are all oriented in the same direction.

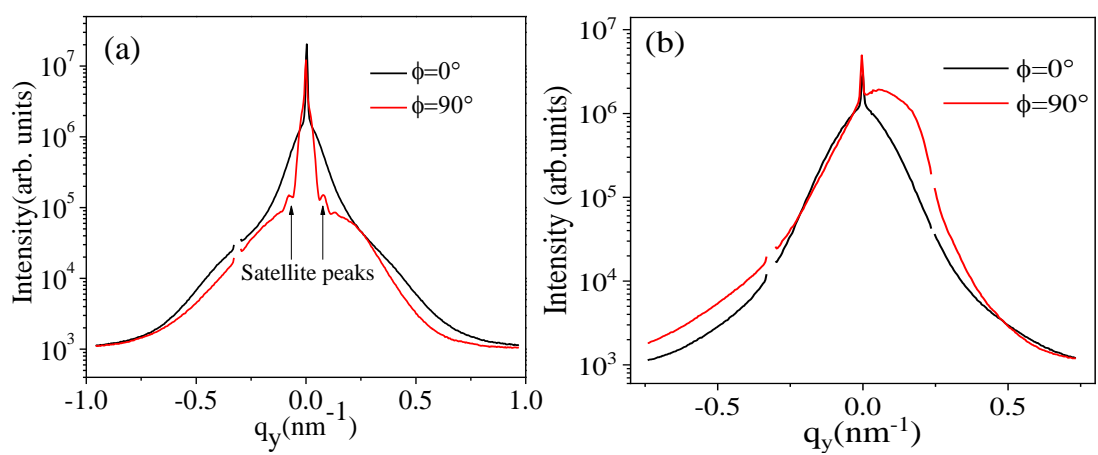

**Figure A.4:** 1-d in-plane intensity profile extracted around Yoneda peak from GISAXS images of (a) IBED$_{10°}$ and (b) IBED$_{20°}$ sample for both orientations, $\phi = 0°$ (black line) and $\phi = 90°$ (red line).

The crystalline structure of both films is probed by GIWAXS measurement. In Fig. A.5, representative 2D GIWAXS data are shown for $\phi=0°$ and $\phi=90°$ direction. It may be noted that due to the geometric constraints and also shadowed by the substrate, only the fourth parts of the diffraction pattern are visible. For IBED$_{10°}$ sample, the scattering pattern is primarily dominated by one intense and one dim Debye ring, as shown in Fig A.5(a,b). The rings are identified to be the (110) and (200) planes of b.c.c. Fe structure. The (110) ring is textured at 58° from the surface normal with an angle of dispersion ± 5°. For IBED$_{20°}$ sample, (110)$_{Fe}$ peak of relatively weak intensity has been observed, as shown in Fig A.5c. In this case, the peak is textured at 48° from the surface normal. The difference in texture axis between IBED$_{10°}$ and IBED$_{20°}$ samples may be due to the difference in deposition angle with respect to Fe target and relative angle (α) between the ion beam and sample surface. Thus, it opens up the possibility of tailoring the texture axis along a predefined direction by choosing a suitable combination of the above parameters. The relatively narrow and sharp rings, as observed in Fig A.5d, may arises from the edges of the samples due to finite misalignment.

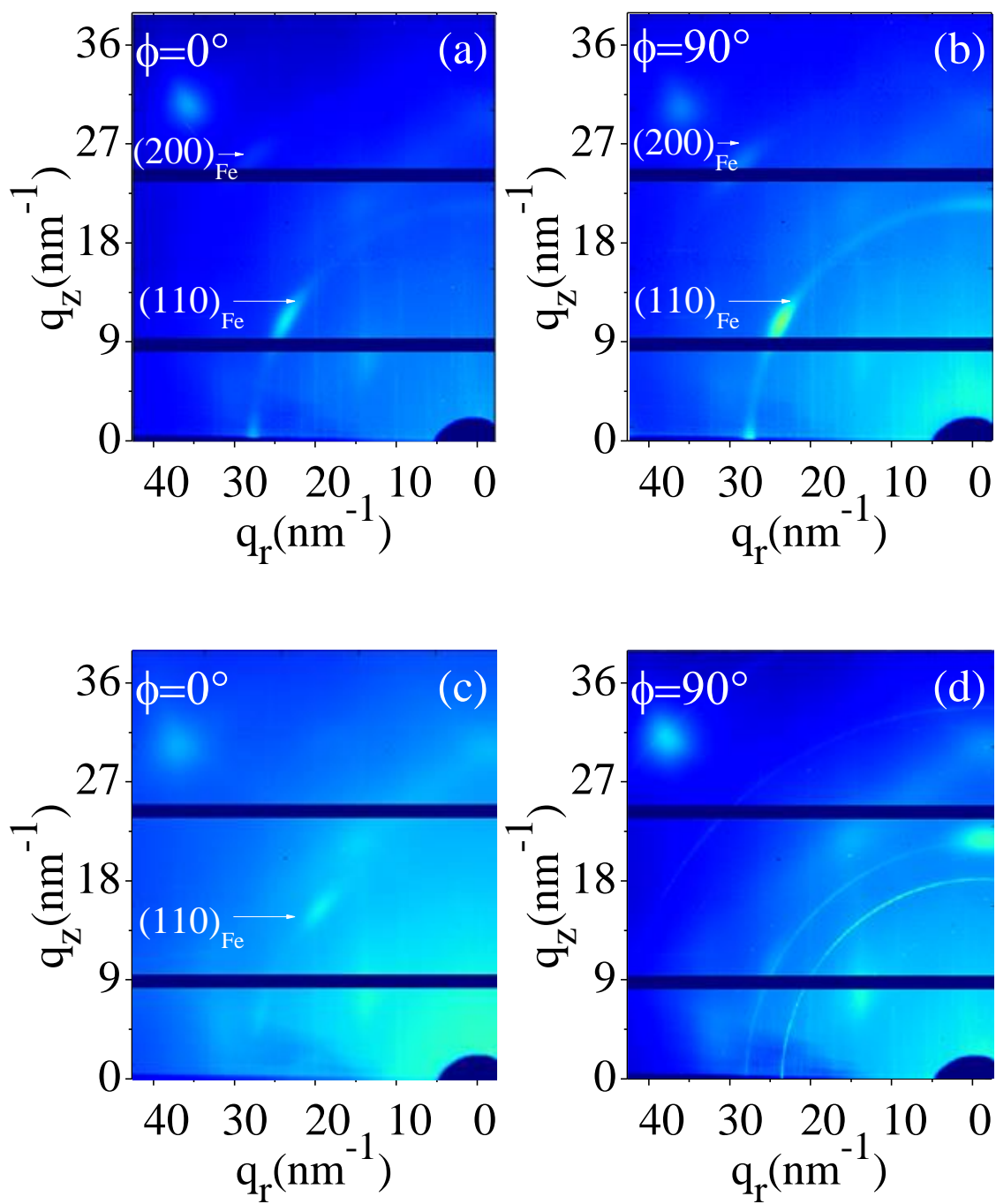

**Figure A.5:** 2D GIWAXS images of the (a,b) IBED$_{10°}$ and (c,d) IBED$_{20°}$ samples taken (a,c) along ($\phi=0°$) and (b,d) across ($\phi=90°$) to the projection of ion beam on the sample surface direction, respectively.

Appendix

The magnetic characterization of the sample was done by MOKE measurement. Figure A.6(a,b) represents the longitudinal MOKE hysteresis loop of IBED$_{10°}$ and IBED$_{20°}$ samples taken by applying an external magnetic field parallel ($\phi=0°$) and perpendicular ($\phi=90°$) to the projection of the ion beam on the film plane. The hysteresis loop of the IBED$_{10°}$ sample along $\phi=0°$ direction exhibits a regular rectangular shape. However, the loop along $\phi=90°$ rounds off. It suggests that an UMA is present in the film with an easy axis of magnetization parallel to the projection of the IBE direction. The saturation field equal to 95Oe has been obtained from the hard axis hysteresis loop. The hysteresis loop of the IBED$_{20°}$ sample is presented in Fig A.6b. We observe that along $\phi=90°$ direction hysteresis loop exhibits a regular rectangular shape, and the loop along $\phi=0°$ rounds off. Thus, the present film shows an UMA; however, its anisotropy axis is rotated by 90° compared to the IBED$_{10°}$ sample. The hard axis saturation field was determined to be ≈ 200Oe. Therefore, compared to the IBED$_{10°}$ film, almost double the strength of UMA has been observed in the IBED$_{20°}$ film. The reason behind the switching of the anisotropy axis is yet to be understood. One possible reason may be the direction of shape anisotropy originating from the orientation and aspect ratio of the nanostructured object decides the anisotropy axis.

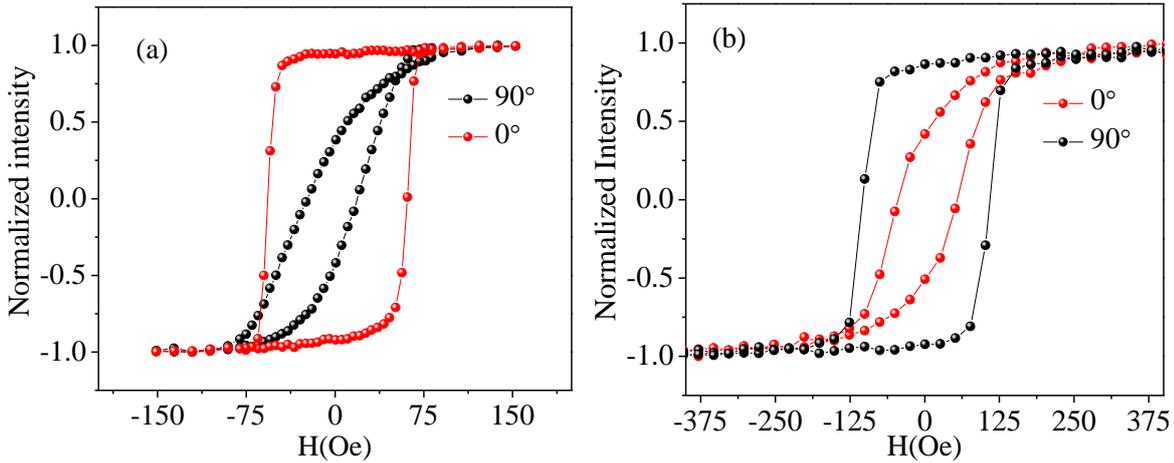

**Figure A.6:** Longitudinal MOKE hysteresis loop of (a) IBED$_{10°}$ and (b) IBED$_{20°}$ sample taken along ($\phi=0°$, red circle) and across ($\phi=90°$, black circle) to the projection of ion beam on the sample.

The surface morphology of IBED$_{10°}$ and IBED$_{20°}$ samples as imaged by AFM is shown below in Fig A.7(a,c). The surface morphology consists of aligned rows of island/elongated grains oriented in the direction parallel to the projection of the ion beam. The degree of ordering of the nanopatterned morphology in Fig. A.7 was quantified from the height-height correlation function (HHCF) C: $2\sigma^2\left\{1 - \exp\left[-\left(\frac{r}{\xi}\right)^{2h}\right]\right\}$ and plotted in Fig. A.7b,d. The correlation length ($\xi$) is 76nm and 113nm for IBED$_{10°}$ and IBED$_{20°}$ samples, respectively. Therefore, the average size of the ordered domain in the IBED$_{20°}$ sample is relatively large compared to IBED$_{10°}$ sample.

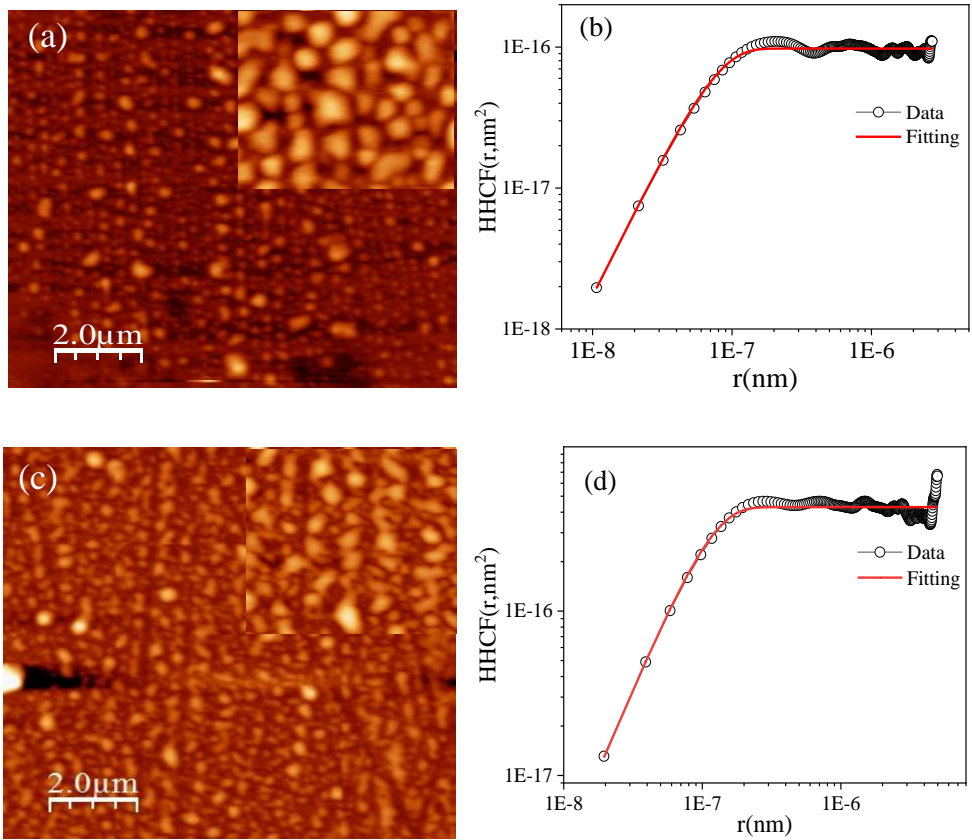

**Figure A.7:** AFM image of (a) IBED$_{10°}$ and (b) IBED$_{20°}$ sample, respectively. (b,d) HHCF obtained from the AFM image in (a,c).

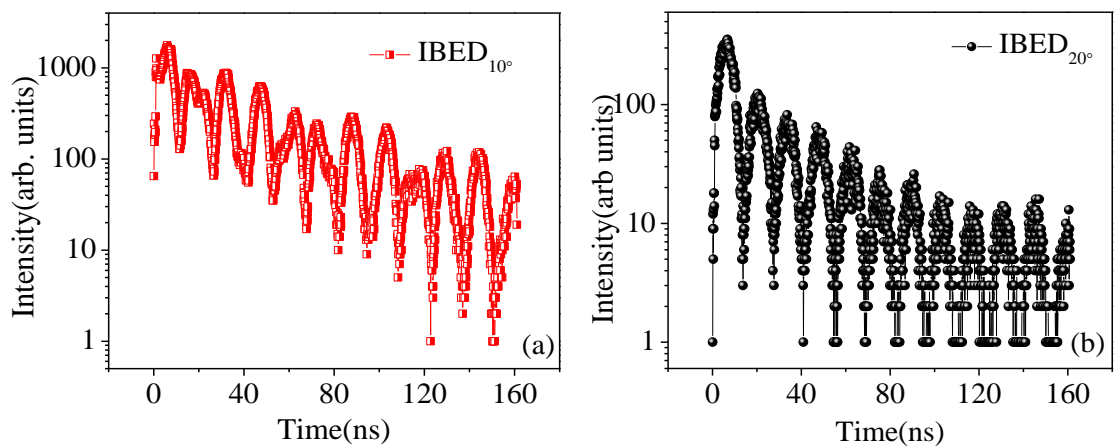

**Figure A.8:** Nuclear forward scattering time spectra for the Si/$^{57}$Fe structure of (a) IBED$_{10°}$ and (b) IBED$_{20°}$ samples, respectively.

Figure A.8(a,b) illustrates the Nuclear forward scattering (NFS) measurements (time spectra) for both the samples $IBED_{10°}$ and $IBED_{20°}$, respectively. NFS experiments were carried out using X-ray synchrotron radiation source at the P01, Dynamics Beamline at PETRA III, DESY, Hamburg, Germany. The x-ray energy was tuned to 14.41 KeV, the nuclear transition energy corresponding to the $Fe^{57}$ isotope. The beat pattern arises due to the interference of different components of the hyperfine field. However, the pattern observed in both samples is different, which suggests different multiplets of hyperfine field and orientation of the spins are exists in the samples.

## A.4 Conclusions

A detailed study on structure, morphology and magnetic anisotropy of nanostructured Fe thin films prepared by ion beam irradiation with co-deposition of material has been performed. The nanostructures grow in a textured polycrystalline state and exhibit different morphological ordering and symmetry. An UMA is also present, whose strength and symmetry depend on experimental parameters. Thus, we have emphasized how the experimental setup in this method of surface nano-structuring influences the pattern qualities in the case of ion beam irradiation with concurrent metal co-deposition. Nonetheless, more controlled studies, as well as thorough and extensive theoretical modelling, are required to better understand how the experimental setup of surface nano-structuring influences the pattern qualities, the mechanics behind the improved order that these patterns might exhibit, etc. Furthermore, more investigation is required to examine the long-term development of these patterns. As significant concerns still need to be resolved, it is worthwhile to do more study in this aera.

# Abbreviations

| | |
|---|---|
| Reflection High Energy Electron Diffraction | RHEED |
| Magneto Optical Kerr Effect | MOKE |
| X-ray reflectivity | XRR |
| X-ray diffuse scattering | XDS |
| Grazing incidence small angle X-ray Scattering | GISAXS |
| Grazing incidence wide angle x-ray scattering | GIWAXS |
| Nuclear Forward Scattering | NFS |
| Oblique angle deposition | OAD |
| Uniaxial Magnetic Anisotropy | UMA |
| Magnetic Anisotropy | MA |
| Magneto crystalline anisotropy | MCA |
| Ion Beam Sputtering | IBS |
| Ion Beam Erosion | IBE |

# List of Publications

**Referred Journal Publications**

1. <u>Anup Kumar Bera</u> and Dileep Kumar *"In-Situ Investigation of Ion Beam-Induced Crossover in Uniaxial Magnetic Anisotropy of Polycrystalline Fe Films"*, ACS Appl. Electron. Mater., **2** (11), 3686–3694, (2020). https://doi.org/10.1021/acsaelm.0c00720

2. <u>Anup Kumar Bera,</u> Pooja Gupta, Debi Garai, Ajay Gupta, Dileep Kumar, *"Effect of surface morphology on magnetization dynamics of cobalt ultrathin films: An in-situ investigation". Applied Surface Science Advances* **6** 100124 (2021). https://doi.org/10.1016/j.apsadv.2021.100124

3. <u>Anup Kumar Bera</u> and Dileep Kumar, *"Self-Smoothening of an Ion-Beam-Sputtered Ag (100) Surface: Evolution of Surface Morphology Using RHEED",* Metall Mater Trans A, **49**, 5205–5210 (2018). https://doi.org/10.1007/s11661-018-4809-7

4. <u>Anup Kumar Bera</u>, Sadhana Singh, Md. Shahid Jamal, Zainab Hussain, V. R. Reddy and Dileep Kumar. *"Growth and in-situ characterization of magnetic anisotropy of epitaxial Fe thin film on ion-sculpted Ag (001) substrate"*, J. Magn. Magn. Mater**. 544**, 168679 (2022). https://doi.org/10.1016/j.jmmm.2021.168679

5. <u>Anup Kumar Bera</u>, Arun Singh Dev, Manik Kuila, Mukesh Rajan, Pallavi Pandit, Matthias Schwartzkopf, Stephan V. Roth, V.R. Reddy and Dileep Kumar, *"Morphology induced large magnetic anisotropy in obliquely grown polycrystalline thin film on nanopatterned substrate." Applied Surface Science* **581**, 152377 (2022). https://doi.org/10.1016/j.apsusc.2021.152377

6. Zainab Hussain, <u>Anup Kumar Bera</u>, Arun Singh Dev, Dileep Kumar and V. Raghavendra Reddy, *"Exchange bias effect in Fe/LaAlO$_3$: An interface induced effect"*, J. Alloys Compd., **849**, 156484 (2020). https://doi.org/10.1016/j.jallcom.2020.156484.

7. <u>Anup Kumar Bera</u>, Arun Singh Dev and Dileep Kumar, *"Enhancing the limit of uniaxial magnetic anisotropy induced by ion beam erosion"*. (Under review) *arXiv id: https://doi.org/10.48550/arXiv.2208.05249*

8. Arun Singh Dev, <u>Anup Kumar Bera</u>, Pooja Gupta, Velaga Srihari, Pallavi Pandit, Marie Betker, Matthias Schwartzkopf, Stephan V. Roth, and Dileep Kumar "*Oblique angle deposited FeCo multi-layered nanocolumnar structure: magnetic anisotropy and its thermal stability in polycrystalline thin films*" **590**, 153056, (2022). https://doi.org/10.1016/j.apsusc.2022.153056

9. Monika Saxena, <u>Anup Kumar Bera</u>, R. Venkatesh. "*Wet chemical assisted synthesis of self-assembled tellurium nanostructures with enhanced stability*", Materials Letters, **301**, 130300 (2021). https://doi.org/10.1016/j.matlet.2021.130300

10. Monika Saxena, Tarachand, <u>Anup Kumar Bera</u>, Gunadhor S. Okram. "*Impact of non-stoichiometry in the thermoelectric performance of polyol method prepared $Cu_{1+x}In_{1-x}S_2$ (x = −0.3, −0.2, −0.1, 0, 0.1, 0.2) nanowires.*" Journal of Alloys and Compounds **881** 160517 (2021). https://doi.org/10.1016/j.jallcom.2021.160517

11. <u>Anup Kumar Bera</u>, Md. Shahid Jamal, Avinash G. Khanderao, Manik Kuila, Mukul Gupta, Matthias Schwartzkopf, Stephan V. Roth, V.R. Reddy and Dileep Kumar, *"Inducing magnetic anisotropy in Fe thin film by concurrent ion sputtering and simultaneous metal co-deposition"*, (Under Preparation).

12. <u>Anup Kumar Bera</u> and Dileep Kumar, "*In-situ investigation of surface reconstruction and uniaxial magnetic anisotropy in ultrathin epitaxial Fe film induced by low energy ion beam erosion.*" (Under review)

## Publications in Conference Proceedings

1. <u>Anup Kumar Bera</u> and Dileep Kumar, *"Effect of Ar+ beam sculpting on in-plane magnetic anisotropy and morphology of epitaxial Fe/MgO(001) film"*, 64th DAE Solid State Physics Symposium (DAE-SSPS 2019), Indian Institute of Technology Jodhpur. AIP Conference Proceedings 2265, 030315 (2020); https://doi.org/10.1063/5.0017794

# Conferences and workshops attended

- Poster presentation at 62$^{nd}$ DAE Solid State Physics Symposium (DAE SSPS-2017) 26-30 December 2017, BARC, Mumbai, India.
- Poster presentation at 64$^{th}$ DAE Solid State Physics Symposium (DAE SSPS-2019) 18-22 December 2019, Indian Institute of Technology Jodhpur, Rajasthan.
- Poster presentation at 4$^{th}$ international Conference on Nano Structuring by Ion Beam (ICNIB2017) Conference 11-13th October 2017, DAVV, Indore, Madhya Pradesh, India.
- Poster presentation at 4$^{th}$ International Conference on Advanced Materials (ICAM-2019) 12-14th Jun 2019 Kannur, Kerala, India.
- Oral at 6$^{th}$ International Conference on Nanoscience and Nanotechnology (ICONN 2021) 1-3 Feb 2021 at SRM Institute of Science and Technology, Chennai.
- Attended Winter School on Synchrotron Techniques in Materials Science 25-31 Oct 2018 at S. N. Bose National Centre for Basic Sciences, Kolkata.
- Poster presentation at International Conference on Magnetic Materials and Applications (ICMAGMA-2018) 9-13$^{th}$ Dec 2018 at NISER, Bhubaneswar.
- Poster presentation at International Workshop on Nano/Micro 2D & 3D fabrication and manufacturing of Electronic & Biomedical Devices and Applications 1-3 Nov 2018 (IWNEBD-2018), IIT Mandi.

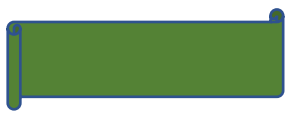